\documentclass[12pt,twoside,a4paper]{report}
\usepackage{graphicx}
\usepackage{latexsym}
\usepackage{amsmath}
\usepackage{amssymb}
\usepackage{graphicx}
\usepackage{dcolumn}
\usepackage{bm}
\usepackage{color}
\usepackage{url}
\usepackage{amsmath,amssymb,graphicx}
\usepackage{amsmath}
\usepackage{amssymb}
\usepackage{mathrsfs}
\usepackage{accents}
\usepackage{epsfig}
\usepackage[lofdepth,lotdepth,caption=false]{subfig}
\usepackage{hyperref}
\usepackage{mathrsfs}
\usepackage{slashed}
\usepackage{braket}

\renewcommand{\O}{\mathcal{O}}

\newcommand{\Z}{\mathbb{Z}}
\newcommand{\keV}{{\rm keV}}
\newcommand{\MeV}{{\rm MeV}}
\newcommand{\GeV}{{\rm GeV}}
\newcommand{\TeV}{{\rm TeV}}

\newcommand{\vev}[1]{\langle #1 \rangle}

\begin{document}

\marginparsep 0pt
\textwidth 15.0 truecm


\setlength{\baselineskip}{0.780cm}
\begin{center}
\includegraphics[width=0.4\textwidth]{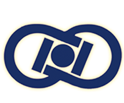}
\end{center}

\vspace{0.1 mm}
\begin{center}
{\Large \textbf{Institute for Research in Fundamental Sciences}}

Institute for Studies in Theoretical Physics and Mathematics (IPM)
\end{center}

\vspace{1 mm}
\pagestyle{empty}
\begin{center}
\large{ School of Particles and Accelerators  }
\end{center}

\vspace{1 mm}
\pagestyle{empty}
\begin{center}
\large{ Ph.D Thesis }
\end{center}

\vspace{0.9 mm}
\pagestyle{empty}
\begin{center}
\Large{ Phenomenology of the Sterile Neutrinos  }
\end{center}

\vspace{10mm}
\begin{center}
\Large{  By: Zahra Khajeh Tabrizi}\\
\end{center}

\vspace{10mm}
\begin{center}
\large {Supervisor: Dr. Orlando L. G. Peres}
\end{center}

\vspace{10mm}
\begin{center}
July, 2015
\end{center}

\newpage
\setlength{\baselineskip}{0.780cm}
\begin{center}
\includegraphics[width=0.4\textwidth]{oxlogo}
\end{center}

\vspace{0.1 mm}
\begin{center}
{\Large \textbf{Institute for Research in Fundamental Sciences}}

Institute for Studies in Theoretical Physics and Mathematics (IPM)
\end{center}

\vspace{1 mm}
\pagestyle{empty}
\begin{center}
\large{ School of Particles and Accelerators  }
\end{center}

\vspace{1 mm}
\pagestyle{empty}
\begin{center}
\large{ Ph.D Thesis }\\
\large{Particle Physics}
\end{center}

\vspace{0.9 mm}
\pagestyle{empty}
\begin{center}
\Large{ Phenomenology of the Sterile Neutrinos  }
\end{center}

\vspace{10mm}
\begin{center}
\Large{  By: Zahra Khajeh Tabrizi}\\
\end{center}

\vspace{10mm}
\begin{center}
\large {Supervisor: Dr. Orlando L. G. Peres}
\end{center}

\vspace{10mm}
\begin{center}
July, 2015
\end{center}

\newpage
\setcounter{page}{1}
\pagestyle{myheadings}
\markright{}

\noindent
{\Huge {\bf Acknowledgments}}
\vspace{1.2cm}

\noindent

To IPM and FAPESP for the financial support.

To Prof. Orlando L. G. Peres, whom I admire, not only for the excellent guidance, not only by the unique opportunities that he offered me, nor for the time and patience he spent with me, but for the confidence he gave me, and for all his kindness.

To Dr. Arman Esmaili, for our collaborations and teaching me how to be a physicist.

To Prof. Renata Z. Funchal, Dr. Pedro Machado, Yuber F. Perez and Olcyr Sumensari for our collaborations, for being so kind to me and for all the useful discussions.

To UNICAMP, for its kind hospitality. 

To Prof. H. Arfaei, for providing this opportunity for me to do my research in the field of my interest.

To my friends and colleagues Sara Khatibi, Behrooz Eslami, Felipe Campos Penha, Renan Pocireti, Gabriela Stenico and Najme Mirian.

\newpage
{\large {\bf Abstract }}

\vspace{5mm}

In this thesis we investigate several topics in neutrino physics, with an emphasis on the phenomenology of the sterile neutrinos. We study the existence of a light sterile neutrino within the so called $3+1$ scenario using the data of the medium baseline reactor experiments. We will also probe the parameters of the Large Extra Dimension model with the high energy atmospheric data of the IceCube experiment, and will find an equivalence between the Kaluza Klein modes and the sterile neutrinos. We will study the secret interaction of the sterile neutrinos which is proposed to solve the tension between cosmology and the sterile neutrino hypothesis. In addition to these, we will show that a minimal 2-Higgs-Doublet-Model extended with a $U(1)$ or $\Z_2$ symmetry cannot explain the smallness of the neutrino masses.\\


\newpage
{\large {\tableofcontents}}

\newpage
	\addcontentsline{toc}{chapter}{\listfigurename}
	\setcounter{lofdepth}{2}
	\listoffigures
	
	\newpage
	\addcontentsline{toc}{chapter}{\listtablename}
	\listoftables

\addcontentsline{toc}{chapter}{Publications}
\chapter*{Publications}

This thesis is based on my articles: 

\begin{enumerate}
\item
Z.~Tabrizi and O.~L.~G.~Peres,\\
  {\it Hidden Interactions of Sterile Neutrinos As a Probe For New Physics},\\
 \href{http://dx.doi.org/10.1103/PhysRevD.93.053003}{Phys.\ Rev.\ D {\bf 93}, no. 5, 053003 (2016)
  [arXiv:1507.06486 [hep-ph]].}
 
\item
P.~A.~N.~Machado, Y.~F.~Perez, O.~Sumensari, Z.~Tabrizi and R.~Z.~Funchal,\\
 {\it On the Viability of Minimal Neutrinophilic Two-Higgs-Doublet Models},\\
\href{http://link.springer.com/article/10.1007%2FJHEP12%282015%29160}{JHEP {\bf 1512}, (2015) 160 
  [arXiv:1507.07550 [hep-ph]].}
  \item  
  A.~Esmaili, O.~L.~G.~Peres and Z.~Tabrizi,\\
  {\it Probing Large Extra Dimensions With IceCube},\\
  \href{http://iopscience.iop.org/1475-7516/2014/12/002/}{JCAP {\bf 1412}, no. 12, 002 (2014)
  [arXiv:1409.3502 [hep-ph]].}
  
  \item A.~Esmaili, E.~Kemp, O.~L.~G.~Peres and Z.~Tabrizi, \\
{\it Probing light sterile neutrinos in medium baseline reactor experiments},  \\ \href{http://prd.aps.org/abstract/PRD/v88/i7/e073012}{Phys.\ Rev.\ D {\bf 88}, 073012 (2013)
  [arXiv:1308.6218 [hep-ph]].}\\
  \end{enumerate}

\addcontentsline{toc}{chapter}{Introduction}
\chapter*{Introduction}

 
 Neutrinos are the lightest particles known in the universe with non-zero mass. The upper limit on the mass of the heaviest neutrino is less than 4 million times lighter than the mass of electrons. They are electrically-neutral particles; however, are affected by the weak force. Neutrinos are the second most abundant particle in the universe.  
  
The energy of neutrinos vary from $10^{-4}$~eV (neutrinos originating from the Big Bang) to EeV scale (with the possible extra-galactic sources). The dependence of the cross section of neutrinos to their energy as well as their sources is
shown in Fig. \ref{fig:NeutrinoEnergies}. Although there are many different sources of neutrinos, some of them are not accessible by today's neutrino detectors. The big bang neutrinos (sometimes known as the relic neutrinos) have extremely low cross section and cannot be detected directly. On the other hand, since neutrinos only participate in the weak force, they need giant detectors for their observation. The biggest available neutrino detector is the IceCube experiment in Antarctica, with a volume of $1~\rm{km}^3$, which is able to detect  neutrinos in GeV-EeV range of energy.  
\begin{figure}[!hHtb]
\centering
\includegraphics[width=0.95\textwidth]{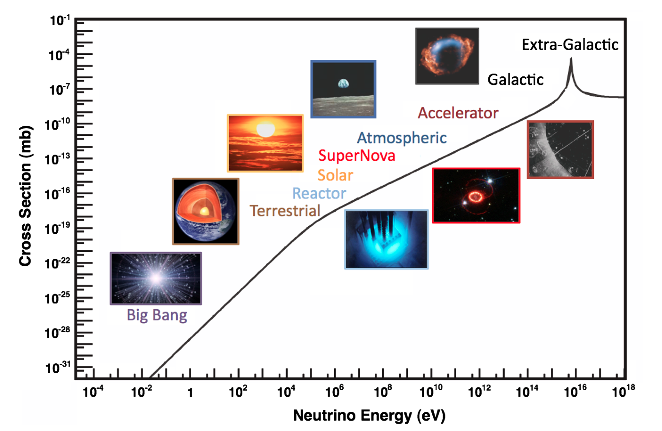}
\label{fig:NeutrinoEnergies}
\caption{\label{fig:NeutrinoEnergies}Various sources of neutrinos as a function of energy. Plot taken from \cite{Formaggio:2013kya}.}
\end{figure}

Three distinct types of neutrinos have been discovered, which are named after their Charged Current (CC) partners: the electron $(e)$-, muon ($\mu$)- and tau ($\tau$)- neutrinos. The most precise measurement on the number of light neutrinos (the active neutrinos in the SM) comes from the studies of $Z$ production in $e^-e^+$ collisions, specially in the LEP experiment. By calculating the invisible width of the $Z$ boson, in which they subtract the visible partial widths of the $Z$ decay (corresponding to the decay into quarks and charged leptons) from the total $Z$ decay, and assuming that all the invisible width is due to the light neutrinos, they have found that the number of light neutrinos with mass below $m_Z/2$ is $N_\nu=2.984\pm0.008\simeq3$ \cite{ALEPH:2005ab}. 

Plenty of neutrino experiments performed in the last two decades have confirmed that the three types of neutrinos in the SM are massive particles, and their mass eigenstates $(\nu_1,\nu_2,\nu_3)$ do not coincide with their flavor eigenstates $(\nu_e,\nu_\mu,\nu_\tau)$, which enter into the charged current interactions. These experiments have shown that neutrinos \textbf{\textit{change}} there flavor after they propagate in a finite distance. This change depends on the energy of neutrinos $E_\nu$, and the baseline of propagation $L$. The only possible explanation of all the data collected from neutrino experiments is that neutrinos have distinct masses, and they mix.

The mixing and flavor oscillation phenomena of neutrinos can be described by the mass-squared differences $\Delta m_{ij}^2\equiv m_i^2-m_j^2$ (where $m_i$ is the mass of $\nu_i$ state), and the elements of the so-called Pontecorvo-Maki-Nakagawa-Sakata (PMNS) unitary mixing matrix, which is parametrized by 3 mixing angles $(\theta_{12},\theta_{23},\theta_{13})$ and 1 CP-violating phase $\delta$. The experimental values of these parameters and the details of neutrino oscillation will be presented in Chapter \ref{chap1}. \\

  

A lot of neutrino properties can be explained through the current rich data of neutrino experiments, yet there are still some un-answered questions in neutrino sector of the SM:
\begin{itemize}

\item How many neutrino species do we have? Are there any sterile neutrinos? 

Although most of the data collected from the neutrino oscillation experiments are in agreement with 3-neutrino hypothesis, there exists some anomalies which cannot be explained in this paradigm. The existence of a sterile neutrino state was motivated by the LSND~\cite{Aguilar:2001ty}, MiniBooNE~\cite{miniboone} and reactor anomalies~\cite{reactoranomalie}. The most popular way to clarify these anomalies is to assume that there exists 1 (or more) neutrino state which does not have any interaction in the SM (therefore is sterile), but can mix with the active neutrinos and change their oscillation behavior pattern. In order to explain these anomalies, there must be at least one heavier mass eigenstate, with a mass $\sim 1~{\rm eV}$, the so-called $3+1$ model. Plenty of experiments have been designed to check this scenario (see~\cite{Abazajian:2012ys} and references therein; see also~\cite{Esmaili:2013vza,Esmaili:2013cja,Esmaili:2012nz}), yet the data collected to date, present an incomplete picture of sterile neutrinos, some in favor and some against sterile hypothesis. Clarifying the existence/absence of the sterile neutrinos is one of the biggest questions in the neutrino physics, also the main emphasis of this thesis.

\begin{figure}[!hHtb]
\centering
\includegraphics[width=0.8\textwidth]{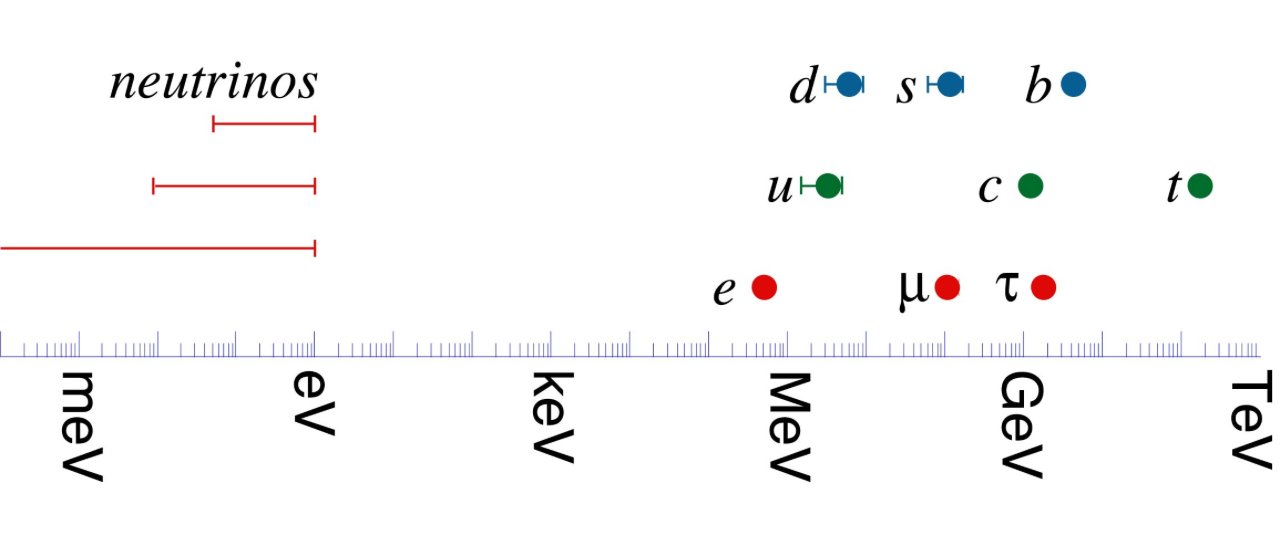}
\label{fig:fermion_masses}
\caption{\label{fig:fermion_masses}Neutrino masses compared to the masses of other fermions. Plot taken from \cite{fig:Fermion_Masses}.}
\end{figure}

\item What is the scale of neutrino masses? Why are neutrinos so light?

From the neutrino oscillation experiments we know that neutrinos are massive, and their mass and flavor eigenstates do not coincide; while from cosmological \cite{Hinshaw:2012aka,Planck:2015xua} and terrestrial \cite{Aseev:2011dq,Kraus:2004zw} experiments, we know that the masses of neutrinos should be below the eV scale (See Fig. \ref{fig:fermion_masses}) (The bounds from cosmology on the sum of neutrinos will be presented in Section \ref{neutrinos_cosmology}.). It means that if the same Higgs mechanism which is responsible for the masses of other fermions is also responsible for the masses of neutrinos, the Yukawa couplings of neutrinos have to be up to 6 orders of magnitude smaller than the one for electron, which produces a non-pleasant hierarchy between the Yukawa couplings. The so-called seesaw mechanism is an alternative which clarifies the masses of neutrinos \cite{Minkowski:1977sc}. Also, the large extra dimension (LED) model, which is primarily motivated as a solution to the hierarchy problem of Higgs, can explain the smallness of neutrino masses \cite{Dienes:1998sb,ArkaniHamed:1998vp}. Within the next few years the KATRIN experiment \cite{KATRIN} will investigate the beta decay of Tritium to study the absolute neutrino masses.

\item What is the hierarchy of neutrino masses? Or in other words, is $m_1$, the mass of the mostly $\nu_e$ state, smaller than $m_3$, the mass of the mostly $\nu_\tau$ state ("Normal Hierarchy", $\Delta m^2_{31}>0$), or bigger than that ("Inverted Hierarchy", $\Delta m^2_{31}<0$) (see Fig. \ref{fig:mass_normal_inverted})? 

Roughly speaking, most of the information we have about $\theta_{23}$ and $\Delta m^2_{31}$ comes from the atmospheric neutrino experiments (mainly MINOS \cite{Adamson:2013whj}). In these experiments, the matter effects on neutrino oscillation is negligible (due to the baseline of experiment), and the main oscillation probability they measure is the survival probability of muon neutrinos:
\begin{equation}\label{eq0.1}
P(\nu_\mu\to\nu_\mu)=1-\sin^22\theta_{23}\sin^2\big(\frac{\Delta m^2_{31}L}{4E}\big)+\rm{subleading~terms}.
\end{equation}
As can be seen, this oscillation probability is not sensitive to the sign of $\Delta m^2_{31}$. Therefore, these experiments are blind to the order of $1-3$ masses. However, the presence of matter effects modify the calculations, and hence the oscillation probability is different for normal and inverted hierarchies\footnote{The details of neutrino oscillation in matter will be explained in the Chapter \ref{chap3}.}.  
\begin{figure}[!hHtb]
\centering
\includegraphics[width=0.8\textwidth]{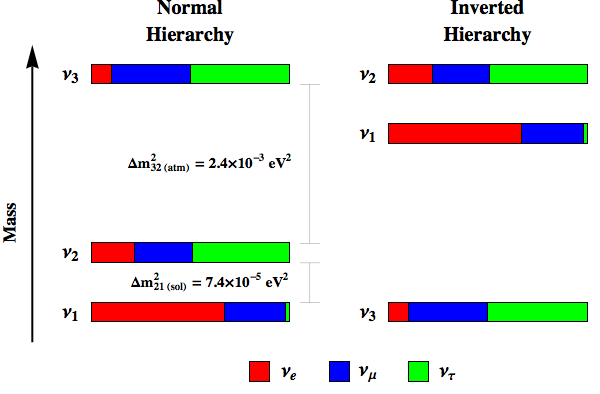}
\label{fig:mass_normal_inverted}
\caption{\label{fig:mass_normal_inverted}Neutrino mass hierarchies.}
\end{figure}
The IceCube experiment detects the atmospheric neutrinos coming from the north pole passing through the earth. Therefore, the matter effects cannot be neglected in this experiment, and the mass hierarchy can be determined using the data of the atmospheric neutrinos in PINGU ("Precision IceCube Next Generation Upgrade"). The data taking of this experiment is expected to start in 2017, and the mass hierarchy would be known by 2020, by up to $3\sigma$ C.L. \cite{Winter:2013ema}.

\item Is there $CP$-violation in the neutrino sector? If yes, what is the value of the $CP$-violating phase?

Leptonic $CP$-violation \textbf{\textit{might}} be the reason why we see asymmetry between matter and anti-matter in the universe (Leptogenesis). $CP$-violation in neutrinos means the oscillation probability for neutrinos and anti-neutrinos would be different. This asymmetry arises if the transition probability of neutrinos is different with the antineutrinos:
\begin{equation}\label{eq0.2}
A^{CP}=P(\nu_\alpha\to\nu_\beta)-P(\bar{\nu}_\alpha\to\bar{\nu}_\beta).
\end{equation}
As it will be explained in Chapter \ref{chap1}, in the case $\alpha=\beta$, the survival probability of neutrinos and antineutrinos become equal. Therefore, this asymmetry can be only measured in the appearance experiments, in which $\alpha\neq\beta$, such as in the T2K experiment \cite{Abe:2014ugx}, where they measure the transition oscillation of muon neutrinos to electron neutrinos. The future large underground detectors are supposed to determine this asymmetry, in addition to the neutrino mass hierarchy. These experiments measure the asymmetry by comparing the events of neutrinos and anti-neutrinos (direct measurement of $CP$-violation). 

\item Is neutrino its own antiparticle? Are neutrinos Dirac or Majorana? 

If neutrinos are Majorana particles, then double beta decay can proceed without emission of any neutrino. Thus observing a neutrinoless double beta decay ("$0\nu\beta\beta$") process is a proof that neutrinos are their own antiparticles, and hence Majorana. Also, observing such "$0\nu\beta\beta$" process means that the lepton number is not conserved anymore. The decay rate of the "$0\nu\beta\beta$" process is proportional to the absolute neutrino masses. Therefore, observing such process not only solves the Dirac vs Majorana mystery, but also gives information about neutrino masses. \\

\end{itemize}

The main goal of this thesis is investigating the phenomenology of sterile neutrinos. To do this, we have used the data of various neutrino oscillation experiments to probe different sterile models and find constraints on the related parameters. The thesis is organized as follows: In Chapter \ref{chap1}, we will present a brief review on the properties of the standard 3-neutrino model as well as the sterile neutrino hypothesis. 

In Chapter \ref{chap2}, we will analyze the data of the medium baseline reactor experiments, Double Chooz, Daya Bay and RENO, in the search of light-sterile neutrinos and also test the robustness of $\theta_{13}$ determination in the presence of sterile neutrinos. We will show that the existence of a light sterile neutrino state improves the fit to these data moderately.

In Chapter \ref{chap3}, we will study the phenomenological consequences of the Large Extra Dimension (LED) model for the high energy atmospheric neutrinos. For this purpose we will construct a detailed equivalence between a model with a large extra dimension and a $(3+n)$ scenario consisting of $3$ active and $n$ extra sterile neutrino states, which provides a clear intuitive understanding of the Kaluza-Klein modes. Then, we will analyze the collected data of high energy atmospheric neutrinos by the IceCube experiment to obtain the bounds on the radius of the extra dimension.

In Chapter \ref{chap4}, we will investigate the possibility of the sterile neutrino state interacting with a new gauge boson $X$, with mass $\sim$ MeV. This new interaction of the sterile neutrinos (the hidden interaction) produces neutral current matter potential for the sterile states, and hence changes the oscillation probability of neutrinos drastically. We will explain how the oscillation of neutrinos will be affected in this model. Then we will use the current data of the MINOS experiment to perform an analysis on the hidden interaction and will find the best fit values of the parameters. Finally, we will constrain the mass of the light gauge boson using the data of the MINOS neutrino experiment.

In addition to the phenomenology of the sterile neutrinos, in this thesis we will explain the smallness of neutrino masses via neutrinophilic 2 Higgs Doublet Models (2HDMs). To do this, in Chapter~\ref{chap5} we will study the phenomenology of neutrinophilic 2HDMs. The extra Higgs, via its very small vacuum expectation value, is the sole responsible for the smallness of neutrino masses. We will show that these models accompanied with an extra $Z_2$ or a softly broken $U(1)$ symmetry are either strongly disfavored by the electroweak precision data or very constrained. 

We will summarize our work in this thesis in the last Chapter.

\chapter{The Framework of Neutrino Oscillations }\label{chap1}

\newpage

{\Large Cosmic Gall}\\

Neutrinos they are very small.

They have no charge and have no mass

And do not interact at all.

The earth is just a silly ball

To them, through which they simply pass,

Like dustmaids down a drafty hall

Or photons through a sheet of glass.

They snub the most exquisite gas,

Ignore the most substantial wall,

Cold-shoulder steel and sounding brass,

Insult the stallion in his stall,

And, scorning barriers of class,

Infiltrate you and me! Like tall

And painless guillotines, they fall

Down through our heads into the grass.

At night, they enter at Nepal

And pierce the lover and his lass

From underneath the bed--you call

It wonderful; I call it crass.\\

~~~~John Updike\\

\newpage
In this Chapter, we will provide a mini-review on the framework of neutrino oscillations. We will first introduce the standard picture of 
3-neutrinos in Section \ref{Standard_Picture_of_Neutrinos}. Then in Section \ref{Sterile_Neutrinos}, we will present the experimental 
anomalies which led to the hypothesis of the sterile neutrinos, and explain the framework of the so-called $3+1$ scenario. The neutrino experiments and neutrinos in cosmology  will be briefed in Sections \ref{neutrino_experiments} and \ref{neutrinos_cosmology}, 
 respectively.
\section{The Standard Picture of Neutrinos}\label{Standard_Picture_of_Neutrinos}

\subsection{The Neutrino Lagrangian}

The phenomenological properties of neutrino interactions and oscillations is contained in the following Lagrangian:
\begin{eqnarray}\label{eq1.1}
\mathcal{L}&=&\sum_{\alpha=e,\mu,\tau}i\bar{\nu}^\alpha_L\slashed{\partial}\nu^\alpha_L~~~~~~~~~~~~~~~~~~~~~~~~~~~~~~~~~~~~~~~~(\rm{neutrino~Kinetic~term})\nonumber\\
&-&\frac{g}{2\sqrt{2}}\sum_{\alpha=e,\mu,\tau}\bar{\nu}^{\alpha}_L\gamma^\rho(1-\gamma^5)l_\alpha W_\rho+h.c.~~~~~~~~~~~(\rm{Charged~Current~interaction})\nonumber\\
&-&\frac{g}{4\cos\theta_w}\sum_{\alpha=e,\mu,\tau}\bar{\nu}^{\alpha}_L\gamma^\rho(1-\gamma^5)\nu^\alpha_L Z_\rho~~~~~~~~~~~~~~~(\rm{Neutral~Current~interaction})\nonumber\\
&-&\sum_{\alpha=e,\mu,\tau}\bar{l}^\alpha_LM^ll^\alpha_R+h.c.~~~~~~~~~~~~~~~~~~~~~~~~~~~~~~~~(\rm{leptonic~mass~term})\nonumber\\
&-&\sum_{\alpha=e,\mu,\tau}\bar{\nu}^{\alpha}_R M^\nu_D\nu^\alpha_L+h.c..~~~~~~~~~~~~~~~~~~~~~~~~~~~~~(\rm{Dirac~mass~term~for~neutrinos})\nonumber\\
\end{eqnarray}
Here $g$ is the weak coupling constant, $\theta_w$ is the Weinberg angle, $W$ and $Z$ are the weak gauge boson fields, $l_{\alpha}$'s
 are the charged lepton fields $(e,\mu,\tau)$, $\gamma^i$'s are the Dirac matrices and $M^l$ is the leptonic Dirac mass term. Neutrinos 
 are strictly massless in the SM, since it does not contain right handed neutrinos $\nu_R$. Therefore, to build a Dirac mass term for 
 neutrinos similar to the last term of Eq. (\ref{eq1.1}), we require not only the left handed field $\nu_L$, but also the right handed field 
 $\nu_R$ \footnote{For simplicity, we have only considered the Dirac mass terms for neutrinos, since it can be shown that the oscillation 
 phenomenology of Dirac and Majorana neutrinos are the same \cite{Akhmedov:1999uz}.}.

To diagonalize the mass matrices, we need to go from the weak basis to mass eigenstates. To do this, we have to make the following 
changes:
\begin{eqnarray}\label{eq1.2}
l^\alpha_{L(R)}&=&\sum_{j=1}^3{\big(V^l_{L(R)}\big)}^\alpha_{j}l^j_{L(R)},\nonumber\\
\nu^\alpha_{L(R)}&=&\sum_{j=1}^3{\big(U^\nu_{L(R)}\big)}^\alpha_{j}\nu^j_{L(R)},
\end{eqnarray}
where ${V^l_L}^\dagger M^lV^l_R=\rm{diag}\big(m_e,m_\mu,m_\tau\big)$ and ${U^\nu_L}^\dagger M^\nu U^\nu_R=\rm{diag}
\big(m_1,m_2,m_3\big)$, in which $m_\alpha$'s~$(\alpha=e,\mu,\tau)$ are the masses of the charged leptons, while $m_i$'s~$(i=1,2,3)$ 
are the masses of neutrinos in the mass basis. Please note that $V^l_{L(R)}$ and $U^\nu_{L(R)}$ are unitary matrices. From now on, the 
latin indices $i,j,\cdots=1,2,3\cdots$ correspond to the mass eigenstates, while the greek indices $\alpha,\beta,\cdots=e,\mu,\tau,\cdots$ correspond 
to the flavor (weak) eigenstates. 

The relation between the flavor and mass eigenstates of neutrinos is given by the so-called Pontecorvo-Maki-Nakagawa-Sakata (PMNS) matrix which appears in the Charged Current Lagrangian. This matrix is defined as 
$$
U_{\rm{PMNS}}\equiv {U^\nu_L}{U^l_L}^\dagger.
$$
Therefore
\begin{equation}\label{eq1.3}
\begin{pmatrix}
\nu_e\\
\nu_\mu\\
\nu_\tau
\end{pmatrix}
=U_{\rm{PMNS}}\begin{pmatrix}
\nu_1\\
\nu_2\\
\nu_3
\end{pmatrix}.
\end{equation}
Various parameterizations of the PMNS matrix exist. However, the oscillation properties of neutrinos is independent of different 
parameterizations. The $3\times3$ PMNS matrix is most commonly parameterized by 3 mixing angles $\theta_{12},\theta_{13}$ and 
$\theta_{23}$, and a $CP$-violating phase $\delta$ \footnote{An $n\times n$ unitary matrix has $n^2$ independent parameters, $n(n-1)/2$ 
of these are mixing angles, while $(n-1)(n-2)/2$ are the number of physical phases. There are also $2n-1$ non-physical phases
 which can be absorbed by appropriate rotations. }, in which the matrix can be written as:
\begin{eqnarray}\label{eq1.4}
U_{\rm{PMNS}}&=&R^{23}(\theta_{23})R^{13}_\delta(\theta_{13},\delta)R^{12}(\theta_{12})\nonumber\\
&=&\begin{pmatrix}
c_{12}c_{13}&s_{12}c_{13}&s_{13}e^{-i\delta}\\
-s_{12}c_{23}-c_{12}s_{23}s_{13}e^{i\delta}&c_{12}c_{23}-s_{12}s_{23}s_{13}e^{i\delta}&s_{23}c_{13}\\
s_{12}s_{23}-c_{12}c_{23}s_{13}e^{i\delta}&-c_{12}s_{23}-s_{12}c_{23}s_{13}e^{i\delta}&c_{23}c_{13}
\end{pmatrix},
\end{eqnarray}
where $s_{ij}\equiv\sin\theta_{ij}$ and $c_{ij}\equiv\cos\theta_{ij}$. The matrix $R^{ij}(\theta_{ij})~(i,j=1,2,3;~i<j)$ is the rotation matrix in 
the $ij-$plane. For an arbitrary dimension, the rotation matrix $R_{\delta}^{ij}(\theta_{ij},\delta)$ includes the $CP$-violating phase, and it 
is obtained by changing $s_{ij}\to s_{ij}e^{-i\delta}$ and $-s_{ij}\to -s_{ij}e^{i\delta}$ in the rotation matrix, for non-successive $i$ and
 $j$ indices. 

The analysis of the solar and long baseline reactor neutrinos lead to the best-fit values~\cite{GonzalezGarcia:2012sz}
\begin{eqnarray}
\sin^2 \theta_{12}&=&0.3,\nonumber\\
\Delta m_{21}^2&=&7.4\times 10^{-5}~{\rm eV}^2;\nonumber
\end{eqnarray}
while the data from atmospheric and long baseline accelerator experiments result in 
\begin{eqnarray}
\sin^2\theta_{23}&=&0.4,\nonumber\\
|\Delta m_{31}^2|\approx|\Delta m_{32}^2|&=&2.4\times10^{-3}~{\rm eV}^2.\nonumber
\end{eqnarray}
The last mixing angle has been measured recently with the new generation of medium baseline reactor experiments, including 
Double Chooz~\cite{DCmainpaper}, Daya Bay~\cite{An:2012eh} and RENO~\cite{Ahn:2012nd} experiments, with the best-fit value 
$$\sin^2\theta_{13}=0.023.$$
Please note that from the solar experiments, the sign of the solar mass squared difference is known ($\Delta m^2_{21}>0$, therefore $m_2>m_1$), while the atmospheric neutrino experiments have not verified the sign of the atmospheric mass squared difference $\Delta m^2_{31}$ yet, which is the source of hierarchy problem in neutrino masses. 

\subsection{Neutrino Oscillation in Vacuum}
To obtain the evolution equation of neutrinos in vacuum, one has to solve the following Schr\"{o}dinger-like equation
 \cite{Giunti:book}: 
\begin{equation}\label{eq1.5}
i\frac{d}{dt}\Ket{\nu_i}=\mathcal{H}\Ket{\nu_i}=E_i\Ket{\nu_i},
\end{equation}
where $\mathcal{H}$ is the Hamiltonian, and $E_i=\sqrt{m_i^2+p^2}$ is the energy of neutrinos, which since neutrinos are 
ultra relativistic particles, it can be written as $E_i\sim |\vec{p}|+\frac{m_i^2}{2|\vec{p}|}$. Therefore, 
\begin{equation}\label{eq1.6}
\Ket{\nu_i(t)}=e^{-iE_it}\Ket{\nu_i},
\end{equation}
where $\Ket{\nu_i}\equiv\Ket{\nu_i(0)}$. By using Eq. (\ref{eq1.3}), we can rewrite the evolution equation in the flavour basis:
\begin{eqnarray}\label{eq1.7}
\Ket{\nu_\alpha(t)}&=&\sum_{i=1}^3U_{\alpha i}e^{-iE_it}\Ket{\nu_i}\nonumber\\
&=&\sum_{\beta=e,\mu,\tau}\sum_{i=1}^3U_{\alpha i}e^{-iE_it}U^*_{\beta i}\Ket{\nu_\beta},
\end{eqnarray}
where $U\equiv U_{\rm{PMNS}}$. We can calculate the amplitude of the transition $\nu_\alpha\to\nu_\beta$ as a function of time $t$ by
\begin{eqnarray}\label{eq1.8}
A_{\nu_\alpha\to\nu_\beta}(t)\equiv \Braket{\nu_\beta|\nu_\alpha(t)}=\sum_{i=1}^3U_{\alpha i}U^*_{\beta i}e^{-iE_it}.
\end{eqnarray}
Hence, the probability that a neutrino with flavor $\alpha$ oscillates to a neutrino with flavor $\beta$ after the time $t$ is given by
\begin{eqnarray}\label{eq1.9}
P_{\nu_\alpha\to\nu_\beta}(t)=|A_{\nu_\alpha\to\nu_\beta}(t)|^2=\sum_{i,j=1}^3U_{\alpha i}U^*_{\beta i}U^*_{\alpha j}U_{\beta j}e^{-i(E_i-E_j)t}.
\end{eqnarray}
Since neutrinos are ultra relativistic particles, we can approximate the power of the exponential by 
$(E_i-E_j)t \simeq \big(\frac{m_i^2-m_j^2}{2E}\big)t=\frac{\Delta m^2_{ij}t}{2E}$, where we have assumed the momentum of all neutrino states is the same, and $|\vec{p}|\simeq E$. The next step is changing the $t$ dependence to the distance, $L$ dependence, as what we know 
well is the distance between the source of neutrinos and where they are detected. Again, as neutrinos are ultrarelativistic particles and 
their speed is very close to the speed of light, we can simply assume $t=L$, and write the oscillation probability as
\begin{eqnarray}\label{eq1.10}
P_{\nu_\alpha\to\nu_\beta}(L)=\sum_{i,j=1}^3U_{\alpha i}U^*_{\beta i}U^*_{\alpha j}U_{\beta j}e^{-i\frac{\Delta m^2_{ij}L}{2E}}.
\end{eqnarray}
Finally, using the unitarity condition for the PMNS matrix $U$, we can write the oscillation probability in a more convenient form:
\begin{eqnarray}\label{eq1.11}
P_{\nu_\alpha\to\nu_\beta}(L)=\delta_{\alpha\beta} &-&4 \sum_{i>j}^3\Re \big[U_{\alpha i}U^*_{\beta i}U^*_{\alpha j}U_{\beta j}\big]\sin^2\Big(\frac{\Delta m^2_{ij}L}{4E}\Big)\nonumber\\
&+&2 \sum_{i>j}^3\Im \big[U_{\alpha i}U^*_{\beta i}U^*_{\alpha j}U_{\beta j}\big]\sin\Big(\frac{\Delta m^2_{ij}L}{2E}\Big).
\end{eqnarray}
The same expression holds for anti-neutrinos if we change $U\to U^*$. 

It should be noted that for calculating the survival probability, in which $\alpha=\beta$, the amplitude $U_{\alpha i}U^*_{\alpha i}U^*_{\alpha j}U_{\alpha j}=|U_{\alpha i}|^2|U_{\alpha j}|^2$ is real, and therefore, the last term of Eq. (\ref{eq1.11}) vanishes, and we can write:
\begin{eqnarray}\label{eq1.12}
P_{\nu_\alpha\to\nu_\alpha}(L)=1 &-&4 \sum_{i>j} |U_{\alpha i}|^2|U_{\alpha j}|^2\sin^2\Big(\frac{\Delta m^2_{ij}L}{4E}\Big).
\end{eqnarray}
We can also define the useful parameter oscillation length, which is the distance at which the oscillation given by $\Delta m^2_{ij}$ becomes $2\pi$, and is defined by
$$
L^{\rm{osc}}_{ij}=\frac{4\pi E}{\Delta m^2_{ij}}.
$$
\subsubsection{The Case of 2 Neutrinos}\label{2-neutrino-model}

In this part we consider the case of only 2 neutrino flavors, e.g. $\nu_e$ and $\nu_\mu$, which in many cases is a very 
good approximation. Therefore, the oscillation would be described by the PMNS matrix, which now only has 1 mixing angle $\theta$, and includes no $CP$-violating phase:
\begin{eqnarray}\label{eq1.13}
U=\begin{pmatrix}
\cos\theta&\sin\theta\\
-\sin\theta&\cos\theta
\end{pmatrix},
\end{eqnarray}
and 1 mass squared difference $\Delta m^2_{21}\equiv\Delta m^2$. In this case the transition probability becomes
\begin{eqnarray}\label{eq1.14}
P_{\nu_\alpha\to\nu_\beta}\Big|_{\alpha\neq\beta}=\sin^22\theta\sin^2\big(\frac{\Delta m^2L}{4E}\big),
\end{eqnarray}
while the survival probability is $P_{\nu_\alpha\to\nu_\alpha}=1-P_{\nu_\alpha\to\nu_\beta}\Big|_{\alpha\neq\beta}$.

\begin{figure}[!hHtb]
\centering
\includegraphics[width=0.8\textwidth]{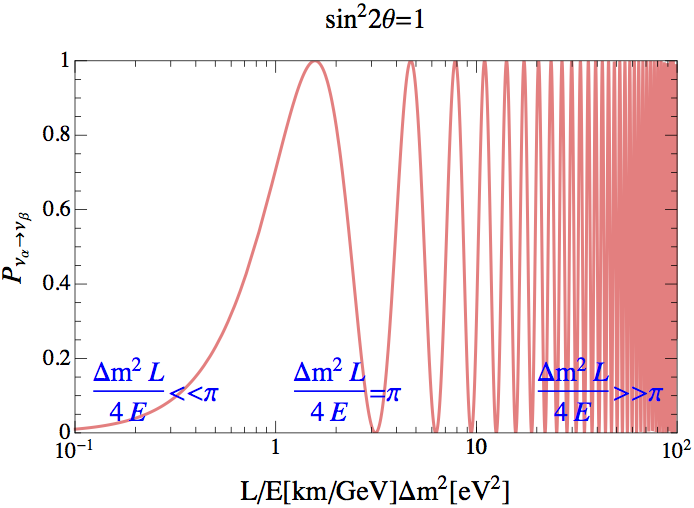}
\label{fig:2flavor_prob}
\caption{\label{fig:2flavor_prob}The transition probability for a 2-neutrino case. }
\end{figure}

In Fig. \ref{fig:2flavor_prob} we have shown the transition probability for a 2-neutrino case, putting the amplitude of the oscillation equal 
to 1. As can be seen, for the case $\Delta m^2L/4E\ll\pi$ (or $L\ll L^{\rm{osc}}$), the oscillation is suppressed, while for
 $\Delta m^2L/4E\gg\pi$ (or $L\gg L^{\rm{osc}}$), it averages out. The oscillation length defined after Eq. (\ref{eq1.12}) corresponds to 
 the first dip in Fig. \ref{fig:2flavor_prob}, where the argument in the sine function in Eq. (\ref{eq1.14}) is equal to $\pi$. This is the 
 sensitivity of the neutrino oscillation experiments. Therefore, the energy range of the experiment or the location of the detectors is 
 determined depending on the parameters an experiment plans to measure. For example, for measuring the atmospheric mass squared 
 difference $\Delta m^2_{31}=0.0024$ eV$^2$ in reactor experiments, which their energy range is around $\sim$ a few MeV, the 
 detectors have to be almost $1$ km far from the reactors to be sensitive to the oscillation. Hence, in this range of energy and baseline, 
 the oscillation induced by the solar mass squared difference $\Delta m^2_{21}=7.4\times10^{-5}$ eV$^2$ can be ignored. In general, the oscillation length for the mass squared difference 31 is $\sim1$ km and for the mass squared difference 21 is $\sim30$ km. 
 
\subsection{Neutrino Oscillation in Matter}\label{Neutrino_Oscillation_in_Matter}

\begin{figure}[!hHtb]
\centering
\includegraphics[width=0.8\textwidth]{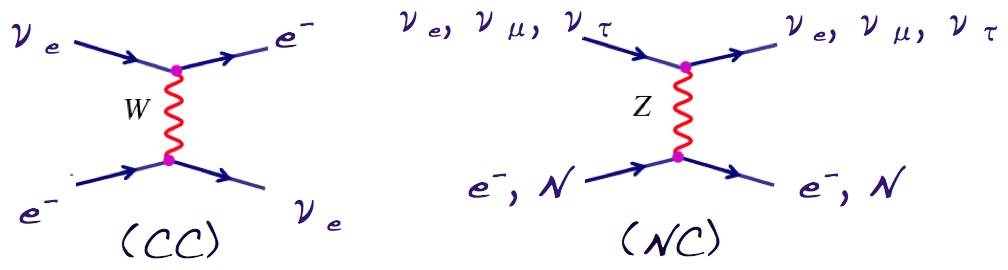}
\label{fig:coherent_scattering}
\caption{\label{fig:coherent_scattering}Coherent CC and NC forward scattering of neutrinos. }
\end{figure}
When neutrinos pass through matter, their evolution equation would be modified by the coherent scattering from particles they collide 
along their way. This coherent scattering results in an effective potential which is due to the charged current (CC) and neutral current 
(NC) interaction of neutrinos with matter. The electron neutrinos scatter from the electrons in the medium and experience the CC 
potential $V_{CC}$, through the exchange of $W^\pm$, while all the active neutrinos, $\nu_e$, $\nu_{\mu}$ and $\nu_\tau$ can 
experience the NC potential $V_{NC}$, due to the exchange of the $Z$ boson with $e^-$, $p$ and $n$ in the medium. 
These potentials are position dependent, and are given by \cite{Giunti:book}:
\begin{eqnarray}
V_{CC}(r)&=&\sqrt{2}G_FN_e(r),\\\label{eq1.15}
V_{NC}(r)&=&-\frac{\sqrt{2}}{2}G_FN_n(r), \label{eq1.16}
\end{eqnarray}
in which $G_F$ is the Fermi constant, while $N_e(r)$ and $N_n(r)$ are the electron and neutron number densities of the earth, respectively. For all practical reasons we assume that $N_e(r)$ and $N_n(r)$ are the same in our calculations. 

To calculate the evolution of neutrinos in the presence of matter one has to solve the following Schr\"{o}dinger-like equation:
\begin{equation}\label{eq1.17}
i\frac{d}{dr}\nu_\alpha=\Big[\frac{1}{2E_\nu}UM^2U^\dagger+V(r)\Big]_{\alpha\beta}\nu_\beta,
\end{equation}
where $M^2=\rm{diag}\Big(m_1^2,m_2^2,m_3^2\Big)$ is the mass squared matrix, $U$ is the $3\times3$ PMNS matrix and $E_{\nu}$ is the energy of neutrinos. The matrix $V(r)$ is the matter potential as a function of the distance:
\begin{equation}\label{eq1.18}
V(r)=
\begin{pmatrix}
V_{CC}(r)+V_{NC}(r)&0&0\\
0&V_{NC}(r)&0\\
0&0&V_{NC}(r)
\end{pmatrix}.
\end{equation}
The same equations hold for anti-neutrinos after we change $V(r)\to-V(r)$.
\subsubsection{The MSW Effect}

In principle, Eq. (\ref{eq1.17}) is a distance dependent Schr\"{o}dinger equation, and one can only solve it numerically. However, for 
the case of constant density we can solve the equations analytically. For simplicity we will consider the 2-neutrino case (See Section
 \ref{2-neutrino-model}). We assume the mixing between $\nu_e$ and $\nu_\mu$ with $\nu_1$ and $\nu_2$ \footnote{The case of 
 $\nu_e$ and $\nu_\tau$ is similar, since the potential of $\nu_\tau$ is the same as $\nu_\mu$.}. In this case the PMNS matrix is the 
 same as Eq. (\ref{eq1.13}). After subtracting from the mass squared and the potential matrices the constants
  $m_1^2 \times\mathbb{I}$ and $(V_{\rm{NC}}+\frac{V_{\rm{CC}}}{2})\times \mathbb{I}$, respectively, we can write the effective 
  Hamiltonian in matter for the 2-neutrino case as  \cite{Giunti:book}:
\begin{eqnarray}\label{eq1.19}
\mathcal{H}_M&=&\frac{1}{2E_\nu}
\begin{pmatrix}
\cos\theta&\sin\theta\\
-\sin\theta&\cos\theta
\end{pmatrix}
\begin{pmatrix}
0&0\\
0&\Delta m^2
\end{pmatrix}
\begin{pmatrix}
\cos\theta&-\sin\theta\\
\sin\theta&\cos\theta
\end{pmatrix}+
\begin{pmatrix}
V_{\rm{CC}}/2&0\\
0&-V_{\rm{CC}}/2
\end{pmatrix}\nonumber\\
&=&\frac{1}{4E_\nu}
\begin{pmatrix}
-\Delta m^2\cos2\theta+A_{\rm{CC}}&\Delta m^2\sin2\theta\\
\Delta m^2\sin2\theta&\Delta m^2\cos2\theta-A_{\rm{CC}}
\end{pmatrix},
\end{eqnarray}
where $\Delta m^2 \equiv m_2^2-m_1^2$ and $A_{\rm{CC}}=2E_{\nu}V_{\rm{CC}}$. The Hamiltonian can be diagonalized using a $2\times2$ rotation matrix $U_M$ with angle $\theta_M$, as ${U_M}^{T}\mathcal{H}_MU_M=\mathcal{H}_M^{\rm{diag}}$, where $\mathcal{H}_M^{\rm{diag}}=\frac{1}{4E_\nu}\rm{diag}\big(-\Delta m^2_M,\Delta m^2_M\big)$. The effective mass squared difference $\Delta m^2_M$ is given by
\begin{equation}\label{eq1.20}
\Delta m^2_M=\sqrt{\big(\Delta m^2\cos2\theta-A_{\rm{CC}}\big)^2+\big(\Delta m^2\sin2\theta\big)^2},
\end{equation}
and the effective mixing angle $\theta_M$ is
\begin{equation}\label{eq1.21}
\tan2\theta_M=\frac{\tan2\theta}{1-\frac{A_{\rm{CC}}}{\Delta m^2\cos2\theta}}.
\end{equation}
Therefore, the transition probability of $\nu_e\to\nu_\mu$ becomes
\begin{equation}\label{eq1.22}
P_{\nu_e\to\nu_\mu}=\sin^22\theta_M\sin^2\Big(\frac{\Delta m^2_M L}{4E}\Big).
\end{equation}

An interesting phenomenon which was discovered by Mikhaev, Smirnov and Wolfenstein in 1985, is that when the first term in 
(\ref{eq1.20}) becomes 0, a resonance happens:
\begin{equation}\label{eq1.23}
A^R_{\rm{CC}}=\Delta m^2\cos2\theta. 
\end{equation}
At the resonance, the effective mixing angle in (\ref{eq1.21}) becomes $\pi/4$, which is the maximal mixing angle, and means that the 
amplitude of the transition probability at the resonance is $\sin^22\theta_M=1$, and a total transition happens between the 2 flavors. This 
phenomenon is called the MSW effect, and was first proposed in  \cite{MSW-effect}.

\subsection{Seesaw Mechanism}\label{seesaw_mechanism}
The Dirac neutrino mass terms are described by 
\begin{equation}\label{eq1.24}
-\mathcal{L}_D=m_D(\bar{\nu}_L\nu_R+\bar{\nu}_R\nu_L)=m_D\bar{\nu}_D\nu_D,
\end{equation}
in which $\nu_D\equiv\nu_L+\nu_R$. The Dirac mass term $m_D$ can be generated through the Higgs mechanism by vacuum expectation value (vev) of the Higgs doublet:
\begin{equation}\label{eq1.25}
m_D=yv/\sqrt{2},
\end{equation} 
in which $y$ is the Yukawa coupling of neutrinos, and $v=246$ GeV is the weak scale. For $m_D\sim0.1$ eV, the Yukawa coupling would be $y\sim10^{-12}$, which is 6 orders of magnitude smaller than the Yukawa coupling of electrons: $y_e=3\times10^{-6}$. 

The Majorana mass term is written as
\begin{equation}\label{eq1.26}
-\mathcal{L}_T=\frac{m_L}{2}(\bar{\nu}_L\nu_R^c+\bar{\nu}_R^c\nu_L)=\frac{m_L}{2}(\bar{\nu}_L\mathcal{C}\nu_L^T+\nu_L^T\mathcal{C}\nu_L)=
\frac{m_L}{2}\bar{\nu}_M\nu_M,
\end{equation}
where $\nu^c_{L(R)}=\mathcal{C}{\gamma^0}^T\nu_{R(L)}^*=i\gamma^2\gamma^0{\gamma^0}^T\nu_{R(L)}^*$, and $\nu_M\equiv \nu_L+\nu_R^c=\nu_M^c$ is a 2-component Majorana field. A right handed neutrino $\nu_R$ (which in many models is considered as sterile neutrino) can also obtain Majorana mass term:
\begin{equation}\label{eq1.27}
-\mathcal{L}_S=\frac{m_R}{2}(\bar{\nu}_L^c\nu_R+\bar{\nu}_R\nu_L^c)=\frac{m_R}{2}(\bar{\nu}_L^c\mathcal{C}{\bar{\nu}_L^c}^T+{\nu_L^c}^T\mathcal{C}\nu_L^c)=
\frac{m_R}{2}{\bar{\nu}_M}_s{\nu_M}_s,
\end{equation}
where ${\nu_M}_s\equiv \nu_L^c+\nu_R={\nu_M^c}_s$.

When Dirac and Majorana mass terms are both present at the same time, we need to diagonalize the mass
matrix in order to obtain the mass-eigenstates; which in general, will be linear combinations of $\nu_L$ and $\nu^c_R$. In this case, for one right handed and one left handed neutrino, the effective mass Lagrangian becomes
\begin{equation}\label{eq1.28}
-\mathcal{L}=\frac{1}{2}
\begin{pmatrix}
\bar{\nu}_L&\bar{\nu}_L^c
\end{pmatrix}
\begin{pmatrix}
m_L&m_D\\
m_D&m_R
\end{pmatrix}
\begin{pmatrix}
\nu_R^c\\
\nu_R
\end{pmatrix}
+\rm{h.c.}.
\end{equation}
This mass matrix can be easily diagonalized by a $2\times2$ matrix $\mathcal{U}$. In following we list a number of special cases and limits:
\begin{itemize}
\item The pure Majorana case $m_D=0$: There is no mixing between the active neutrino $\nu_L$ and the sterile neutrino $\nu_R$.
\item The pure Dirac case $m_L=m_R=0$:  Leads to two degenerate Majorana neutrinos, which in principle can be combined to form a Dirac neutrino.
\item The seesaw limit $m_R\gg m_{D,L}$: There is one, mainly sterile, state with $m_2 \simeq m_R$, which decouples at low energy, and one light, mainly active state, with mass $m_1\simeq m_L - m^2_D/m_R$. If one of the eigenvalues goes up, the other goes down, and vice versa. This is the point of the name "seesaw" of the mechanism. In this case, if $m_L=0$, an elegant explanation is obtained for why $|m_1|\ll m_D$.
\end{itemize}

\section{Sterile Neutrinos}\label{Sterile_Neutrinos}

Although the huge amount of the information collected from the solar, atmospheric, reactor and accelerator neutrino oscillation experiments 
are in agreement with 3-neutrino hypothesis, there are a number of neutrino oscillation experiments the results of which cannot be 
explained within the framework of 3-neutrino oscillations. The most popular way to explain these anomalies is assuming the existence of 1 
(or more) sterile neutrino state(s), which do not have any interactions, hence are singlets of the SM (therefore are \textit{sterile}), and can 
only be detected through their mixing with the active neutrinos. In this Section we will explain the experimental anomalies which need to be
 clarified within the sterile neutrino hypothesis. Then, we will explain the oscillation of neutrinos in the presence of the sterile states.

\subsection{Experimental Motivation}\label{Experimental_Motivation}

\subsubsection{The LSND Anomaly}

The Liquid Scintillator Neutrino Detector (LSND) was primarily designed to check neutrino oscillations by measuring the number of neutrino 
events produced by an accelerator source \cite{Aguilar:2001ty}. The baseline of the experiment was $L\sim20$ m, and it detected 
neutrinos in the energy range of $E_\nu\sim20-200$ MeV. In this experiment, they observed electron anti-neutrino $\bar{\nu}_e$ events in
 a pure $\bar{\nu}_\mu$ beam. The most straightforward interpretation of this result is that $\bar{\nu}_\mu$ is oscillating to $\bar{\nu}_e$; 
 however, for this to happen, the corresponding mass squared difference has to be $\sim~1$ eV$^2$, which is in conflict with the solar and
  atmospheric mass squared differences, which are respectively $\Delta m^2_{21~(\rm{sol})}=7.4\times10^{-5}~$eV$^2$ and 
  $\Delta m^2_{31~(\rm{atm})}=2.4\times10^{-3}~$eV$^2$. Therefore, there has to be a forth neutrino state with mass $\sim~1$ eV. Since
   from the LEP results the number of the active neutrinos which participate in the weak interaction is 3, this forth neutrino state has to be a
    gauge singlet of the SM.

\subsubsection{The MiniBooNE Anomaly}
The MiniBooNE experiment at Fermilab has been designed to test the LSND result. The baseline of the experiment was $L\sim540$ m, 
and it detected neutrinos in the energy range of $475~\rm{MeV<E_\nu<3~\rm{GeV}}$ . The experiment has reported oscillation results for
 both neutrino and anti-neutrino channels. They have fit the 2-neutrino model (see Section \ref{2-neutrino-model}) to the data, and have found the
  allowed region in ($\sin^22\theta,\Delta m^2$) plane. They have shown that the $\nu_\mu\to\nu_e$ and $\bar{\nu}_\mu\to\bar{\nu}_e$ 
  oscillations with the mass squared difference in the $0.01$ eV$^2\leq\Delta m^2\leq1$ eV$^2$ range is consistent with the allowed region
   reported by the LSND experiment \cite{miniboone}. They report an unexplained excess of events in the low energy range of the neutrino 
   and anti-neutrino events. However, in the higher energy ranges, no significant excess has been seen. 

\subsubsection{The Reactor Anomaly}
A recent reevaluation of the flux of $\bar{\nu}_e$'s produced in the reactors show a $3\%$ increase of the flux. At the same time, the 
experimental value of the lifetime of neutrons is smaller than what it was expected. Therefore, the cross section of the Inverse Beta Decay
 (IBD) process becomes larger, and hence, the expected number of $\bar{\nu}_e$  events from the reactor experiments with baseline 
 $L < 100$ m, is $6\%$ higher than the observed data. This deficit can be explained if a sterile neutrino exists with $\Delta 
m^2_{\rm{sterile}}>1$ eV$^2$ \cite{reactoranomalie}.\\ 

\subsubsection{The Galium Anomaly}
In the radioactive source experiments GALLEX and SAGE, a $2.7\sigma$ deficit was seen in the expected number of events \cite{Giunti:2010zu}. This anomaly which is usually known as the "gallium anomaly", also hints towards a new mass squared difference of the order of the reactor anomaly.

\subsubsection{The Dark Radiation Anomaly}
An analysis done on the cosmological data from the cosmic microwave background (CMB) and Large Scale Structure (LSS) shows a tendency in favor of extra relativistic degrees of freedom, when the decoupling of CMB happens \cite{Hamann:2010pw}. This new source of radiation is often called as the "dark radiation", and allows for 1 sterile neutrino state. \\

In conclusion, there are different indications on the existence of the sterile neutrinos from different neutrino oscillation experiments. However, although these experiments have different sources and detector techniques, yet none of them can claim a discovery.

\subsection{The Framework of Sterile Neutrinos: "3+1" model}

As indicated in the previous Section, there are plenty of experimental anomalies which can be explained if we extend the SM with one 
(or more) neutrino state. Because of the fact that from the LEP results the number of light neutrinos which
 couple to the $Z$ boson should be $3$, hence this
 new state has to be a singlet fermion of the SM, i.e. \textbf{\textit{sterile}}. The sterile neutrino states do not have any charge under the 
 SM; therefore, they do not have any weak interaction, and can only be detected indirectly through their mixing with the active neutrinos. 

The SM singlets, i.e. the sterile neutrinos appear in many extensions of the SM. The most attractive scenario which explains the neutrino masses, the seesaw mechanism, includes heavy sterile neutrinos \cite{Minkowski:1977sc,Mohapatra:1979ia,Yanagida:1979as,Schechter:1980gr}. Also, in the Large Extra Dimension (LED) models which were proposed to explain the hierarchy problem in Higgs sector, the smallness of neutrino masses could be explained as well assuming that the sterile neutrinos could propagate in the bulk, and result in naturally small neutrino Yukawa couplings which are suppressed by the volume of the extra dimensions \cite{Dienes:1998sb,ArkaniHamed:1998vp} \footnote{We will study the phenomenology of the LED model in Chapter \ref{chap3}.}. In this thesis we are specifically concerned with relatively light sterile neutrinos which have significant mixing with the active neutrinos. 

The minimal scheme which can explain the experimental anomalies in neutrino sector is the so called $"3+1"$ model, which includes 3 active neutrinos $\nu_e$, $\nu_\mu$ and $\nu_\tau$, extended with a forth sterile state $\nu_s$. In this model the mass spectrum of neutrino sector consists of three mostly active neutrino mass eigenstates with masses $(m_1,m_2,m_3)$ and one mostly sterile neutrino mass eigenstate with mass $m_4$. To be able to explain the experimental anomalies, the mass of the mostly sterile state has to be much bigger than the other masses, in such a way that
\begin{equation}\label{eq1.29}
m_1,m_2,m_3<<m_4.
\end{equation}
\begin{figure}[!hHtb]
\centering
\includegraphics[width=0.58\textwidth]{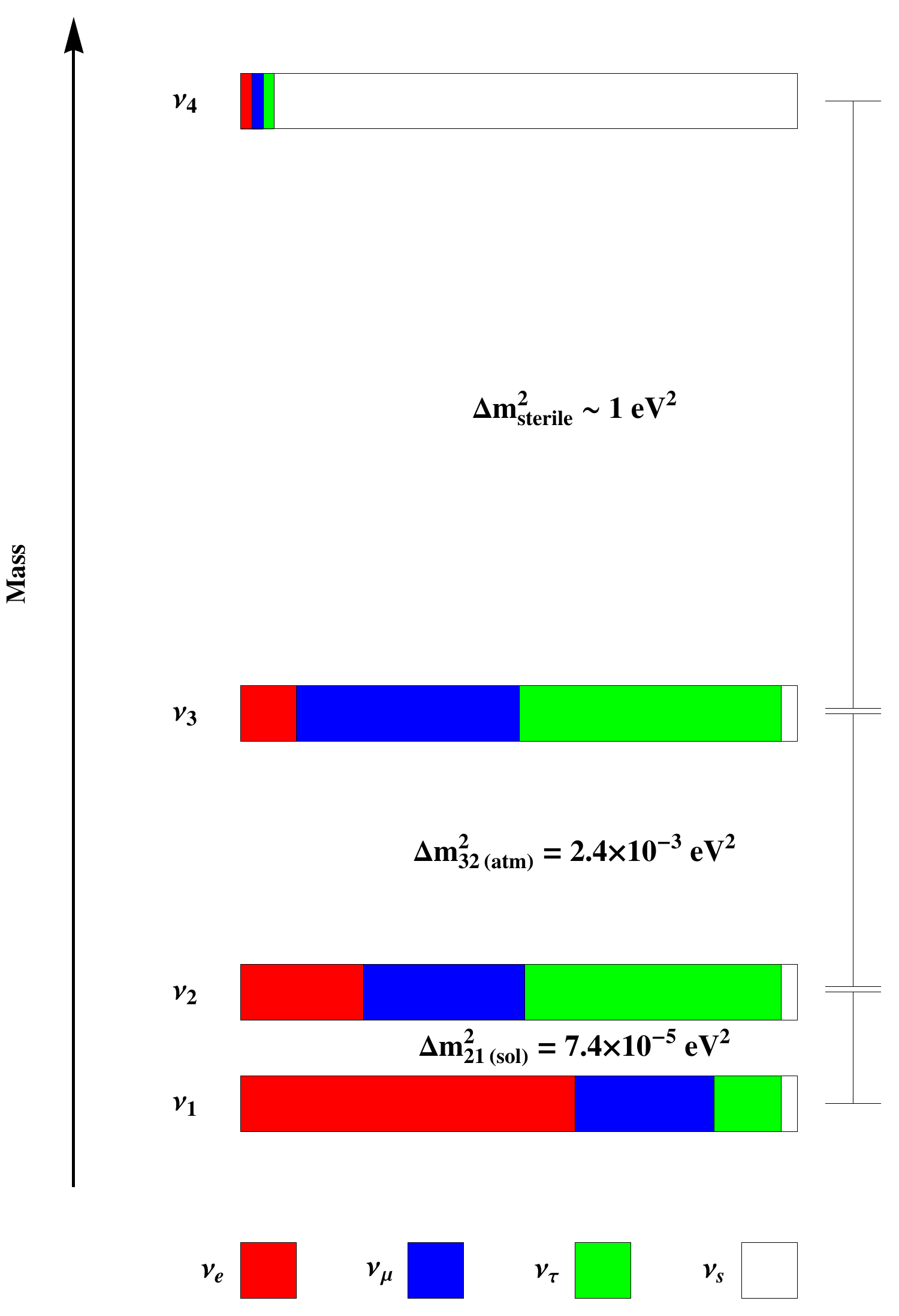}
\label{fig:mass_hierarchy_3+1}
\caption{\label{fig:mass_hierarchy_3+1}The mass hierarchy of neutrinos for $3+1$ framework}
\end{figure}

The mixing between the flavor and mass eigenstates is described by generalizing the PMNS matrix to a $4\times4$ unitary matrix, $U^{(4)}
$, which can be parametrized by six mixing angles: 3 active-active mixing angles ($\theta_{12},\theta_{13},\theta_{23}$), as well as  3 
active-sterile mixing angles ($\theta_{14},\theta_{24},\theta_{34}$), and 3 CP-violating phases: ($\delta_{13},\delta_{14},\delta_{24}$). To describe $U^{(4)}$ we use the following parametrization:
\begin{eqnarray}\label{eq1.30}
U^{(4)}= R^{34}(\theta_{34})R^{24}_{\delta_{24}}(\theta_{24},\delta_{24})R^{14}_{\delta_{14}}(\theta_{14},\delta_{14})R^{23}(\theta_{23})R^{13}_{\delta_{13}}(\theta_{13},\delta_{13})R^{12}(\theta_{12}),
\end{eqnarray}
where $R^{ij}(\theta_{ij})$ (i,j = 1,...,4 and $i < j$) is the $4\times4$ rotation matrix in the $ij$-plane with angle $\theta_{ij}$\footnote{For the
 inclusion of the CP-phases in the rotation matrices see the explanation after Eq. (\ref{eq1.4}).}. In total, 6 new mixing parameters are 
introduced in the $(3+1)$ model: 3 mixing angles $(\theta_{14},\theta_{24},\theta_{34})$, which quantify the $\nu_s-\nu_e$, $\nu_s-\nu_
\mu$ and $\nu_s-\nu_\tau$ mixings, respectively; one new mass-squared difference which we choose as $\Delta m_{41}^2\equiv m_4^2-
m_1^2$; and 2 CP-violating phases $\delta_{14}$ and $\delta_{24}$. The two extra mass-squared differences $\Delta m_{42}^2$ and $
\Delta m_{43}^2$ are not independent and can be written as: $\Delta m^2_{42}=\Delta m^2_{41}-\Delta m^2_{21}$ and $\Delta m^2_{43}=
\Delta m^2_{41} -\Delta m^2_{31}$. In Fig. \ref{fig:mass_hierarchy_3+1} we have schematically shown the relation between the mass and 
flavor eigenstates for the $3+1$ model. 

The sterile neutrino hypothesis should not spoil the results of the 3-neutrino framework. Therefore, the mass squared differences in the 
$3+1$ model should satisfy this condition:
$$|\Delta m^2_{21}|\ll|\Delta m^2_{31}|\ll|\Delta m^2_{41}|.$$
This condition guaranties that the oscillations induced by the new mass squared difference is averaged out in the experiments sensitive to
 $\Delta m^2_{21}$ or $\Delta m^2_{31}$. Also, the assumption that the new mass state is mostly sterile indicates that the elements of 
 $U^{(4)}$ should satisfy this condition:
$$|U^{(4)}_{e4}|^2,|U^{(4)}_{\mu4}|^2,|U^{(4)}_{\tau4}|^2\ll1,$$
so that the standard oscillation mixing angles are untouched. 
The oscillation probability of neutrinos in vacuum for a general $3+1$ framework is similar to Eq. (\ref{eq1.11}), replacing the PMNS matrix
 with $U^{(4)}$:
\begin{eqnarray}\label{eq1.31}
P_{\nu_\alpha\to\nu_\beta}(L)=\delta_{\alpha\beta} &-&4 \sum_{i>j}^4\Re \big[U^{(4)}_{\alpha i}{U^{(4)}}^*_{\beta i}{U^{(4)}}^*_{\alpha j}U^{(4)}_{\beta j}\big]\sin^2\Big(\frac{\Delta m^2_{ij}L}{4E}\Big)\nonumber\\
&+&2 \sum_{i>j}^4\Im \big[U^{(4)}_{\alpha i}{U^{(4)}}^*_{\beta i}{U^{(4)}}^*_{\alpha j}U^{(4)}_{\beta j}\big]\sin\Big(\frac{\Delta m^2_{ij}L}{2E}\Big).
\end{eqnarray}\\

In $\nu_e$ $(\bar{\nu}_e)$ disappearance experiments, in which the survival probability of $\nu_e\to\nu_e$ ($\bar{\nu}_e\to\bar{\nu}_e$) is measured, for the short baseline (SBL) limit, where the baseline of the experiment is $L<100$ m, we can neglect the $1-2$ and $1-3$ frequencies (since the oscillation lengths for these 2 mass squared differences are $30$ km and $1$ km respectively, which are much bigger than the baseline of these experiments), and write the effective survival probability in vacuum in the following form 
(See Eq. (\ref{eq1.12})):
\begin{eqnarray}\label{eq1.32}
P^{3+1}_{\nu_e(\bar{\nu}_e)\to\nu_e(\bar{\nu}_e)}&=&1-4|U^{(4)}_{e4}|^2\big(1-|U^{(4)}_{e4}|^2\big)\sin^2\big(\frac{\Delta m^2_{41}L}{4E}\big)\nonumber\\
&=&1-\sin^22\theta_{14}\sin^2\big(\frac{\Delta m^2_{41}L}{4E}\big),
\end{eqnarray}
where for writing the last term, we have used the parametrization of $U^{(4)}$ in Eq. (\ref{eq1.25}). The SBL experiments are not sensitive to the standard $3-$neutrino scheme, but they can observe the oscillatory behavior of $\Delta m^2_{41}$ and $\theta_{14}$ parameters. On the other hand, in medium baseline experiments ($L\sim1$ km), oscillations due to ($\Delta m^2_{31}, \theta_{13}$) are the relevant parameters, and oscillations due to eV$^2$-scale mass-squared differences are averaged out.

In $\nu_\mu$ $(\bar{\nu}_\mu)$ disappearance experiments with long baselines ($L~\sim$ a few hundred-a few thousand km), the effective survival probability in vacuum takes the following form:
\begin{eqnarray}\label{eq1.33}
P^{3+1}_{\nu_\mu(\bar{\nu}_\mu)\to\nu_\mu(\bar{\nu}_\mu)}=1-4|U^{(4)}_{\mu4}|^2\big(1-|U^{(4)}_{\mu4}|^2\big)\sin^2\big(\frac{\Delta m^2_{41}L}{4E}\big),
\end{eqnarray}
where $U^{(4)}_{\mu4}=\sin\theta_{24}\cos\theta_{14}$. Therefore, these experiments are sensitive to both of these mixing angles. However, in most of these experiments $\cos\theta_{14}$ can be approximated to 1. On the other hand, it was shown in \cite{Esmaili:2013vza} that due to the importance of the matter effect in the IceCube experiment, the survival probability of $\nu_\mu$ $(\bar{\nu}_\mu)$ in matter is sensitive to $U^{(4)}_{\tau4}$ as well, where $U^{(4)}_{\tau4}=\sin\theta_{34}\cos\theta_{14}\cos\theta_{24}$.

In $\nu_\mu\to\nu_e$ ($\bar{\nu}_\mu\to\bar{\nu}_e$) appearance searches, the effective transition oscillation is written as 
\begin{eqnarray}\label{eq1.34}
P^{3+1}_{\nu_\mu(\bar{\nu}_\mu)\to\nu_e(\bar{\nu}_e)}=4|U^{(4)}_{\mu4}|^2|U^{(4)}_{e4}|^2\sin^2\big(\frac{\Delta m^2_{41}L}{4E}\big),
\end{eqnarray}
and hence these experiments are sensitive to both $U^{(4)}_{e4}$ and $U^{(4)}_{\mu4}$ at the same time.

The framework of the sterile neutrinos has been considered for a long time, see e.g. \cite{Peres:2000ic,Giunti:2000wt}. The prospect of the $3+1$ model has been studied in a number of papers. In a recent global analysis on the data of short baseline ($L<100$ m) and medium baseline ($L\sim1$ km) reactor and accelerator experiments, as well as atmospheric and solar neutrino data, they have found that \cite{Kopp:2013vaa}:
\begin{eqnarray}\label{eq1.35}
\Delta m^2_{41}&=&0.93~\rm{eV}^2,\nonumber\\
|U^{(4)}_{e4}|&=&0.15,\nonumber\\
|U^{(4)}_{\mu4}|&=&0.17,\nonumber\\
|U^{(4)}_{\tau4}|^2&\leq&0.2;
\end{eqnarray}
 in which by using the parametrization of Eq. (\ref{eq1.25}), the active-sterile mixing angles become (neglecting all the CP-violating phases)
 \begin{eqnarray}\label{eq1.36}
\sin^2\theta_{14}&=&0.022,\nonumber\\
\sin^2\theta_{24}&\simeq&0.029,\nonumber\\
\sin^2\theta_{34}&\leq&0.19.
\end{eqnarray}


\subsubsection{"3+1" model in matter}\label{3plus1matter}

To calculate the evolution of neutrinos in $3+1$ scenario in matter, we have to solve the following Schr\"{o}dinger-like equation \cite{Esmaili:2012nz}:
\begin{equation}\label{eq1.37}
i\frac{d}{dr}
\begin{pmatrix}
\nu_e\\\nu_\mu\\\nu_\tau\\\nu_s
\end{pmatrix}
=\Big[\frac{1}{2E_\nu}U^{(4)}M^2{U^{(4)}}^\dagger+V(r)\Big]
\begin{pmatrix}
\nu_e\\\nu_\mu\\\nu_\tau\\\nu_s
\end{pmatrix},
\end{equation}
in which $M^2=\rm{diag}\Big(0,\Delta m^2_{21},\Delta m^2_{31},\Delta m^2_{41}\Big)$ is the mass squared difference matrix, $U^{(4)}$ is the $4\times4$ PMNS matrix and $E_{\nu}$ is the energy of neutrinos. The potential matrix $V(r)$ is 
\begin{equation}\label{eq1.33}
V(r)=\sqrt{2}G_F\rm{diag}\Big(N_e(r)-N_n(r)/2,-N_n(r)/2,-N_n(r)/2,0\Big),
\end{equation}
where $G_F$ is the Fermi constant, and $N_e(r)$ ($N_n(r)$) is the electron (neutron) number density of the earth (See the discussion in Section \ref{Neutrino_Oscillation_in_Matter} for $3$-neutrino case in matter). The same equation holds for anti-neutrinos if we change $V(r)\to-V(r)$. 

Eq. (\ref{eq1.37}) is in principle a distance dependent Schr\"{o}dinger equation and one can only solves it numerically. The method on how to solve these equations will be described in Chapter \ref{chap3}. 


\section{Neutrino Experiments}\label{neutrino_experiments}

\subsection{Oscillation Experiments }
There are 2 kinds of neutrino oscillation experiments: the \textit{\textbf{disappearance}} and the \textit{\textbf{appearance}} experiments. In disappearance experiments, there is usually a source of neutrinos with flavor $\alpha$ and 2 near and far detectors. The near detector is very close to the source, in a range in which the flavor oscillation does not happen. Both near and far detectors measure the flux of $\nu_\alpha$. In these experiments, they compare the flux in both detectors to look for a loss of events. The loss of neutrino events in the far detector is due to neutrino oscillation. 

In the appearance experiments, there usually exists a source of neutrinos with flavor $\alpha$, with a far detector which measures neutrino events with flavor $\beta$, for $\alpha\neq\beta$. Detecting neutrinos with flavor different from the source neutrinos indicate there is neutrino oscillation. 

In what follows we describe different sources of neutrinos, and the related experiments. 

\subsubsection{Solar Neutrinos }

The sun is a huge source of electron neutrinos with energy $\sim1$~MeV to 20 MeV. Most neutrinos which pass through the earth originate from the 
sun ($\sim6.5\times10^{10}$ $\rm{particles}/\rm{cm}^2\rm{s}$). The flux of solar neutrinos on the earth was first calculated by John N. 
Bahcall in 1960s~\cite{Bahcall_1964}. After Bahcall calculated the flux, Raymond Davis designed an experiment to measure this flux, the Homestake (or usually  
called Davis) experiment \cite{Davis_1964}. The purpose of this experiment was counting neutrinos which were emitted by the nuclear fusion inside the sun. 
The results of this experiment was the first big surprise in neutrino physics, as the measured number of neutrinos from the experiment was approximately one 
third of the expected number of events calculated by Bahcall. This problem is the so called "\textit{solar neutrino problem}". This discrepancy between the 
measured and the expected number of events was later found to be due to the "flavour oscillation" of neutrinos. 

\begin{figure}[!hHtb]
\centering
\includegraphics[width=0.8\textwidth]{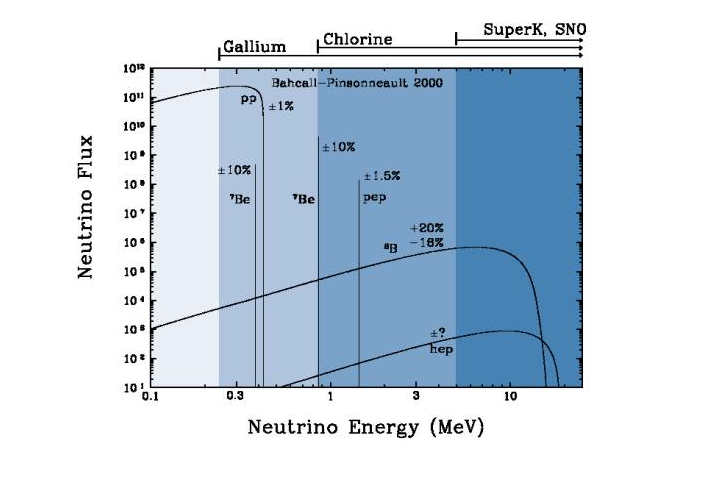}
\label{fig:solar_flux}
\caption{\label{fig:solar_flux}Solar neutrino flux as a function of energy. Plot taken from \cite{solar_flux}. }
\end{figure}

After the surprising results of the Davis experiment, the GALLEX/GNO and SAGE experiments (the gallium experiments) were established 
to detect solar neutrinos in the beginning of 1990s. These experiments measured the low energy electron neutrinos produced in the 
sun. Since the end of 1990s, the Super-Kamiokande and SNO experiments have been providing high precession data on the high energy 
range of the solar neutrinos. Finally it was the SNO experiment which solved the solar mystery. The Sudbury Neutrino Observatory (SNO) experiment has measured the flux of electron neutrinos, as well as the combined flux of all neutrino flavours. The SNO experiment has determined that $\phi_e/\phi_{\rm{total}}=0.306\pm0.05$ \cite{Ahmad:2002ka}, which is a proof that the electron neutrinos produced in the sun oscillate to muon and tau neutrinos in their path to the earth. The combined rate of these fluxes showed an agreement with the Bahcall calculations. The no-oscillation hypothesis has been ruled out by more than $17\sigma$ C.L. since then. A combined analysis on all solar neutrino data as well as the data from the KamLAND reactor experiment gives the best fit values for the solar mass squared difference and mixing angle~\cite{GonzalezGarcia:2012sz}:
\begin{eqnarray}
\Delta m_{21}^2&=&7.4\times 10^{-5}~{\rm eV}^2,\nonumber\\
\sin^2 \theta_{12}&=&0.3,~(\theta_{12}\sim28~\rm{degrees}).\nonumber
\end{eqnarray}

\begin{figure}[!hHtb]
\centering
\includegraphics[width=1\textwidth]{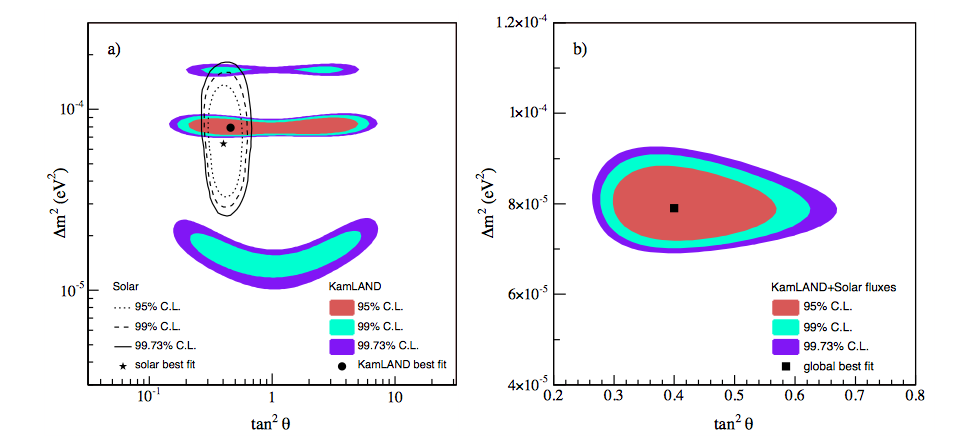}
\label{fig:KamLAND_combined_solar}
\caption{\label{fig:KamLAND_combined_solar}Neutrino oscillation parameter allowed region from KamLAND anti-neutrino data (shaded regions) and solar neutrino experiments. Here $\Delta m^2\equiv\Delta m^2_{21}$ is the solar mass squared difference, while $\theta\equiv\theta_{12}$ is the solar mixing angle. Plot taken from \cite{Araki:2004mb}.}
\end{figure}

\subsubsection{Atmospheric Neutrinos }
Atmospheric neutrinos are produced by the interactions of the primary cosmic rays with the atmosphere. These interactions produce pions in particular. The pions decay into muons and muon neutrinos. The muons also decay to electrons, electron neutrinos, as well as secondary muon neutrinos. Therefore the flux of muon atmospheric neutrinos is 2 times more than the flux of electron atmospheric neutrinos (this is only true for energies below GeV, i.e. $E<1$~GeV.).

Atmospheric neutrinos are not yet completely explored. They have a huge energy range, from $100$ MeV to $100$ TeV. The baseline of the atmospheric experiments can vary from a few km, to the diameter of the earth. The neutrinos are produced in the atmosphere, in a thin layer around the earth. They cross various layers of the earth and are detected in the underground/under-water/under-ice detectors. The atmospheric neutrinos which pass through the earth can experience matter effect with density $\sim ~2.5-15~\rm{gr}/\rm{cm}^3$. 

The Kamiokande \cite{Hirata:1988uy} and IMB~\cite{BeckerSzendy:1992ym} experiments began to observe the atmospheric neutrinos in the second half of 1980's. These experiments observed a number of events significantly smaller than what was predicted. This was the source of "\textit{atmospheric neutrino anomaly}". The solution of this anomaly in favor of neutrino oscillation came from the observations of the Super-Kamiokande water Cherenkov detector \cite{Fukuda:1998mi}. A combined analysis on the data from atmospheric and long baseline accelerator experiments results in 
\begin{eqnarray}
|\Delta m_{31}^2|\approx|\Delta m_{32}^2|&=&2.4\times10^{-3}~{\rm eV}^2,\nonumber\\
\sin^2\theta_{23}&=&0.4,~(\theta_{23}\sim33~\rm{degrees}).\nonumber
\end{eqnarray}
The IceCube neutrino observatory located in Antarctica is the hugest atmospheric neutrino experiment built so far. Its main goal is detecting the highest energy neutrinos possibly originating from extra-galactic sources. The IceCube detector is able to probe the atmospheric neutrinos from 10 GeV-400 TeV, as well as the very high energy neutrinos with energy up to several EeV. The experiment can also observe neutrino oscillation induced by 2-3 mass squared difference. Searching for the sterile neutrinos is also from the goals of IceCube \cite{Abbasi:2010ie,Nunokawa:2003ep}.

\begin{figure}[!hHtb]
\centering
\includegraphics[width=0.8\textwidth]{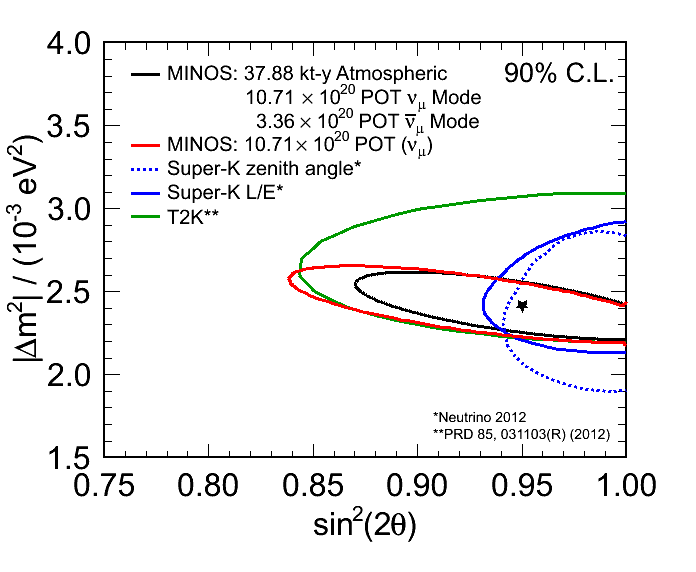}
\label{fig:MINOS_combined_atmospheric}
\caption{\label{fig:MINOS_combined_atmospheric}The allowed regions for $\Delta m^2_{23}$ and $\sin^22\theta_{23}$, from MINOS, T2K and SuperK experiments. Plot taken from \cite{Adamson:2013whj}. }
\end{figure}

\subsubsection{Reactor Neutrinos }

Fission reactors are a huge source of electron anti-neutrinos (about $10^{20}~\rm{s}^{-1}$ per nuclear core). The reactor neutrinos were first detected in the Cowan-Reines experiment, in 1956 \cite{Cowan:1992xc}. In this experiment as well as the other reactor experiments, the neutrinos are detected through the Inverse Beta Decay (IBD) process:
$$
\bar{\nu}_e+p\rightarrow e^++n.
$$
The electron anti-neutrinos with threshold energy above $1.8$ MeV interact with protons in water, and produce positrons and neutrons in the final products. The positrons annihilate with electrons and produce $0.5~$MeV photons, while the neutrons are captured by the nucleons of the detector liquid and produce secondary photons a few mili-seconds later, with energy around $3-11$ MeV. 

\begin{figure}[!hHtb]
\centering
\includegraphics[width=0.5\textwidth]{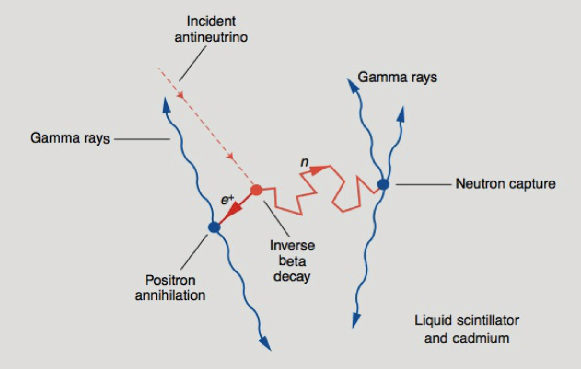}
\label{fig:IBD_process}
\caption{\label{fig:IBD_process}IBD process \cite{Maiani:2014vqa}. }
\end{figure}

The reactor neutrino experiments can detect electron antineutrinos with energy around a few MeV. The medium baseline (MBL) reactor experiments with baseline $\sim0.1-1$~km are sensitive to the atmospheric mass squared difference $\Delta m^2_{31}$, while the long baseline (LBL) reactor experiments with baseline $\sim10-100$~km are able to measure the solar mass squared difference $\Delta m^2_{21}$. The Kamioka Liquid Scintillator Antineutrino Detector (KamLAND) experiment is surrounded by 55 Japanese nuclear power reactors with average distance of 150 km \cite{Eguchi:2002dm}. The experiment started to take data in 2002. KamLAND measures the solar mass squared difference most precisely, while the solar experiments are more accurate in measuring $\theta_{12}$; therefore, to have the most precise measurement for these parameters, a combined analysis is performed.

\begin{figure}[!hHtb]
\centering
\includegraphics[width=1\textwidth]{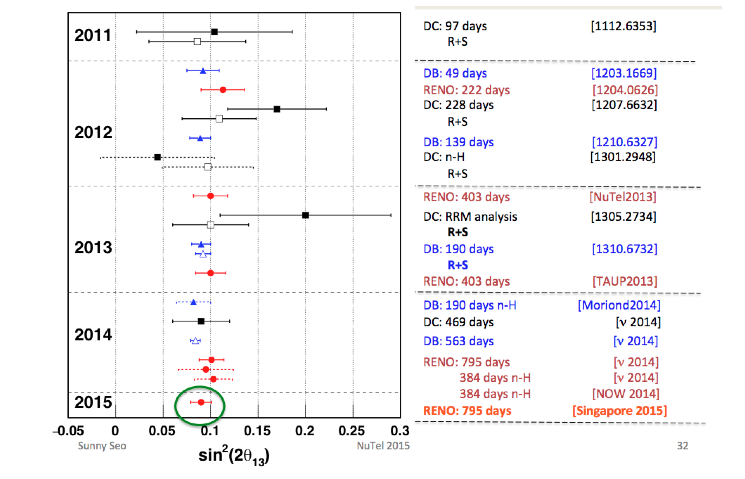}
\label{fig:reactor_combined}
\caption{\label{fig:reactor_combined}Summary of $\theta_{13}$ measurements by the reactor neutrino experiments of RENO, Daya Bay and Doube-Chooz. Plot and Table are taken from \cite{Kim:2015dag}. }
\end{figure}

Recently in 2012, the SBL rector experiments Double Chooz~\cite{DCmainpaper}, Daya Bay~\cite{An:2012eh} and RENO~\cite{Ahn:2012nd} experiments measured the value of the last unknown mixing angle:
$$\sin^2\theta_{13}=0.023,~(\theta_{13}\simeq9~\rm{degrees}).$$ 
The measurement of the small mixing angle $\theta_{13}$ in these experiments was achieved thanks to the highly controlled systematic errors and efficient background rejection down to $\sim10\%$ of signal. Therefore, in principle, the data of these experiments can be also used to discover/constrain new physics in the neutrino sector.

\subsubsection{Accelerator Neutrinos }

\begin{figure}[!hHtb]
\centering
\includegraphics[width=0.8\textwidth]{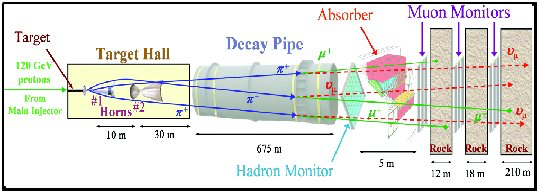}
\label{fig:numi_beamline}
\caption{\label{fig:numi_beamline} NuMI beamline at Fermilab \cite{numi_beamline}. }
\end{figure}

The idea of producing accelerator neutrino beams was primarily introduced by Schwartz \cite{Schwartz:1960hg} and Pontecorvo \cite{pontecorvo:accelerator}. Conventional neutrino beams are produced very similar to atmospheric neutrinos: high energy protons hit a target, pions and kaons are produced, they decay into muons and neutrinos. Then the muons also decay into electrons and neutrinos. The accelerator neutrinos have energy between $E_\nu\sim0.1-100$~GeV, and baseline $L\sim1-1000$~km.  Accelerator neutrino beams are always a mixture of both neutrinos and antineutrinos.

 The NuMI ("Neutrinos at Main Injector") beamline at Fermilab is an example of accelerator neutrino beams (See Fig. \ref{fig:numi_beamline}). The MINOS (Main Injector Neutrino Oscillation Search) experiment uses the neutrinos produced in the NuMI accelerator to search for neutrino oscillations \cite{Adamson:2013whj}. The K2K accelerator experiment \cite{Ahn:2002up} also looks for evidence for muon neutrino disappearance. The T2K experiment which is the second generation of K2K is also an accelerator based experiment which measures the oscillation of $\nu_\mu$ to $\nu_e$ and the value of $\theta_{13}$ mixing angle \cite{Abe:2013xua}.

\subsubsection{Measuring $CP$-violation}
The $CP$-violation in neutrino sector happens if the oscillation probability for neutrinos is different with anti-neutrinos. This difference is defined by (See Eq. (\ref{eq0.2}) and the discussion after that.)
\begin{equation}
A_{\alpha\beta}=P(\nu_\alpha\to\nu_\beta)-P(\bar{\nu}_\alpha\to\bar{\nu}_\beta).
\end{equation}
The invariance of $CPT$ implies that $A_{\alpha\beta}=-A_{\beta\alpha}$; hence $A_{\alpha\alpha}=0$. This behavior can also be seen from the probability relation in Eq. (\ref{eq1.12}), in which it is shown that the survival probability for neutrinos and anti-neutrinos is the same; therefore, $A_{\alpha\alpha}=0$. This means that $CP$-violation can only be measured in the appearance experiments, e.g. in the T2K experiment in which it measures the oscillation of $\nu_\mu$ to $\nu_e$.

The conservation of the probability implies that $\sum_{\alpha\neq\beta}A_{\alpha\beta}=0$, which means $A_{e\mu}=A_{\mu\tau}=A_{\tau e}=-A_{\mu e}=-A_{\tau\mu}=-A_{e\tau}\equiv A^{\rm{CP}}$. Therefore, $CP$-violation is the same in all experimental channels up to a sign. For 3-neutrino oscillation, we have
\begin{equation}
A^{\rm{CP}}=-4s_{12}c_{12}s_{13}c_{13}^2s_{23}c_{23}\sin\delta\Big[\sin\big(\frac{\Delta m^2_{12}L}{2E}\big)+\sin\big(\frac{\Delta m^2_{23}L}{2E}\big)+\sin\big(\frac{\Delta m^2_{31}L}{2E}\big)\Big],
\end{equation}
which is maximal for $\delta=\pi/2$ or $\delta=3\pi/2$. A recent global analysis of solar, atmospheric, reactor and accelerator neutrino data in the framework of three-neutrino oscillations show that the favored values of $\delta$ are around $3\pi/2$, while values around $\pi/2$ are disfavored by more than $2\sigma$ \cite{Gonzalez-Garcia:2014bfa}. Also, it is shown in \cite{Machado:2013kya} that if the mass hierarchy is known, then the T2K experiment can exclude about $50\%-60\%$ of the $\delta$ space by about $90\%$ C.L., by 10 years running of the experiment.

\section{Neutrinos in Cosmology}\label{neutrinos_cosmology}



In the temperatures higher than $10^{10}$ K in the early universe, the matter which was mainly composed of neutrinos, electrons and positrons, was in thermal equilibrium with the photons. At this temperature processes like 
\begin{eqnarray}
e^++e^-&\rightleftharpoons&\nu+\bar{\nu},\nonumber\\
\overset{(-)}{\nu}+e^{\pm}&\rightleftharpoons&\overset{(-)}{\nu}+e^{\pm}\nonumber
\end{eqnarray}
 happened rapidly by weak interaction. When the temperature fell below $10^{10}$ K (which is less than the rest energy of electrons/positrons), since there was not enough energy to produce $e^--e^+$ pairs from neutrinos, they decoupled from the rest. However, due to the electromagnetic interaction, electrons/positrons were still coupled to photons. At this time there were 2 different temperatures: that of neutrinos, which decreased like the inverse of the scale parameter: $T_\nu\propto1/a$, as neutrinos were relativistic particles; and that of $\gamma-e^--e^+$, which decreased at a slower rate. This goes on until the recombination time, in which the electrons combine with protons to form the neutral atoms. Hence, there was no electric charge in the universe, and photons decoupled from the rest. After decoupling of photons, their temperature decreased like inverse of the scale parameter: $T_\gamma\propto1/a$. Therefore, after the decoupling of the  photons, the ratio of the temperature of the photons to that of neutrinos remain constant, although $T_\gamma$ is bigger than $T_\nu$, as until before the decoupling of the photons, $T_\gamma$ was decreasing with a slower rate. This ratio is $R\equiv\frac{T_\gamma}{T_\nu}$.
 
 The temperature of the relic photons from cosmic microwave background (CMB) has been measured extremely well: $T^0_\gamma=2.72$~K, while the temperature of the relic neutrinos is related to the temperature of photons. At the present time this temperature can be calculated as: $T^0_\nu=(\frac{4}{11})^{1/3}T^0_\gamma\sim0.71T^0_\gamma$. Hence, the temperature of the relic neutrinos is: $T^0_\nu=1.96$ K~($1.7\times10^{-4}$~eV). 
 
The cross section of relic neutrinos is extremely low: 
$$\sigma\sim G_F^2 m_\nu^2\sim10^{-56}(\frac{m_\nu}{1\rm{eV}})^2~\rm{cm}^2,$$
therefore the direct measurement of the relic neutrinos is extremely impossible with the present experimental techniques. However, if the relic neutrinos be observed, our knowledge of the universe would be expanded to almost 1 second after the big bang (Please note that the CMB dates back to when the universe is $\sim380,000$ years, while the decoupling of neutrinos happens at $t\sim1$ s). 

Since the relic neutrinos were decoupled before photons, they were present at the time photons were decoupled. Therefore, we can calculate the number of light neutrinos from CMB. The density of the matter and radiation contents of the universe are given by 
\begin{equation}
\rho_m=\rho^0_m(\frac{a_0}{a})^3~~\rm{and}~~\rho_r=\rho^0_r(\frac{a_0}{a})^4,
\end{equation}
where $a$ is the scale factor and is related to the redshift as $a=\frac{1}{1+z}$. The contribution of the neutrinos to the total radiation content of the universe which is composed of photons and relativistic neutrinos can be parametrized as a function of the effective number of neutrinos, $N_{\rm{eff}}$
\begin{equation}
\rho_r=\rho_\gamma+\rho_\nu=\big[1+\frac{7}{8}(\frac{4}{11})^{\frac{4}{3}}N_{\rm{eff}}\big]\rho_\gamma,
\end{equation}
where $\rho_{r(\gamma,\nu)}$ is the energy density of radiation (photons, neutrinos). The factor $7/8$ accounts for the fermionic number of degrees of freedom. In the standard model of cosmology, $N_{\rm{eff}}= 3.046$, which accounts for 3 active neutrinos at the time of thermalization.

Nonrelativistic massive neutrinos have a significant contribution to the energy density of the universe. This energy density is proportional to neutrinos masses $m_i$ and their number density $n_\nu$: $\rho_\nu\propto m_in_\nu$. It is usually common to define dimensionless energy density parameter in cosmology, which is defined by the ratio of the energy density to the critical energy density of the universe which is a constant factor. The density fraction for neutrinos is then proportional to the total masses, and the reduced Hubble parameter  $h$: 
$$
\Omega^0_\nu=\frac{\sum_in^0_{\nu_i}m_i}{\rho^0_ch^2}\simeq\frac{\sum_im_i}{93~ \rm{eV}~h^2},
$$
where $n^0_{\nu_i}$ is the number density of neutrinos which for each flavor we have a density of $112$~cm$^{-3}$, $\rho^0_c = 3H^2/(8\pi G)$ is the critical density today and $h=0.742 \pm 0.036$. Therefore from the Planck observation of temperature and polarization anisotropies of the CMB, it is possible to find the upper bound on the sum of neutrinos masses \cite{Ade:2015xua}:
$$
\sum_im_i<0.23~\rm{eV}.
$$
Although cosmological experiments are sensitive to the sum of neutrino masses, they are blind to the oscillation parameters.


\chapter{Probing light sterile neutrinos in medium baseline reactor experiments }\label{chap2}
\newpage

{\Large Without experimentalists, theorists tend to drift.

Without theorists, experimentalists tend to falter.\\

~~~~T.D. Lee}
\newpage

\footnote{This chapter is prepared based on my work published in \cite{Esmaili:2013yea}.} Medium baseline reactor experiments (Double Chooz, Daya Bay and RENO) provide a unique opportunity to test the presence of very light sterile neutrinos. In this chapter, we analyze the data of these experiments in the search of sterile neutrinos and also test the robustness of $\theta_{13}$ determination in the presence of sterile neutrinos. 

The chapter is organized as follows: we present the introduction of light sterile neutrinos in section \ref{sec:int}. In section~\ref{sec:standard3nu}, we analyze the Double Chooz, Daya Bay and RENO data in the standard $3\nu$ framework. We will show that our results are consistent with theirs within the error budget. In section~\ref{sec:lightsterile}, we discuss the phenomenology of light sterile neutrinos and derive the $\bar{\nu}_e$ survival probability in the $(3+1)_{\rm light}$ model. Section~\ref{sec:3p1lightsterile} is devoted to the analysis of data in the $(3+1)_{\rm light}$ model. In section~\ref{analysis3p1dc} we analyze the data of Double Chooz, and in section~\ref{combined} we present the results of combined analysis of Double Chooz, Daya Bay and RENO data. We summarize our conclusions in Section~\ref{sec:con_ch2}.

\section{Introduction: Light sterile neutrinos in medium baseline experiments}\label{sec:int}

Double Chooz, Daya Bay and RENO experiments with $L/E\sim \left( 10-10^3 \right)~{\rm m}/{\rm MeV}$, where $L$ and $E$ are the baseline and neutrino energy respectively, are sensitive to $\Delta m_{31}^2$-induced flavor oscillation with the amplitude $\sin^2 2\theta_{13}$. In fact, the measurement of the small mixing angle $\theta_{13}$ in these experiments was achieved thanks to the highly controlled systematic errors and efficient background rejection down to $\sim 10\%$ of signal. Thus, in principle, the data of these experiments can be used also to discover/constrain new physics in the neutrino sector. 

The set up and baseline to energy ratio of Double Chooz, Daya Bay and RENO experiments make them sensitive to small admixture of a new sterile neutrino state with electron anti-neutrinos, with mass-squared difference $\sim \left(10^{-3}-10^{-1}\right)~{\rm eV}^2$. The existence of a sterile neutrino state with mass $\sim 1~{\rm eV}$, the so-called $3+1$ model, was motivated by LSND~\cite{Aguilar:2001ty}, MiniBooNE~\cite{miniboone} and reactor anomalies~\cite{reactoranomalie}; and several experiments have been proposed to check this scenario (see~\cite{Abazajian:2012ys} and references therein; see also~\cite{Esmaili:2013vza}). However, from the phenomenological point of view, it is worthwhile to probe the existence of a \textbf{{\textit {light}}} sterile neutrino, a model which we call it $(3+1)_{\rm light}$. In this regard, we perform a detailed analysis of Double Chooz, Daya Bay and RENO data for the $(3+1)_{\rm light}$ model. In this model, in addition to the $\theta_{13}$ and $\Delta m_{31}^2$ parameters, the active-sterile mixing parameters $(\theta_{14},\Delta m_{41}^2)$ contribute to the $\bar{\nu}_e\to\bar{\nu}_e$ oscillation probability. We discuss the correlation among these parameters and the constraints that can be derived on them. Also, we discuss the robustness of $\theta_{13}$ determination in the presence of the sterile neutrinos. We show that the reported value of $\theta_{13}$ holds also in the presence of the sterile neutrinos and the data of Daya Bay and RENO play a crucial role in this robustness. 

The prospect of sterile neutrino search in medium baseline reactor experiments has been studied in a number of papers. In~\cite{Bandyopadhyay:2007rj}, by calculating the sensitivity of Double Chooz to the sterile neutrinos with $\Delta m^2\sim 1~{\rm eV}^2$ in the $3+2$ model (3 active and 2 sterile neutrinos), it has been concluded that with only the far detector data, the $\theta_{13}$ angle can be confused with the active-sterile neutrino mixing angles. In~\cite{deGouvea:2008qk} the interplay between a sterile neutrino with $\Delta m^2_{41} \sim (10^{-2}-1)~{\rm eV}^2$ and $\theta_{13}$ determination has been studied by computing the sensitivity of Double Chooz and Daya Bay, with the conclusion that disentangling these parameters requires information about the positron recoil energy distortions. A simulation of medium baseline experiments in search of light sterile neutrinos has been performed in~\cite{Bora:2012pi} and the dependence of limits on the systematic errors has been studied. Also, in~\cite{Ciuffoli:2012yd}, the correlation between $\theta_{13}$ and the active-sterile mixing parameters have been studied with an emphasize on the reactor anomaly and its connection to the cosmological data. The effect of the sterile neutrinos on $\theta_{13}$ determination in both medium and long baseline experiments has been studied in~\cite{Bhattacharya:2011ee}. In~\cite{Kang:2013gpa} the data of Daya Bay and RENO is analyzed in $3+1$ framework; the obtained limits are consistent with the limits of this work in the same range of $\Delta m_{41}^2$. Constraining the sterile neutrino scenario with the solar and KamLAND data has been studied in~\cite{Palazzo:2012yf} which we will discuss it in section~\ref{combined}. 

In this chapter we extend for the first time the previous searches to $\Delta m^2_{41}\sim (10^{-3}-10^{-1})~{\rm eV}^2$ and perform an analysis of the available data from the medium baseline experiments. We will show that due to the slight mismatch of data and $3\nu$ prediction at Double Chooz at $E_{\rm prompt}\sim (3-4)$~MeV, the $(3+1)_{\rm light}$ model is favored by $\sim 2.2\sigma$ significance; however, incorporating the Daya Bay and RENO data decreases the significance to $\sim1.2\sigma$ C.L..


\section{Standard analysis in the $3\nu$ framework}\label{sec:standard3nu}

In this section we reproduce the results of the medium baseline reactor experiments (Double Chooz~\cite{DCmainpaper}, Daya Bay~\cite{An:2012eh} and RENO~\cite{Ahn:2012nd}) in the $3\nu$ framework. The survival probability of the electron anti-neutrinos $\bar{\nu}_e$ takes the following form (see Eq. (\ref{eq1.12})):     
\begin{eqnarray}\label{eq2.1}
P({{\bar{\nu}_e\to\bar{\nu}_e}})&=&1-4\Big[|U_{e2}|^2|U_{e1}|^2\sin^2\big(\frac{\Delta m^2_{21}L}{4E}\big)+|U_{e3}|^2|U_{e1}|^2\sin^2\big(\frac{\Delta m^2_{31}L}{4E}\big)\nonumber\\
&+&|U_{e3}|^2|U_{e2}|^2\sin^2\big(\frac{\Delta m^2_{32}L}{4E}\big)\Big],
\end{eqnarray}
where $L\sim1~{\rm km}$ is the distance of the detector from the source, and $E\sim{\rm a~few~MeV}$ is the energy of the reactor neutrinos. Due to the moderately short baseline of these experiments, the oscillation induced by $\Delta m^2_{21}$ can be ignored, and using the definition of the PMNS matrix in Eq. (\ref{eq1.4}), the $\bar{\nu}_e$ survival oscillation probability can be casted as following:
\begin{eqnarray}\label{eq2.2}
P({{\bar{\nu}_e\to\bar{\nu}_e}})&\simeq&1-4|U_{e3}|^2\big(|U_{e1}|^2+|U_{e2}|^2\big)\sin^2\big(\frac{\Delta m^2_{31}L}{4E}\big)\nonumber\\
&=&1-\sin^2 2\theta_{13}\sin^2\left( \frac{\Delta m^2_{31}L}{4E} \right),
\end{eqnarray}
in which we have used the fact that $\Delta m^2_{32}\simeq\Delta m^2_{31}$. As discussed in~\cite{DCmainpaper,An:2012eh}, the effect of the uncertainty in the value of $\Delta m_{31}^2$ in the extraction of the $\theta_{13}$ parameter is quite small; therefore, we fix it to the best-fit value $2.4\times10^{-3}~{\rm eV}^2$ measured by the MINOS experiment~\cite{Adamson:2011ig}.

The detection of the reactor antineutrinos is through the Inverse Beta Decay (IBD) process, $\bar{\nu}_e +{\rm p} \to e^+ + {\rm n}$. Neutrino energy can be reconstructed by measuring the prompt positron energy $E\sim E_{{\rm prompt}}+0.78~{\rm MeV}$ (neglecting the neutron recoil energy). Thus, by reconstructing the spectrum of the observed $\bar{\nu}_e$ events, a fit of Eq.~(\ref{eq2.2}) to the data can give information about the value of $\theta_{13}$. In the following we discuss each of these experiments in detail.   

\subsection{Double Chooz}\label{dc3analysis}

The Double Chooz experiment has detected 8,249 candidates of electron antineutrino events 
using a $10.3~{\rm m}^3$ detector which is located at $L=1050$~m far from the 2 reactor cores. The total livetime of the experiment is $227.93$ days. The expected number of events in the case of no-oscillation ({\it i.e.}, $\theta_{13}=0$) are 8,937 (including background events). From a rate plus spectral shape analysis they have found $\sin^2 2\theta_{13}=0.109\pm 0.055$, which excludes the no-oscillation hypothesis at $99.8\%$ C.L. ($2.9~\sigma$)~\cite{DCmainpaper}. 

\begin{figure}[t!]
\centering
\includegraphics[width=1\textwidth]{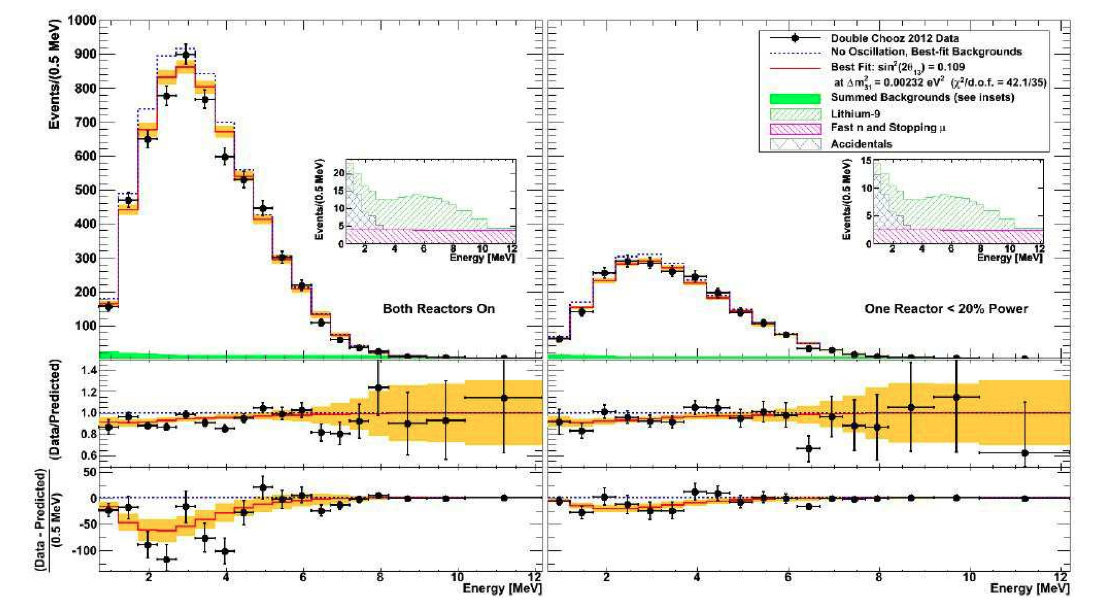}
\caption{\label{fig:DC_data}Summary of the Double Chooz experiment. 
Plot taken from \cite{DCmainpaper}.}
\end{figure}

To reproduce the results of Double Chooz we follow the method described in~\cite{DCmainpaper}. The observed events in Double Chooz are separated in 18 prompt energy bins, between 0.7~MeV and 12.2~MeV (Fig.~\ref{fig:DC_data}). The analysis is performed by defining two different data-taking periods: {\it i}) both reactors on (139.27 days) with 6,088 total IBD candidates; {\it ii}) one reactor with less than $20\%$ of nominal power (88.66 days) with 2,161 total IBD candidates. 

The numbers of expected events without oscillation ({\it i.e.} $\theta_{13}=0$), background events and observed events in each bin of energy is published in~\cite{DCmainpaper} (See Fig.~\ref{fig:DC_data}). The expected number of $\bar{\nu}_e$ events in a detecter with detection efficiency $\epsilon$ and at distance $L$ from the source can be calculated as \cite{Ahn:2010vy}
\begin{equation}\label{eq2.3}
N^{\rm osc}=\frac{N_p}{4\pi L^2}\int\int\epsilon P(\bar{\nu}_e\to\bar{\nu}_e)\frac{d\sigma}{dE_{e^+}}\frac{d\phi_{\bar{\nu}_e}}{dE}dE_{e^+}dE,
\end{equation}
where $N_p$ is the number of free protons in the detector target, $\frac{d\sigma}{dE_{e^+}}$ is the differential cross section of the IBD process, $\frac{d\phi_{\bar{\nu}_e}}{dE}$ is the differential energy distribution at the reactor, and $E_{e^+}$ and $E$ are the energy of the prompt positrons and the electron anti-neutrinos, respectively. It is quite difficult to calculate the differential cross sections and the accurate detector efficiency for phenomenology point of view; however, in these experiments, they usually report the no-oscillation number of  events as well. Therefore, one can use Eq. (\ref{eq2.3}) to calculate $N^{\rm {no-osc}}$, noticing that in this case the probability is 1, and write 
\begin{equation}\label{eq2.4}
\frac{N^{\rm {osc}}}{N^{\rm {no-osc}}}= \left<P(\bar{\nu}_e\to\bar{\nu}_e)\right>,
\end{equation}
in which $ \left<P(\bar{\nu}_e\to\bar{\nu}_e)\right>$ is the averaged $\bar{\nu}_e$ survival probability in Eq.~(\ref{eq2.2}). Hence, the expected number of events for $\theta_{13}\neq 0$ in the $i$-th bin of energy can be calculated by   
\begin{equation}\label{eq2.5}
N^{{\rm osc}}_i(\theta_{13})=N^{\rm no-osc}_i\times \left<P_{{\rm sur}}(\theta_{13})\right>_i,
\end{equation}
where $N^{\rm no-osc}_i$ is the expected number of events for $\theta_{13}=0$ in the $i$-th energy bin (obtained from Fig.~\ref{fig:DC_data}, after subtracting background events), and $\left<P_{{\rm sur}}(\theta_{13})\right>_i$ is the averaged $\bar{\nu}_e$ survival probability as a function of $\theta_{13}$ in the $i$-th energy bin. 

To analyze the data of Double Chooz we define the following $\chi^2$ function 
\begin{eqnarray}\label{eq2.6}
\chi^2_{\rm DC}(\sin^2 2\theta_{13};\alpha,b)&=&\sum_{i=1}^{36}\frac{\Big( N^{{\rm obs}}_i-\big[(1+\alpha)N^{{\rm osc}}_i(\theta_{13})+(1+b)B_i\big]\Big)^2}{(\sigma^{{\rm obs}}_{i})^2+(\sigma^{{\rm osc}}_i)^{2}}\nonumber\\
&+&\frac{\alpha^2}{\sigma^2_{\alpha}}+\frac{b^2}{\sigma^2_b},
\end{eqnarray}
where $i$ runs over the 36 bins of energy (18 for each period of data-taking); $N^{{\rm obs}}_i$, $B_i$ and $N^{\rm osc}_i$ are the observed, background and expected number of events in the $i$-th bin, respectively. The ${\sigma^{{\rm obs}}_{i}}=\sqrt{N^{{\rm obs}}_i}$ and ${\sigma^{{\rm osc}}_{i}}=\sqrt{N^{{\rm osc}}_i}$ represent the statistical errors of the observed events and the expected events with oscillation, respectively. The systematic uncertainties in the normalization of the reactor neutrino flux and the background events are taken into account by the $\alpha$ and $b$ pull terms, with $\sigma_{\alpha}=0.02$ and $\sigma_b=0.27$.
\begin{table}
\centering
\begin{tabular}{|c|c|c|c|c|}
\hline
{\rm Experiment}     & $\chi^2_{{\rm no-osc}}$ & $\chi^2_{{\rm osc}}$ &$\delta \chi^2$ &$\sin^2 2\theta_{13}$ \\
\hline
Double Chooz analysis~\cite{DCmainpaper}         & $\sim$ 52 & 42.1& 9.9 &  0.109$\pm$ 0.055 \\ 
Double Chooz (our analysis) & 35.2 & 26.2 &  9.0 & 0.115$\pm$ 0.037 \\
\hline
Daya Bay analysis~\cite{An:2012eh}                  & $\sim$ 31 & 4 & 27 & 0.092 $\pm$ 0.021\\
Daya Bay (our analysis)           &  31.8 & 3.5 & 28.3 & 0.091 $\pm$ 0.014\\
\hline
RENO analysis~\cite{Ahn:2012nd}  & $\sim$ 22 & 0 & 22 & 0.113 $\pm$ 0.032\\
RENO (our analysis) & 19.0   & 0 & 19 & 0.110  $\pm$ 0.024 \\
\hline
\end{tabular}
\caption{Comparison between our analysis and the analyses reported by Double Chooz, Daya Bay and RENO experiments. The quantity $\delta \chi^2\equiv \chi^2_{{\rm no-osc}}-\chi^2_{{\rm osc}}$ shows the improvement in the fit of data due to nonzero $\theta_{13}$.}
\label{table2.1}
\end{table}

\begin{figure}[t!]
\centering
\subfloat[both reactors on]{
\includegraphics[width=0.5\textwidth]{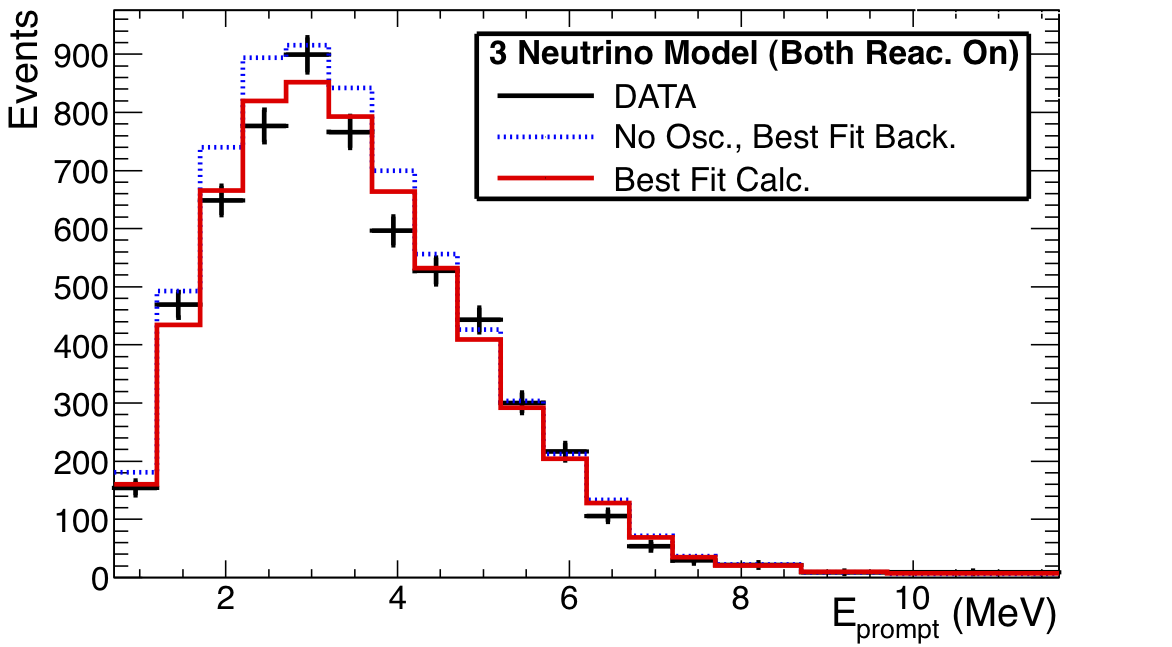}
\label{fig:dchis3,1}
}
\subfloat[one reactor off]{
\includegraphics[width=0.5\textwidth]{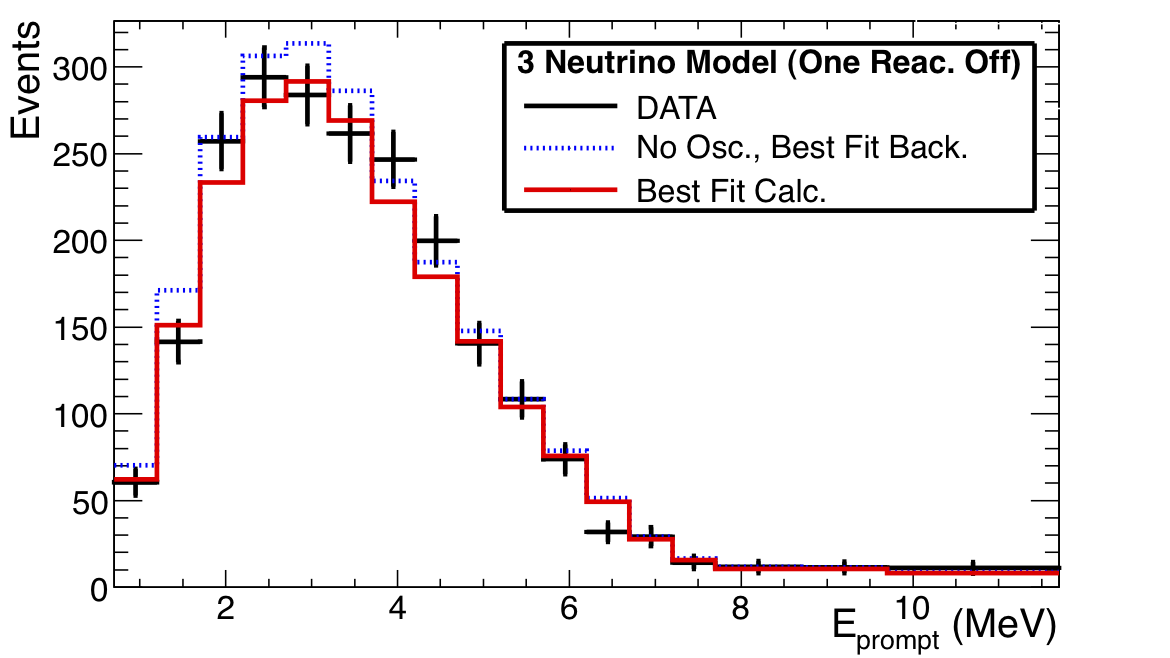}
\label{fig:dchis3,2}
}
\caption{\label{fig:dchis3}Prompt energy distribution of events in the Double Chooz experiment (data points), compared with distributions for the best-fit value of $\theta_{13}$ (red solid curve) and $\theta_{13}=0$ (blue dashed curve). The left and right panels correspond to the two data-taking periods with both reactors on and one reactor off, respectively.}
\end{figure}

By marginalizing $\chi^2_{\rm DC}(\sin^2 2\theta_{13};\alpha,b)$ with respect to $\alpha$ and $b$, we obtain the best-fit value of the mixing angle $\sin^2 2\theta_{13}=0.115$, with the normalized (to the number of degrees of freedom) $\chi^2$ value of $\chi^2_{{\rm DC}}/{\rm d.o.f.}=26.2/35$, which excludes the no-oscillation hypothesis at 2.7$\sigma$ C.L. (compare to the reported 2.9$\sigma$ C.L. in~\cite{DCmainpaper}). Our $1\sigma$ range of mixing parameter $\sin^2 2\theta_{13}$ from Double Chooz is given in the first row of Table~\ref{table2.1}, which is similar to the range reported by the Double Chooz collaboration. Also, in Fig.~\ref{fig:dchis3} we compare the best-fit energy spectrum of events from our analysis in the $3\nu$ framework (including the background events) with Double Chooz data and the spectrum for $\theta_{13}=0$, for each integration period. The contribution of the background events is $\sim15$ events in the first couple of energy bins and reduces to $\sim5$ events for higher energy bins. The left and right panels of Fig.~\ref{fig:dchis3} are for the cases where both reactors are on and one reactor runs with less than $20\%$ of power; which clearly the former plays the main role in the analysis due to the higher statistics. The preference to nonzero $\theta_{13}$ can be easily recognized by comparing the red solid curve for best-fit value of $\theta_{13}$ with the blue dotted curve which shows the distribution for vanishing $\theta_{13}$.

\subsection{Daya Bay}\label{daya_bay}

The Daya Bay reactor neutrino experiment has measured the best-fit value of the mixing angle $\sin^2 2\theta_{13}=0.092$, excluding the zero value at 5.2$\sigma$ C.L.~\cite{An:2012eh}. In 55 days of livetime, 10,416 (80,376) electron anti-neutrino candidates have been detected in the far hall (near halls). The ratio of the observed to the expected number of events is $R=0.940$. From this deficit, $\sin^2 2\theta_{13} = 0.092$ has been determined, based on a rate-only analysis ({\it i.e.} comparing the total number of the observed events with the total number of the expected events).

\begin{figure}[t!]
\centering
\includegraphics[width=0.7\textwidth]{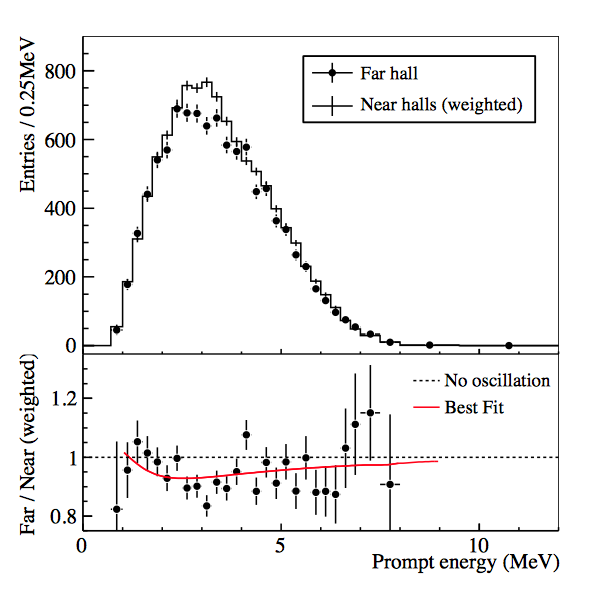}
\caption{\label{fig:DB_data}Summary of the Daya Bay experiment. Plot taken from \cite{An:2012eh}.}
\end{figure}

\begin{figure}[t!]
\centering
\includegraphics[width=0.8\textwidth]{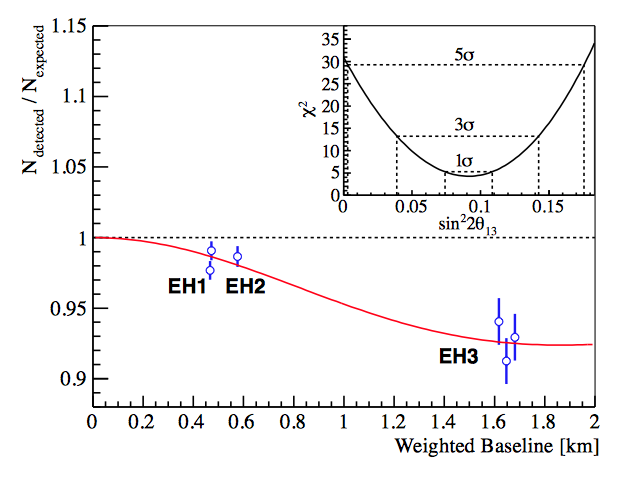}
\caption{\label{fig:DB_ratio}Daya Bay experiment: Ratio of the measured versus expected signal in each detector, assuming no oscillation. 
Plot taken from \cite{An:2012eh}.}
\end{figure}

The Daya Bay experiment consists of three underground experimental halls (EH1, EH2 and EH3), where two Antineutrino Detectors (AD) are located in EH1 and one AD in EH2 (near halls). Three ADs are located at the far hall (EH3) at a distance where $\bar{\nu}_e$ survival oscillation probability deviates maximally from one. The Daya Bay collaboration has published the observed number of events in the near and far halls as a function of the prompt energy (Fig.~\ref{fig:DB_data}). However, due to the low energy resolution of the Daya Bay experiment, it is not possible to perform a bin by bin analysis similar to the case of Double Chooz. Nevertheless, we can do a rate only analysis, which means comparing the \textbf{\textit{total}} observed and expected number of events. Daya Bay has published the ratio of the observed to the expected number of events (assuming no-oscillation) for each AD (Fig.~\ref{fig:DB_ratio}); {\it i.e.}, 
\begin{equation}\label{eq2.7}
R_i\equiv\frac{\#{\rm ~of~observed~events~in~}i{\rm -th~AD}}{\#{\rm ~of~expected~events~in~}i{\rm -th~AD}}~,
\end{equation}
where $i=1,\ldots, 6$. Using $R_i$s and the number of observed IBD candidates in each AD (given in Table~\ref{table2.2}, also in Table II of~\cite{An:2012eh}), the expected number of events without oscillation can be calculated by: $N^{\rm no-osc}_i=N_i^{\rm obs}/R_i$. The averaged oscillation probability for each AD of Daya Bay can be written as (see~\cite{Ciuffoli:2012yd})
\begin{equation}\label{eq2.8}
\langle P_{\rm sur}(\theta_{13})\rangle_i=1-\sin^2 2\theta_{13} \int \sin^2 \left(\frac{\Delta m^2_{31}d_i}{4E}\right)\rho(E) {\rm d}E~,
\end{equation} 
where $d_i$ is the weighted distance of the $i$-th AD from the reactors (the weighted baseline given in Fig.~\ref{fig:DB_ratio}) and $\rho(E)$ is the fractional energy distribution of neutrinos. Using the top panel of Fig.~\ref{fig:DB_data} for the energy distribution of events, the expected number of events including the oscillation for the $i$-th AD becomes
\begin{equation}\label{eq2.9}
N^{\rm exp}_i(\theta_{13})=\left(\dfrac{N_i^{{\rm obs}}}{R_i}\right)\times \langle P_{\rm sur}(\theta_{13})\rangle_i~.
\end{equation}
\begin{table}
\small
\centering
\begin{tabular}{|c|c|c|c|}
\hline
\hline
   &\rm{AD}1~~~~~ \rm{AD}2&\rm{AD}3&\rm{AD}4~~~~~~~\rm{AD}5~~~~~~~\rm{AD}6\\
   \hline
   \hline
\rm{IBD~candidates}&28935~~~~28975&22466&3528~~~~~~~3436~~~~~~~3452\\
\hline
\rm{total background}&$694$~~~~~~~$697$&$517$&$182$~~~~~~~~~$184$~~~~~~~~~$174$\\
\hline
\rm{total background uncertainty}&107~~~~~~107&89.6&19.7~~~~~~~~19.7~~~~~~~~19.7\\
\hline
\rm{fraction of the neutrino flux}&0.188~~~~~~0.202&0.109 &0.124 ~~~~~~0.188 ~~~~~0.186\\
\hline
\hline
\end{tabular}
\caption{Signal and background summary of the Daya Bay experiment \cite{An:2012eh}.}
\label{table2.2}
\end{table}

We analyze the Daya Bay data by defining the following $\chi^2$ function (which includes only the rate information):
 \begin{eqnarray} \label{eq2.10}
 \chi^2_{{\rm DB}}(\sin^2 2\theta_{13};\epsilon,\epsilon_i,\alpha_r,\eta_i)&=&\sum_{i=1}^6 \frac{\left[N_i^{{\rm obs}}-N^{{\rm exp}}_i(\theta_{13})\left\{1+\epsilon+\sum_{r=1}^6 \omega_r^i \alpha_r+\epsilon_i\right\}+\eta_i\right]^2}{
\left(\sigma_i^{{\rm stat}}\right)^2} \nonumber \\
  &+&\sum_{r=1}^6\frac{\alpha_r^2}{\sigma_r^2}+\sum_{i=1}^6\left(\frac{\epsilon^2_i}{\sigma^2_d}+\frac{\eta^2_i}{(\sigma^2_B)_i}\right)~,
 \end{eqnarray}
where $N_i^{{\rm obs}}$ and $N_i^{{\rm exp}}$ are respectively the total number of observed and expected IBD candidate events in the $i$-th AD; with $\sigma_i^{{\rm stat}}$ representing the statistical error of the observed number of events which is defined as $(\sigma_i^{{\rm stat}})^2=N_i^{{\rm obs}}+B_i$, where $B_i$ is the number of background events in the $i$-th AD. The $\omega^i_r$ is the fraction of the neutrino flux from $r$-th reactor at the $i$-th AD (given in the last row of Table~\ref{table2.2}). The systematic uncertainties of the reactor flux, detection efficiency and background events are taken into account by pull terms with pull parameters $\alpha_r$, $\epsilon_i$ and $\eta_i$ respectively; with $\sigma_r=0.8\%$, $\sigma_d=0.2\%$ and $(\sigma_B)_i$ presented in the third row of Table~\ref{table2.2}. The parameter $\epsilon$ accommodates the uncorrelated flux normalization uncertainty which we marginalize without any pull term compensation. All the values used in $\chi^2_{\rm DB}$ function of Eq.~(\ref{eq2.10}) are listed in Table~\ref{table2.2} \cite{An:2012eh}.

After minimizing $\chi^2_{\rm DB}$ with respect to all parameters, we find the best-fit value of the mixing parameter $\sin^2 2\theta_{13}=0.091$, which is consistent with Daya Bay result. The $\chi^2$ value and $1\sigma$ range of $\sin^2 2\theta_{13}$ are shown in the second row of Table~\ref{table2.1}.

\subsection{RENO}

\begin{figure}[t!]
\centering
\includegraphics[width=0.8\textwidth]{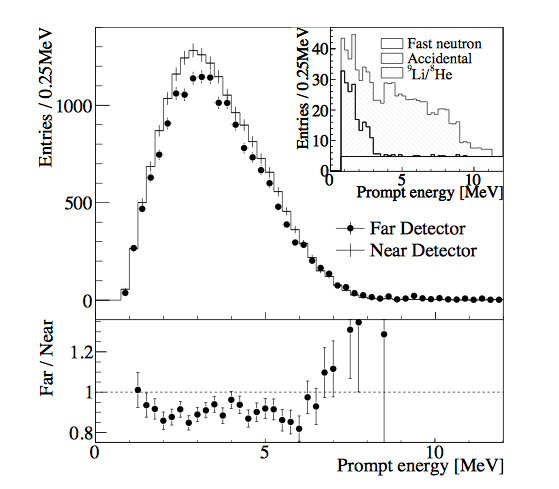}
\caption{\label{fig:RENO_data}Summary of the RENO experiment. Plot taken from \cite{Ahn:2012nd}.}
\end{figure}

The RENO experiment has observed disappearance of reactor $\bar{\nu}_e$ with $4.9\sigma$ of significance. In 229 days of data-taking period, the number of observed neutrinos at far (near) detector has been 17,102 (154,088). The ratio of the number of observed neutrinos to the number of expected neutrinos (for $\theta_{13}=0$) is $R=0.920$. From this deficit, the RENO collaboration has obtained $\sin^2 2\theta_{13} = 0.113$, based on a rate-only analysis~\cite{Ahn:2012nd}.

\begin{table}
\small
\centering
\begin{tabular}{|c|c|c|c|c|c|c|}
\hline
\hline
\tiny{Fraction of $\bar{\nu}_e$ flux}&reactor 1&reactor 2&reactor 3&reactor 4&reactor 5&reactor 6\\
\hline
\hline
detector 1&0.0678& 0.1493& 0.3419& 0.2701& 0.115&0.0558\\
\hline
detector 2& 0.1373&0.1574&0.1809&0.1856&0.178&0.1608\\
\hline
\end{tabular}
\caption{Fraction of $\bar{\nu}_e$ flux from different detectors on different reactors for RENO experiment \cite{An:2012eh}.}
\label{table2.3}
\end{table}

\begin{table}
\small
\centering
\begin{tabular}{|c|c|c|c|c|c|c|}
\hline
\hline
\tiny{Weighted distance (m)}&reactor 1&reactor 2&reactor 3&reactor 4&reactor 5&reactor 6\\
\hline
\hline
detector 1&667.9&451.8& 304.8& 336.1&513.9&739.1\\
\hline
detector 2&1556.5&1456.2& 1395.9&1381.3&1413.8& 1490.1\\
\hline
\end{tabular}
\caption{Weighted distance of different detectors from different reactors for RENO experiment \cite{An:2012eh}.}
\label{table2.4}
\end{table}

\begin{figure}[t!]
\centering
\includegraphics[width=0.8\textwidth]{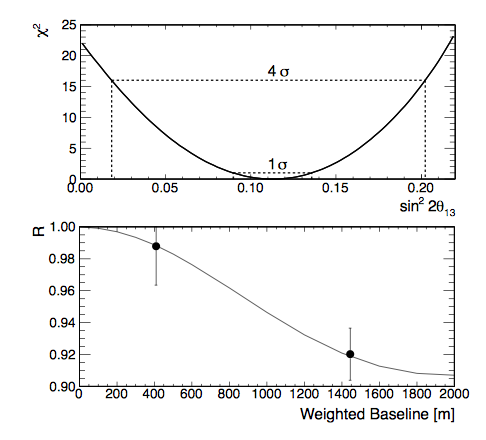}
\caption{\label{fig:RENO_ratio}RENO experiment: The $\chi^2$ distribution as a function of $\sin^2 2\theta_{13}$. Bottom: Ratio of the measured reactor neutrino events relative to the expected number of events with no oscillation. The curve represents the oscillation survival probability at the best fit, as a function of the flux-weighted baselines. Plot taken from \cite{Ahn:2012nd}.}
\end{figure}

The RENO experiment consists of two detectors, near and far, detecting $\bar{\nu}_e$ emission form   6 reactors. The average distance of the near (far) detector from the center of the reactor array is 294 m (1383 m). The details of the experiment can be found at~\cite{Ahn:2012nd}. The RENO collaboration has published the observed number of events in the near and far detectors as a function of the prompt energy (Fig. \ref{fig:RENO_data}). The analysis of RENO is similar to the analysis of Daya Bay. Following the method described in~\cite{Ciuffoli:2012yd}, we calculate the averaged survival probability of neutrinos in $i$-th detector ($i=$~near,far) in the following way:
\begin{eqnarray}\label{eq2.11}
\langle P_i\rangle=\sum_{j=1}^6f_{ij}\left[1-\sin^2 2\theta_{13}\int \sin^2\left(\frac{\Delta m^2_{31}d_{ij}}{4E}\right)\rho_j(E) {\rm d}E\right],~~~~~
\end{eqnarray}

where $f_{ij}$ is the fraction of antineutrino flux at the $i$-th detector coming from the $j$-th reactor, given in Table~\ref{table2.3}; $d_{ij}$ is the weighted distance of the $i$-th detector from the $j$-th reactor, given in Table~\ref{table2.4} and $\rho_j$ is the fractional energy distribution of neutrinos emitted from the $j$-th reactor. Using the ratio of observed to expected number of events (as defined in Eq.~(\ref{eq2.5})) from the bottom panel of Figure~\ref{fig:RENO_ratio}, the expected number of events without oscillation can be derived. The expected number of events including oscillation can be calculated form Eq.~(\ref{eq2.7}).

We use the same $\chi^2$ function defined by the collaboration~\cite{Ahn:2012nd}:
\begin{eqnarray}\label{eq2.12}
\chi^2_{{\rm RENO}}(\sin ^2 2\theta_{13};\alpha,b_i,\xi_i,f_r)=\sum_{i=N,F} \frac{\left[N_i^{{\rm obs}}+b_i-(1+\alpha+\xi_i)\sum_{r=1}^6(1+f_r)N_{i,r}^{\rm exp}\right]^2}{\left(\sigma_i^{{\rm obs}}\right)^2}\nonumber\\
+\sum_{i=N,F}\left(\frac{\xi^2_i}{\sigma_d^{\xi^2}}+\frac{b^2_i}{\sigma^{b^2}_i}\right)+\sum_{r=1}^6\left(\frac{f_r}{\sigma_r}\right)^2,~~~~~~~~~~~~~~~~~~~~~~~~~
\end{eqnarray}
where $i$ denotes to either Near or Far detectors, $N_i^{{\rm obs}}$ is the total number of neutrinos observed in each detector (after background subtraction), $r=1,\ldots,6$ runs over the reactors and $\alpha$ takes into account the global flux normalization uncertainty. $\sigma_i^{{\rm obs}}$ is the statistical error of observed events; and uncorrelated systematic error of reactor flux and detection efficiency are $\sigma_r=0.9\%$ and $\sigma _d^\xi=0.2\%$ respectively (taken from Table II of~\cite{Ahn:2012nd}). $\sigma^b_{N(F)}$ is the background uncertainty at the near (far) detector: $\sigma^b_N=1140.93$ ($\sigma^b_F=166.545$) \cite{Ahn:2012nd}. The corresponding pull parameters are respectively $f_r$, $\xi_i$ and $b_i$.

\begin{figure}[t!]
\centering
\includegraphics[width=0.9\textwidth]{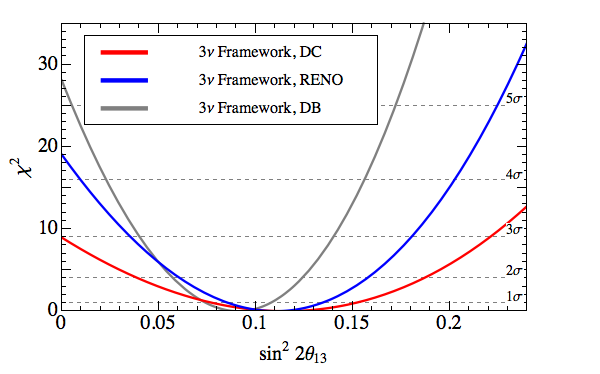}
\caption{\label{fig:chi2,all,3nu}$\Delta \chi^2\equiv \chi^2 - \chi^2_{\rm min}$ versus $\sin^2 2\theta_{13}$, for Double Chooz (red solid curve), RENO (blue dot-dashed curve) and Daya Bay (black dashed curve) experiments.}
\end{figure}

Minimizing $\chi^2_{{\rm RENO}}$ of Eq.~(\ref{eq2.12}) with respect to all the pull parameters, we find the best-fit value $\sin^2 2\theta_{13}=0.110$, which is consistent with RENO result reported in~\cite{Ahn:2012nd}. The summary of all our results about RENO can be seen in the third row of Table~\ref{table2.1}.\\

In Fig.~\ref{fig:chi2,all,3nu} we show $\Delta\chi^2\equiv \chi^2-\chi^2_{\rm min}$ versus $\sin^2 2\theta_{13}$ for the three discussed experiments. Comparing our results with the results of Double Chooz, Daya Bay and RENO collaborations in Table~\ref{table2.1} show that: $i$) our best-fit values for $\sin^2 2\theta_{13}$ are fairly close to the reported values by the collaborations; $ii$) our exclusion of $\theta_{13}=0$ (quantified by $\delta \chi^2$) and also $1\sigma$ allowed interval of $\sin^2 2\theta_{13}$ are compatible with the corresponding values reported by the collaborations. In the rest of this chapter, we will use the data of these experiments to probe the existence of the sterile neutrino state with mass-squared difference $\Delta m^2\sim (10^{-3}-10^{-1})\,{\rm eV}^2$.

\section{The Framework of light sterile neutrino: $(3+1)_{\rm light}$}\label{sec:lightsterile}

Although the majority of data from neutrino oscillation experiments can be interpreted consistently in the $3\nu$ framework, persisting anomalies, including MiniBooNE, LSND, reactor and Gallium anomalies (see section \ref{Experimental_Motivation}) motivate existence of a sterile neutrino state with mass $\sim \mathcal{O}(1)$~eV. However, although several experiments have been proposed to check the existence of $\sim 1$~eV sterile neutrinos (see~\cite{Abazajian:2012ys,Esmaili:2013vza}), the possibility of the existence of a light sterile neutrino ($\Delta m^2_{41}\ll 1~$eV$^2$) is neither excluded strongly nor planned to be explored substantially\footnote{The potential of upcoming reactor experiments with larger baseline to probe ``super light" sterile neutrinos ($\Delta m_{41}^2\simeq10^{-5}~{\rm eV}^2$) is studied in \protect{\cite{Bakhti:2013ora}}.}. In this chapter we probe this possibility in the light of Double Chooz, Daya Bay and RENO published data.

For simplicity, we extend the neutrino sector of the Standard Model by adding one light sterile neutrino: the $(3+1)_{\rm light}$ model. In this model the mass spectrum of neutrino sector consists of three mostly active neutrino mass eigenstates with masses $(m_1,m_2,m_3)$ and one mostly sterile neutrino mass eigenstate with mass $m_4$ such that: 
\begin{equation}\label{eq2.13}
m_1<m_2 \ll m_3<m_4~.
\end{equation}
The mixing of these states can be described by the $4\times4$ PMNS matrix $U^{(4)}$ described in Eq.~(\ref{eq1.30}). 


\begin{figure}[t!]
\centering
\setcounter{subfigure}{0}
\subfloat{
\includegraphics[width=0.5\textwidth]{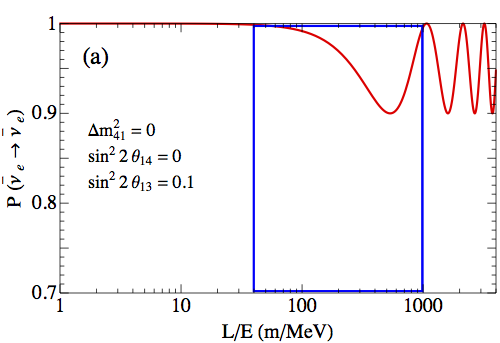}
\label{fig:prob11}
}
\subfloat{
\includegraphics[width=0.5\textwidth]{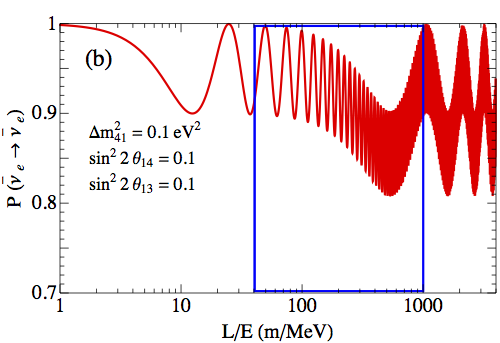}
\label{fig:prob2}
}
\quad
\subfloat{
\includegraphics[width=0.5\textwidth]{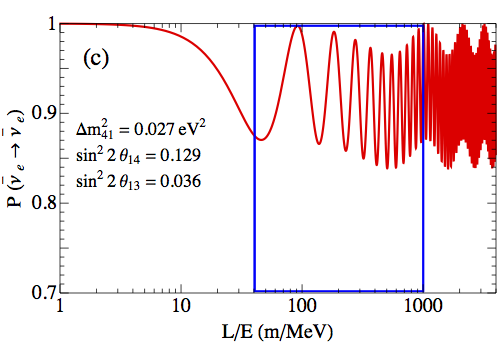}
\label{fig:prob3}
}
\subfloat{
\includegraphics[width=0.5\textwidth]{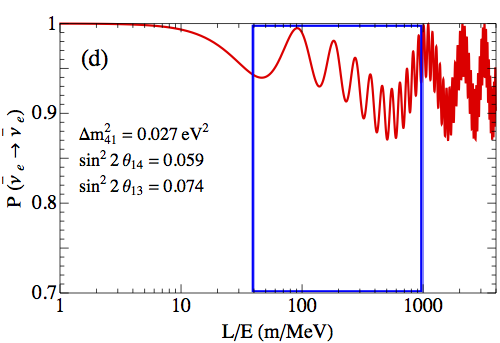}
\label{fig:prob4}
}
\caption{\label{fig:prob,3+1}The $\bar{\nu}_e$ survival probability versus $L/E$ for different values of $\sin^2 2\theta_{13}$, $\sin^2 2\theta_{14}$ and $\Delta m_{41}^2$. The blue box represents the relevant $L/E$ region for medium baseline experiments. Panel (a) shows the probability in the $3\nu$ framework with the best-fit value of $\theta_{13}$. Panel (b) is for nonzero active-sterile mixing parameters, with a rather large $\Delta m_{41}^2$. Panels (c) and (d) show the probability for the best-fit values obtained in the analysis of the $(3+1)_{\rm light}$ model, for Double Chooz and the combined data, respectively.}
\end{figure}

In the medium baseline reactor experiments (Double Chooz, Daya Bay and RENO) which we are considering in this chapter, the distances of the near and far detectors to the sources are a few hundred meters and $\sim1$~km, respectively; and the energy of neutrinos emitted from reactors is $\sim$ few MeV. In this energy and baseline ranges the oscillation phase induced by $\Delta m_{21}^2$ can be ignored and the $\bar{\nu}_e$ survival probability is given by:
\begin{eqnarray}\label{eq2.14}
P({\bar{\nu}_e\to\bar{\nu}_e})&=&1-4\left(|U_{e1}^2|+|U_{e2}|^2\right)\times\left[|U_{e3}|^2\sin^2
\left(\frac{\Delta m^2_{31}L}{4E}\right) \right.\nonumber\\
 & + &\left. |U_{e4}^2|\sin^2\left(\frac{\Delta m^2_{41}L}{4E}\right)\right] -4|U_{e3}|^2|U_{e4}|^2\sin^2\left(\frac{\Delta m^2_{43}L}{4E}\right)~,
\end{eqnarray}
where $L$ and $E$ are the baseline and energy respectively. In terms of the parametrization of Eq.~(\ref{eq1.30}), the oscillation probability is
\begin{eqnarray}\label{eq2.15}
P(\bar{\nu}_e\to\bar{\nu}_e)&=&1-\sin^22\theta_{13} \cos^4\theta_{14}\sin^2\left(\frac{\Delta m^2_{31}L}{4E}\right) - \cos^2\theta_{13}\sin^2 2\theta_{14}\sin^2\left(\frac{\Delta m^2_{41}L}{4E}\right) \nonumber \\
 & - & \sin^2\theta_{13}\sin^2 2\theta_{14}\sin^2\left(\frac{\Delta m^2_{43}L}{4E}\right)~.
\end{eqnarray}
In the analysis of the $(3+1)_{\rm light}$ scenario, we treat $\theta_{13}$ , $\theta_{14}$ and $\Delta m_{41}^2$ in Eq.~(\ref{eq2.15}) as free parameters and fix $\Delta m_{31}^2$ to its best-fit value from the MINOS experiment~\cite{Adamson:2011ig}. In Fig.~\ref{fig:prob,3+1} we show the $\bar{\nu}_e$ survival probability for different values of $\sin^2 2\theta_{14}$, $\sin^2 2\theta_{13}$ and $\Delta m_{41}^2$, versus $L/E$. The blue box shows the relevant values of $L/E$ for medium baseline experiments for both near and far detectors. The upper left panel shows the probability for the best-fit value of $\theta_{13}$ in the $3\nu$ framework. The upper right panel depicts the probability for nonzero active-sterile mixing parameters shown in the legend; and as can be seen, for relatively large $\Delta m_{41}^2=0.1~{\rm eV}^2$ in this panel the effect of sterile neutrino is averaged out especially for the far detector. The two lower panels show the probability for best-fit values of mixing parameters obtained in our analysis of the $(3+1)_{\rm light}$ (see section~\ref{sec:3p1lightsterile}).

Some symmetries in Eq.~(\ref{eq2.15}) can be recognized. For example, there is a degeneracy in Eq.~(\ref{eq2.15}) where in the limit $\Delta m_{43}^2\to0$ (or $\Delta m_{41}^2\to\Delta m_{31}^2$), when $\theta_{13}=0$ the angle $\theta_{14}$ imitates the role of $\theta_{13}$ when $\theta_{14}=0$. Thus, it is always possible to obtain a fit to the data in the $(3+1)_{\rm light}$ model as good as the fit in $3\nu$ framework by setting $\Delta m_{41}^2=\Delta m_{31}^2$ and exchanging $\theta_{14}$ with $\theta_{13}$. However, it would be possible that deviation of $\Delta m_{41}^2$ from $\Delta m_{31}^2$ (that is, nonzero $\Delta m_{43}^2$), which leads to a shift of extrema positions in Fig.~\ref{fig:prob,3+1}, results in a better fit than the $3\nu$ fit. Notice that varying $\theta_{13}$ in the $3\nu$ scheme leads to change in the depth of minima while leaving the positions of minima unchanged. So, generally we expect to obtain better fits by extending the $3\nu$ framework to the $(3+1)_{\rm light}$ model. The possibility of having two independent measurements in near and far detectors (as in Daya Bay and RENO) helps in lifting this degeneracy (see section~\ref{combined}), while with one detector only (Double Chooz) the degeneracy will be manifested (see section~\ref{analysis3p1dc}).

\section{Analysis of the $(3+1)_{\rm light}$ model}\label{sec:3p1lightsterile}
In this section we confront the data from Double Chooz, Daya Bay and RENO experiments with the prediction of the $(3+1)_{\rm light}$ model. In section~\ref{analysis3p1dc}, we consider the Double Chooz data and in section~\ref{combined}, we perform the combined analysis including the data of the other two experiments.   

\subsection{Probing the $(3+1)_{\rm light}$ model with Double Chooz}
\label{analysis3p1dc}

For the analysis of Double Chooz data for the $(3+1)_{\rm light}$ model, we follow the method of section~\ref{dc3analysis}, with the modification that the number of expected events in $(3+1)_{\rm light}$ model is given by 
\begin{equation}\label{eq2.16}
N^{{\rm osc,3+1}}_i=N^{{\rm no-osc}}_i \times \left\langle P(\theta_{13},\theta_{14},\Delta m^2_{41} )\right\rangle_i~,
\end{equation}
where $\left\langle P\right\rangle_i$ is the average of survival probability in Eq.~(\ref{eq2.15}) in the $i$-th energy bin.

Using the same $\chi^2_{\rm DC}$ as in Eq.~(\ref{eq2.6}), we find the following best-fit values: 
\begin{equation}\label{eq2.17}
\sin^2 2\theta_{13}=0.036\quad , \quad\sin^2 2\theta_{14}=0.129\quad , \quad\Delta m^2_{41}=0.027~{\rm eV}^2~,
\end{equation}
with the minimum value $\chi^2_{{\rm DC}}/{\rm d.o.f}=19.1/33$, which shows improvement with respect to the $3\nu$ analysis with minimum of $\chi^2_{{\rm DC}}/{\rm d.o.f}=26.2/35$ (see the first row of Table~\ref{table2.1}). The main feature of the $(3+1)_{\rm light}$ analysis with Double Chooz data is the significantly different best-fit value for $\sin^2 2\theta_{13}$ in Eq.~(\ref{eq2.17}) with respect to the $3\nu$ best-fit value $\sin^2 2\theta_{13}=0.115$. In Fig.~\ref{fig:chi2,dc3+1,q13}, we compare the $\chi^2_{\rm DC}$ as a function of $\sin^2 2\theta_{13}$ for $3\nu$ and $(3+1)_{\rm light}$ models. The red dashed curve for the $(3+1)_{\rm light}$ is calculated by marginalizing $\chi^2_{\rm DC}$ with respect to $\Delta m_{41}^2$ and $\sin^2 2\theta_{14}$. The shift in the best-fit value of $\sin^2 2\theta_{13}$ in $(3+1)_{\rm light}$ is clear from Fig.~\ref{fig:chi2,dc3+1,q13}. Also, from the best-fit value $\sin^22\theta_{13}=0.036$ down to zero, the $\chi^2$ value is nearly constant which shows the negative impact of $\theta_{14}$ on establishing a nonzero value for $\theta_{13}$ from Double Chooz data. Thus, as we expect, for the Double Chooz data active-sterile mixing parameters can mimic the effect of $\theta_{13}$ (see section~\ref{sec:lightsterile}).

\begin{figure}[t!]
\centering
\includegraphics[width=0.8\textwidth]{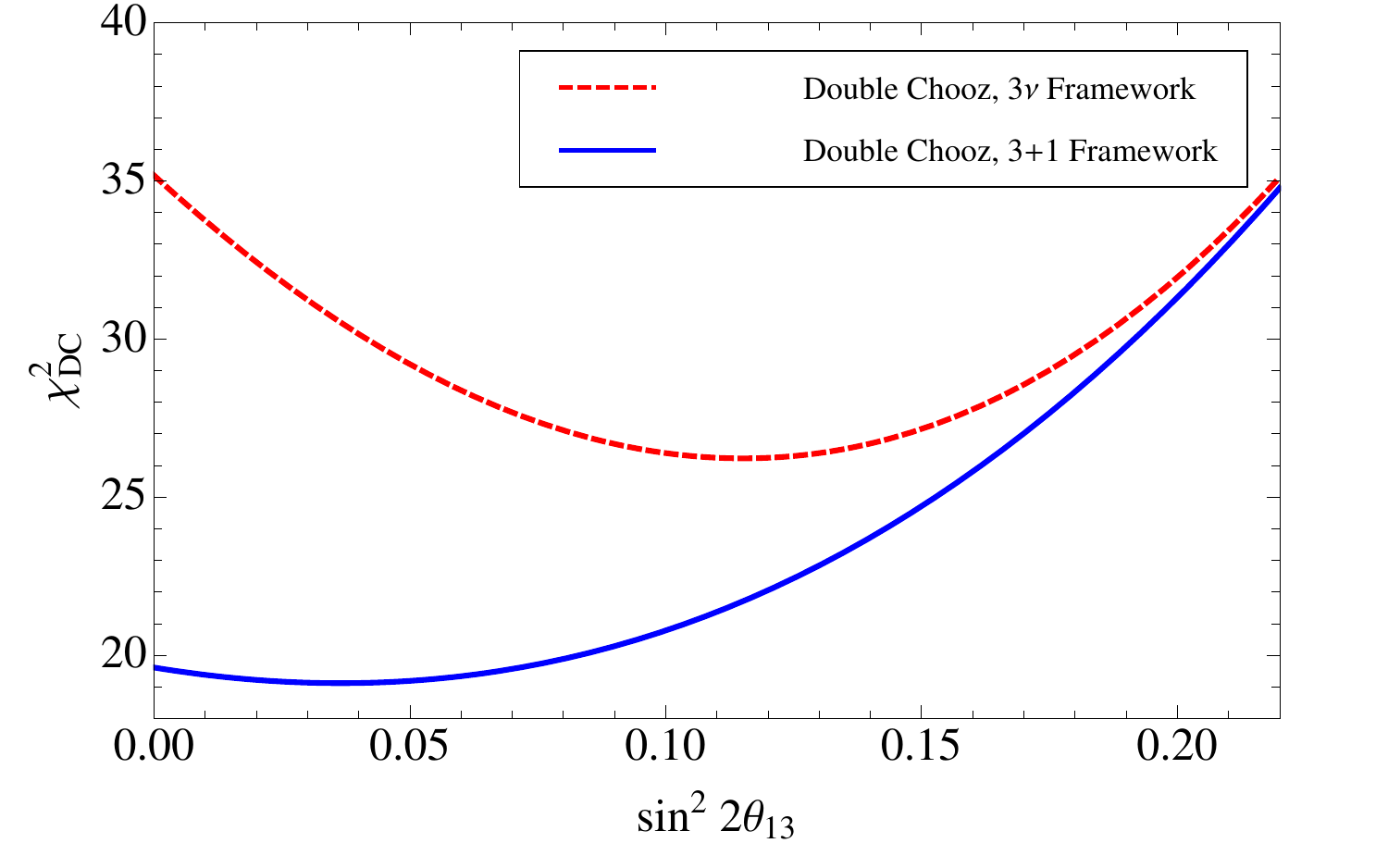}
\label{fig:prob1}
\caption{\label{fig:chi2,dc3+1,q13}$\chi^2_{{\rm DC}}$ versus $\sin^2 2\theta_{13}$ for the $(3+1)_{\rm light}$ model (red dashed curve) and the $3\nu$ framework (blue solid curve), for the Double Chooz data.}
\end{figure}

\begin{figure}[t!]
\centering
\subfloat[]{
\includegraphics[width=0.7\textwidth]{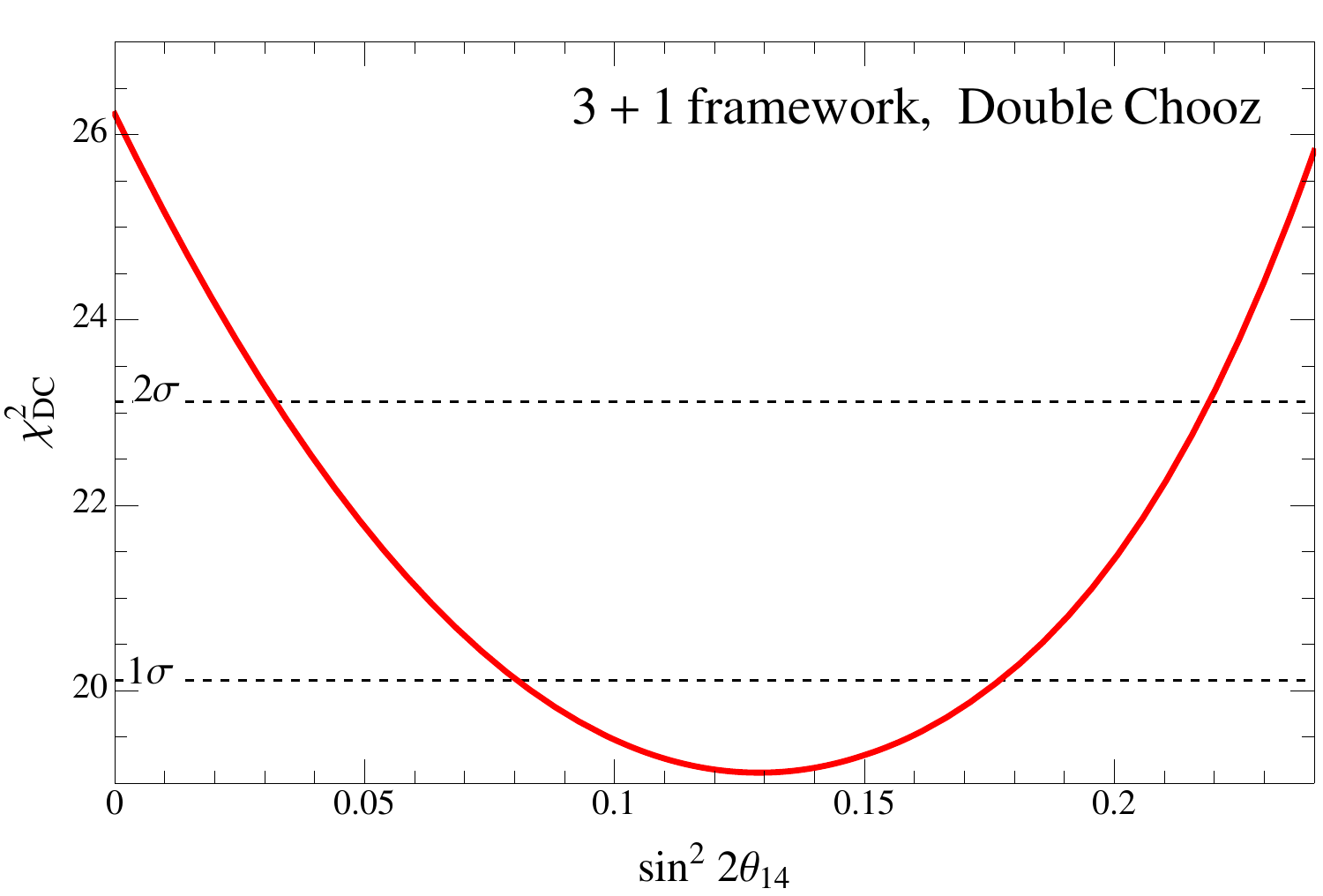}
\label{fig:chi2,dc3+1,1}
}
\quad
\subfloat[]{
\includegraphics[width=0.7\textwidth]{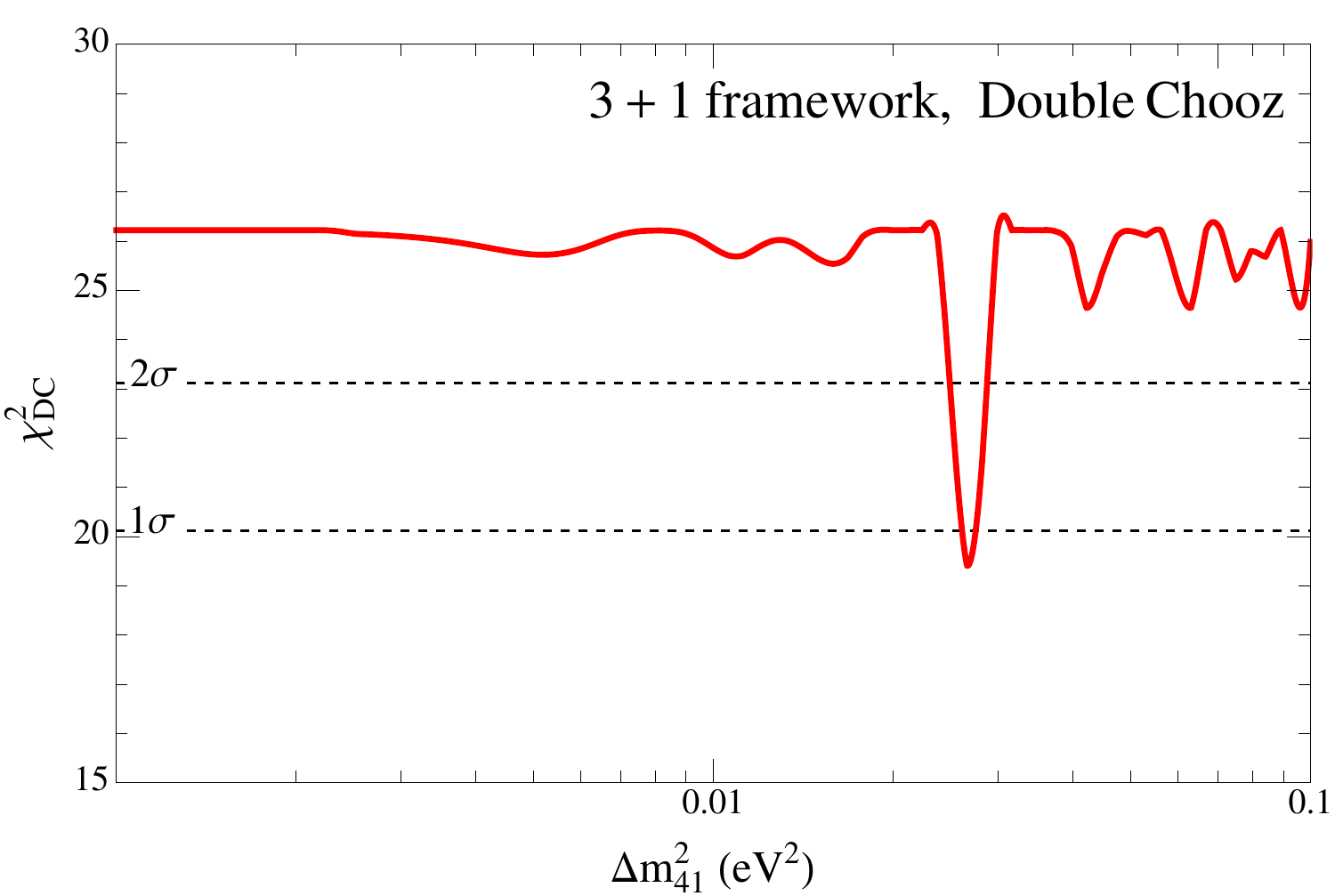}
\label{fig:chi2,dc3+1,2}
}
\caption{\label{fig:chi2,dc3+1}$\chi^2_{{\rm DC}}$ with respect to $\sin^2 2\theta_{14}$ (left panel) and $\Delta m_{41}^2$ (right panel), which are calculated by marginalizing $\chi^2_{\rm DC}$ with respect to $\Delta m_{41}^2$ and $\sin^2 2\theta_{13}$ for the left panel, and $\sin^2 2\theta_{13}$ and $\sin^2 2\theta_{14}$ for the right panel.}
\end{figure}

In the left and right panels of Fig.~\ref{fig:chi2,dc3+1} we show the $\chi^2_{\rm DC}$ values as a function of $\sin^2 2\theta_{14}$ and $\Delta m_{41}^2$, respectively; where the $\chi^2_{\rm DC}$ values are obtained by marginalization over other parameters. As can be seen from Fig.~\ref{fig:chi2,dc3+1,1}, with the Double Chooz data, zero value of $\sin^2 2\theta_{14}$ can be excluded by more than $\sim $ 2.2$\sigma$. Comparing Fig.~\ref{fig:chi2,dc3+1,1} with the red curve in Fig.~\ref{fig:chi2,all,3nu} shows that, as we anticipated, for Double Chooz data extending the $3\nu$ framework to $(3+1)_{\rm light}$ effectively corresponds to exchanging $\theta_{13}$ with $\theta_{14}$. In Fig.~\ref{fig:chi2,dc3+1,2} the best-fit value $\Delta m_{41}^2=0.027~{\rm eV}^2$, shows up as a minimum in $\chi^2_{\rm DC}$. The $\bar{\nu}_e$ survival probability for the best-fit values in Eq.~(\ref{eq2.17}), is shown in Fig.~\ref{fig:prob3}. To understand qualitatively the choice of the best-fit values, in Fig.~\ref{fig:his,dc3+1} we plot the energy distribution of events at Double Chooz detector. In this figure the red and green curves correspond respectively to the distribution of the events for the best-fit values in the $3\nu$ and $(3+1)_{\rm light}$ models. The improvement of the fit to data for $(3+1)_{\rm light}$ is fairly visible, specially for the prompt energies $\sim (3 - 4)$~MeV (due to the higher statistics, the main contribution comes from the left panel of Fig.~\ref{fig:his,dc3+1}). Let us take a closer look at this energy range. For example, counting the energy bins from the left side, in the seventh bin ($E_{\rm prompt}\sim (3.7-4.2)$~MeV) clearly the $(3+1)_{\rm light}$ model matches to the data better. This energy bin corresponds to $L/E\sim (210-233)$~m/MeV. Comparing Figs.~\ref{fig:prob11} and \ref{fig:prob3} in this range of $L/E$ shows that $\bar{\nu}_e$ survival probability decreases in $(3+1)_{\rm light}$ model with respect to $3\nu$ framework. Numerically, for the average of the probability in this bin we obtain: $\langle P \rangle_{3\nu}=0.964$ and $\langle P \rangle_{3+1}=0.883$; where the ratio of these two ($\sim 1.1$) coincides with the ratio of the red to green curves in Fig.~\ref{fig:his,dc3+1,1} for the 7th bin of energy. Conversely, for the fifth bin of energy in Fig.~\ref{fig:his,dc3+1,1}, we obtain $\langle P \rangle_{3\nu}=0.947$ and $\langle P \rangle_{3+1}=0.966$ which leads to an increase in the number of events in the $(3+1)_{\rm light}$ model and again better fit to data. The same improvement occurs for the other neighbor energy bins which eventually leads to a better fit in the $(3+1)_{\rm light}$ model with $\Delta m_{41}^2=0.027\,{\rm eV}^2$. By changing the value of $\Delta m_{41}^2$, this distortion in the number of events moves to the higher or lower energies where already the prediction of $3\nu$ model matched with data points and consequently the fit deteriorates. Notice that although a significant difference exists between Figs.~\ref{fig:prob11} and \ref{fig:prob3} in the range $L/E\sim(40 -100)$~m/MeV, since $L=1050$~m in Double Chooz experiment, the contribution of this range of $L/E$, corresponding to $E_{\rm prompt} \gtrsim10$~MeV, is quite negligible.     

In Figs.~\ref{fig:allowed,dc3+1,2} and \ref{fig:allowed,dc3+1,3} we show the 2-dimensional allowed regions from Double Chooz data in the $(\sin^2 2\theta_{13},\Delta m_{41}^2)$ and $(\sin^2 2\theta_{14},\Delta m_{41}^2)$ planes, respectively. As can be seen from Fig.~\ref{fig:allowed,dc3+1,2} for $\Delta m_{41}^2 \sim 0.027\,{\rm eV}^2$, even vanishing $\theta_{13}$ is allowed at 95\% C.L.. For lower values of $\Delta m_{41}^2$ ($\lesssim 0.005\,{\rm eV}^2$) also $\theta_{13}=0$ is allowed at $3\sigma$ C.L., which is a consequence of $\theta_{14}-\theta_{13}$ degeneracy mentioned in section~\ref{sec:lightsterile}. On the other hand, in the same range of $\Delta m_{41}^2$, larger values of $\sin^22\theta_{14}$ are allowed (see Fig.~\ref{fig:allowed,dc3+1,3}). At lower confidence levels, the closed allowed regions appear in both Figs.~\ref{fig:allowed,dc3+1,2} and \ref{fig:allowed,dc3+1,3} for $\Delta m_{41}^2\sim 0.027\,{\rm eV}^2$ which is a result of the minimum in Fig.~\ref{fig:chi2,dc3+1,2}.

To summarize, extending the standard $3\nu$ framework by adding one light sterile neutrino state improves the fit to the Double Chooz data and weakens the lower limit on $\sin^2 2\theta_{13}$ such that the zero value is allowed at less than $\sim1\sigma$ C.L.. Also, in the $(3+1)_{\rm light}$ model, the allowed region and the best-fit of $\sin^2 2\theta_{13}$ shift to smaller values. From the Double Chooz data, the $(3+1)_{\rm light}$ model is favored at $\sim$ 2.2$\sigma$ C.L. with respect to the $3\nu$ framework. However, it should be noticed that even in the $3\nu$ framework, the signal of nonzero $\sin^2 2\theta_{13}$ from the Double Chooz experiment is rather weak ($2.9\sigma$), and as we saw in section~\ref{sec:standard3nu}, the inclusion of Daya Bay and RENO data significantly enhances the signal. In this regard, we perform a combined analysis of all data for the $(3+1)_{\rm light}$ model in the next section.

\begin{figure}[t!]
\centering
\subfloat[both reactors on]{
\includegraphics[width=0.5\textwidth]{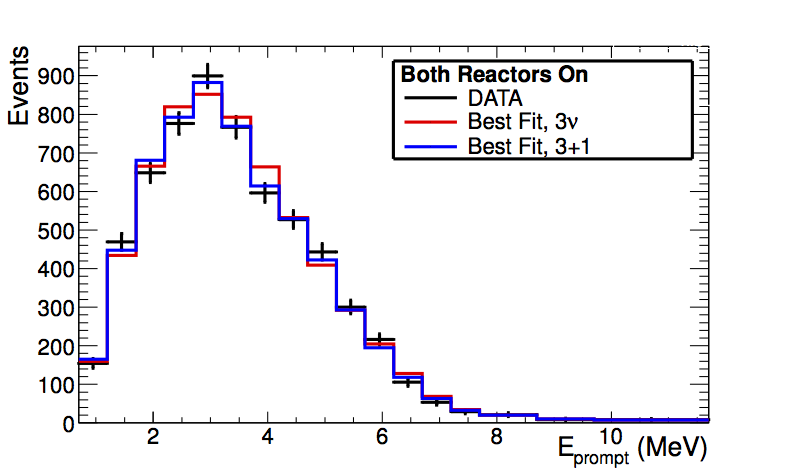}
\label{fig:his,dc3+1,1}
}
\subfloat[one reactor off]{
\includegraphics[width=0.5\textwidth]{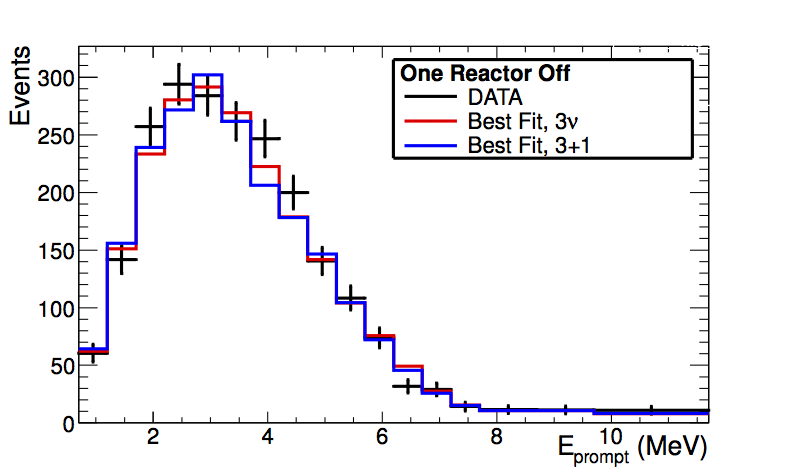}
\label{fig:his,dc3+1,2}
}
\caption{\label{fig:his,dc3+1}Prompt energy distribution of events in the Double Chooz experiment (data points), compared with the predictions of the $3\nu$ (red curve) and $(3+1)_{\rm light}$ (blue curve) models. Left and right panels correspond to the two data-taking periods with both reactor on and one reactor off, respectively.}
\end{figure}

\begin{figure}[t!]
\centering
\subfloat[]{
\includegraphics[width=0.5\textwidth]{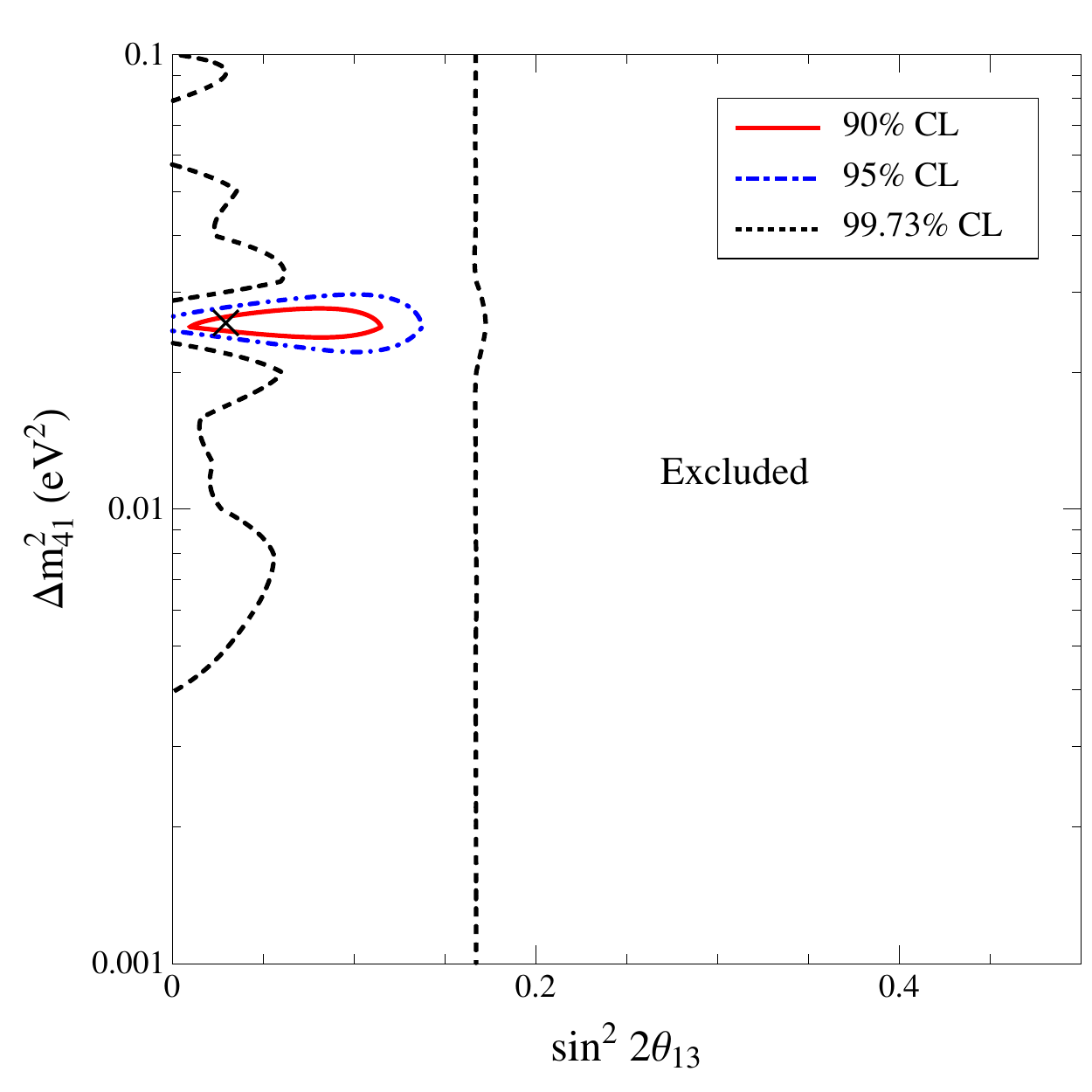}
\label{fig:allowed,dc3+1,2}
}
\subfloat[]{
\includegraphics[width=0.5\textwidth]{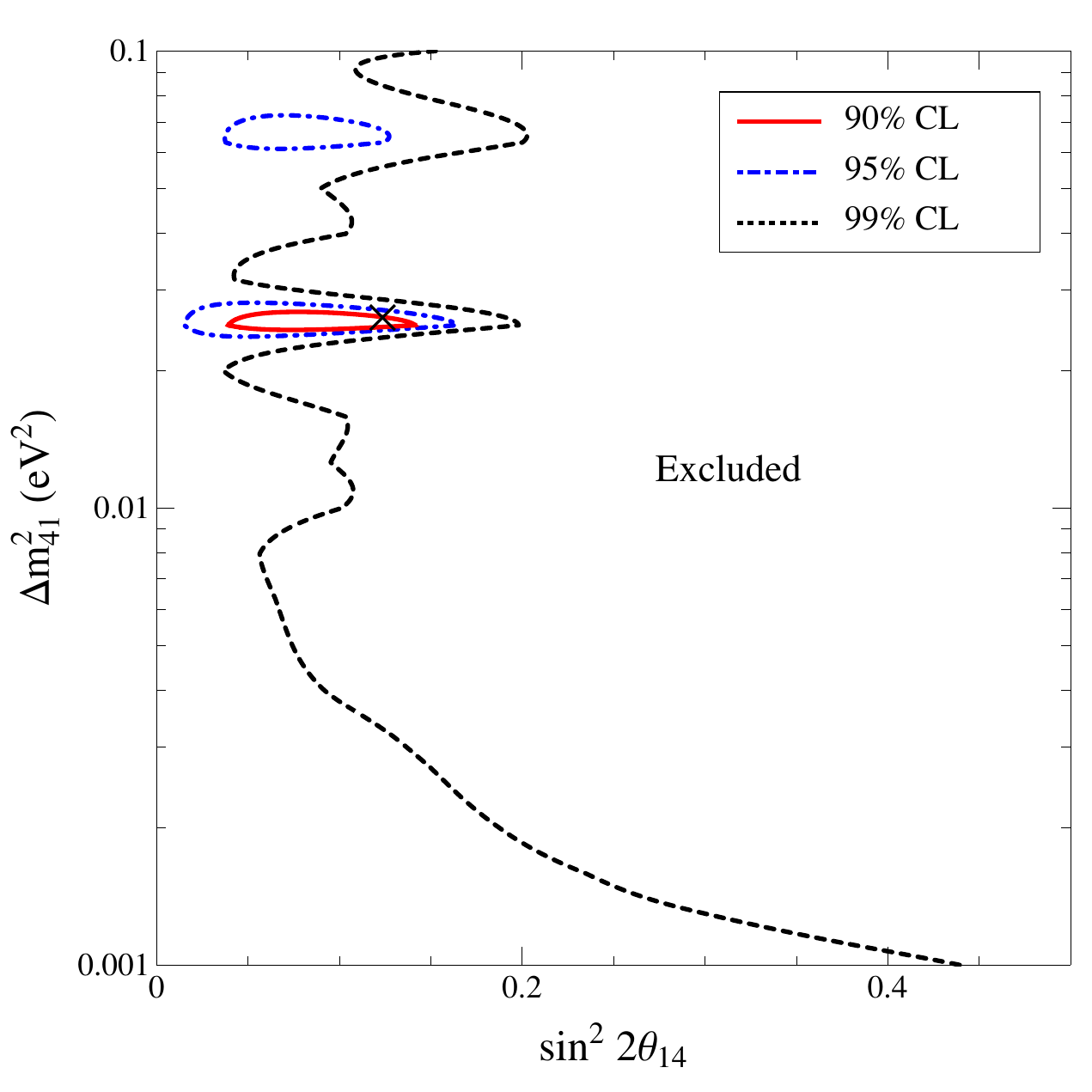}
\label{fig:allowed,dc3+1,3}
}
\caption{\label{fig:allowed,dc3+1}Allowed regions in (a) $(\sin^2 2\theta_{13},\Delta m^2_{41})$, (b) $(\sin^2 2\theta_{14},\Delta m^2_{41})$ plane from the Double Chooz data for different confidence levels. The best-fit value is shown by a cross.}
\end{figure}

\subsection{Probing the $(3+1)_{\rm light}$ model with the combined data of Double Chooz, Daya Bay and RENO}
\label{combined}

In this section we probe the $(3+1)_{\rm light}$ model with the combined analysis of Double Chooz (shape and rate), Daya Bay and RENO (rate only) data. The global $\chi^2_{\rm all}$ is defined by:
\begin{equation}\label{eq2.18}
\chi^2_{{\rm all}}\left(\sin^2 2\theta_{13},\sin^2 2\theta_{14},\Delta m^2_{41}\right)=\chi^2_{{\rm DC}}+\chi^2_{{\rm DB}}+\chi^2_{{\rm RENO}}~,
\end{equation}
as described in section~\ref{sec:standard3nu} in Eqs.~(\ref{eq2.6}), (\ref{eq2.10}) and (\ref{eq2.12}). After minimizing $\chi^2_{{\rm all}}$ with respect to all the pull parameters, we find the following best-fit values
\begin{equation}\label{eq2.19}
\sin^2 2\theta_{13}=0.074\quad ,\quad \sin^2 2\theta_{14}=0.059\quad ,\quad \Delta m^2_{41}=0.027~{\rm eV}^2~,
\end{equation}
with the minimum value $\chi^2_{{\rm min}}/{\rm d.o.f.}=26.7/35$. Comparing with the $3\nu$ framework $\chi^2_{\rm min}/{\rm d.o.f.}=29.7/37$, the $(3+1)_{\rm light}$ leads to an improvement of the fit to global data; however, the significance of the improvement is small. 

The following comments are in order about the combined analysis: 

\begin{figure}[t!]
\centering
\subfloat[]{
\includegraphics[width=0.5\textwidth]{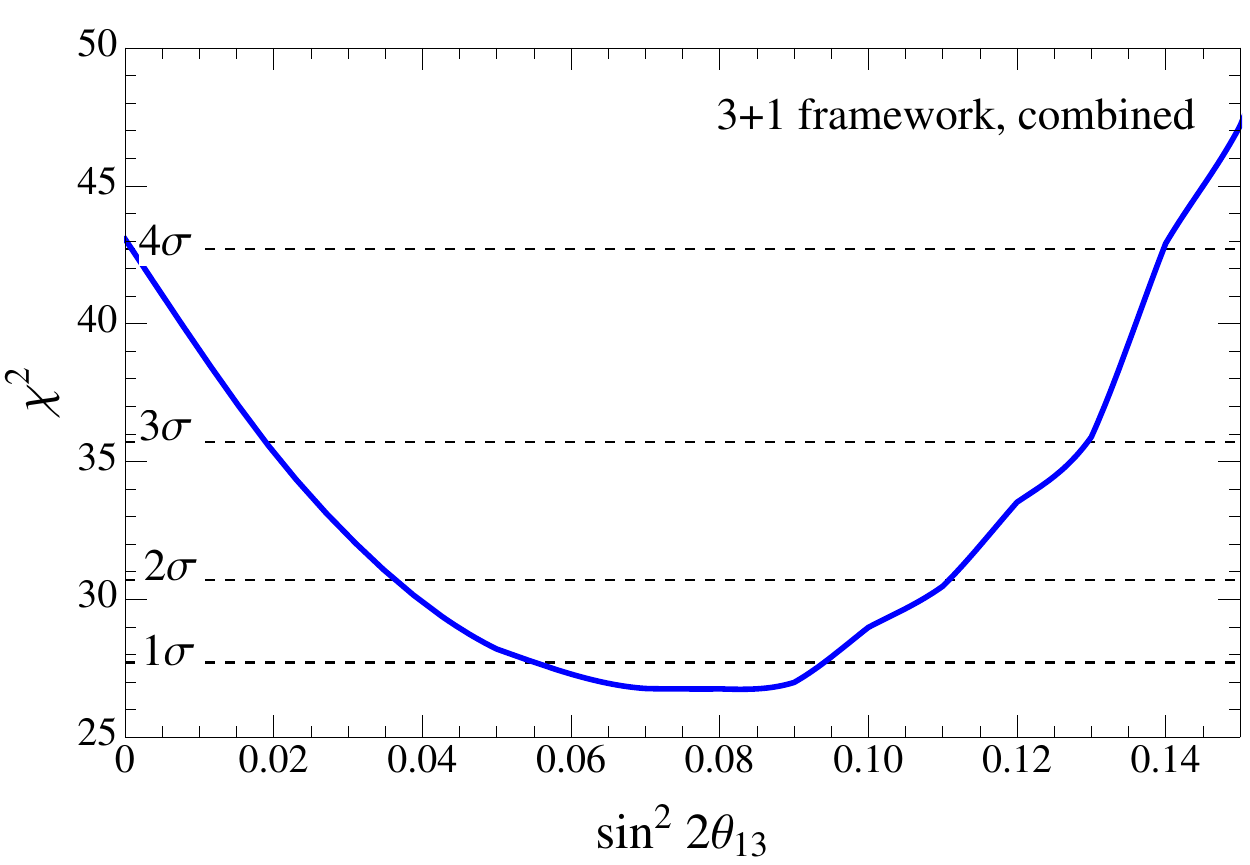}
\label{fig:chi2,combined,q13}
}
\subfloat[]{
\includegraphics[width=0.5\textwidth]{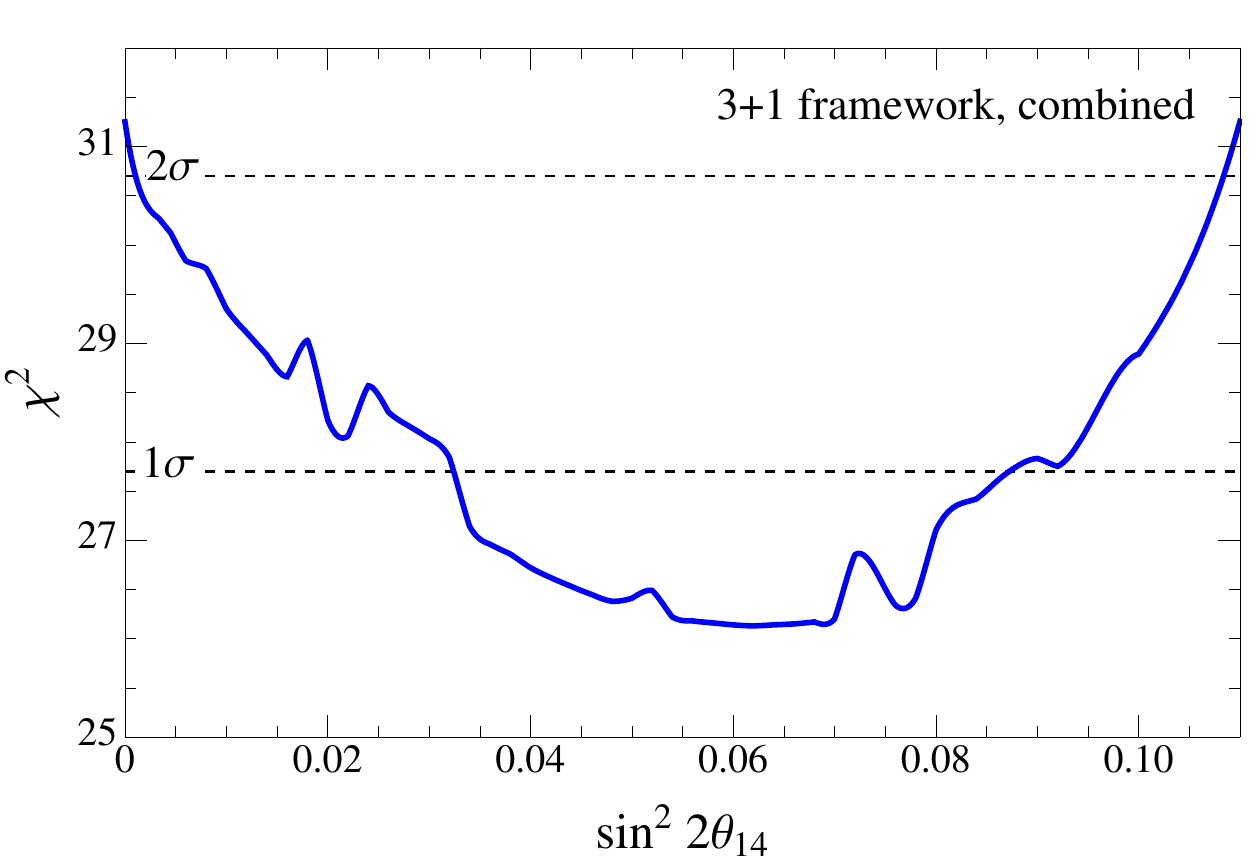}
\label{fig:chi2,combined,q14}
}
\caption{\label{fig:chi2,combined}The $\chi^2_{{\rm all}}$ versus (a) $\sin^2 2\theta_{13}$ and (b) $\sin^2 2\theta_{14}$ in the $(3+1)_{\rm light}$ model for the combined data of Double Chooz, Daya Bay and RENO experiments.}
\end{figure}

\begin{itemize}

\item In Fig.~\ref{fig:chi2,combined,q13} we show $\chi^2_{\rm all}$ as a function of $\sin^2 2\theta_{13}$ after marginalizing over $\Delta m_{41}^2$ and $\sin^2 2\theta_{14}$. The $1\sigma$ range of the value of the 13-mixing angle is $\sin^2 2\theta_{13}=0.074^{+0.017}_{-0.013}$. As can be seen, inclusion of Daya Bay and RENO data increases the best-fit value of $\sin^2 2\theta_{13}$, such that the zero value of $\theta_{13}$ can be excluded by $\sim 4\sigma$. The best-fit value of $\theta_{13}$ and exclusion of $\theta_{13}=0$ in the combined analysis of the $(3+1)_{\rm light}$ model is comparable with the case of $3\nu$ framework, although a bit smaller (see Table~\ref{table2.1}).

\item Fig.~\ref{fig:chi2,combined,q14} shows $\chi^2_{\rm all}$ versus $\sin^2 2\theta_{14}$. Comparison with Fig.~\ref{fig:chi2,dc3+1,1} shows that inclusion of Daya Bay and RENO data shifts the best-fit value of $\sin^2 2\theta_{14}$ to lower values, and also the significance of nonzero $\theta_{14}$ decreases to $\sim2\sigma$. The $1\sigma$ range is $\sin^2 2\theta_{14}~=~0.059^{+0.021}_{-0.016}$.

\begin{figure}[t!]
\centering
\includegraphics[width=0.7\textwidth]{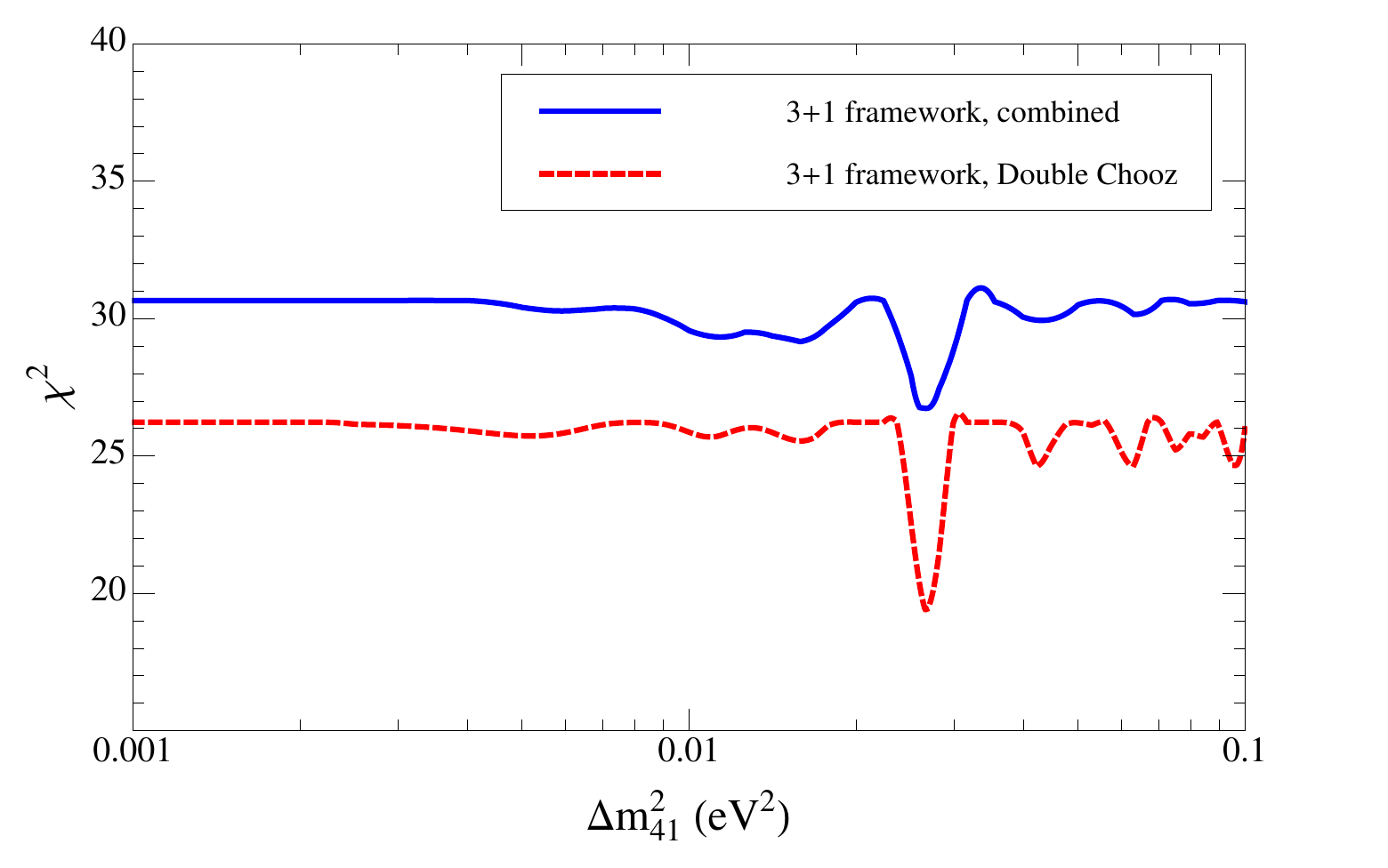}
\label{fig:combinedw}
\caption{\label{fig:combinedw}The $\chi^2$ versus $\Delta m_{41}^2$ for the $(3+1)_{\rm light}$ model, from the combined data of Double Chooz, Daya Bay and RENO experiments (blue curve). For comparison the curve for only Double Chooz data is shown (red dashed curve).}
\end{figure}

\item For the mass-squared difference $\Delta m^2_{41}$, the best-fit value of the Double Chooz and combined analysis is the same (see Eqs.~(\ref{eq2.17}) and (\ref{eq2.19}), see also Fig.~\ref{fig:combinedw}). The best-fit value $\Delta m_{41}^2=0.027\,{\rm eV}^2$ for the combined analysis originates from Double Chooz data for the same reason discussed in section~\ref{analysis3p1dc} about Fig.~\ref{fig:his,dc3+1}; namely since the position of the extrema in the $\bar{\nu}_e$ oscillation probability depends only on $\Delta m_{41}^2$ value. It is straightforward to check that for the mixing parameters of Eq.~(\ref{eq2.19}) still the improvement of fit to the Double Chooz data in $(3+1)_{\rm light}$ holds to some extend in the range $E_{\rm prompt}\sim (3-4)$~MeV.  

\begin{figure}[t!]
\centering
\subfloat[]{
\includegraphics[width=0.5\textwidth]{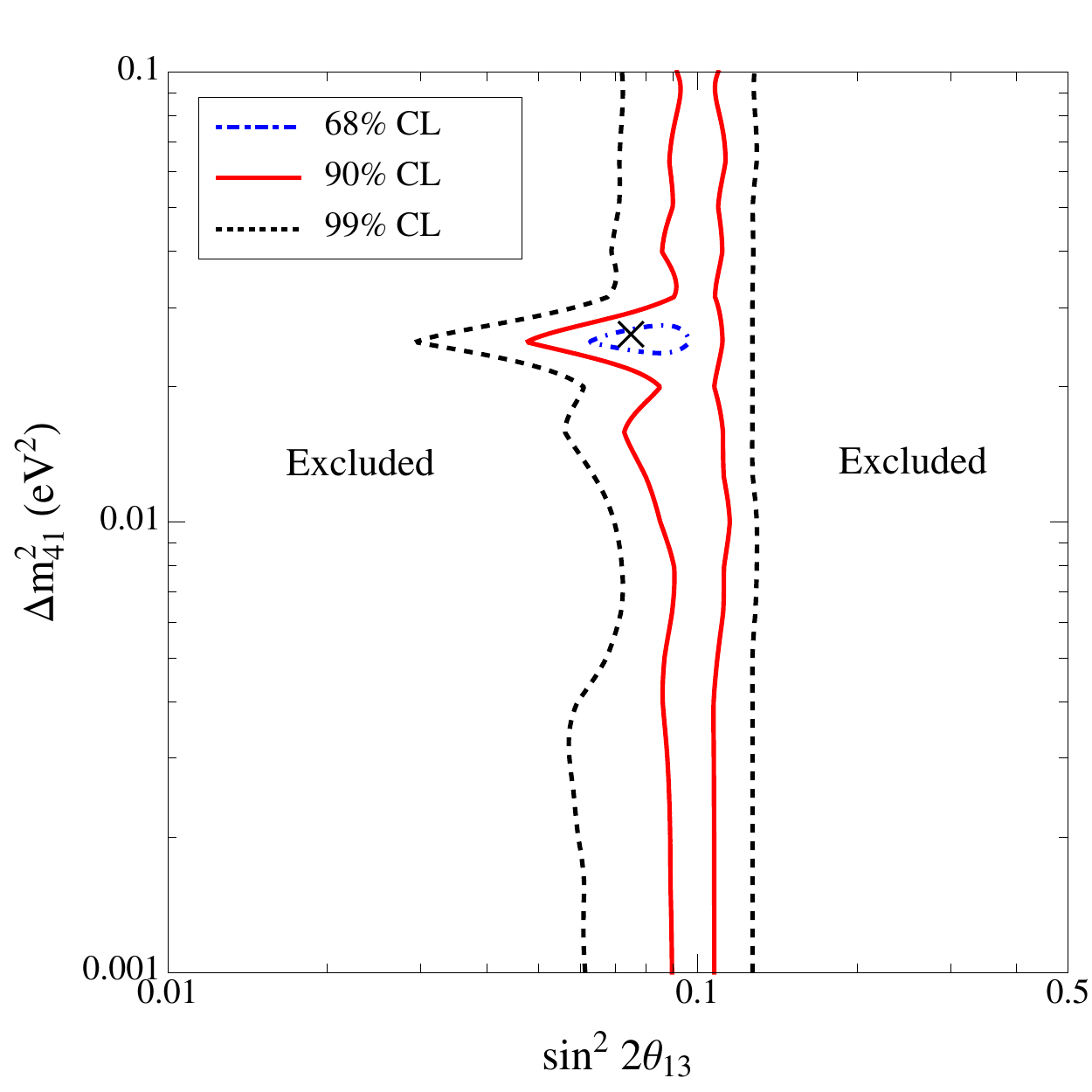}
\label{fig:allowed,combined,q13dms14}
}
\subfloat[]{
\includegraphics[width=0.5\textwidth]{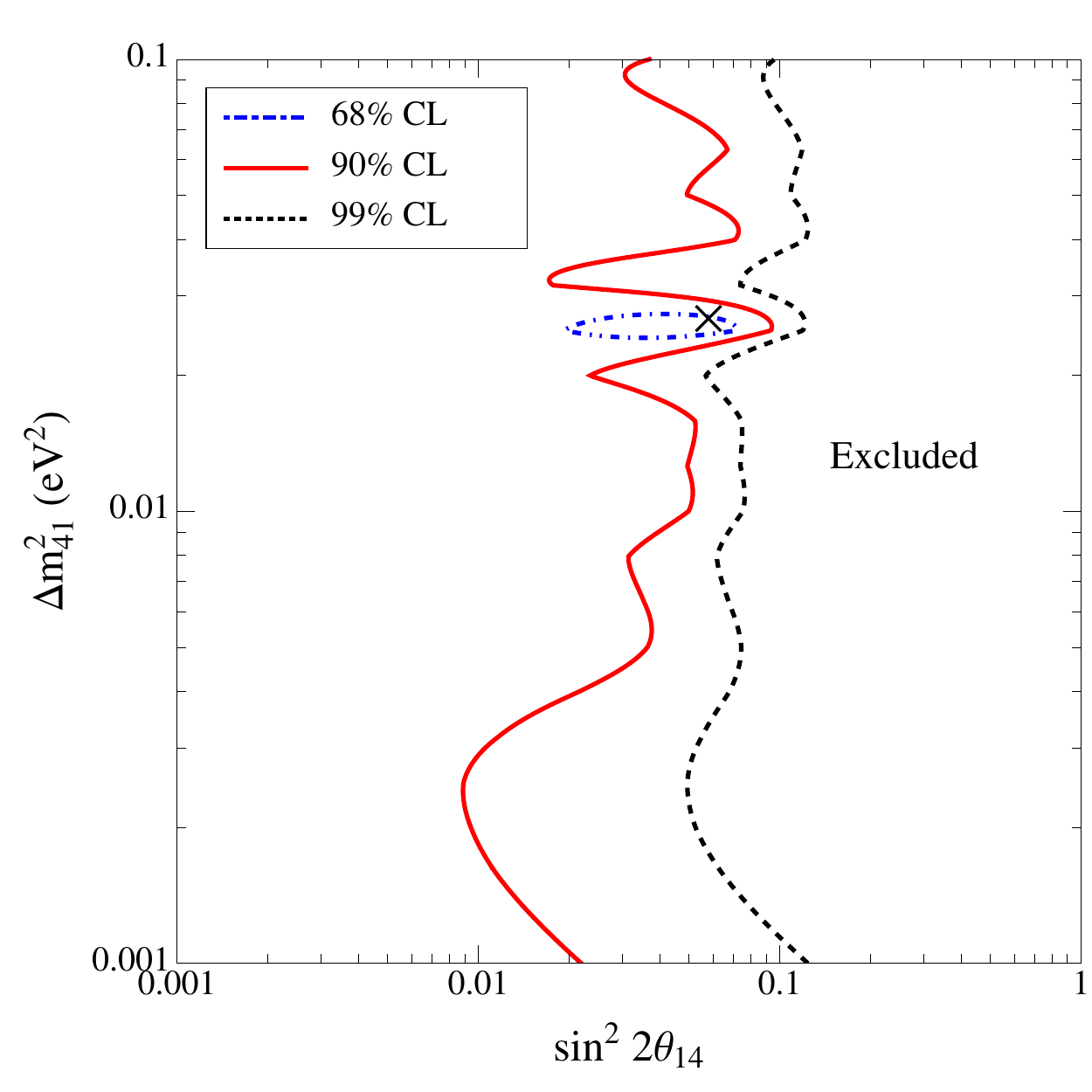}
\label{fig:allowed,combined,q14dms14}
}
\caption{\label{fig:allowed,combined,dms14}Allowed regions in (a) $(\sin^2 2\theta_{13},\Delta m^2_{41})$, (b) $(\sin^2 2\theta_{14},\Delta m^2_{41})$ plane; for the combined analysis of Double Chooz, Daya Bay and RENO data. The best-fit value is shown by a cross.}
\end{figure}

\item In Figs.~\ref{fig:allowed,combined,q13dms14} and \ref{fig:allowed,combined,q14dms14} we show the allowed region in $(\sin^2 2\theta_{13},\Delta m^2_{41})$ and $(\sin^2 2\theta_{14},\Delta m^2_{41})$ planes respectively. From Fig.~\ref{fig:allowed,combined,q13dms14} we see that the global analysis of Double Chooz, Daya Bay and RENO data in the $(3+1)_{\rm light}$ model excludes $\theta_{13}=0$ with high confidence level ($\sim4.1\sigma$). Also, the best-fit value of $\sin^22\theta_{13}$ is fairly close to the value in the $3\nu$ framework. Thus, the measured value of $\sin^2 2\theta_{13}$ is robust with respect to the existence of a light sterile neutrino; and the Daya Bay and RENO data play an important role in this robustness (compare Fig.~\ref{fig:allowed,dc3+1,2} and \ref{fig:allowed,combined,q13dms14}). In both Figs.~\ref{fig:allowed,combined,q13dms14} and \ref{fig:allowed,combined,q14dms14} a closed allowed region appears at $\Delta m_{41}^2\sim 0.027\,{\rm eV}^2$ in low confidence levels which stem from the Double Chooz data. Particularly, in Fig.~\ref{fig:allowed,combined,q14dms14} the closed region indicates that $\theta_{14}=0$ can be excluded at $\sim 68\%$ C.L.; however, by increasing the significance the closed region transforms to upper limit and $\theta_{14}=0$ is allowed.

\item In Fig.~\ref{fig:allowed,q13q14} we show the allowed region in $(\sin^2 2\theta_{13},\sin^2 2\theta_{14})$ plane for Double Chooz only (left panel) and for the combined analysis (right panel).  In Fig.~\ref{fig:allowed,dc,q13q14} (which is for Double Chooz), there is an anti-correlation between the allowed values of $\theta_{13}$ and  $\theta_{14}$: for smaller values of $\theta_{13}$ larger values of $\theta_{14}$ are favored and vice-versa. This anti-correlation is a manifestation of $\theta_{13}-\theta_{14}$ degeneracy discussed in section~\ref{sec:lightsterile}. The break of degeneracy in low confidence levels (red and blue curves in Fig.~\ref{fig:allowed,dc,q13q14}) is due to the mismatch of the Double Chooz data and the $3\nu$ prediction in $E\sim(3-4)$~MeV which favors larger $\theta_{14}$. However, by including the Daya Bay and RENO data in Fig.~\ref{fig:allowed,combined,q13q14} (with the advantage of making two independent measurements in near and far detectors of each experiment) the anti-correlation and degeneracy break and we end up in two islands the both of them favoring nonzero $\theta_{13}$. For the first island: $\sin^2 2\theta_{14}<\sin^2 2\theta_{13}$, while for the second one: $\sin^2 2\theta_{14}>\sin^2 2\theta_{13}$; where the latter originates from the Double Chooz contribution. In both panels of Fig.~\ref{fig:allowed,q13q14} the green dashed curve represent the limit from solar and KamLAND data~\cite{Palazzo:2012yf}.

\begin{figure}[t!]
\centering
\subfloat[]{
\includegraphics[width=0.5\textwidth]{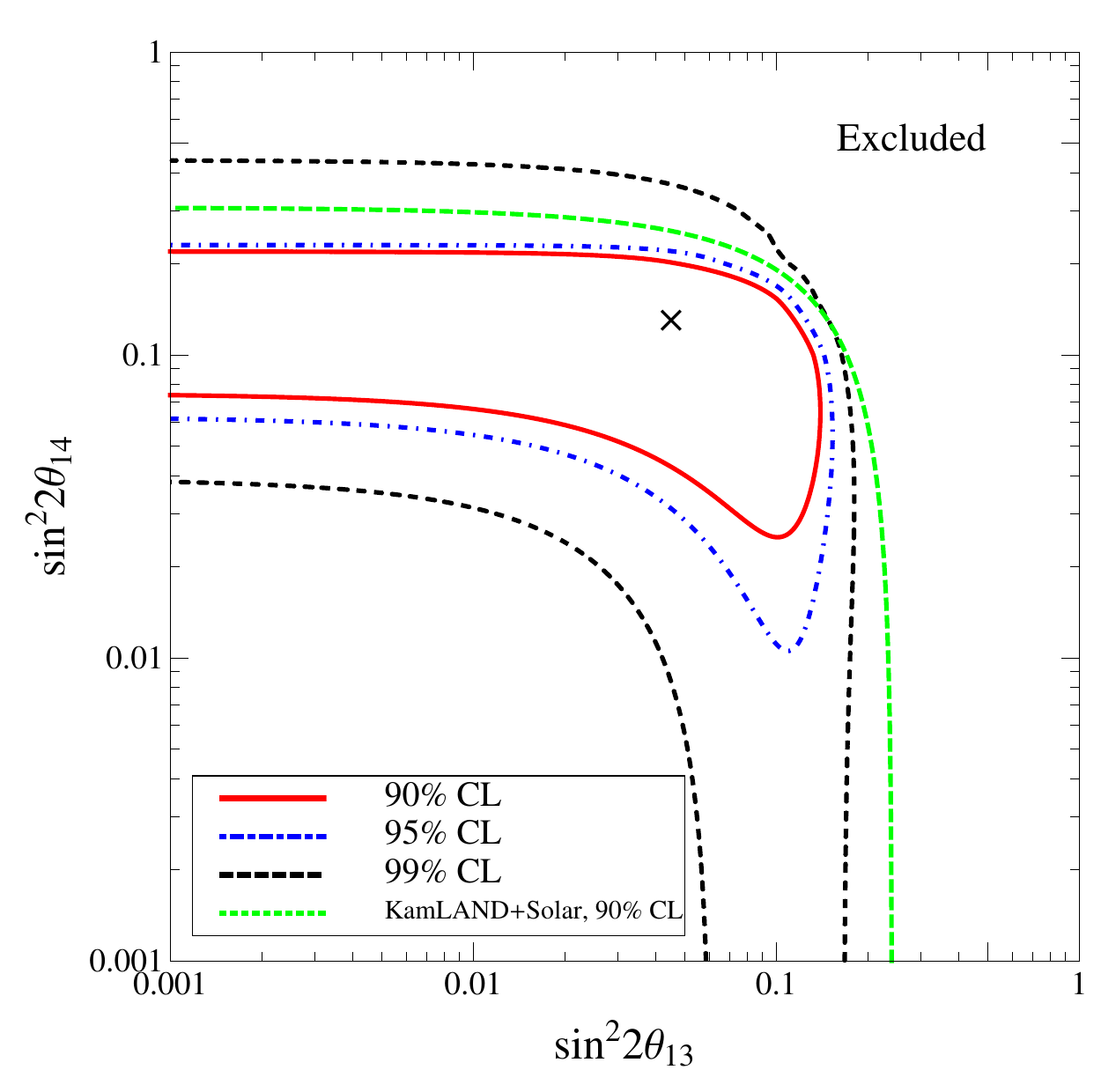}
\label{fig:allowed,dc,q13q14}
}
\subfloat[]{
\includegraphics[width=0.5\textwidth]{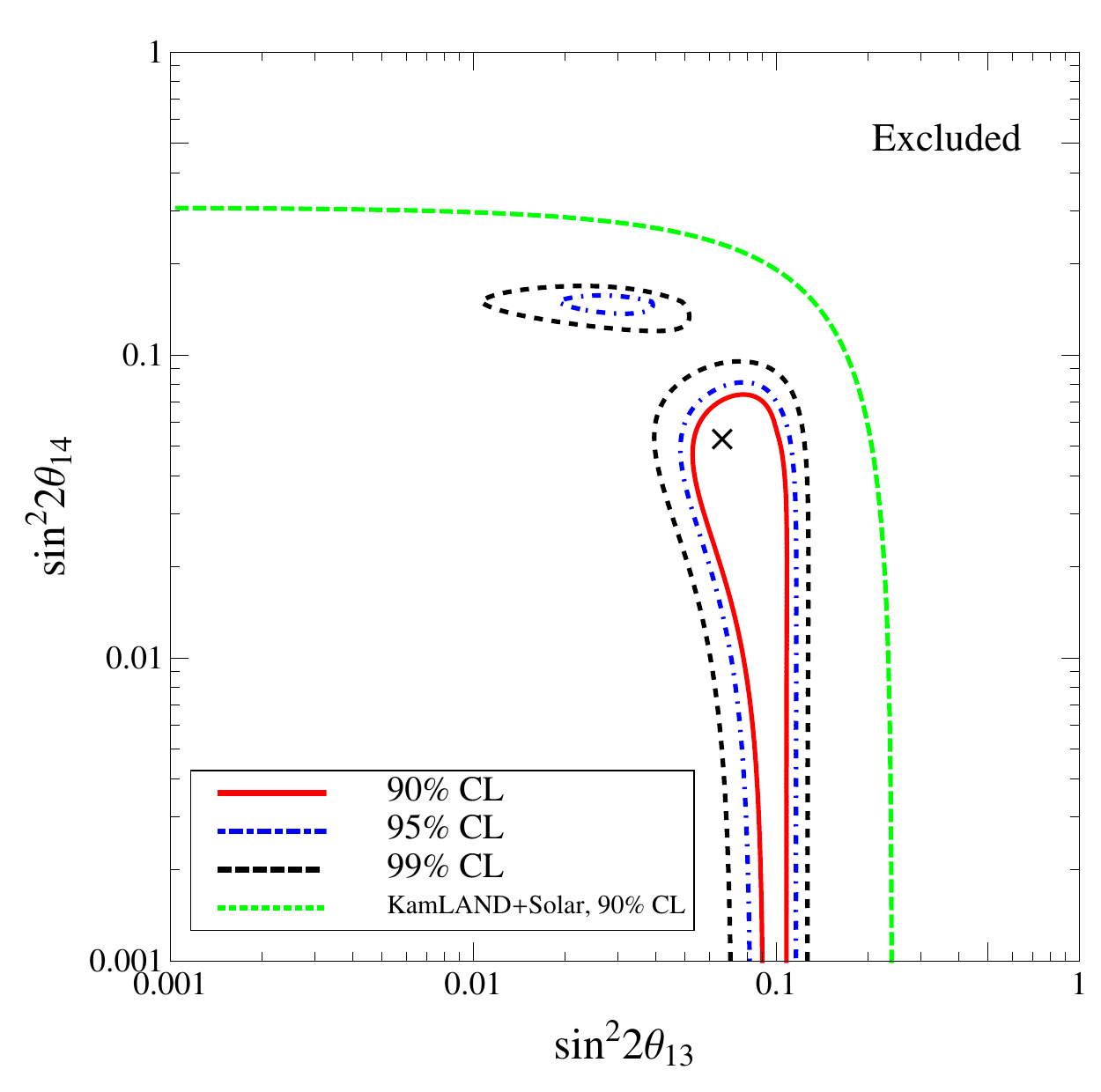}
\label{fig:allowed,combined,q13q14}
}
\caption{\label{fig:allowed,q13q14}Allowed regions in $(\sin^2 2\theta_{13},\sin^2 2\theta_{14})$ plane from the Double Chooz data (left panel) and the combined data of Double Chooz, Daya Bay an RENO experiments (right panel). The best-fit value is shown by a cross. The green dashed curve shows the upper limit at $90\%~$C.L. from the KamLAND and Solar data \cite{Palazzo:2012yf}.}
\end{figure}

\item Finally we would mention that our results are robust with respect to the uncertainty in the value of $\Delta m^2_{31}$, since varying this parameters will change just the positions of extrema in Fig.~\ref{fig:prob11} while the improvement to data requires additional extrema as in Fig.~\ref{fig:prob4}. As a cross check, we have tested the stability of our results by varying $\Delta m^2_{31}$ within its $1\sigma$ uncertainty ranges from MINOS experiment~\cite{Adamson:2011ig}. The negligible change in our results justifies the initial assumption of fixed $\Delta m_{31}^2$.

\end{itemize}

\section{Conclusions}\label{sec:con_ch2}

Searches for sterile neutrinos and investigating its impact on experimental results obtained, or planned to be obtained, is one of the cutting edge questions in neutrino physics. Although the initial motivation was interpretation of LSND anomaly by sterile neutrinos with mass ~$\sim\mathcal{O}(1)$~eV, further anomalies such as cosmological hints and low energy solar data stimulated existence of lighter sterile neutrinos. In the work presented in this chapter, we investigated the consequence of the existence of a light sterile neutrino, the $(3+1)_{\rm light}$ model, on medium baseline reactor experiments: Double Chooz, Daya Bay and RENO. The baseline and energy of these experiments provide the opportunity to probe active-sterile oscillation with $\Delta m_{41}^2\sim (10^{-3}-10^{-1})\,{\rm eV}^2$.  

Among these three experiments, the Double Chooz consists of one detector and presents both rate and shape (in energy) information of observed events; while the Daya Bay and RENO experiments, each equipped by near and far detectors, provide the deficit in total number of events in the far detector(s) with respect to the near detector(s). The energy distribution of events in Double Chooz is in good agreement with the $3\nu$ framework prediction except in the range $E_{\rm prompt}\sim(3-4)$~MeV. This discrepancy leads to a better fit in the $(3+1)_{\rm light}$ such that we obtained the best-fit values $\sin^2 2\theta_{13}=0.036$, $\sin^2 2\theta_{14}=0.129$ and $\Delta m^2_{41}=0.027~{\rm eV}^2$ for mixing parameters. With the Double Chooz data the $3\nu$ framework can be excluded at $\sim 2.2\sigma$ C.L.. Also, the best-fit value of $\theta_{13}$ angle is significantly different than the reported value in the $3\nu$ framework and $\theta_{13}=0$ is allowed in less than $1\sigma$ C.L.. 

Inclusion of the rate information from the Daya Bay and RENO experiments alters the conclusion such that with the combined data of Double Chooz, Daya Bay and RENO we obtain the best-fit values
\begin{equation}\label{eq2.20}
\sin^2 2\theta_{13}=0.074 \quad , \quad \sin^2 2\theta_{14}=0.059 \quad ,\quad \Delta m^2_{41}=0.027~{\rm eV}^2~. 
\end{equation}
With the combined data the $(3+1)_{\rm light}$ model is favored at $\sim1.2\sigma$ C.L.. The value of $\theta_{13}$ angle is close to the reported value in the $3\nu$ framework and so robustness of $\theta_{13}$ determination can be claimed. The persisting $\theta_{13}-\theta_{14}$ degeneracy in $(3+1)_{\rm light}$, which exists in the limit $\Delta m_{41}^2\to\Delta m_{31}^2$, can be lifted mainly by the data of Daya Bay and RENO. Despite the preference for the $(3+1)_{\rm light}$ model, a large part of the parameter space of this model is excluded in our analysis, better than the constraints from the other analyses by a factor of 2.

The planned near future data from these experiments can significantly exclude/strengthen the favored nonzero active-sterile mixing parameters found in this work. Especially, the energy spectrum of the data in the Daya Bay and RENO experiments can decisively rule out/confirm it. Also, installation of the near detector in the Double Chooz experiment can provide valuable information about the observed anomaly in $E_{\rm prompt}\sim(3-4)$~MeV, wether supporting it or contradicting it.

\chapter{Probing Large Extra Dimensions With IceCube }\label{chap3}

\newpage

{\Large Only neutrinos, with their extremely small interaction cross-sections, can enable us to see into the interior of a star, and thus verify directly the hypothesis of nuclear energy generation in stars.\\

~~~~John N. Bahcall}

\newpage

\footnote{This chapter is prepared based on my work published in \cite{Esmaili:2014esa}.} In models with Large Extra Dimensions the smallness of neutrino masses can be naturally explained by introducing gauge singlet fermions which propagate in the bulk. The Kaluza-Klein modes of these fermions appear as towers of the sterile neutrino states on the brane. In this chapter we study the phenomenological consequences of this picture for the high energy atmospheric neutrinos. For this purpose, we construct a detailed equivalence between a model with large extra dimensions and a $(3+n)$ scenario consisting of 3 active and $n$ extra sterile neutrino states, which provides a clear intuitive understanding of the Kaluza-Klein modes. Finally, we analyze the collected data of high energy atmospheric neutrinos by IceCube experiment and obtain bounds on the radius of the extra dimensions.  

The chapter is organized as follows:  We will briefly introduce the Large Extra Dimension (LED) model in section~\ref{sec:intt_cha3}. In Section~\ref{sec:formalism_ch3} we explain the formalism of the Large Extra Dimension (LED) model and the matter effects of the Earth on the KK modes. In Section~\ref{sec:osc} we calculate the flavor oscillation probabilities of high energy atmospheric neutrinos in the LED model. Then in Section~\ref{sec:3+n} we establish the equivalence between the LED and $(3+n)$ models. Section~\ref{sec:icecube} is devoted to the analysis of the data of IceCube. We summarize our conclusions in Section~\ref{sec:conc_ch3}.

\section{Introduction: The Large Extra Dimension model\label{sec:intt_cha3}}

The large extra dimension (LED) model has been introduced and motivated as a solution to the hierarchy problem~\cite{ArkaniHamed:1998rs,ArkaniHamed:1998nn,Antoniadis:1998ig}, which is the huge difference between the Planck scale $M_{\rm{pl}}\simeq1.2\times10^{19}$~GeV and the electroweak scale $v=246$~GeV. The basic idea is that if a singlet of the SM exists (such as Graviton
), it can propagate into the bulk (i.e. the space including the extra dimensions), while the Standard Model (SM) particles are localized in a 4-dimensional space-time embedded in the balk~\cite{Rubakov:1983bb}. In this scenario the \textit{fundamental} Planck scale in the bulk is suppressed down to the weak scale by the volume of the extra dimension space and so there is no hierarchy problem anymore. The relation between the observed Planck scale in 4-dimensions and the fundamental Planck scale $M_F$ is given by
\begin{equation}\label{eq3.1}
M^2_{\rm{Pl}}=(2\pi)^D(R_1R_2\cdots R_D)M_{\rm{F}}^{2+D}=V_DM_{\rm{F}}^{2+D},
\end{equation}
where the extra $D$ dimensions are compactified in tori with radii  $R_i$'s and volume $V_D$. 

In the same scenario, the same idea has been proposed to explain the smallness of neutrino masses~\cite{Dienes:1998sb,ArkaniHamed:1998vp}. In fact, the mechanism of confinement of SM particles on the brane relies on the gauge flux conservation which necessitates that just singlets under the SM gauge symmetry can propagate into the bulk. Thus, in principle, in addition to the graviton, the hypothesized right handed neutri- nos can also live in the bulk and consequently the volume suppression explains the small neutrino masses. However, the Kaluza-Klein (KK) expansion of the right handed neutrinos after the compactification of the extra dimensions manifest towers of the sterile neutrinos from the brane point of view which can dramatically affect the oscillation phenomenology of the active neutrinos. This behavior has been studied extensively in the literature\footnote{For possible signatures of bulk KK modes at colliders and also their impact on the lepton number violating processes see~\cite{Cao:2004tu}. A review of the collider signatures is given in~\cite{Gingrich:2009az}.}~\cite{Dvali:1999cn,Machado:2011jt}. Although the majority of studies derive more and more stringent upper bound on the radius of the extra dimensions, still, interestingly, with the current upper limit on the size of the extra dimensions, the first KK mode sterile neutrino can have a mass $\mathcal{O}(1)$~eV, which is in the ballpark of what is required for the interpretation of the recently observed anomalies in the short baseline neutrino experiments and LSND/MiniBooNE experiments~\cite{Kopp:2013vaa}. For instance, in this line, it is proposed in~\cite{Machado:2011kt} that the reactor and gallium anomalies can be interpreted within the LED model.   

In this chapter we study an independent probe of the LED model by the use of the high energy atmospheric neutrinos. During the past few years, the completed IceCube detector at the south pole has collected a high statistics sample of the atmospheric neutrino data with energies $>10$~GeV, which actually plays the role of \textit{background} for the astrophysical/cosmic neutrino searches that IceCube is intended to do. However, these background data provide a unique opportunity to probe the new physics scenarios with unprecedented precision. The atmospheric neutrino data of IceCube has been used to probe the sterile neutrinos~\cite{Esmaili:2012nz,Esmaili:2013vza,Esmaili:2013cja,Razzaque:2011ab}, violation of the equivalence principle~\cite{Esmaili:2014ota}, the non-standard neutrino interactions~\cite{Esmaili:2013fva} and the matter density profile of the Earth~\cite{Agarwalla:2012uj}. In this chapter we study the signature of the LED model in the high energy atmospheric neutrinos and, by analyzing the data sets of IC-40~\cite{Abbasi:2010ie} and IC-79~\cite{Aartsen:2013jza}, we show that it is possible to constrain the radius of the extra dimension to $<4\times10^{-5}$~cm (at $2\sigma$ C.L.). Also, we estimate the sensitivity of IceCube to the LED model after taking into account the energy information of the collected data and show that the favored region of the parameter space by reactor and gallium anomalies~\cite{Machado:2011kt} can be excluded by the IceCube data.  

From the brane point of view the KK modes of the LED model resemble a series of the sterile neutrino states with increasing masses. The Earth's matter density induce resonant conversion of the active neutrinos to these sterile states, which the rate of conversion depends on the energy and zenith angle ($\theta_z$) of the atmospheric neutrinos. Phenomenologically, these signatures are similar to the signatures of the $(3+n)$ scenarios consisting of 3 active neutrinos and $n$ sterile states with mixing pattern determined by various mixing angles. We elaborate on this similarity and establish a detailed equivalence between them.

\section{Matter effects on the neutrino propagation in the Large Extra Dimension model\label{sec:formalism_ch3}}

In this section we study the propagation of neutrinos in matter in the LED model. Our aim is to investigate the Earth's matter effects on the propagation of the high energy atmospheric neutrinos in the presence of the Kaluza-Klein modes. The collected data of the high energy atmospheric neutrinos by the IceCube detector provides a unique opportunity to search for these effects and so to probe the LED model.

The number of LEDs should be $D\geq 2$, where the $D=1$ case is excluded by the observed $1/r^2$ behavior of the gravitational force at the scale of the solar system (If there were only one extra dimension, the radius of this extra dimension had to be $R\sim10^{10}$~km.). The factor suppressing the 4-dimensional Planck scale down to $\sim$ TeV scale is the volume of the $D$-dimensional space, where for the case that LEDs are compactified on tori with radii $R_j$ ($j=1,\ldots,D$), the volume is given by $(2\pi)^D R_1\cdots R_D$ (See Eq.~(\ref{eq3.1})). It should be noticed that all the radii $R_j$ are not necessarily equal, and in fact, assuming an {\it asymmetrical} compactification in the $D$-dimensional space, a hierarchical pattern of $R_j$ elevates the existing bounds on the size of the LED radii from supernovae cooling and the cosmological considerations~\cite{Kaloper:2000jb}. A $(4+D)$-dimensional space with hierarchical radii of compactification in the $D$-dimensional space of LED effectively is equivalent to a $5$-dimensional bulk space with the LED radius given by the largest $R_j$ which will be denoted by $R_{\rm ED}$ hereafter. The LED scenario explains the smallness of the active neutrino masses through the volume suppression of the Yukawa couplings between the Higgs field $H$, the active left-handed neutrinos $\nu_{iL}$ and the 5-dimensional fermions $\Psi_i$ (singlet under the SM gauge group) where $i=1,2,3$ correspond to the number of the active flavors. The action of interaction between the active neutrinos and $\Psi_i$ fields is given by $S= \sum_{i=1}^3 S_i$, where~\cite{Dienes:1998sb}
\begin{eqnarray}\label{eq3.2}
S_i=\int d^4x~dy~i\overline{\Psi}_i\Gamma^A\partial_A\Psi_{i}+\int d^4x~\left[ i\bar\nu_{iL}\gamma^{\mu}\partial_{\mu}\nu_{iL}+\lambda_{ij}H\bar\nu_{iL}\psi_{jR}(x,y=0)\right]+\rm{h.c.}.\nonumber\\
\end{eqnarray}
In this equation $\Gamma_A ~(A=0,\ldots,4)$ are the Dirac matrices and $(\psi_{iL},\psi_{iR})$ are the Weyl components of the fermion $\Psi_i$ living in the 5-dimensional space $(x^\mu,y)$. The first term of Eq.~(\ref{eq3.2}) is the kinetic term of $\psi_{iL}$ and $\psi_{iR}$ fields and the first term in bracket is the kinetic term of the active neutrino fields $\nu_{iL}$. The last term is the Yukawa term with the coupling constant $\lambda_{ij}$, which gives the interaction of $\Psi_i$ fields in the bulk with the active neutrinos living on the brane $y=0$ (we are assuming compactification on a $Z_2$ orbifold where $\psi_{iL}$ and $\psi_{iR}$ are odd and even under its $Z_2$ action, respectively; and so $\psi_{iL}(x,y=0)$ vanishes.). Please note that the Yukawa couplings $\lambda_{ij}$ are not dimensionless parameters. The mass dimension of the fields in the action of Eq.~(\ref{eq3.2}) are: $[\psi_{R}]_M=\frac{D+3}{2}$, $[\nu_L]_M=\frac{3}{2}$ and $[H]_M=1$; therefore the mass dimension of the Yukawa couplings would be $[\lambda_{ij}]_M=-\frac{D}{2}$. At this point the only mass scale in the theory is $M_F$, hence we can define the dimensionless Yukawa coupling as \cite{Davoudiasl:2002fq}
\begin{eqnarray}\label{eq3.3}
h_{ij}=\lambda_{ij}M_F^{D/2},
\end{eqnarray}
in which we assume $h_{ij}\sim\mathcal{O}(1)$.
The mixings of the active neutrinos are parametrized with the PMNS matrix $U$ through
\begin{equation}\label{eq3.4}
\nu_{\alpha L} = \sum_{i=1}^3 U_{\alpha i} \nu_{iL}~,
\end{equation}
where $U\to U^\ast$ for antineutrinos. Without loss of generality, the Yukawa coupling matrix $\lambda_{ij}$ can be diagonalized by the above field redefinition and a corresponding redefinition of the bulk fields. After electroweak symmetry breaking and expansion of the $\psi_{iR}$ and $\psi_{iL}$ fields in terms of the Kaluza-Klein modes, the mass terms of action in Eq.~(\ref{eq3.2}) take the following form~\cite{Dienes:1998sb,Dvali:1999cn}
 \begin{equation}\label{eq3.5}
\sum_{n=-\infty}^{\infty}m_i^D \bar{\nu}_{iL} \psi_{iR}^{(n)}+ \sum_{n=1}^{\infty} \frac{n}{R_{\rm ED}} \left( \overline{\psi_{iL}^{(n)}} \psi_{iR}^{(n)} - \overline{\psi_{iL}^{(-n)}} \psi_{iR}^{(-n)}\right) + \rm{h.c.},
\end{equation}
where $\psi_{iR}^{(n)}$ and $\psi_{iL}^{(n)}$ are the $n^{\rm th}$ KK mode of the bulk fermions $\psi_{iR}$ and $\psi_{iL}$, respectively. The $m_i^D$ are the three mass parameters that form the diagonal Dirac mass matrix $m_{\rm diag}^D$ in this basis, which in turn results from the diagonalization of the matrix 
\begin{equation}\label{eq3.6}
m^D_{ij}=\frac{h_{ij}vM_F^{-D/2}}{\sqrt{V_D}}=h_{ij}v\frac{M_F}{M_{\rm{pl}}},
\end{equation}
where for writing the las term we have used the relation in Eq.~(\ref{eq3.1}). (Please note that the KK modes have a pre factor proportional to $V_D^{-1/2}$.) Therefore the coupling in the above equation becomes 
\begin{equation}\label{eq3.7}
\frac{M_F}{M_{\rm{pl}}}\simeq10^{-16}\frac{M_F}{1~\rm{TeV}},
\end{equation}
and neutrinos can get a naturally small Dirac mass (For $M_F\sim1~$TeV, $m^D_i\sim10^{-5}$~eV). Let us define the following basis of fields:
\begin{eqnarray}\label{eq3.8}
\nu^{(0)}_{iR} & = &  \psi^{(0)}_{iR},\nonumber\\
\nu^{(n)}_{iR} & = &  \frac{\psi^{(n)}_{iR}+\psi^{(-n)}_{iR}}{\sqrt 2},~~~~n=1,\ldots,\infty,\nonumber\\
\nu^{(n)}_{iL} & = &  \frac{\psi^{(n)}_{iL} -\psi^{(-n)}_{iL}}{\sqrt 2},~~~~n=1,\ldots,\infty,
\end{eqnarray}
and the combinations orthogonal to $\nu^{(n)}_{iR}$ and $\nu^{(n)}_{iL}$ which since they decouple from the system we ignore them. In this basis the mass terms in Eq.~(\ref{eq3.5}) can be written as $\overline{L}_i M_i R_i$, where $L_i^T=\left(\nu_{iL},\nu^{(n)}_{iL}\right)$, $R_i^T=\left(\nu^{(0)}_{iR},\nu^{(n)}_{iR}\right)$ and
\begin{equation}\label{eq3.9}
M_i=\lim_{n\to\infty}
\begin{pmatrix}
m_i^D & \sqrt{2}m_i^D & \sqrt{2}m_i^D & \sqrt{2}m_i^D & \ldots & \sqrt{2}m_i^D\\
0 & 1/R_{\rm{ED}} & 0 & 0 & \ldots & 0\\
0 & 0 & 2/R_{\rm{ED}} & 0 & \ldots & 0\\
\vdots & \vdots & \vdots & \vdots& \ddots & \vdots\\
0 & 0 & 0 & 0 & \cdots & n/R_{\rm{ED}}
\end{pmatrix}.
\end{equation}
As can be seen the mass matrix $M_i$ is not diagonal and so we will call the basis of $L_i$ and $R_i$ the ``pseudo-mass" basis.

The Schr\"{o}dinger-like evolution equation of the whole physical states, that is the active neutrinos and the KK modes $\nu^{(n)}_{iL}$, including the matter potentials (which in our case are induced by the Earth's matter) can be written in the pseudo-mass basis as ($k=1,2,3$)\\
\begin{equation}\label{eq3.10}
\left[ i \frac{d}{dr} L_k = \frac{1}{2E_\nu} M_k^\dagger M_k L_k+ \sum_{j=1}^3
\begin{pmatrix}
X_{kj} & 0_{1\times n} \\
0_{n\times 1} & 0_{n\times n}
\end{pmatrix} L_j \right]_{n\to\infty}~,
\end{equation} 
where $X_{kj} = \sum_{\alpha} U_{\alpha k}^\ast U_{\alpha j} V_\alpha$, and
\begin{equation}\label{eq3.11}
V_\alpha = \delta_{e\alpha} V_{\rm CC} + V_{\rm NC} = \sqrt{2} G_F \left( \delta_{e\alpha} N_e  - \frac{N_n}{2} \right)~,
\end{equation}
where $n_e$ and $n_n$ are the electron and neutron number density profiles, respectively. The same evolution equation applies to antineutrinos with the replacement $X_{kj} \to -X_{kj}$. Eq.~(\ref{eq3.10}) has the following explicit form \cite{Machado:2011jt}:
\begin{eqnarray}\label{eq3.12}
{\tiny i \frac{d}{dr}
\begin{pmatrix}
\nu_{1L}\\
\nu_{2L}\\
\nu_{3L}\\
\nu_{1L}^{(1)}\\
\nu_{2L}^{(1)}\\
\nu_{3L}^{(1)}\\
\nu_{1L}^{(2)}\\
\nu_{2L}^{(2)}\\
\nu_{3L}^{(2)}\\
\vdots\\
\nu_{1L}^{(N)}\\
\nu_{2L}^{(N)}\\
\nu_{3L}^{(N)}\\
\end{pmatrix}
=\frac{1}{2ER^2_{\rm{ED}}}
\begin{pmatrix}
\begin{array}{@{}*{13}{c}@{}}
\eta_1+V_{11}&V_{12}&V_{13}&\xi_1&0&0&2\xi_1&0&0&\cdots&N\xi_1&0&0\\
V_{21}&\eta_2+V_{22}&V_{23}&0&\xi_2&0&0&2\xi_2&0&\cdots&0&N\xi_2&0\\
V_{31}&V_{32}&\eta_3+V_{33}&0&0&\xi_3&0&0&2\xi_3&\cdots&0&0&N\xi_3\\
\xi_1&0&0&1&0&0&0&0&0&\cdots&0&0&0\\
0&\xi_2&0&0&1&0&0&0&0&\cdots&0&0&0\\
0&0&\xi_3&0&0&1&0&0&0&\cdots&0&0&0\\
2\xi_1&0&0&0&0&0&4&0&0&\cdots&0&0&0\\
0&2\xi_2&0&0&0&0&0&4&0&\cdots&0&0&0\\
0&0&2\xi_3&0&0&0&0&0&4&\cdots&0&0&0\\
\vdots&\vdots&\vdots&\vdots&\vdots&\vdots&\vdots&\vdots&\vdots&\ddots&\vdots&\vdots&\vdots\\
N\xi_1&0&0&0&0&0&0&0&0&\cdots&N^2&0&0\\
0&N\xi_2&0&0&0&0&0&0&0&\cdots&0&N^2&0\\
0&0&N\xi_3&0&0&0&0&0&0&\cdots&0&0&N^2\\
\end{array}
\end{pmatrix}}\nonumber\\
\times
{\tiny \begin{pmatrix}
\nu_{1L}\\
\nu_{2L}\\
\nu_{3L}\\
\nu_{1L}^{(1)}\\
\nu_{2L}^{(1)}\\
\nu_{3L}^{(1)}\\
\nu_{1L}^{(2)}\\
\nu_{2L}^{(2)}\\
\nu_{3L}^{(2)}\\
\vdots\\
\nu_{1L}^{(N)}\\
\nu_{2L}^{(N)}\\
\nu_{3L}^{(N)}\\
\end{pmatrix}}_{N\to\infty},
~~~~~~~~~~~~~~~~~~~~~~~~~~~~~~~~~~~~~~~~~~~~~~~~~~~~~~~~~~~~~~~~~~~~~~~~~~~~~~~~~~~~~
\end{eqnarray}
where $\xi_i=\sqrt{2}m_iR_{\rm{ED}}$, $V_{ij}=2ER_{\rm{ED}}^2X_{ij}$, and $\eta_i=(N+\frac{1}{2})\xi_i^2$.
An immediate interpretation of the set of evolution equations in Eq.~(\ref{eq3.10}) is that, from the brane ($y=0$) point of view, the KK modes $\nu^{(n)}_{iL}$ (for each $i$, and $n=1,2,\ldots$) constitute a tower of sterile neutrinos which their masses (and also the masses of the active states $\nu_{iL}$) can be obtained by the diagonalization of the matrix $M_i^\dagger M_i$. The matrices $M_i^\dagger M_i$ can be diagonalized by changing the basis from pseudo-mass basis $L_i=(\nu_{iL},\nu_{iL}^{(n)})^T$ to the ``true" mass basis $L^\prime_i=(\nu_{iL}^\prime,\nu_{iL}^{\prime(n)})^T$, where $L^\prime_i = S^\dagger_i L_i$ and $S^\dagger_i M_i^\dagger M_i S_i = (M_i^\dagger M_i)_{\rm diag}$. The active flavor neutrino states $\nu_{\alpha L}$ can be expanded in terms of the ``true" mass basis as
\begin{equation}\label{eq3.13}
\nu_{\alpha L} = \sum_{i=1}^3 U_{\alpha i} \nu_{iL}  = \sum_{i=1}^3 U_{\alpha i} \sum_{n=0}^\infty S_i^{0n} \nu_{iL}^{\prime(n)}~,
\end{equation}
where $S_i^{0n}$ is the $0n$ element of the matrix $S_i$ and we defined $\nu_{iL}^{\prime(0)}\equiv\nu_{iL}^{\prime}$. The eigenvalues $\left(\lambda_i^{(n)}\right)^2$ of the matrices $R_{\rm ED}^2 M_i^\dagger M_i$ are the roots of the following transcendental equation~\cite{Dienes:1998sb}
\begin{equation}\label{eq3.14}
\lambda_i-\pi \left(m_i^DR_{\rm ED}\right)^2 \cot(\pi\lambda_i)=0~.
\end{equation}
So the mass\footnote{These are the masses in vacuum. The matter potentials will modify these masses in the usual way.} of each state $\nu^{\prime(n)}_{iL}$ in $L_i^\prime$ is $\lambda_i^{(n)}/R_{\rm ED}$. The matrix elements $S_i^{0n}$ are given by~\cite{Dienes:1998sb}
\begin{equation}\label{eq3.15}
\left(S^{0n}_i\right)^2=\frac{2}{1+\pi^2 \left(m_i^DR_{\rm ED}\right)^2 + \left(\lambda^{(n)}_i\right)^2 /\left(m_i^DR_{\rm ED}\right)^2}~.
\end{equation}
It can be shown that Eq.~(\ref{eq3.14}) has infinite number of solutions $\lambda^{(n)}_i$ where $n < \lambda^{(n)}_i < n+0.5$. Thus, the masses of the KK modes $\nu^{\prime(n)}_{iL}$ ($n\neq0$) are increasing roughly as $\sim n/R_{\rm ED}$, while the contribution of the KK modes to the active flavor states (that is $S_i^{0n}$) decreases by increasing $n$. The decrease of the active-sterile mixings by the increase of $n$ means that the higher KK modes gradually decouple from the evolution equation in Eq.~(\ref{eq3.10}), and so for an experimental setup sensitive to a known energy range we need to consider only a finite number of the KK modes. 
It is easy to find the solutions of Eq.s~(\ref{eq3.14}) and (\ref{eq3.15}) for $m_iR_{\rm{ED}}\ll1$ and $m_iR_{\rm{ED}}\gg1$. In these limits we find:
\begin{eqnarray}\label{eq3.16}
m_i^DR_{\rm{ED}}\ll1&:&\nonumber\\
\lambda_i^{(0)}&=&m_i^DR_{\rm{ED}}\Big(1-\frac{\pi^2{(m_i^{D})^2}R_{\rm{ED}}^2}{6}+\cdots\Big),\nonumber\\
S^{00}_i&=&1-\frac{\pi^2}{6}{(m_i^{D})^2}R_{\rm{ED}}^2+\cdots,\nonumber\\
\lambda_i^{(k)}&=&k+\frac{{(m_i^{D})^2}R_{\rm{ED}}^2}{k}+\cdots,~~~~~~~(k=1,2,\cdots),\nonumber\\
S^{0k}_i&=&\frac{\sqrt{2}m_i^DR_{\rm{ED}}}{k}+\cdots,
\end{eqnarray}
\begin{eqnarray}\label{eq3.17}
m_i^DR_{\rm{ED}}\gg1&:&\nonumber\\
\lambda^{(n)}&=&(n+\frac{1}{2})\Big(1-\frac{1}{\pi^2{(m_i^{D})^2}R_{\rm{ED}}^2}+\cdots\Big),~~~~~~~~~(n=0,1,2,\cdots)\nonumber\\
S^{0n}_i&=&\frac{\sqrt{2}}{\pi m_i^DR_{\rm{ED}}}.
\end{eqnarray}
In the following we discuss the number of KK modes that should be considered for the analysis of the IceCube atmospheric neutrino data.

In the high energy range ($E_\nu \gtrsim 0.1$~TeV) the Earth's matter effects dramatically change the oscillation pattern of atmospheric neutrinos in the LED model. The matter potentials modify the oscillation phases which lead to resonant conversion of the active neutrinos to the KK modes comprising the tower of sterile neutrinos with increasing masses. The resonance condition in the $2\nu$ approximation of $\nu_{iL}^{(0)}-\nu_{iL}^{\prime(n)}$ system with the effective mixing angle denoted by $\vartheta_n$ is (See Eq.~(\ref{eq1.23}))
\begin{equation}\label{eq3.18}
\frac{\Delta m^2}{2E_\nu}\cos2\theta_n=\frac{\left(\lambda^{(n)}_i\right)^2 - \left(\lambda^{(0)}_i\right)^2}{2E_\nu R_{\rm ED}^2} \cos2\vartheta_n = V_\alpha~.
\end{equation}  
Due to the sign of $V_\alpha$ for the Earth's matter ($V_e>0$, while $V_\mu, V_\tau < 0$), the resonance condition in Eq.~(\ref{eq3.18}) can be fulfilled for $\nu_e$, $\bar{\nu}_\mu$ and $\bar{\nu}_\tau$; which means that at energies satisfying the condition in Eq.~(\ref{eq3.18}) the $\nu_e$ ($\bar{\nu}_{\mu/\tau}$) converts to the sterile flavor KK mode $\nu_{sL}^{(n)}$ ($\bar{\nu}_{sL}^{(n)}$). The atmospheric neutrino flux at high energies is dominated by $\nu_\mu$ and $\bar{\nu}_\mu$ with the $\nu_e$ and $\bar{\nu}_e$ components suppressed at least by a factor of ~$\sim20$~\cite{Honda:2006qj}. On the other hand, in this work we analyze the so-called muon-track events in IceCube which originate from the charged current interactions of $\nu_\mu$ and $\bar{\nu}_\mu$ with the nuclei in the detector. Thus, the main signature of the LED model in the high energy atmospheric neutrinos is in the muon-flavor survival probabilities. Before passing, let us mention two points. Firstly, in the LED model for each flavor of the active neutrinos (or equivalently for each mass eigenstate) there is a tower of KK modes. So, just by considering the first mode ($n=1$) three different mass-squared differences can be inserted in Eq.~(\ref{eq3.18}):
$$
\frac{\left(\lambda^{(1)}_1\right)^2 - \left(\lambda^{(0)}_1\right)^2}{R_{\rm ED}^2}~~,~~\frac{\left(\lambda^{(1)}_2\right)^2 - \left(\lambda^{(0)}_2\right)^2}{R_{\rm ED}^2}~~\rm{and}~~\frac{\left(\lambda^{(1)}_3\right)^2 - \left(\lambda^{(0)}_3\right)^2}{R_{\rm ED}^2}, 
$$
which lead to three different resonance energies. However, for $R_{\rm ED} \lesssim 10^{-4}$~cm the first KK mode masses are large enough (for reasonable values of $m_i^D$) such that all the active-sterile mass-squared differences are almost equal and effectively there is just one mass-squared difference for each $n$. The current upper limit on $R_{\rm ED}$ from oscillation experiments is $\sim10^{-4}$~cm~\cite{Machado:2011jt} and so the three mass-squared differences for each $n$ are degenerate. Secondly, although for the numerical calculations in sections~\ref{sec:osc} and \ref{sec:icecube} we use the exact position-dependent mass density profile of the Earth from the PREM model~\cite{1981PEPI...25..297D}, in the analytical description of the oscillation pattern we assume a constant average density $\bar{\rho} = 5.5~{\rm g~cm}^{-3}$ for the core-crossing atmospheric neutrinos. The resonances described in Eq.~(\ref{eq3.18}) are constant density MSW resonances and the variability of matter density is not playing a significant role except for the core crossing trajectories where the castle wall configuration of mantle-core-mantle leads to the parametric resonances~\cite{Liu:1997yb}.

Let us study the series of resonance energies from Eq.~(\ref{eq3.18}). By increasing $n$, ${\cos2\vartheta_n \to 1}$ and $\left(\lambda^{(n)}_i\right)^2 \propto n^2$; so for the resonance energy of conversion to the $n^{\rm th}$ KK mode we obtain $E_\nu^{{\rm res},(n)} \simeq\frac{ n^2}{2VR^2_{\rm{ED}}}$. For $\left(\left(\lambda^{(n)}_i\right)^2 - \left(\lambda^{(0)}_i\right)^2\right)/R_{\rm ED}^2 = 1~{\rm eV}^2$ the resonance energy for core crossing trajectories of $\bar{\nu}_\mu$ (that is $\cos\theta_z=-1$) is\footnote{The MSW resonance energy from Eq.~(\ref{eq3.18}) is $\sim4$~TeV. However, for trajectories passing through the core of Earth the parametric resonance dominates at $\sim2.5$~TeV~\cite{Liu:1997yb}.} $\sim 2.5$~TeV. Thus, the series of resonance energies for the atmospheric $\bar{\nu}_\mu$ conversion to the KK modes (assuming $\cos\theta_z=-1$) are
\begin{equation}\label{eq3.19}
E_{\nu}^{{\rm res},(n)} \simeq 10n^2 ~{\rm TeV} \left( \frac{10^{-5}~{\rm cm}}{R_{\rm ED}} \right)^2~.
\end{equation} 
For the neutrinos passing just the mantle ($\cos\theta_z \gtrsim-0.8$) the resonance energies are $\simeq~16n^2 ~{\rm TeV} \left( 10^{-5}~{\rm cm}/R_{\rm ED} \right)^2$. At high energies ($E_\nu\gtrsim 0.1$~TeV) in the standard $3\nu$ framework the muon-flavor survival probability is $P(\bar{\nu}_\mu\to\bar{\nu}_\mu)=1$; while, qualitatively from Eq.~(\ref{eq3.19}), in the LED model a series of dips exist at energies $E_\nu^{{\rm res},(n)}$ ($n=1,2,\ldots$), which reflect the conversion of $\bar{\nu}_\mu$ to the $n^{\rm th}$ KK sterile states. The infinite number of resonance energies can be truncated at some $n$ for two reasons: 1) by increasing $n$ the resonance energy increases while the flux of atmospheric neutrinos decreases by the increase of energy as $\propto E_\nu^{-2.7}$. So, the statistics at higher KK modes resonance energies are low and IceCube (or in general any neutrino telescope) would not be sensitive to these KK modes. 2) By the increase of $n$ the mixing between the active and the $n^{\rm th}$ KK mode states decreases ($\sin\vartheta_n\simeq\sqrt{2}m_i^DR_{\rm ED}/n$) which leads to less intense active to sterile conversion. So the depth of resonance dips decrease by the increase of energy and for the large values of $n$ it is beyond the sensitivity reach of the detector. 

In this chapter we analyze the atmospheric neutrino data collected during two phases of IceCube construction IC-40~\cite{Abbasi:2010ie} and IC-79~\cite{Aartsen:2013jza} (the numbers mean that at the period of data collection 40 and 79 strings bearing DOMs were deployed, out of the final 86 strings). The energy range of IC-40 and IC-79 data sets are $(0.1-400)$~TeV and $(0.1-10)$~TeV respectively\footnote{The IC-79 data set consists of two high energy and low energy subsets~\cite{Aartsen:2013jza}. For our analysis the high energy subset is relevant which its energy range is $(0.1-10)$~TeV.}. Taking 100~TeV as the energy where above it the statistics are too low, from Eqs.~(\ref{eq3.18}) and (\ref{eq3.19}) the resonance energies are within the energy range of IC-40 and IC-79 for $n\lesssim3~(R_{\rm ED}/10^{-5}~{\rm cm})$. By inserting the current upper limit $R_{\rm ED}\lesssim10^{-4}$~cm it means that at least $\sim30$ KK modes should be taken into account in the calculation of oscillation probabilities. On the other hand, IceCube is sensitive\footnote{This is the sensitivity of IceCube from the analysis of zenith distribution of muon-track events. Adding the energy information improves the sensitivity by a factor of few for resonance energies $\lesssim10$~TeV~\cite{Esmaili:2013vza}. We elaborate more on this in section~\ref{sec:icecube}.} to the active-sterile mixing angles $\sin^22\vartheta_n\gtrsim0.1$~\cite{Esmaili:2012nz}. From Eq.~(\ref{eq3.15}) this sensitivity can be translated to (assuming $m_i^D R_{\rm ED} \ll 1$)
\begin{equation}\label{eq3.20}
n \lesssim 4.5 \left( \frac{R_{\rm ED}}{10^{-5}~{\rm cm}} \right) \left( \frac{\max\left[m_i^D,\sqrt{\Delta m_{\rm atm}^2}\right]}{{\rm eV}} \right)~.
\end{equation}       
The ``max" function in the above relation comes from the fact that if $m_1^D\to 0$, although the mixing between $\nu_{1L}^{(0)}$ and $\nu_{1L}^{(1)}$ vanishes, but the mixing between $\nu_{3L}^{(0)}$ and $\nu_{3L}^{(1)}$ is still sizable because $\lambda_3^{(0)}=\sqrt{\left(\lambda_1^{(0)}\right)^2+R_{\rm ED}^2\Delta m_{\rm atm}^2}$ is not zero. Plugging the current bounds on $m_i^D$ and $R_{\rm ED}$ from~\cite{Machado:2011jt} into Eq.~(\ref{eq3.20}) we obtain $n\lesssim3$. Thus, practically very few KK modes contribute substantially to the oscillation pattern of the atmospheric neutrinos. In the above discussion we assumed that all the sensitivity of IceCube to the sterile neutrinos originate from the resonance region; while the interference terms in lower energies are also important and so a few more KK modes should be taken into account. As a conservative assumption, in the numerical calculations of the next section we consider $n=5$ KK modes in the evolution equations.

\section{Numerical calculation of the oscillation probabilities\label{sec:osc}}

The oscillation probabilities of the active neutrinos can be found by solving the set of evolution equations in Eq.~(\ref{eq3.10}). In the case of zero matter potential (in vacuum), the oscillation probability of ${\nu_\alpha}\to{\nu_\beta}$ is given by
\begin{equation}\label{eq3.21}
P_{\nu_{\alpha}\to\nu_{\beta}}=\Bigg|\sum_{i=1}^3U^{\alpha i}{U^*}^{\beta i}\sum_{n=0}^{\infty}(S_i^{0n})^2\exp (-i\frac{{\lambda_i^{(n)}}^2L}{2E_{\nu}R^2_{\rm{ED}}})\Bigg|^2,
\end{equation}
where $U$ is the $3\times3$ PMNS matrix, $L$ is the baseline and $E_\nu$ is the energy of neutrinos. In the rest of this section we find the solutions of Eq.~(\ref{eq3.10}) for nonzero matter potential. As we mentioned before, for the high energy atmospheric neutrinos which is our interest in this work, the relevant channel is the survival probability $P(\bar{\nu}_\mu\to\bar{\nu}_\mu)$, or more generally the oscillation probabilities of $\nu_\mu\to\nu_\alpha$ and $\bar{\nu}_\mu\to\bar{\nu}_\alpha$.  

As we discussed and justified in section~\ref{sec:formalism_ch3}, we consider $n=5$ KK modes in our numerical calculations. The initial conditions for the calculation of $\bar{\nu}_\mu$ oscillation probabilities in the pseudo-mass basis of Eq.~(\ref{eq3.10}) are $L_i^j=\delta^{j}_{0} U_{\mu i}^\ast$ ($i=1,2,3$ are the indices of the 3 neutrino generations and $j=0,1,\cdots,n$), where $L_i^j$ is the $j^{\rm th}$ component of $L_i$ (defined before Eq.~(\ref{eq3.9})) and elements of the PMNS matrix $U$ are fixed to their best-fit values (see Eq.~(\ref{eq1.4}) and the discussion after that)~\cite{GonzalezGarcia:2012sz}. The values of $m_i^D$ depend on the mass hierarchy of the active neutrinos. For normal hierarchy (NH) $m_2^D= \sqrt{(m_1^{D})^2+\Delta m_{\rm 21~(sol)}^2}$ and $m_3^D= \sqrt{(m_1^D)^2+\Delta m_{\rm 31~(atm)}^2}$; and so $m_1^D$ and $R_{\rm ED}$ are the free parameters of the model. For inverted hierarchy (IH) $m_1^D\simeq m_2^D\simeq\sqrt{(m_3^D)^2+\Delta m_{\rm 31~(atm)}^2}$; and so $m_3^D$ and $R_{\rm ED}$ are the free parameters\footnote{A technical note: To be precise, the relations between masses should be applied to the eigenvalues $\lambda_i^{(0)}$, since the masses of the active neutrinos in the LED model is given by $\lambda_i^{(0)}/R_{\rm ED}$. Therefore, for example for the inverted hierarchy case, we would write $\lambda_1^{(0)} \simeq \lambda_2^{(0)} \simeq \sqrt{\left(\lambda_3^{(0)}\right)^2+R_{\rm ED}^2 \Delta m_{\rm 31~(atm)}^2}$. Then, by knowing the values of $\lambda_1^{(0)}$ and $\lambda_2^{(0)}$ we can calculate $m_1^D$ and $m_2^D$ from Eq.~(\ref{eq3.14}), which can be used to calculate $\lambda_i^{(n)}$ by the same equation. This procedure have been discussed in detail in~\cite{BastoGonzalez:2012me}. However, in the region of parameter space where we are interested in (and also taking into account the current bounds), it can be shown that applying the mass relations to $m_i^D$ lead to the same results and we can ignore this technical point.}. However, in the high energy range ($E_\nu\gtrsim 0.1$~TeV), since the oscillations driven by $\Delta m_{\rm atm}^2$ and $\Delta m_{\rm sol}^2$ are suppressed and the first KK mode is much heavier than the active neutrino states, the oscillation pattern is the same for both NH and IH and so we show the oscillation probabilities just for the NH case. Finally, in our numerical calculation, for the matter potential $X_{kj}$ in Eq.~(\ref{eq3.10}) we used the PREM model~\cite{1981PEPI...25..297D}. The Preliminary reference Earth model (PREM) is a one-dimensional model representing the average Earth properties as a function of the radius of the earth. The density of the different layers of the earth from the PREM model is shown in Table~\ref{table3.1}. 
\begin{table}
\begin{tabular}{|c|c|c|}
\hline
\rm{Region}&\rm{Radius~(km)}&\rm{Density~}(gr~cm$^{-3}$)\\
\hline
\rm{Inner core}&$0-1221.5$&$13.08-8.83x^2$\\
\rm{Outer~core}&$1221.5-3480.0$&$12.58-1.26x-3.64x^2-5.53x^3$\\
\rm{Lower mantle}&$3480.0-5701.0$&$7.96-6.47x+5.53x^2-3.08x^3$\\
\rm{Transition zone}&$5701.0-5771.0$&$5.32-1.48x$\\
~~~~~~~~&$5771.0-5971.0$&$11.25-8.03x$\\
~~~~~~~~&$5971.0-6151.0$&$7.11-3.80x$\\
\rm{LVZ}&$6151.0-6291.0$&$2.69+0.69x$\\
\rm{LID}&$6291.0-6346.6$&$2.69+0.69x$\\
\rm{Crust}&$6346.6-6356.0$&$2.90$\\
&$6356.0-6368.0$&$2.60$\\
\rm{Ocean}&$6368.0-6371.0$&$1.02$\\
\hline
\end{tabular}
\caption{The density of different layers of the earth based on the PREM model~\cite{1981PEPI...25..297D}. Here the variable $x$ is the normalized radius $x=R/R_{\oplus}$, where $R_{\oplus}=6371.0$~km. }
\label{table3.1}
\end{table}

\begin{figure}[t!]
\centering
\includegraphics[width=0.6\textwidth]{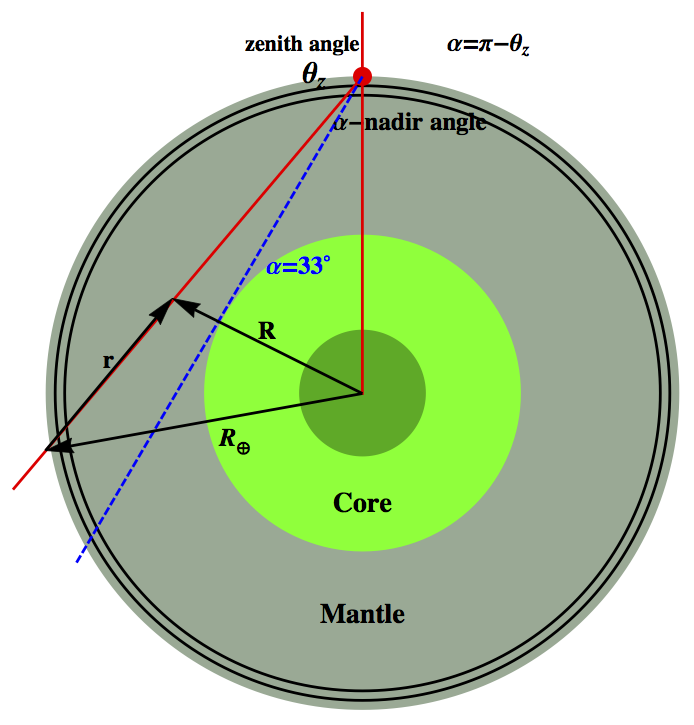}
\label{fig:matter_in_earth}
\caption{\label{fig:matter_in_earth}The trajectory of neutrinos inside the earth }
\end{figure}

As it was mentioned before, to find the evolution of neutrinos for a varying density, we have to solve Eq. (\ref{eq3.10}) numerically.  Atmospheric neutrinos have a huge energy range, from $10$ GeV to $100$ TeV. The baseline of the atmospheric experiments can vary from a few km, to the diameter of the earth. The neutrinos are produced in the atmosphere, in a thin layer around the earth. They cross various layers of the earth and are detected in the underground/under-water/under-ice detectors. Therefore, we deal with oscillations in multi-medium layer, and the density is a function of the radius of different layers; therefore, a function of the zenith angle of the earth:
$$
r=\sqrt{R^2+R_{\oplus}^2+2RR_{\oplus}\cos\theta_z},
$$ 
where $r$ is the distance neutrinos are traveling inside the earth after they are produced in the atmosphere, $R$ is the radius of different layers, and $\theta_z$ is the zenith angle (See Fig. \ref{fig:matter_in_earth}) (Here we are neglecting the depth of the detector inside the earth the the hight the atmospheric neutrinos are produced.). The total distance neutrinos travel to get to the underground detector is $L=-2R_\oplus\cos\theta_z$. The zenith angle $\theta_z=\pi$ corresponds to the directions where neutrinos are coming exactly from down, crossing the centre of the earth, hence passing through all layers of the earth, while $\theta_z=\pi/2$ is for the horizontal direction. The atmospheric neutrinos passing through the earth can experience matter effect $\sim ~1-15~\rm{gr}/\rm{cm}^3$.

\begin{figure}[t!]
\centering
\subfloat[$P(\bar{\nu}_\mu\to\bar{\nu}_\mu)$]{
\includegraphics[width=0.5\textwidth]{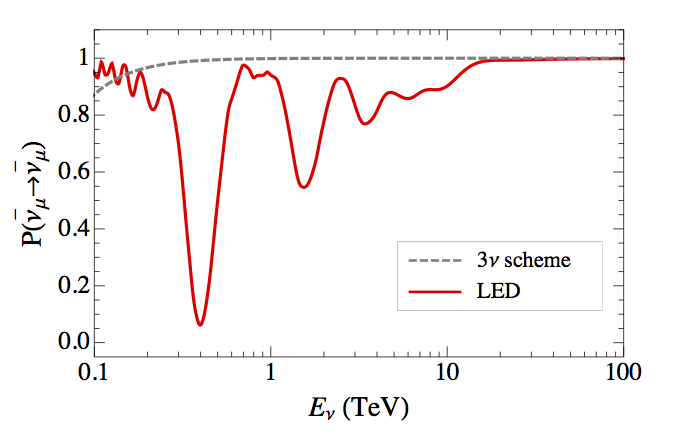}
\label{fig:LEDprob-mubar}
}
\subfloat[$P(\nu_\mu\to\nu_\mu)$]{
\includegraphics[width=0.5\textwidth]{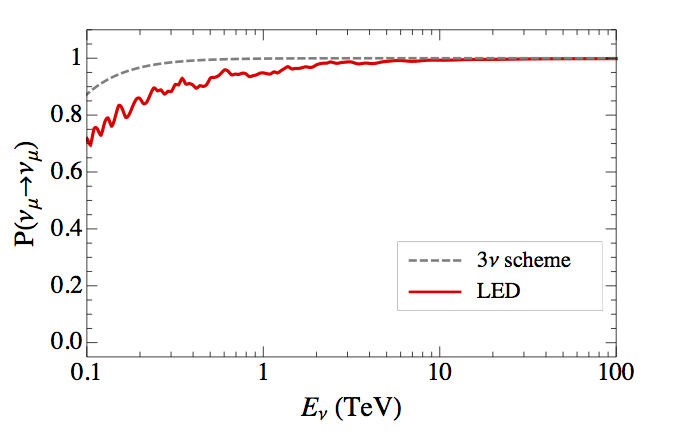}
\label{fig:LEDprob-mu}
}
\quad
\subfloat[$P(\bar{\nu}_\mu\to\bar{\nu}_\tau)$]{
\includegraphics[width=0.5\textwidth]{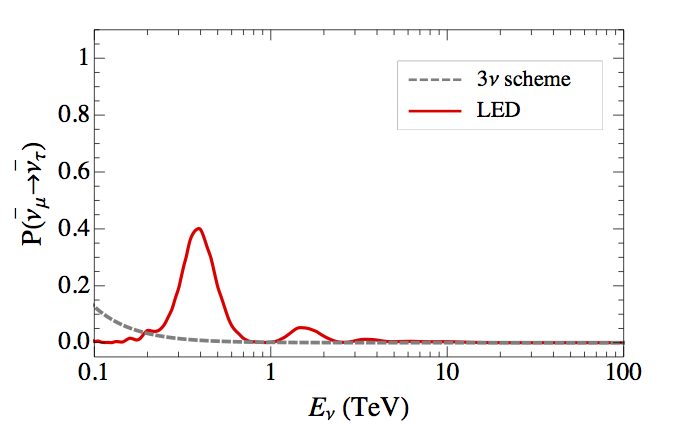}
\label{fig:LEDprob-taubar}
}
\subfloat[$P(\nu_\mu\to\nu_\tau)$]{
\includegraphics[width=0.5\textwidth]{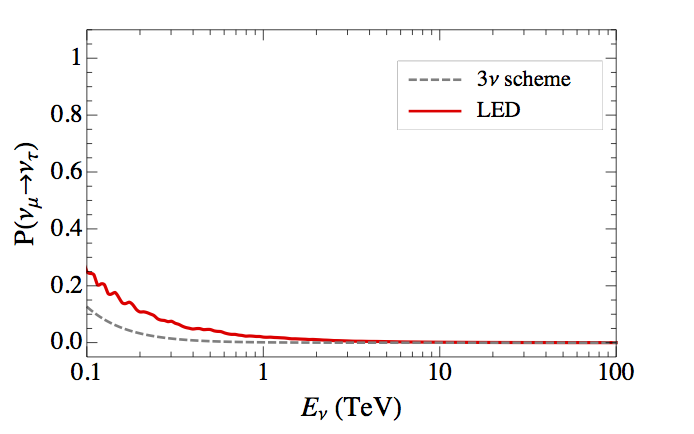}
\label{fig:LEDprob-tau}
}
\caption{\label{fig:LEDprob}The oscillation probabilities as function of neutrino energy $E_\nu$ for $\cos\theta_z=-1$. In all the panels $m^D_1=0.01~\rm{eV}$ and $R_{\rm{ED}}=5\times10^{-5}~\rm{cm}$. The oscillation channel is denoted in each subcaption. In all panels, the gray dashed and red solid curves are for the standard $3\nu$ scheme and the LED model, respectively.}
\end{figure}

Figures~\ref{fig:LEDprob-mubar} and \ref{fig:LEDprob-taubar} show the oscillation probabilities of $\bar{\nu}_\mu\to\bar{\nu}_\mu$ and $\bar{\nu}_\mu\to\bar{\nu}_\tau$, respectively. Correspondingly, Figures~\ref{fig:LEDprob-mu} and \ref{fig:LEDprob-tau} are for $\nu_\mu$ oscillation probabilities. In all the figures we assumed $m_1^D=0.01$~eV and $R_{\rm ED}=5\times10^{-5}$~cm, and the plots are for neutrinos passing the diameter of the Earth, which is $\cos\theta_z=-1$. The gray dashed and red solid curves are for the standard $3\nu$ scheme and the LED model, respectively. The resonances discussed in Eq.~(\ref{eq3.18}) can be seen in Figure~\ref{fig:LEDprob-mubar}. As we expected, the resonances exist just for $\bar{\nu}_\mu$. For $R_{\rm ED}=5\times10^{-5}$~cm, Eq.~(\ref{eq3.19}) gives $$E^{\rm res}_\nu=(0.4,1.6,3.6,6.4,10)~\rm{TeV}$$ for the first five resonance energies which match the position of dips in Figure~\ref{fig:LEDprob-mubar}. The decreasing depth of the dips for the higher KK modes is a consequence of the decreasing mixing angle between $\nu_{iL}^{(0)}$ and $\nu_{iL}^{(n)}$ ($\sin\vartheta_{n}\propto1/n$). The $\nu_\mu\to\nu_e$ and $\bar{\nu}_\mu\to\bar{\nu}_e$ oscillation probabilities are not shown since in both $3\nu$ scheme and the LED model the matter potential $V_e$ suppresses oscillation and the oscillation probability is zero for $E_\nu\gtrsim 0.1$~TeV. The nonzero oscillation probability $\bar{\nu}_\mu\to\bar{\nu}_\tau$ in Figure~\ref{fig:LEDprob-taubar}, showing as peaks at the resonance energies, are due to the $\bar{\nu}_\tau - \bar{\nu}_{s}^{(n)}$ mixings (we will discuss it in section~\ref{sec:3+n}, see also~\cite{Esmaili:2013cja,Choubey:2007ji}).

The oscillation probabilities for the trajectories passing the mantle ($\cos\theta_z\gtrsim-0.8$) are qualitatively similar to Figure~
\ref{fig:LEDprob}, while the resonances are at $$E^{\rm res}_\nu=(0.64,2.56,5.76,10.24,16)~\rm{TeV} $$(for the same values of $m_1^D$ and $R_{\rm ED}$ as in 
Figure~\ref{fig:LEDprob}) and the dips are less profound due to the absence of the parametric resonance for these trajectories.    

In Figure~\ref{fig:LEDprob} the oscillation probabilities are shown for fixed values of $m_1^D$ and $R_{\rm ED}$. However, to confront the IceCube data with the expectation from the LED model, we would scan all the parameter space of $(m_1^D,R_{\rm ED})$. We will report the result of this analysis in section~\ref{sec:icecube}. In the next section we elaborate on the interpretation of Figure~\ref{fig:LEDprob} in terms of the $(3+n)$ scenario.

\section{The equivalence between the LED and $(3+n)$ models\label{sec:3+n}}

The KK modes in the LED model resemble a tower of sterile neutrinos from the brane point of view. There is a tower of sterile neutrinos for each flavor of the active neutrinos (or equivalently for each mass eigenstates of the active neutrinos), so an LED model with $n$ KK modes can be considered as a $(3+3n)$ model consisting of 3 active neutrinos and $3n$ sterile neutrinos. The translation of the LED model to a $(3+3n)$ model provides a better intuitive understanding of the results presented in the previous section, especially since there are already a rich literature on the oscillation pattern of the high energy atmospheric neutrinos in the $(3+1)$ model~\cite{Esmaili:2012nz,Esmaili:2013vza,Esmaili:2013cja,Razzaque:2011ab,Choubey:2007ji,Nunokawa:2003ep,Peres:2000ic}, which can be easily generalized to the $(3+3n)$ model. 

Let us briefly summarize the active-sterile mixing in the $(3+3n)$ model. The mixing matrix in this scenario is a $(3+3n)\times(3+3n)$, unitary matrix $W_{3+3n}$ which can be parametrized by $(3n+3)(3n+2)/2$ mixing angles. Among these mixing angles, there are $3$ active-active ($\theta_{12}$, $\theta_{13}$, $\theta_{23}$) and $3\times3n$ active-sterile ($\theta_{i,j}$, $i=1,2,3$ and $4\le j\le3n$) mixing angles, while $3n(3n-1)/2$ angles quantify the sterile-sterile mixings of the $3n$ sterile states ($\theta_{l,m}$,  $4\le l< m\le3n$). Since the sterile states do not enter the charged current interactions, these angles are not relevant to the phenomenology of the active neutrinos on the brane. We parametrize this mixing angle as following (we assume CP symmetry in lepton sector):
\begin{equation}\label{eq3.22}
W_{3+3n} = \prod_{j=2}^{3+3n} \left( \prod_{i=1}^{j-1} R_{ij} (\theta_{ij}) \right)~,
\end{equation}
where the ordered product is defined as $\prod_{i=1}^k A_i = A_k A_{k-1} \ldots A_1$, and $R_{ij} (\theta_{ij})$ is the rotation matrix in the $ij$ plane by the angle $\theta_{ij}$. The active flavor states $\nu_{\alpha L}$ are related to the mass eigenstates $\nu_j$ by
\begin{equation}\label{eq3.23}
\nu_{\alpha L} = \sum_{j=1}^{3+3n} \left( W_{3+3n} \right)_{\alpha j} \nu_j~.
\end{equation}
By identifying the mass eigenstates $\nu^{\prime(q)}_{iL}\equiv\nu_{3q+i}$ (where $q$ is the KK index, and $i=1,2,3$), the comparison of Eq.~(\ref{eq3.23}) and Eq.~(\ref{eq3.13}) enables us to derive the values of the elements of mixing matrix $W_{3+3n}$ in terms of the LED model parameters, which are $R_{\rm ED}$ and $m_1^D$. Let us calculate the active-eterile mixing angles for the case of only 1 KK mode. In this case the mixing matrix becomes
\begin{eqnarray}\label{eq3.24}
W_{3+3}&=&R_{56}(\theta_{56})\cdots R_{16}(\theta_{16})R_{45}(\theta_{45})\cdots R_{15}(\theta_{15})\nonumber\\&\times&R_{34}(\theta_{34})R_{24}(\theta_{24})R_{14}(\theta_{14})R_{23}(\theta_{23})R_{13}(\theta_{13})R_{12}(\theta_{12}).
\end{eqnarray}
Therefore, the relation between the flavor and mass eigenstates in Eq.~(\ref{eq3.23}) for $n=1$ is given by
\begin{eqnarray}\label{eq3.25}
\nu_{\alpha L}&=&\sum_{j=1}^{6} \left( W_{3+3} \right)_{\alpha j} \nu_j\nonumber\\
&=& \left( W_{3+3} \right)_{\alpha 1}\nu^{\prime(0)}_{1L}+\left( W_{3+3} \right)_{\alpha 2}\nu^{\prime(0)}_{2L}+\left( W_{3+3} \right)_{\alpha 3}\nu^{\prime(0)}_{3L}\nonumber\\
&+&\left( W_{3+3} \right)_{\alpha 4}\nu^{\prime(1)}_{1L}+\left( W_{3+3} \right)_{\alpha 5}\nu^{\prime(0)}_{2L}+\left( W_{3+3} \right)_{\alpha 6}\nu^{\prime(0)}_{3L}.
\end{eqnarray}
On the other hand, from Eq.~(\ref{eq3.13})  we have
\begin{eqnarray}\label{eq3.26}
\nu_{\alpha L} &=& U_{\alpha 1}S_1^{00}\nu^{\prime(0)}_{1L}+U_{\alpha 2}S_2^{00}\nu^{\prime(0)}_{2L}+U_{\alpha 3}S_3^{00}\nu^{\prime(0)}_{3L}\nonumber\\
&+&U_{\alpha 1}S_1^{01}\nu^{\prime(1)}_{1L}+U_{\alpha 2}S_2^{01}\nu^{\prime(1)}_{2L}+U_{\alpha 3}S_3^{01}\nu^{\prime(1)}_{3L}.
\end{eqnarray}
By comparing Eq.~(\ref{eq3.25}) and Eq.~(\ref{eq3.26}), we find that $\left( W_{3+3} \right)_{\alpha~ 3m+p}=U_{\alpha p}S_p^{0m}$, or equivalently
\begin{eqnarray}\label{eq3.27}
\left( W_{3+3} \right)_{\alpha 1}&=&U_{\alpha 1}S_1^{00},\nonumber\\
\left( W_{3+3} \right)_{\alpha 2}&=&U_{\alpha 2}S_2^{00},\nonumber\\
\left( W_{3+3} \right)_{\alpha 3}&=&U_{\alpha 3}S_3^{00},\nonumber\\
\left( W_{3+3} \right)_{\alpha 4}&=&U_{\alpha 1}S_1^{01},\nonumber\\
\left( W_{3+3} \right)_{\alpha 5}&=&U_{\alpha 2}S_2^{01},\nonumber\\
\left( W_{3+3} \right)_{\alpha 6}&=&U_{\alpha 3}S_3^{01}.\nonumber\\
\end{eqnarray}
Finally using the parametrization of $W_{3+3}$ in Eq.~(\ref{eq3.24}) we can explicitly find the relation of the active-sterile mixing angles as a function of the LED parameters for 1 KK mode:
\begin{eqnarray}
s_{16}&=&U_{e3}S_3^{01},\label{eq3.28}\\
 s_{15} &=&\frac{ U_{e2}S_2^{01}}{c_{16}},\label{eq3.29}\\
 s_{14} &=&\frac{U_{e1}S_1^{01}}{c_{16}c_{15}},\label{eq3.30}\\
  s_{26} &=& \frac{U_{\mu3}S_3^{01}}{c_{16}},\label{eq3.31}\\
 s_{25} &=&  \frac{s_{15} s_{16} s_{26} + U_{\mu2}S_2^{01}}{c_{15}c_{26}},\label{eq3.32}\\
s_{24} &=&\frac{s_{14}(-c_{26} s_{15} s_{25}-c_{15} s_{16} s_{26}) - U_{\mu1}S_1^{01}}{c_{14}c_{25}c_{26}},\label{eq3.33}\\
  s_{36} &=&\frac{U_{\tau3}S_3^{01}}{c_{16}c_{26}},\label{eq3.34}\\
 s_{35} &=&\frac{ -c_{26}s_{15}s_{16}s_{36} - c_{15} s_{25}s_{26}s_{36} - U_{\tau2}S_2^{01}}{c_{15}c_{25} c_{36}},\label{eq3.35}\\
 s_{34} &=&(-c_{25}c_{36} s_{14} s_{15}s_{35} - 
   c_{14}c_{36} s_{24}s_{25}s_{35} -c_{15} c_{26}s_{14}s_{16}s_{36} - 
  c_{14}c_{25}s_{24}s_{26}s_{36}\nonumber\\
   &+&s_{14}s_{15}s_{25}s_{26}s_{36} - U_{\tau1}S_1^{01})/(-c_{14}
   c_{24}c_{35}c_{36} ),\label{eq3.36}
\end{eqnarray}
where $s_{ij}~(c_{ij})\equiv\sin\theta_{ij}~(\cos\theta_{ij})$.\\

\begin{figure}[t!]
\centering
\subfloat[$P(\bar{\nu}_\mu\to\bar{\nu}_\mu)$]{
\includegraphics[width=0.5\textwidth]{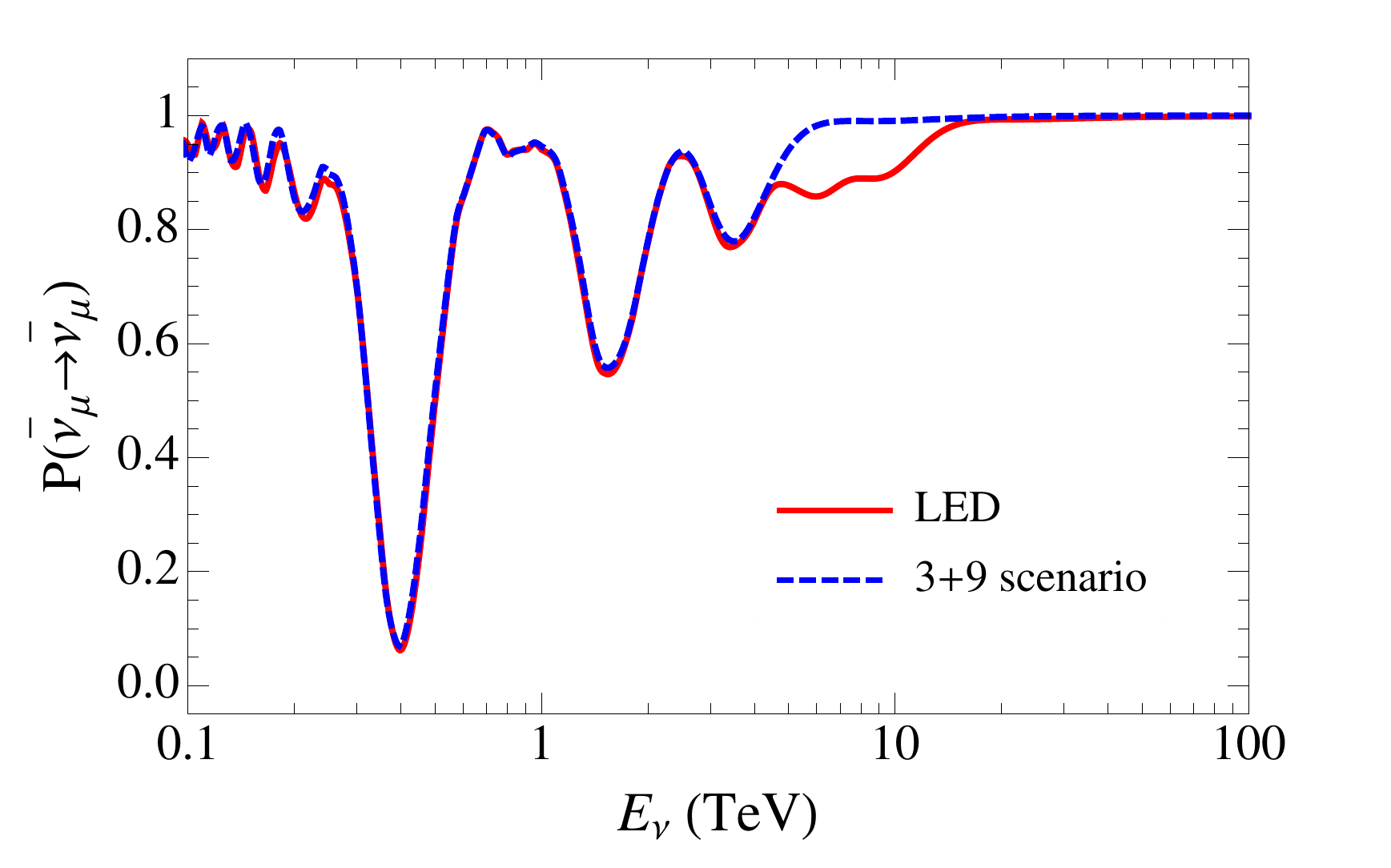}
\label{fig:compare-mu}
}
\subfloat[$P(\bar{\nu}_\mu\to\bar{\nu}_\tau)$]{
\includegraphics[width=0.5\textwidth]{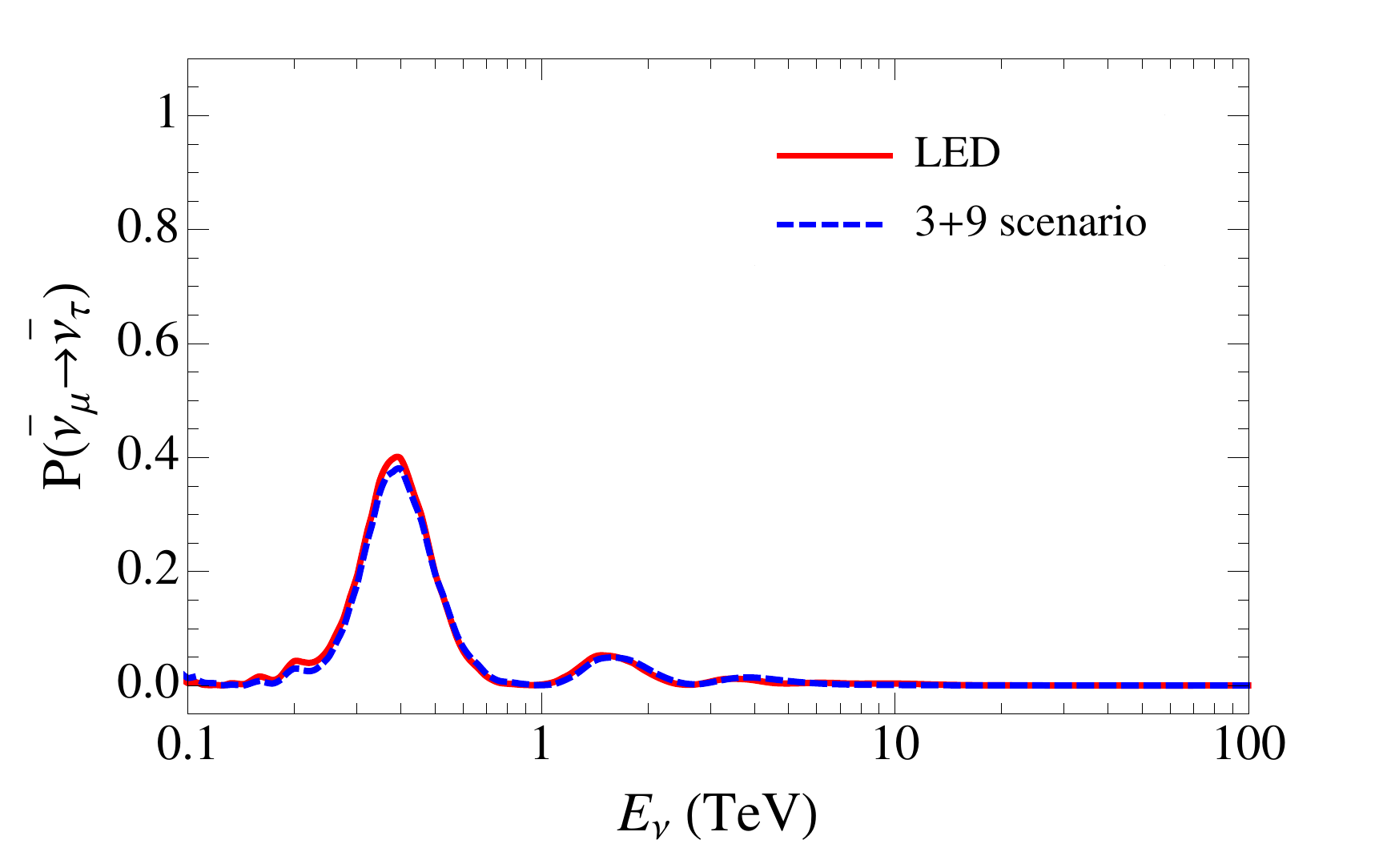}
\label{fig:compare-tau}
}
\caption{\label{fig:compare}Comparison of the oscillation probabilities calculated in the LED and the $(3+3n)$ models. For the LED model, we assume $5$ KK modes, $m_1^D=0.01$~eV and $R_{\rm ED}=5\times10^{-5}$~cm. For the $(3+3n)$ scenario we assume $n=3$ sterile neutrino states. Oscillation probabilities are for neutrinos passing the diameter of Earth ($\cos\theta_z=-1$). As can be seen up to the third KK mode the calculation in both models agree. By considering $(3+3n)$ scenario with larger $n$ this agreement extends to higher KK modes.}
\end{figure}

In order to elaborate on this equivalence between the LED model and the $(3+3n)$ model, in Figure~\ref{fig:compare} we compare the oscillation probabilities calculated in both models. In both panels of Figure~\ref{fig:compare} the red solid curve is for the LED model, the same as the one shown in Figure~\ref{fig:LEDprob}, with 5 KK modes. The dashed blue line corresponds to $(3+3n)$ scenario with $n=3$. The dashed blue line is obtained by solving the following evolution equation     
\begin{equation}\label{eq3.37}
i\frac{d\nu_{\alpha}}{dr} = \left[ \frac{1}{2E_\nu} W_{3+3n} \mathbf{M}^2 W_{3+3n}^\dagger + \mathbf{V}(r) \right]_{\alpha\beta} \nu_{\beta}~,
\end{equation}
where $\alpha,\beta=e,\mu,\tau,s_1,\ldots,s_{3n}$ (the $s_i$ is the $i^{\rm th}$ sterile flavor eigenstate). The elements of $W_{3+3n}$ are obtained the same way we calculated $W_{3+3}$; and $\mathbf{M}^2$ is a $(3+3n)\times (3+3n)$ diagonal matrix where the elements are mass-squared differences 
$$\mathbf{M}^2 = {\rm diag} \left( 0, \Delta m_{21}^2,\Delta m_{31}^2, \Delta m_{41}^2,\ldots, \Delta m_{3+3n,1}^2 \right)~,$$ 
where for the $q^{\rm{th}}$ KK mode ($q\geq1$ and we are assuming $m_1^D R_{\rm ED}\ll1$) 
$$\Delta m_{3+q,1}^2=\Delta m_{3+q+1,1}^2=\Delta m_{3+q+2,1}^2 = \frac{q^2}{R_{\rm ED}^2}~.$$
The potential matrix in Eq.~(\ref{eq3.37}) is given by 
$$\mathbf{V}(r) = \sqrt{2}G_F{\rm diag} \left( N_e(r), 0,0, \frac{1}{2}N_n(r), \ldots, \frac{1}{2}N_n(r) \right)~.$$
Since we are assuming $n=3$ in $(3+3n)$ scenario, in the comparison of the LED model with $n=5$ KK modes the oscillation probabilities in both models should match up to the third KK mode and for higher KK modes deviations should appear.

Figures~\ref{fig:compare-mu} and \ref{fig:compare-tau} show the oscillation probabilities $P(\bar{\nu}_\mu\to\bar{\nu}_\mu)$ and $P(\bar{\nu}_\mu\to\bar{\nu}_\tau)$, respectively. As can be seen in panel (a), the probabilities match up to the third KK mode; the same is in panel (b), although since the peaks are very small the deviation in higher KK modes is not visible. In panel (a) clearly the deviation can be seen for the fourth and fifth KK mode resonances. Now, with the equivalence we are discussing in this section, it is easy to understand the peak in panel (b). It originates from the nonzero value of $(W_{3+3n})_{\tau j}$, that is the mixing between $\nu_\tau$ and the sterile states. This effective conversion of $\bar{\nu}_\mu\to\bar{\nu}_\tau$ when $(W_{3+3n})_{\tau j}\neq0$ has been already studied in the literature~\cite{Esmaili:2013cja,Choubey:2007ji}. In fact this effective conversion is the source of the sensitivity of the cascade events in IceCube to $\theta_{3,3+3n}$ angles, which are poorly constrained by the current experiments (see the discussion in~\cite{Esmaili:2013cja}).

Although we discussed the equivalence between the LED model with $n$ KK modes and $(3+3n)$ scenario, this equivalence can be further simplified to the $(3+n)$ scenario. As we mentioned in section~\ref{sec:formalism_ch3}, for $R_{\rm ED}\lesssim10^{-4}$~cm \footnote{If $R_{\rm ED}\sim10^{-4}$~cm, the related mass squared difference for the first KK mode becomes $\Delta m^2_{41}=1/R_{\rm ED}^2=0.04$~eV$^2$; therefore, the mass of the first mostly sterile state would be $m_4\sim0.2$~eV.} even the first KK mode states are much heavier than the active neutrino states and so effectively the three states $\nu_{1L}^{\prime(n)}$, $\nu_{2L}^{\prime(n)}$ and $\nu_{3L}^{\prime(n)}$ of the $n^{\rm th}$ KK mode are degenerate in mass. Thus, in principle it would be possible to redefine the states in each KK mode in such a way that, in two flavors approximation of the active-sterile oscillation, just one of the new states mixes with the active neutrinos and the other two decouple. By this redefinition of states, the LED model with $n$ KK modes would be equivalent (at two flavors approximation) to the $(3+n)$ model, which has much fewer mixing parameters than the $(3+3n)$ model. In the following we elaborate on this equivalence and derive the corresponding effective mixing parameter values in the $(3+n)$ model.

In the phenomenology of the high energy atmospheric neutrino oscillation in the presence of the sterile neutrinos it is always possible to reduce the active-sterile mixing patterns to the two-flavor systems of $\nu_e-\nu_{s_p}$, $\nu_\mu-\nu_{s_p}$ and $\nu_\tau-\nu_{s_p}$. In this approximation the oscillation of $\nu_e$, $\nu_\mu$ and $\nu_\tau$ flavors to the $p^{\rm th}$ sterile state $\nu_{s_p}$ can be described by the effective mixing angles $\vartheta_{ep}$, $\vartheta_{\mu p}$ and $\vartheta_{\tau p}$ respectively. In the LED model the expansion of the active flavor neutrino states in terms of the mass eigenstates in Eq.~(\ref{eq3.13}) can be written as 
\begin{equation}\label{eq3.38}
\begin{pmatrix}
\nu_{eL} \\
\nu_{\mu L} \\
\nu_{\tau L}
\end{pmatrix}
= \sum_{n=0}^\infty U\mathcal{S}^{(n)} 
\begin{pmatrix}
\nu_{1L}^{\prime(n)} \\
\nu_{2L}^{\prime(n)} \\
\nu_{3L}^{\prime(n)}
\end{pmatrix},
\end{equation}
where $U$ is the PMNS matrix and $\mathcal{S}^{(n)}$ is a $3\times3$ diagonal matrix with the elements $\mathcal{S}^{(n)}={\rm diag}(S_1^{0n},S_2^{0n},S_3^{0n})$. For a fixed $n\geq1$ we can change the basis $(\nu_{1L}^{\prime(n)},\nu_{2L}^{\prime(n)},\nu_{3L}^{\prime(n)})$ to a new basis $(\tilde{\nu}_{1L}^{(n)},\tilde{\nu}_{2L}^{(n)},\tilde{\nu}_{3L}^{(n)})_\alpha$ such that in this new basis just ${\tilde{\nu}_{1L}^{(n)}}_\alpha$ contributes to $\nu_{\alpha L}$ state and the two states ${\tilde{\nu}_{2L}^{(n)}}_\alpha$ and ${\tilde{\nu}_{3L}^{(n)}}_\alpha$ decouple from the active neutrino $\nu_{\alpha L}$ and just contribute to the sterile flavor states. Let us do these calculations explicitly. We can expand Eq.~(\ref{eq3.38}) for a fixed $n$, similar 
\begin{eqnarray}\label{eq3.39}
\nu_{\alpha L} &=& U_{\alpha 1}S_1^{00}\nu^{\prime(0)}_{1L}+U_{\alpha 2}S_2^{00}\nu^{\prime(0)}_{2L}+U_{\alpha 3}S_3^{00}\nu^{\prime(0)}_{3L}\nonumber\\
&+&U_{\alpha 1}S_1^{01}\nu^{\prime(1)}_{1L}+U_{\alpha 2}S_2^{01}\nu^{\prime(1)}_{2L}+U_{\alpha 3}S_3^{01}\nu^{\prime(1)}_{3L}+\cdots~.
\end{eqnarray}
The new state ${\tilde{\nu}_{1L}^{(n)}}_\alpha$ will be defined as
\begin{eqnarray}\label{eq3.40}
{\tilde{\nu}_{1L}^{(n)}}_\alpha:=\frac{1}{N^{(n)}_\alpha}\sum_{i=1}^3U_{\alpha i}S_i^{0n}\nu^{\prime(n)}_{iL},
\end{eqnarray}
where $N^{(n)}_\alpha=\sqrt{\sum_{i=1}^3|U_{\alpha i}S_i^{0n}|^2}$ is a normalization constant. Thus Eq.~(\ref{eq3.39}) takes the form
\begin{eqnarray}\label{eq3.41}
\nu_{\alpha L} &=& \sum_{i=1}^3U_{\alpha i}S_i^{00}\nu^{\prime(0)}_{iL}+\sum_{n=1}^{\infty}N^{(n)}_\alpha{\tilde{\nu}_{1L}^{(n)}}_\alpha,
\end{eqnarray}
and as far as the evolution of $\nu_\alpha$ is concerned, the model is effectively a $(3+n)$ model. So, in the two-flavor system of $\nu_\alpha-\nu_{s_n}$ the effective mixing angle is given by (for $n\geq1$)
\begin{equation}\label{eq3.42}
\sin \vartheta_{\alpha n} =N^{(n)}_\alpha= \left[ \sum_{i=1}^3 \left| U_{\alpha i} S_{i}^{0n} \right|^2 \right]^{1/2}~.
\end{equation}

\begin{figure}[t!]
\centering
\subfloat[]{
\includegraphics[width=0.5\textwidth]{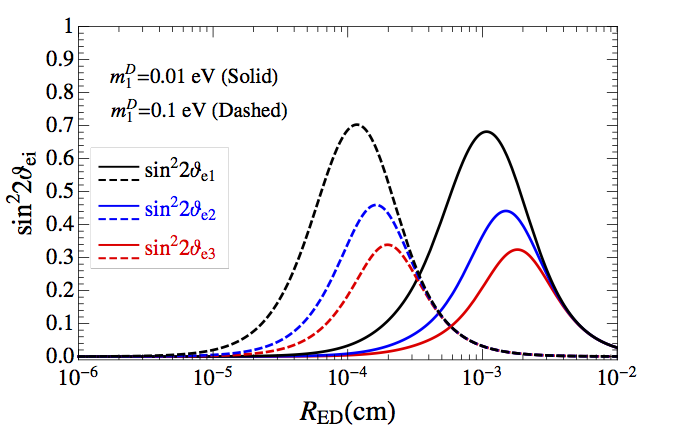}
\label{fig:theta-1n}
}
\subfloat[]{
\includegraphics[width=0.5\textwidth]{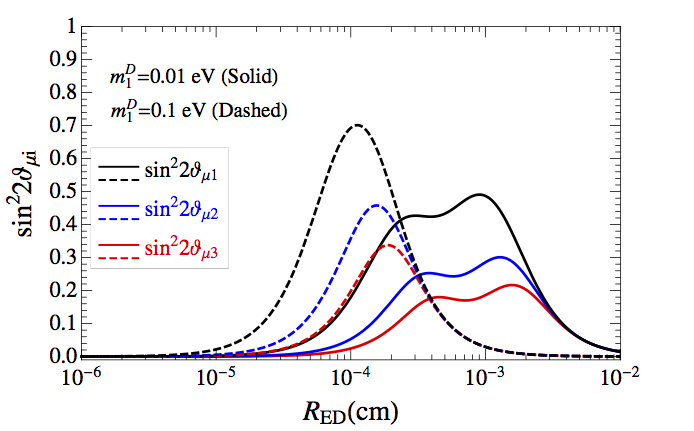}
\label{fig:theta-2n}
}
\quad
\subfloat[]{
\includegraphics[width=0.5\textwidth]{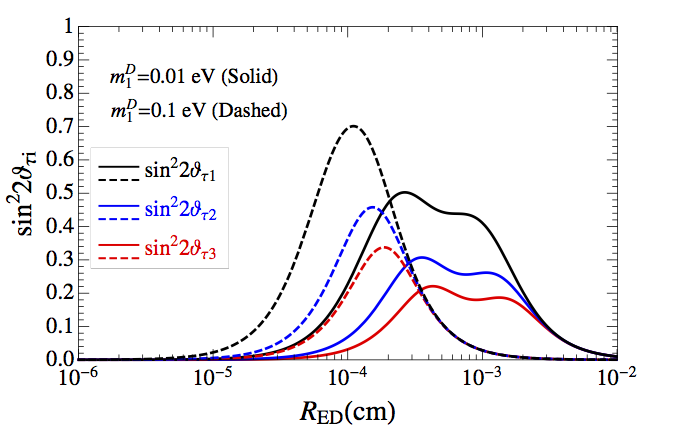}
\label{fig:theta-3n}
}
\caption{\label{fig:theta-in}The effective mixing angles $(\vartheta_{en},\vartheta_{\mu n},\vartheta_{\tau n})$ in the $(3+3)$ scenario which is equivalent to the LED model with 3 KK modes. In all the plots the solid and dashed curves correspond respectively to $m_1^D=0.01$~eV and $0.1$~eV.}
\end{figure}

Figure~\ref{fig:theta-in} shows the corresponding effective active-sterile mixing angles of the $(3+3)$ scenario equivalent to the LED model with 3 KK modes ($n=3$), as function of $R_{\rm ED}$ for $m_1^D=0.1$~eV (dashed curves) and $0.01$~eV (solid curves). The peak-shape behavior of the curves in all the panels originate from the behavior of $S_i^{0n}$. It can be shown from Eq.~(\ref{eq3.15}) that the maxima of $S_i^{0n}$ occur at $m_i^D R_{\rm ED}\simeq \sqrt{n/\pi}$. This condition is easily obtained from Eq.~(\ref{eq3.15}) by putting $\lambda_i^{(n)}\sim n$. We will call it the maximum condition hereafter. In fact in each curve of Figure~\ref{fig:theta-in} there are three peaks at values of $R_{\rm ED}$ derived from\footnote{Notice that there are few percent uncertainties in this computation since we are approximating $\lambda_i^{(n)}\simeq n$; while more accurately $\lambda_i^{(n)}$ is a number between $n$ and $n+1/2$.} $m_i^D R_{\rm ED}\simeq \sqrt{n/\pi}$ for $i=1,2,3$; and the relative heights of these peaks are controlled by the relative size of $U_{\alpha i}$ (where $\alpha=e$, $\mu$ and $\tau$, respectively for $\vartheta_{en}$, $\vartheta_{\mu n}$ and $\vartheta_{\tau n}$). However, as far as $m_1^D\gtrsim\sqrt{\Delta m_{\rm atm}^2}$, since in this case $m_1^D\sim m_2^D\sim m_3^D$, hence $S_1^{0n}\sim S_2^{0n} \sim S_3^{0n}$, and due to the unitarity of the $U$ matrix, Eq.~(\ref{eq3.42}) will become $\sin \vartheta_{\alpha n}=S_1^{0n}$. Therefore, the three peaks for $\alpha=e,\mu\tau$ coincide and effectively one peak can be recognized. This coincidence of the peaks can be seen for the case $m_1^D=0.1$~eV depicted by the black dashed curves in Figure~\ref{fig:theta-in}. For $m_1^D=0.1$~eV and $n=1$, from maximum condition we obtain $R_{\rm ED}\simeq1.1\times10^{-4}$~cm which agrees with the peak's positions of black dashed curves in Figure~\ref{fig:theta-in}. For higher KK modes the peak position slightly moves to the larger $R_{\rm ED}$, proportional to $\sqrt{n}$. The separation of peaks is visible for smaller values of $m_1^D$. Let us consider the case $m_1^D=0.01$~eV depicted by solid curves in Figure~\ref{fig:theta-in}. In this case $m_2^D\simeq m_1^D$ while $m_3^D\simeq\sqrt{\Delta m_{\rm atm}^2}=0.05$~eV. Thus, for $n=1$, the maximum condition leads to two peaks at $R_{\rm ED}\simeq2.3\times10^{-4}$~cm and $1.1\times10^{-3}$~cm which clearly can be identified in the black solid curves of Figures~\ref{fig:theta-2n} and \ref{fig:theta-3n}. The first peak (which is due to $m_3^D$) is not visible in the black solid curve of Figures~\ref{fig:theta-1n} since it is suppressed by the small value of $U_{e3}$. For higher $n$, again the peaks slightly move to larger $R_{\rm ED}$ (compare different colors of solid curves in each panel). For $m_1^D\lesssim\sqrt{\Delta m_{\rm sol}^2}\simeq9\times10^{-3}$~eV a third peak in large values of $R_{\rm ED}$ will develop. However, notice that for $m_1^D\to0$ the position of peaks originating from $m_2^D=\sqrt{\Delta m_{\rm sol}^2}$ and $m_3^D=\sqrt{\Delta m_{\rm atm}^2}$ do not change, which means that always there are two peaks at $R_{\rm ED}=2.3\times10^{-4}$~cm and $1.3\times10^{-3}$~cm for both $\vartheta_{\mu n}$ and $\vartheta_{\tau n}$. Thus, we can immediately conclude that for $m_1^D\lesssim10^{-2}$~eV the sensitivity of IceCube to the LED model is independent of the value of $m_1^D$. By inspecting the black solid curve in Figure~\ref{fig:theta-2n}, it can be seen that $\sin^22\vartheta_{\mu 1}\simeq0.1$ for $R_{\rm ED}\simeq5\times10^{-5}$~cm and so IceCube would be able to constrain $R_{\rm ED}$ at this level for $m_1^D\lesssim10^{-2}$~eV.  

Let us discuss the case of $m_1^D\gtrsim0.1$~eV. In this case, as can be seen also from the dashed curves in Figure~\ref{fig:theta-in}, all the three peaks coincide (since $m_1^D\simeq m_2^D\simeq m_3^D$) at $R_{\rm ED}\simeq \sqrt{n/\pi}/m_1^D$. This means that by increasing $m_1^D$, IceCube will be sensitive to smaller values of $R_{\rm ED}$ such that the sensitivity contour in the log-log plot of $(R_{\rm ED},m_1^D)$ plane will be a straight line with the slope $-1$. The intercept of this line can be estimated from Figure~\ref{fig:theta-in}. From the black dashed curve in Figure~\ref{fig:theta-2n}, it can be seen that $\sin^22\vartheta_{\mu 1}\simeq0.1$ for $R_{\rm ED}\simeq2\times10^{-5}$~cm. From this we conclude that IceCube would be able to constrain LED radius down to $R_{\rm ED} \simeq 2\times10^{-6}({\rm eV}/m_1^D)$~cm for $m_1^D\gtrsim0.1$~eV. We should mention that large values of $m_1^D$ have severe conflicts with the bounds on neutrino mass from cosmological considerations such that $m_1^D\gtrsim1$~eV can be ruled out robustly~\cite{Ade:2013zuv}.

\section{Constraining the LED model with the IceCube data\label{sec:icecube}}

In section~\ref{sec:osc} we calculated the flavor oscillation probabilities of the high energy atmospheric neutrinos in the LED model. In this section we analyze the collected atmospheric data in IceCube to search for the signatures of the LED model in the zenith distribution of events. Although, as we have shown in section~\ref{sec:3+n}, the oscillation probabilities can be calculated in the equivalent $(3+3n)$ or $(3+n)$ scenarios, for the analysis of this section the calculations have been done in the original LED model assuming 5 KK modes. However, for the interpretation of the results obtained in this section, we extensively use the terminology of the $(3+n)$ scenario, that is the effective mixing angles in  Eq.~(\ref{eq3.42}).

We analyze two sets of the IceCube data, IC-40~\cite{Abbasi:2010ie} and IC-79~\cite{Aartsen:2013jza}.
These data sets provide the zenith distribution of events and so in our analysis we would consider just the integrated number of events over the energy. We will discuss later the improvements that can be achieved by adding the energy information of events. The number of the muon-track events in the $i^{\rm th}$ bin of the zenith angle $\Delta_i\cos\theta_z$ can be calculated by
\begin{eqnarray}\label{eq3.43}
N_i= T\Delta\Omega \sum_{\alpha=e,\mu} &\Bigg\{& \int dE \int_{\Delta_i} d\cos\theta_z A^{\nu}_{\rm{eff}}(E,\cos\theta_z) \nonumber\\
&\times&\Phi_{\nu_{\alpha}}(E,\cos\theta_z) P(\nu_{\alpha}\to \nu_{\mu})\Bigg\} +(\nu\to\bar{\nu})~,
\end{eqnarray}
where $T$ is the data-taking period, 359 and 319 days respectively for IC-40 and IC-79; $\Delta\Omega=2\pi$ is the azimuthal acceptance of the IceCube detector, $\Phi_{\nu_\alpha}$ is the atmospheric $\nu_\alpha$ flux and $A_{\rm eff}^\nu$ is the neutrino effective area. Finally, the $P(\nu_\alpha\to\nu_\mu)$ in Eq.~(\ref{eq3.43}) is the neutrino oscillation probability which is discussed in section~\ref{sec:osc}. 
    
\begin{figure}[t!]
\centering
\includegraphics[width=0.8\textwidth]{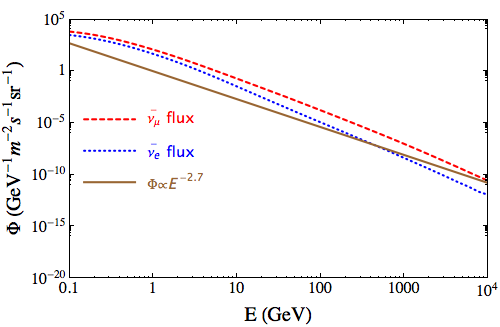}
\caption{\label{fig:atmospheric_neutrino_flux} The conventional flux of the atmospheric neutrinos for $\bar{\nu}_\mu$ (the red dashed curve) and $\bar{\nu}_e$ (the blue dotted curve), taken from~\cite{Honda:2006qj}. The brown solid curve shows an $E^{-2.7}$ power law for comparison.  }
\end{figure}

\begin{figure}[t!]
\centering
\setcounter{subfigure}{0}
\subfloat[IceCube-40 $\nu_\mu$ effective area]{
\includegraphics[width=0.5\textwidth]{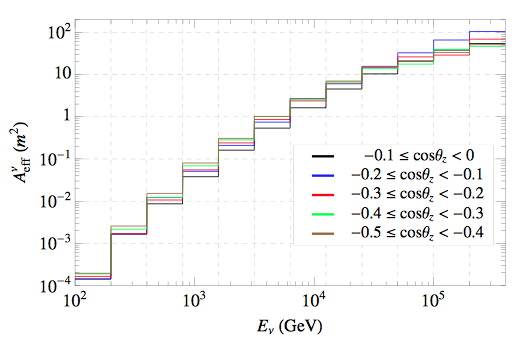}
\label{fig:prob11}
}
\subfloat[IceCube-40 $\bar{\nu}_\mu$ effective area]{
\includegraphics[width=0.5\textwidth]{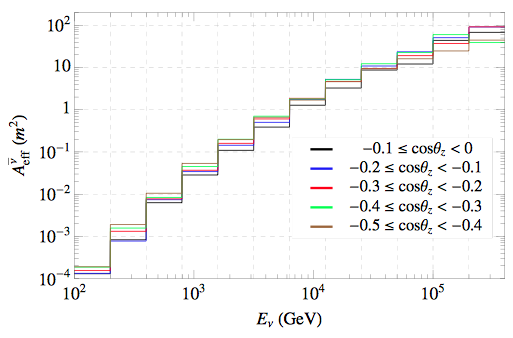}
\label{fig:prob2}
}
\quad
\subfloat[IceCube-40 $\nu_\mu$ effective area]{
\includegraphics[width=0.5\textwidth]{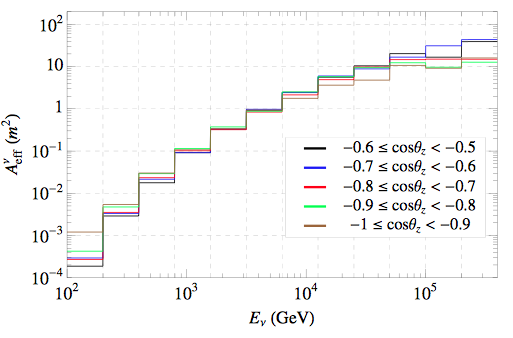}
\label{fig:prob3}
}
\subfloat[[ceCube-40 $\bar{\nu}_\mu$ effective area]{
\includegraphics[width=0.5\textwidth]{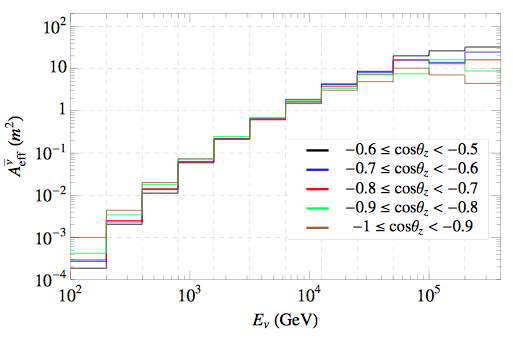}
\label{fig:prob4}
}
\caption{\label{fig:effective_area_IC40} The effective area of IC-40 experiment for  $\nu_\mu$ (panels a and c) and $\bar{\nu}_\mu$ (panels b and d) as a function of energy and zenith angle, taken from \cite{Esmaili:2012nz}. }
\end{figure}

The conventional atmospheric neutrino flux $\Phi_{\nu_\alpha}$ in energy $E>100~$GeV is mainly composed of $\nu_\mu$ and $\bar{\nu}_\mu$ which are produced by the interactions of the primary cosmic rays with the nuclei in the atmosphere of the earth. The interactions of the primary cosmic rays with the nuclei in the atmosphere produce pions and kaons. These mesons decay into muon and muon neutrinos, while muons themselves decay into electrons and electron/muon neutrinos. Therefore, in low energies $(E_{\nu}< 1$~GeV), the flux of muon (anti-) neutrinos is 2 times more than the flux of the electron (anti-)neutrinos. However, in the range of energy we are interested in our work, the flux of $\nu_e$ is more than 1 order of magnitude smaller. Although the $\nu_e$ and $\bar{\nu}_e$ atmospheric fluxes at high energies are quite small, we consider them for the sake of completeness. The flux of the tau neutrinos is negligible in this energy range. The conventional atmospheric flux of $\nu_\mu$, $\bar{\nu}_\mu$, $\nu_e$ and $\bar{\nu}_e$ for energies above $10$~GeV and different bins of zenith angle is published in~\cite{Honda:2006qj}. In Fig.~\ref{fig:atmospheric_neutrino_flux} we have shown the flux of atmospheric neutrinos for $\bar{\nu}_\mu$ (the red dashed curve) and $\bar{\nu}_e$ (the blue dotted curve). As the figure shows, the flux decreases rapidly, approximately $\propto E^{-2.7}$ in GeV-TeV energy range (The solid brown curve). Hence, the calculated neutrino flux decreases rapidly with the increasing energy.

The neutrino effective area $A^{\nu}_{\rm{eff}}$ in Eq.~(\ref{eq3.43}) is defined as the equivalent area of the detector for which the probability of the neutrino detection would be $100\%$. In principle, the concept of the neutrino effective area is used to describe the response of the detector to the flavor, energy and zenith angle distribution of neutrinos. For IC-40 we have taken $A^{\nu}_{\rm{eff}}$ from~\cite{Esmaili:2012nz} (See Fig.~\ref{fig:effective_area_IC40}). Since the effective area of IC-79 is not published yet, we have estimated it using the already known effective area of IC-40, by rescaling it in a way that the simulated expected number of events for IC-79 was produced, as explained in \cite{Esmaili:2013fva}. 

\begin{figure}[t!]
\centering
\includegraphics[width=0.7\textwidth]{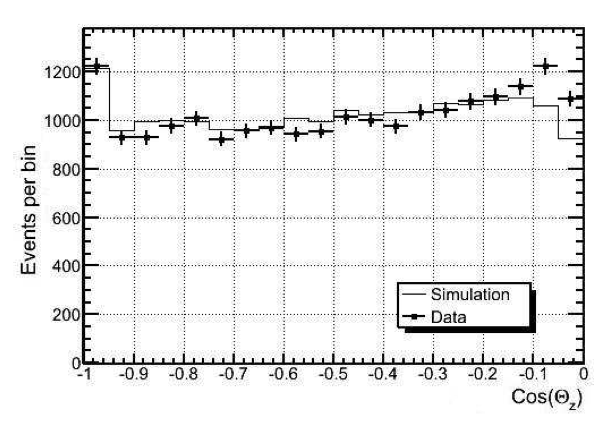}
\caption{\label{fig:IC40_data} IC-40: The $\cos\theta_z$ distribution of data and simulation. Plot taken from \cite{Abbasi:2010ie}. }
\end{figure}

\begin{figure}[t!]
\centering
\includegraphics[width=0.8\textwidth]{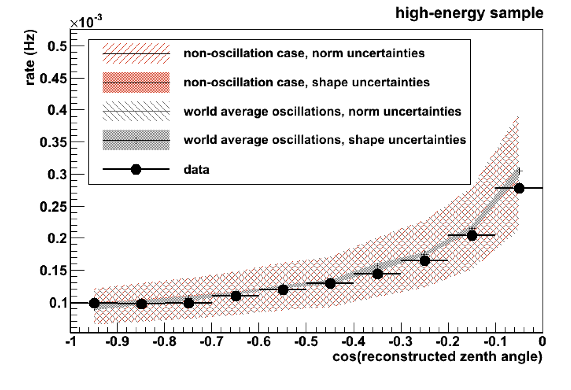}
\caption{\label{fig:IC79_data} IC-79: The Data and Monte Carlo expectation. Plot taken from \cite{Aartsen:2013jza}.}
\end{figure}

We analyze two sets of the IceCube data, IC-40~\cite{Abbasi:2010ie} and IC-79~\cite{Aartsen:2013jza}, consisting of the muon-track events induced by the atmospheric neutrinos. The IC-40 experiment which has measured the atmospheric flux in the energy range $(0.1-400)$~TeV, uses data from 359 days of livetime, while operating on a 40-string configuration of the whole detector. The IC-79 experiment with a livetime of $319$ days has used 79 detector strings to measure atmospheric neutrinos in energy range of $(0.1-10)$~TeV. The data of these experiments as a function of the zenith angle are shown in Fig.~\ref{fig:IC40_data} and Fig.~\ref{fig:IC79_data}.

To confront the LED model with the IceCube data and probing the LED parameters, we define the following $\chi^2$ function:
\begin{eqnarray}\label{eq3.44}
\chi^2 \left( m_1^D,R_{\rm{ED}};\alpha,\beta\right) & = & \sum^{10}_{i=1} \frac{\left\{N^{\rm data}_i - \alpha\left[ 1+\beta\left(0.5+(\cos\theta_z)_i\right)\right] N_i(m_1^D,R_{\rm ED})\right\}^2}{\sigma^2_{i,\rm{stat}}+\sigma^2_{i,\rm{sys}}}\nonumber\\
& + & \frac{(1-\alpha)^2}{\sigma^2_\alpha}+\frac{\beta^2}{\sigma^2_\beta},
\end{eqnarray}
where $N_i^{\rm data}$ is the observed number of events in the $i^{\rm th}$ bin of the zenith angle $\Delta_i\cos\theta_z$, taken from Fig.~\ref{fig:IC40_data} and Fig.~\ref{fig:IC79_data}. For both IC-40 and IC-79 we take 10 equal bins of zenith angle and so the up-going muon-track events are divided to zenith bins with width $\Delta_i\cos\theta_z=0.1$. The $N_i(m_1^D,R_{\rm ED})$ is the expected number of events in the $i^{\rm th}$ bin, given by Eq.~(\ref{eq3.43}), in the LED model with parameters $m_1^D$ and $R_{\rm ED}$. The parameters $\alpha$ and $\beta$ take into account respectively the correlated systematic uncertainties of the normalization and the tilt of the atmospheric neutrino flux, with $\sigma_{\alpha}=0.24$ and $\sigma_{\beta}=0.04$~\cite{Honda:2006qj}. The $\sigma_{i,\rm{stat}}=\sqrt{N_i^{\rm data}}$ is the statistical error and $\sigma_{i,\rm{sys}}=fN_i$ is the uncorrelated systematic uncertainty quantified by the parameter $f$, where $f=7\%$ for IC-40 and $f=2\%$ for IC-79. The marginalization of $\chi^2$ function with respect to $\alpha$ and $\beta$ gives the constraint on the LED parameters. The LED model do not improve the fit to the data as can be seen by comparing the values of $\chi^2$ at best-fit points in the standard $3\nu$ scheme and the LED model reported in Table~\ref{tab:1}. Thus, the data of IceCube can be used to constrain the LED parameters.

\begin{table}
\centering
\vspace{0.5cm}
\begin{tabular}{|c|c|c|}
\hline
data set & $\chi^2_{3\nu,\rm{min}}$ & $\chi^2_{{\rm LED},\rm{min}}$\\
\hline
IC-40 & 10.1 & 9.7 \\
\hline
IC-79 & 8.9 & 9.0 \\
\hline
\end{tabular}
\vspace{0.5cm}
\caption{\label{tab:1}Comparing the goodness of fit between the $3\nu$ scheme and the LED model for IC-40 and IC-79 data sets.}
\end{table}

\begin{figure}[t!]
\centering
\includegraphics[width=0.7\textwidth]{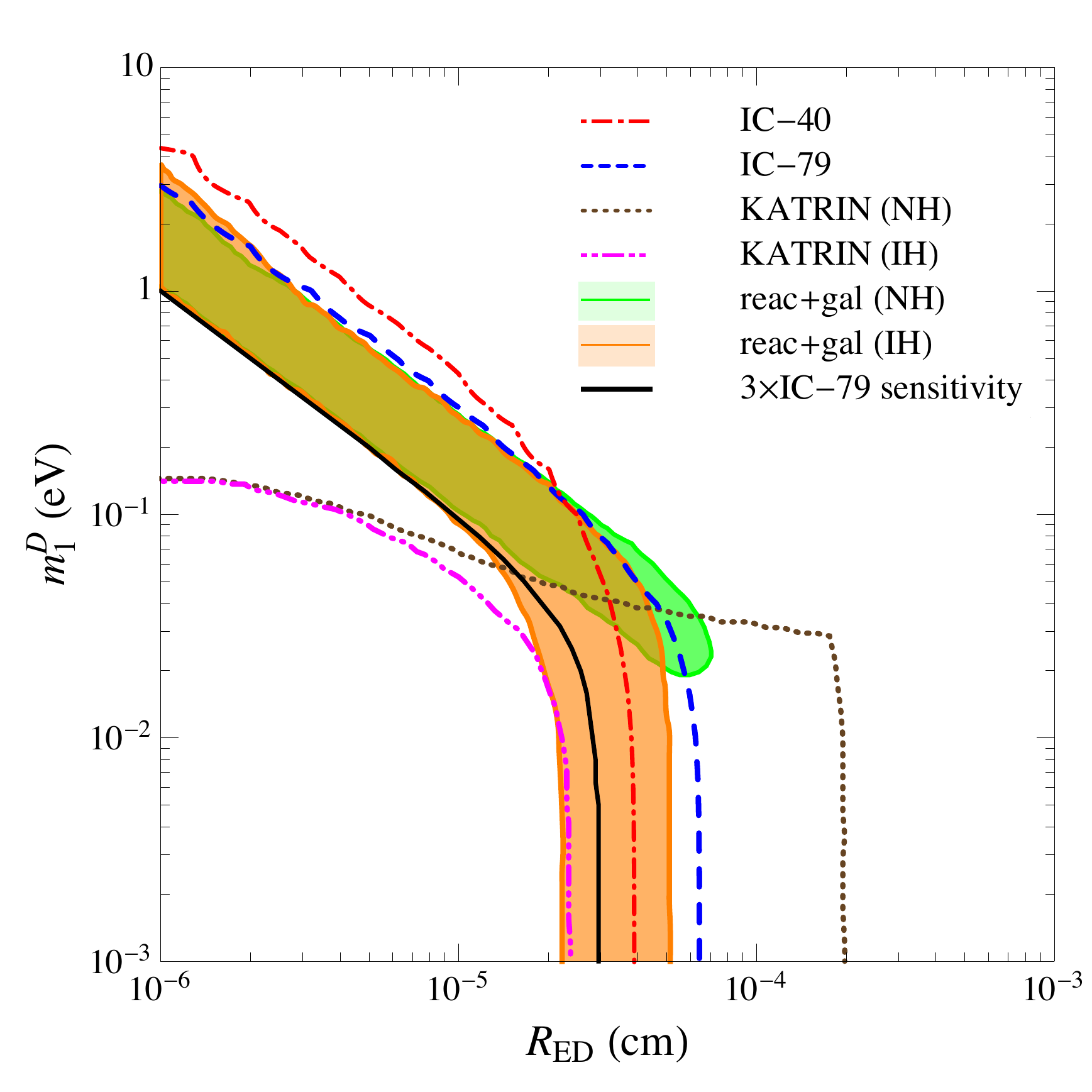}
\caption{\label{fig:limit}The allowed regions for LED model in the plane $(R_{\rm ED},m_1^D)$. The red dashed and blue dot-dashed curves are obtained from the analyses of IC-40 and IC-79 data sets, respectively, at $2\sigma$ C.L.. The green and orange shaded regions are the preferred regions by reactor and gallium anomalies at $2\sigma$ C.L., respectively for NH and IH, taken from~\cite{Machado:2011kt}. The brown dotted and violet dashed curves show the sensitivity of KATRIN, at $90\%$ C.L., respectively for NH and IH, taken from~\cite{BastoGonzalez:2012me}. The black solid curve shows the sensitivity of IceCube to LED model at $99\%$ C.L., assuming 3 times larger statistics than IC-79 and taking into account the energy information of events.}
\end{figure}

Figure~\ref{fig:limit} shows the allowed region in the plane $(R_{\rm ED},m_1^D)$ from the analysis of IceCube data. The red dot-dashed and blue dashed curves show the $2\sigma$ contours obtained from IC-40 and IC-79 data sets respectively. As we discussed in section~\ref{sec:3+n}, these contours consist of two parts: a vertical part for $m_1^D \lesssim 0.1$~eV and a straight line with slope -1 for $m_1^D\gtrsim0.1$~eV. In section~\ref{sec:3+n} we estimated also the position of these parts, that is the intercepts of these lines: $R_{\rm ED} \simeq5\times10^{-5}$~cm for $m_1^D \lesssim 0.1$~eV and $R_{\rm ED}\simeq2\times10^{-6}$~cm for $m_1^D \simeq 1$~eV which are in agreement with Figure~\ref{fig:limit}.  

In Figure~\ref{fig:limit} the green and orange shaded regions show the $2\sigma$ level preferred values of $m_1^D$ and $R_{\rm ED}$ from the reactor and gallium anomalies, respectively for NH and IH, taken from~\cite{Machado:2011kt}. The brown dotted and purple double-dot-dashed curves show the sensitivity of the KATRIN experiment to the LED parameters at $90\%$ C.L., respectively for NH and IH, taken from~\cite{BastoGonzalez:2012me}. Finally, the black solid curve shows the sensitivity of IceCube at $99\%$ C.L., by considering the energy information of events and assuming 3 times of IC-79 data, which is available now. In the following we discuss each of the components in Figure~\ref{fig:limit} and their implications. In fact the equivalence of the LED and the $(3+n)$ models, constructed in section~\ref{sec:3+n}, helps us to easily interpret Figure~\ref{fig:limit}. 

Very concisely, the reactor~\cite{Mueller:2011nm} and gallium~\cite{Giunti:2010zu} anomalies are respectively the deficits in the number of events observed in the short baseline reactor and calibration of the solar neutrino experiments, which point to $P(\nu_e(\bar{\nu_e})\to\nu_e(\bar{\nu_e}))\neq1$ over short distances that obviously cannot be accommodated in the standard $3\nu$ scheme. These deficits can be interpreted in the $(3+n)$ scenario by the $\nu_e-\nu_s$ mixing that leads to the oscillation of $\nu_e$ and $\bar{\nu}_e$ to the sterile neutrino states which escape from detection in the detectors~\cite{Kopp:2013vaa}. Thus, reactor and gallium anomalies require $\vartheta_{en}\neq0$ or in the simplest $(3+1)$ scenario $\vartheta_{e1}\equiv\theta_{14}\neq0$. In a generic $(3+n)$ scenario the mixing angles $\vartheta_{en}$, $\vartheta_{\mu n}$ and $\vartheta_{\tau n}$ are independent parameters that can take any value. On the other hand, the IceCube muon-track data is not sensitive to $\vartheta_{en}$ angles. Also, as it is shown in~\cite{Esmaili:2013cja}, the IceCube cascade data is not sensitive to the values of $\vartheta_{en}$ preferred by reactor and gallium anomalies. Thus, in a generic $(3+n)$ scenario for the interpretation of these anomalies, IceCube cannot provide an independent check. However, this is not the case in the LED model. For the LED model, all the angles in the equivalent $(3+n)$ scenario are inter-related and non-vanishing $\vartheta_{en}$ lead to non-vanishing $\vartheta_{\mu n}$ and $\vartheta_{\tau n}$. Thus, since the IceCube muon-track data can probe $\vartheta_{\mu n}$ and $\vartheta_{\tau n}$, it is possible to probe the LED interpretation of reactor and gallium anomalies which have been proposed in~\cite{Machado:2011kt}. As can be seen from Figure~\ref{fig:limit}, as a result of this work we find that the IC-40 and IC-79 data can exclude a part of the preferred region by these anomalies. 

It is possible to probe the green and orange shaded regions in Figure~\ref{fig:limit} by considering the energy information of IceCube data. Since the energy information of IceCube data is not publicly available we estimate the sensitivity of IceCube assuming a data set 3 times the IC-79 data set (which already are collected). The sensitivity of IceCube to the sterile neutrinos after taking into account the energy information has been calculated in~\cite{Esmaili:2013vza}. From the Figure~10 of~\cite{Esmaili:2013vza} it can be seen that, by considering the energy information, IceCube can probe the sterile neutrino mixing $\sin^22\vartheta_{\mu1}\simeq0.02$ for $\Delta m_{41}^2\lesssim1~{\rm eV}^2$. Using the equivalence constructed in section~\ref{sec:3+n} this sensitivity can be translated to the sensitivity of IceCube to the LED model. From the mixing angles plotted in Figure~\ref{fig:theta-in}, we can check that $\sin^22\vartheta_{\mu1}\simeq0.02$ at $R_{\rm ED}\simeq3\times10^{-5}$~cm for $m_1^D\lesssim0.1$~eV; and at $R_{\rm ED} \simeq10^{-6}({\rm eV}/m_1^D)$~cm for $m_1^D\gtrsim0.1$~eV, which are in agreement with the black solid curve in Figure~\ref{fig:limit}. As can be seen, although the current data exclude only a small part of the region allowed by the reactor and Gallium anomalies, considering the energy information of atmospheric neutrino data can almost exclude all the favored regions (or to confirm the interpretation of these anomalies in terms of LED model). Performing such an analysis (i.e., taking into account the energy binning) requires detailed information of IceCube detector which is not available now. However, with the already collected data the IceCube collaboration can perform this analysis.

The other way of probing the regions preferred by the reactor and gallium anomalies is the KATRIN experiment (the brown dotted and purple double-dot-dashed curves in Figure~\ref{fig:limit}). As can be seen, for both NH and IH cases, the KATRIN can completely exclude the green and orange shaded regions. 

\section{Conclusions\label{sec:conc_ch3}}

An added bonus of the LED model is the explanation of the small neutrino masses which can be achieved by introducing singlet fermions living in the bulk of the extra dimensions. From the brane point of view these fermions constitute towers of the sterile neutrinos with increasing masses (the so-called KK modes) that mix with the active neutrinos and so can affect the phenomenology of the neutrino flavor oscillations. In fact, this picture can be favored due to the recent observed anomalies in the short baseline oscillation experiments which hint on the presence of one (or more) sterile neutrino state(s). On the other hand, the existence of these sterile neutrinos can significantly change the oscillation pattern of the high energy atmospheric neutrinos observed by the IceCube experiment. In this chapter we studied these effects and developed a framework to interpret them. 

The mixing of the KK modes of the bulk fermions with the active neutrinos lead to the resonant conversion of $\bar{\nu}_\mu$ to the undetectable sterile neutrinos at high energies. The resonance originates from the matter effects (constant density MSW resonance) during the propagation of the atmospheric neutrinos through the Earth and would lead to the distortions in the zenith and energy distributions of the muon-track events at the IceCube detector. IceCube has already published two sets of the atmospheric neutrino data (IC-40 and IC-79) and in this chapter we analyzed them in the search of features predicted by the LED model.   

We obtained the limits on the LED parameters (especially the radius of extra dimension $R_{\rm ED}$) by analyzing the zenith distributions of IC-40 and IC-79 data. For $m_1^D\lesssim0.1$~eV the upper limit $R_{\rm ED}\leq4\times10^{-5}$~cm (at $2\sigma$ level) have been set by the IceCube data and is independent of the value of $m_1^D$. For $m_1^D\gtrsim0.1$~eV the limit depends on the value of $m_1^D$ and is stronger: $R_{\rm ED}\lesssim3\times10^{-6}({\rm eV}/m_1^D)$~cm. These bounds can exclude some parts of the parameter space preferred by the reactor and gallium anomalies.  

We have also discussed the prospect of improving the bounds by taking into account the energy distribution of the muon-track events in the IceCube. We have shown that with a sample of data three times larger than the IC-79 data set (which is already collected by the IceCube detector from its completion at December/2010 till now) it would be possible to exclude the $2\sigma$ preferred region by the reactor and gallium anomalies.

As a tool for interpreting the obtained results in this chapter, we developed an equivalence between the LED model and the phenomenological $(3+n)$ scenarios which have been studied extensively in the literature. This equivalence provides a clear and intuitive picture of the oscillation pattern of the atmospheric neutrinos in the LED model and have been used in this chapter to explain the features obtained by the numerical calculations.

\chapter{Hidden Interaction of the Sterile Neutrinos}\label{chap4}

\newpage

{\Large False facts are highly injurious to the progress of science, for they often insure long; but false views, if supported by some evidence, do little harm, for everyone takes a salutary pleasure in proving their falseness.\\

~~~~Charles Darwin}

\newpage

\footnote{This chapter is prepared based on my work published in \cite{Tabrizi:2015bba}.} Recent results from neutrino experiments show evidence for light sterile neutrinos which do not have any Standard Model interactions. In this chapter we study the hidden interaction of sterile neutrinos with an "MeV scale" gauge boson (the $\nu_s$HI model) with mass $M_X$ and leptonic coupling $g^\prime_l$. By performing an analysis on the $\nu_s$HI model using the data of the MINOS neutrino experiment we find that the values above $G_X/G_F=92.4$ are excluded by more than $2\sigma$ C.L., where $G_F$ is the Fermi constant and $G_X$ is the field strength of the $\nu_s$HI model. Using this model we can also probe other new physics scenarios. We find that the region allowed by the $(g-2)_\mu$ discrepancy is entirely ruled out for $M_X\lesssim 100$ MeV. Finally, the secret interaction of sterile neutrinos has been to solve a conflict between the sterile neutrinos and cosmology. It is shown here that such an interaction is excluded by MINOS for $g^\prime_s> 1.6\times10^{-2}$. This exclusion, however, does depend on the value of $g_l^{\prime}$. 

The chapter is organized as follows: we present the introduction of the hidden interaction of the sterile neutrinos in Section~\ref{sec:intt_cha4}. In Section~\ref{sec:formalism_cha4} we discuss the formalism. Section~\ref{sec:analysis_cha4} is devoted to the analysis of the hidden interaction model using the data of the MINOS experiment. We summarize our conclusions in Section~\ref{sec:conclusion_cha4}.

\section{Introduction: The secret interaction of the sterile neutrinos \label{sec:intt_cha4}}

Most of the data collected from the neutrino oscillation experiments are in agreement with the 3-neutrino hypothesis. However, the observation of a deficit of electron anti-neutrinos produced in the reactors~\cite{Mention:2011rk,Huber:2011wv}, together with the results of the MiniBooNE experiment~\cite{AguilarArevalo:2010wv} which shows evidence for $\nu_{\mu}\to \nu_e$ conversion cannot be explained by the usual $3\nu$ scenario~\cite{GonzalezGarcia:2002dzzz}. The most popular way to clarify these anomalies is to assume there exists 1 (or more) neutrino state(s) which does not have any weak interaction (therefore is sterile), but can mix with the active neutrinos in the SM and change their oscillation behavior pattern. In this way the flavor and mass eigenstates of neutrinos are related through the $(3+n)\times(3+n)$ unitary matrix $U$ (with $n$ being the number of sterile states): $\nu_{\alpha}=\sum_{i=1}^{3+n}U^*_{\alpha i}\nu_i$. In the most simple case of only 1 sterile neutrino this matrix would be parametrized by the active-active mixing angles $(\theta_{12},\theta_{13},\theta_{23})$ as well as 3 active-sterile mixing angles $(\theta_{14},\theta_{24},\theta_{34})$. Then the oscillation probability of neutrinos would be described using the active-active and active-sterile mixing angles, as well as the mass squared differences $\Delta m^2_{21}$, $\Delta m^2_{31}$ and $\Delta m^2_{41}$, where $\Delta m^2_{ij}\equiv m_i^2-m_j^2$.

Although most of the anomalies seen in the neutrino sector are in favor of the sterile models with mass $\sim1$ eV, there are conflicts between the sterile hypothesis and cosmology. Such light additional sterile states thermalize in the early universe through their mixing with the active neutrinos; therefore, we effectively have additional relativistic number of neutrinos which can be parametrized by $\Delta N_{\rm eff}$. In the standard model of cosmology we have $\Delta N_{\rm eff}=0$. Massive sterile neutrinos with mass $\sim1~$eV and large enough mixing angles to solve the reactor anomalies imply full thermalization at the early universe. This means that for any additional species of sterile neutrinos, we should have $\Delta N_{\rm eff}=1$. However, this is not consistent with the Big Bang Nucleosynthesis and the Planck results, which state $\Delta N_{\rm eff}<0.7$ with $90\%$ C.L.~\cite{Ade:2015xua}. It was recently proposed in \cite{Hannestad:2013anaaa,Dasgupta:2013zpnnn} that this problem could be solved if the sterile neutrino state interacts with a new gauge boson $X$ with mass $\sim$ a few MeV. This can easily produce a large field strength for the sterile neutrinos. In this way the sterile state experiences a large thermal potential which suppresses the mixing between the active and sterile states in the early universe. Therefore, the abundance of the sterile neutrinos remains small, and its impact on the Big Bang nucleosynthesis (BBN), Cosmic Microwave Background (CMB) and the Large Scale Structure Formation would be negligible, hence the sterile state can be consistent with the cosmological model. 

In this chapters we investigate the possibility of the sterile neutrino states interacting with a new gauge boson $X$, with mass $\sim$ MeV, which has couplings with the sterile neutrinos and the charged leptons in the SM. This new interaction of the sterile neutrinos was first mentioned in~\cite{Pospelov:2011ha}. The "\textbf{\textit{$\nu_s$~Hidden~Interaction}}" ($\nu_s$HI) model produces a neutral current (NC) matter potential for the sterile states proportional to $G_X$, where $G_X$ is the field strength of the new interaction. The NC matter potential in the $\nu_s$HI model changes the oscillation probability of neutrinos and anti-neutrinos drastically. Therefore, using the data of a neutrino oscillation experiment such as the MINOS experiment~\cite{Adamson:2013whj}, we can test the $\nu_s$HI model. 

An advantage of the $\nu_s$HI model is that through it we can use the data of neutrino oscillation experiments to test other new physics scenarios which imply having couplings with a light gauge boson, such as the explanation of the $(g-2)_\mu$ discrepancy with a light gauge boson \cite{Pospelov:2008zw} and the secret interaction of sterile neutrinos proposed in \cite{Hannestad:2013anaaa,Dasgupta:2013zpnnn}  which solves the tension between the sterile hypothesis and cosmology.

\section{The formalism \label{sec:formalism_cha4}}
We enlarge the SM with one extra species of the sterile neutrinos which do not couple with the SM gauge bosons, but have interactions with a new $U_X(1)$ gauge symmetry (the {$\nu_s$HI model}). {The new gauge boson couples to the sterile neutrinos and charged leptons with coupling constants $g^{\prime}_s$ and $g^{\prime}_l$, respectivel}, where for simplicity, we have assumed equal coupling constants for the charged leptons. The strength of this new interaction is given by 
\begin{equation}\label{eq4.1}
\frac{G_X}{\sqrt{2}}=\frac{g^{\prime}_sg^{\prime}_l}{4M_X^2},
\end{equation}
where $M_X$ is the mass of the new gauge boson.

The active neutrinos of the SM have charged and neutral current interactions with the  $W^{\pm}$ and $Z$ bosons. Their matter potential is therefore given by $V_{\alpha}(r)=\delta_{\alpha e}V_{CC}(r)+V_{NC}(r)=\sqrt{2}G_FN_e(r)(\delta_{\alpha e}-1/2)$, where $\alpha=e,\mu,\tau$ and $V_{CC(NC)}$ is the charged (neutral) current potential of the active neutrinos. The factor $G_F$ is the Fermi constant, while $N_e(r)$ is electron number density of the earth given by the PREM model~\cite{1981PEPI...25..297D}. We have assumed that the electron and neutron number densities are equal for our practical purposes.

The sterile neutrinos which couple to the $X$ boson will also have neutral current matter potential which is proportional to the strength field of the new interaction:
\begin{eqnarray}\label{eq4.2}
V_s(r)=-\frac{\sqrt{2}}{2}G_XN_e(r)\equiv\alpha V_{NC}(r),
\end{eqnarray}
where the dimensionless parameter $\alpha$ is defined as
\begin{equation}\label{eq4.3}
\alpha=\frac{G_X}{G_F}.
\end{equation}
For $\alpha\to0$ we recover the minimal $3+1$ sterile neutrino model. In the minimal $"3+1"$ model~\cite{Peres:2000ic} the flavor and mass eigenstates of neutrinos are related through the unitary $(3+1)\times(3+1)$ PMNS matrix $U$: $\nu_{\alpha}=\sum_{i=1}^{3+1}U^*_{\alpha i}\nu_i$. The oscillation probability of neutrinos is described using the active-active and active-sterile mixing angles, as well as the mass squared differences $\Delta m^2_{21}$, $\Delta m^2_{31}$ and $\Delta m^2_{41}$, where $\Delta m^2_{ij}\equiv m_i^2-m_j^2$.

The evolution of neutrinos in the $\nu_s$HI model can be found by solving the following Schr\"{o}dinger-like equation
\begin{eqnarray}\label{eq4.4}
i\frac{d}{dr}
\begin{pmatrix}
\nu_e\\\nu_\mu\\\nu_\tau\\\nu_s
\end{pmatrix}
=\Big[\frac{1}{2E_\nu}UM^2{U}^\dagger+V^{\nu_s{\rm{SI}}}(r)\Big]
\begin{pmatrix}
\nu_e\\\nu_\mu\\\nu_\tau\\\nu_s
\end{pmatrix},
\end{eqnarray}
where $U$ is the $4\times4$ PMNS matrix~\cite{Abazajian:2012ys}, which is parametrized by the active-active mixing angles $(\theta_{12},\theta_{13},\theta_{23})$ as well as 3 active-sterile mixing angles $(\theta_{14},\theta_{24},\theta_{34})$. 
The matrix 
$$M^2=\rm{diag}\Big(0,\Delta m^2_{21},\Delta m^2_{31},\Delta m^2_{41}\Big)$$
 is the matrix of the mass squared differences. Using Eq.~(\ref{eq4.2}), the matter potential matrix in the $\nu_s$HI model will be (after subtracting the constant $V_{NC}(r)\times\mathbb{I}$)
\begin{eqnarray}\label{eq4.5}
V^{\nu_s\rm{HI}}(r)&=&{\rm diag}\Big(V_{\rm{CC}}(r),0,0,V_s(r)-V_{\rm{NC}}(r)\Big)\nonumber\\
&=&\sqrt{2}G_FN_e(r)\rm{diag}\Big(1,0,0,\frac{(1-\alpha)}{2}\Big). 
\end{eqnarray}
The same evolution equation applies to anti-neutrinos with the replacement $V^{\nu_s\rm{HI}}(r)\to-V^{\rm{\nu_sHI}}(r)$. We consider the $\nu_s$HI model with $\alpha>0$.  In an effective 2-neutrino scheme the so called  MSW resonance~\cite{MSW-effect} happens when $\frac{\Delta m^2}{2E_\nu}\cos\theta=V$. Since in the $\nu_s$HI model the sterile states have nonzero matter potential, the potential would be positive in a $\nu_\mu-\nu_s$ system (for $\alpha>1$), which means that at energies where the resonance condition is carried out, $\nu_\mu$ converts to $\nu_s$.

An interesting place to test the $\nu_s$HI model is the MINOS long-baseline neutrino experiment~\cite{Adamson:2013whj}. The MINOS experiment which has a baseline of $735$~km detects both muon and anti-muon neutrinos, and it is one of few experiments that is both sensitive to neutrino and anti-neutrino oscillation probabilities. For the baseline and energy range of the MINOS experiment, the oscillation probabilities of the neutrinos and anti-neutrinos are very similar in the usual 3 neutrino scenario. However, this does not hold in the $\nu_s$HI model anymore. 

\begin{figure}[t!]
\centering
\includegraphics[scale=0.8]{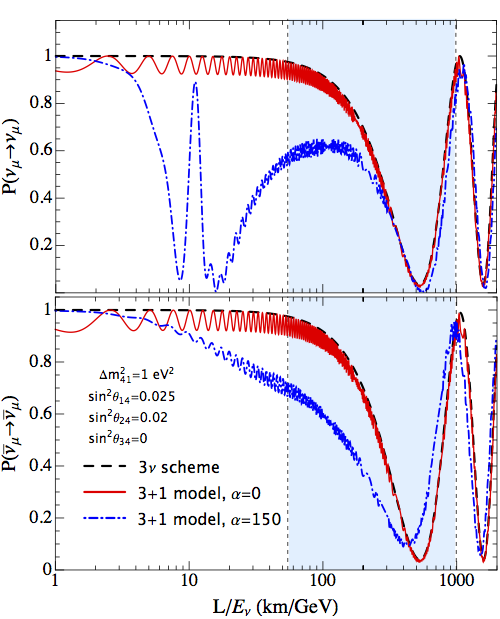}
\caption{\label{probability}
The muon neutrino and anti-neutrino survival probabilities as a function of distance over neutrino energy are shown in the top and bottom, respectively. The black-dashed and the red-solid curves correspond to the 3 and $3+1$ neutrino models, respectively. The blue dot-dashed curves represent the probabilities calculated in the $\nu_s$HI model for $\alpha=150$. The standard 3 neutrino parameters are fixed by the NUFIT best fit values~\cite{GonzalezGarcia:2012sz} and the active-sterile mixing parameters are shown in the plot. The blue shaded area is the range of $L/E_\nu$ for the MINOS experiment~\cite{Adamson:2013whj}.
}
\end{figure}

To see how the $\nu_s$HI model affects the oscillation probability of neutrinos, we compute the full numerical survival probabilities for muon (anti-)neutrino in the case of the standard 3 neutrino scenario and in the $3+1$ and $\nu_s$HI models. We show in Fig.~(\ref{probability}) the survival probability of $\nu_\mu$ (top) and $\bar{\nu}_\mu$ (bottom) for the standard 3 neutrino case (black dashed curve), the $3+1$ model with $\alpha=0$ (red solid curve) and the $\nu_s$HI model with $\alpha=150$ (the blue dot-dashed curve). To calculate the probabilities, we have fixed the 3 neutrino oscillation parameters by the best fit values of NUFIT~\cite{GonzalezGarcia:2012sz}: 
$\Delta m^2_{21}=7.5\times 10^{-5} \hspace{2pt}\mathrm{eV}^2$,
$\Delta m^2_{31}=2.4\times 10^{-3} \hspace{2pt}\mathrm{eV}^2$, $\sin^2\theta_{12}=0.3$,
$\sin^2\theta_{23}=0.6$  and $\sin^2\theta_{13}=0.023$. The values of the active-sterile mixing parameters in the 3+1 and $\nu_s$HI models are listed in Figure~(\ref{probability})~\cite{Abazajian:2012ys}. 
Comparing the 3 neutrino case (the black dashed curve) with the $3+1$ model (the red solid curve), we see that the effect of the sterile neutrino with mass squared difference $\Delta m_{41}^2=1$~eV$^2$ is marginal adding only a very fast oscillation on the top of the oscillation induced by the atmospheric mass squared difference $\Delta m_{31}^2$. However, in the  $\nu_s$HI model we have dramatic effects both for neutrino and anti-neutrino survival probabilities. When the resonance condition is fulfilled, we expect stronger changes for the $\nu_\mu$ survival probability, while for anti-neutrinos the changes are milder. This can be seen in the blue dot-dashed curve at the top and bottom of Fig.~(\ref{probability}).

\section{The analysis \label{sec:analysis_cha4}}
In this section we analyze the collected $\nu_\mu$ and $\bar{\nu}_\mu$ beam data in the MINOS experiment to constrain the $\alpha$ parameter in the $\nu_s$HI model. We calculate the expected number of events in each bin of energy by
\begin{eqnarray}\label{eq4.6}
N^{{\rm osc}}_i
=N^{\rm no-osc}_i\times \left<P_{{\rm sur}}(s_{23}^2,s_{24}^2,\Delta m^2_{31},\Delta m^2_{41};\alpha)\right>_i,
\end{eqnarray}
where $s_{ij}^2\equiv\sin^2\theta_{ij}$ and $N^{{\rm no-osc}}_i$ is the expected number of events for no-oscillation case in the $i$th bin of energy after subtracting background~\cite{Adamson:2013whj};  while $\left<P_{{\rm sur}}\right>_i$ is the averaged $\nu_{\mu}\to\nu_{\mu} ~(\bar{\nu}_{\mu}\to\bar{\nu}_{\mu})$ survival probability in the $i$th energy bin, calculated using Eq.~(\ref{eq4.4}) for the fixed values  $\sin^2\theta_{14}=0.025$ and $\sin^2\theta_{34}=0$ and letting the other parameters to vary.

To analyze the full MINOS data we define the following $\chi^2$ function
\begin{eqnarray}\label{eq4.7}
\chi^2=\sum_i\frac{\Big[(1+a)N^{{\rm osc}}_i+(1+b)N_i^b-N_i^{{\rm obs}}\Big]^2}{(\sigma_i^{\rm obs})^2}+\frac{a^2}{\sigma_a^2}+\frac{b^2}{\sigma_b^2},\nonumber\\
\end{eqnarray}
where $i$ runs over the bins of energy (23 for $\nu_\mu$ events and 12 for $\bar{\nu}_\mu$ events), $N^{{\rm osc}}_i$ is the expected number of events defined in Eq.~(\ref{eq4.6}), $N_i^b$ and $N_i^{\rm{obs}}$ are the background and observed events, respectively. The $\sigma_i^{\rm obs}=\sqrt{N_i^{\rm{obs}}}$ represents the statistical error of the observed events. The parameters $a$ and $b$ take into account the systematic uncertainties of the normalization of the neutrino flux and the background events respectively, with $\sigma_a=0.016$ and $\sigma_b=0.2$. 

\begin{figure}[!hHtb]
\includegraphics[scale=0.7]{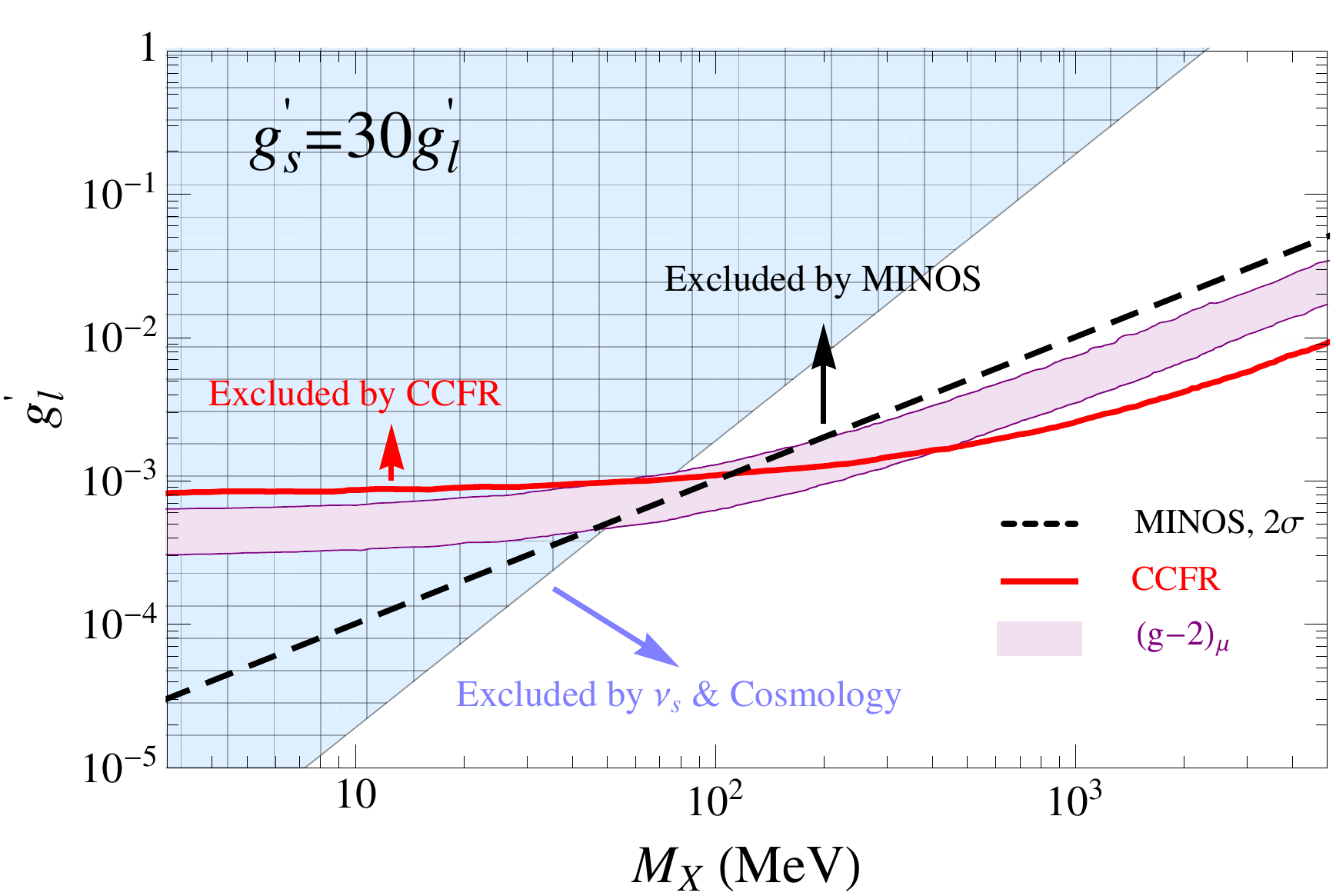}
\caption{\label{gx-mx}
We have shown the region of interest for the $\nu_s$HI model with a light gauge boson with mass $M_X$ and couplings $g^\prime_l$ and $g^\prime_s=\gamma g^\prime_l$. The result of the analysis of the $\nu_s$HI model with the MINOS data is shown by the black dashed curve with $2\sigma$ C.L. (for $\gamma=30$). The purple shaded region is the region favored by the $(g-2)_\mu$ discrepancy, while the red curve is the CCFR~\cite{Mishra:1991bv} measurement of the neutrino trident cross-section~\cite{Altmannshofer:2014pba}. The blue shaded region is where the tension between the sterile neutrino and cosmology is relieved for $f(g^\prime_s,M_X)=100$ and $\gamma=30$ (See Eq.~(\ref{eq4.10}) and the discussion after that). 
}
\end{figure}

After combining the $\chi^2$ function for the $\nu_\mu$ and $\bar{\nu}_\mu$ events and marginalizing over all parameters, we find the following best fit values: 
$\Delta m^2_{31}=2.43\times10^{-3}~\rm{eV}^2$, $\Delta m^2_{41}=4.35~\rm{eV}^2$, $s_{23}^2=0.67$, $s^2_{24}=0.03$,  and $\alpha=19.95$, while the ratio of the $\chi^2$ value over the number of degrees of freedom is $\chi^2/{\rm{d.o.f}}=39.7/30$.  When we increase $\alpha$ from its best fit value we have disagreement between the $\nu_s$HI model and the MINOS data.  From this we can find an upper bound for $\alpha$ at $2\sigma$ C.L.:
\begin{eqnarray}\label{eq4.8}
\alpha<92.4.
\end{eqnarray}
Using Eq.~(\ref{eq4.1}) and Eq.~(\ref{eq4.3}), we can write down the coupling $g^{\prime}_l$ as a function of the gauge boson mass $M_X$ and fixed value of $\alpha$:
\begin{eqnarray}\label{eq4.9}
g^{\prime}_l=\sqrt{\frac{2\sqrt{2}\alpha G_F}{\gamma}}M_X=5.5\times10^{-5}\sqrt{\frac{\alpha}{92.4\gamma}}
(\frac{M_X}{{\rm MeV}} ),
\end{eqnarray}
where we have assumed the two new coupling constants in our model are related as $g^{\prime}_s=\gamma g^{\prime}_l$, in which $\gamma\ge1$. Therefore, we can use the expression above to find an exclusion region in the $(M_X-g^\prime_l)$ plane. Implementing the relation above for the MINOS experiment, we arrive to the the black dashed curve shown in Fig.~(\ref{gx-mx}).

A light gauge boson with mass $\sim$ MeV can be used as a unique explanation for the $3.6~\sigma$ discrepancy between the experimental measurement and the SM prediction of the muon anomalous magnetic moment, $(g-2)_\mu$ \cite{Pospelov:2008zw}. The purple shaded region in Fig.~(\ref{gx-mx}) shows the favored $2\sigma$ region from $(g-2)_\mu$ discrepancy. It is shown in Ref.~\cite{Davoudiasl:2014kua} that nearly the entire $(g-2)_\mu$ band is excluded by various experiments if one assumes that the light gauge boson decays to charged leptons with branching ratio (Br) $\sim1$. However, in the $\nu_s$HI model, the primary decay mode of the light gauge boson is into invisibles (such as the light sterile neutrinos) with Br~$\sim1$. Therefore, all the $(g-2)_\mu$ band in Fig.~(\ref{gx-mx}) will be valid in the $\nu_s$HI model. As Fig.~(\ref{gx-mx}) shows, by comparing our results on the MINOS analysis of the $\nu_s$HI model with the $(g-2)_\mu$ band we can exclude all the masses below $M_X\sim 100 \sqrt{\gamma/30}$~MeV with $2\sigma$ C.L.. 

Another piece of information comes from neutrino trident production: the process in which the $\mu^+\mu^-$ pair is produced from the scattering of $\nu_\mu$ off the Coulomb field of a nucleus.  The red solid curve in Fig.~(\ref{gx-mx}) represents the results of the constrains from CCFR experiment on measurement of  the neutrino trident cross-section~\cite{Mishra:1991bv}. As it can be seen from Fig.~(\ref{gx-mx}), by combining the result of the CCFR experiment with our result from the MINOS analysis, there is only a tiny region in the $(g-2)_\mu$ band which is allowed by all experiments.

The sterile neutrino states with 1 eV mass have dramatic effects in cosmology due to their thermalization in the early universe and are disfavored by the Planck data. This tension could be removed if the sterile states have interactions with a light gauge boson in the so called secret interaction model~\cite{Dasgupta:2013zpnnn,Kopp:2014fha}. This will produce a temperature dependent matter potential for the sterile states which is $V_{\rm{eff}}=-\frac{7\pi^2}{45}\frac{{g^\prime_s}^2}{M_X^4}E_{\nu}T_s^4$~\cite{Dasgupta:2013zpnnn}, where $T_s$ is the temperature of the sterile sector and $E_{\nu}\ll M_X$. Therefore, the oscillation of the active to sterile neutrinos would be suppressed if $|V_{\rm{eff}}|\gg|  \frac{\Delta m^2_{41}}{2E_{\nu}}|$~\cite{Dasgupta:2013zpnnn}. We define the following function: 
\begin{eqnarray}\label{eq4.10}
\dfrac{V_{\rm{eff}}}{\Delta m^2_{41}/2E_\nu} \equiv f(g^\prime_s,M_X)=\frac{14\pi^2 {g^\prime_s}^2 E_\nu^2}{45\Delta m^2_{41}}(\frac{T_s}{M_X})^4.
\end{eqnarray}
Hence, the cosmology condition in the secret interaction model would be satisfied if $f(g^\prime_s,M_X)\gg1$. Similar to Eq.~(\ref{eq4.9}) we can find the values of the coupling constant which satisfy the cosmology condition:
\begin{eqnarray}\label{eq4.11}
g^\prime_s\gg \sqrt{\frac{45}{14\pi^2}}\frac{\sqrt{\Delta m^2_{41}}}{ E_\nu}\Big(\frac{M_X}{T_s}\Big)^2.
\end{eqnarray}
Assuming that the cosmology condition is satisfied for $f(g^\prime_s,M_X)=100$, then using Eq.~(\ref{eq4.9}) and the relation between the 2 coupling constants $g^\prime_s=\gamma g^\prime_l$, the values of $g^\prime_s$ above
\begin{eqnarray}\label{eq4.12}
g^\prime_s=1.6\times10^{-2}\big(\frac{T_s}{\rm{MeV}}\big)^2\big(\frac{E_\nu/\rm{MeV}}{\sqrt{\Delta m^2_{41}/\rm{eV}^2}}\big)\frac{\alpha}{92.4}\frac{\gamma}{30}
\end{eqnarray}
 is excluded by the MINOS analysis.
Therefore, using the MINOS data we find that at the time of BBN ($E_{\nu}\simeq T_s\simeq 1$~MeV) and for $\Delta m^2_{41}=1~$eV$^2$, the values of the coupling constant above $g^\prime_s= 1.6\times10^{-2}$ is excluded with more than $2\sigma$ C.L. (for $\gamma=30$). The blue shaded region in Fig.~(\ref{gx-mx}) shows the cosmology condition for the values mentioned above.

\section{The conclusion \label{sec:conclusion_cha4}}

We have investigated the possibility that the light sterile neutrinos as suggested by the reactor anomaly have hidden interaction with an "MeV scale" gauge boson. In the Hidden Interaction ($\nu_s$HI) model, the sterile neutrinos have neutral current matter potential. Therefore, we can use the data of the neutrino experiments to constrain this model and probe other new physics scenarios. The field strength of this model is described by $G_X$. In this work we studied the $\nu_s$HI model using the MINOS experiment and showed that the values above $G_X/G_F=92.4$ are excluded.

One consequence of the $\nu_s$HI model is constraining other new physics scenarios such as explaining the $(g-2)_\mu$ discrepancy with a light gauge boson. We showed that using the $\nu_s$HI model, the $(g-2)_\mu$ region is entirely ruled out for $M_X\lesssim 100 \sqrt{\gamma/30}$ MeV by the MINOS data.  Also, the secret interaction of sterile neutrinos which is introduced in the literature to solve the tension between the sterile neutrinos and cosmology is excluded by MINOS for $g^\prime_s> 1.6\times10^{-2}\frac{\gamma}{30}$ for any value of $M_X$, where $g^\prime_s$ is the coupling between the sterile states and the light gauge boson. We can use the data of the future neutrino oscillation experiments such as DUNE~\cite{Berryman:2015nua} to further test the $\nu_s$HI model and get a definite answer on the presence of the light gauge boson.
\\

\chapter{Phenomenology of minimal neutrinophilic two Higgs doublet models }\label{chap5}

\newpage

{\Large The most incomprehensible thing about the world is that it is comprehensible.\\

~~~~Albert Einstein}

\newpage

\footnote{This chapter is prepared based on my work published in \cite{Machado:2015sha}.} In this thesis so far we have been studying the phenomenology of the sterile neutrinos. In this chapter we study the models which explain the smallness of neutrinos masses. The smallness of neutrino masses can be explained using the 2 Higgs doublet models (2HDM). In such models, the vacuum expectation value (vev) of the first doublet gives masses to the particles in the SM, while the second Higgs doublet is responsible for the masses of neutrinos and through its small vev generates a small mass for them. We need some extra symmetries such as the $\Z_2$ or a softly broken $U(1)$, to prevent the neutrinos to couple to the first Higgs doublet. In this work we find the constraints that electroweak precision data can now impose on the neutrinophilic two Higgs doublet models which comprise only one extra $SU(2)\times U(1)$ doublet and a new symmetry, namely a spontaneously broken $\Z_2$ or a softly broken global $U(1)$. We find that the model with a $\Z_2$ symmetry is basically ruled out by electroweak precision data, even if the model is modified to include extra right-handed neutrinos, due to the presence of a very light scalar.  While the other model is still perfectly viable, the parameter space is considerably constrained by current data, specially by the $T$ parameter. Particularly, the mass of the new charged scalar has to be similar to the mass of the new neutral scalars.

The chapter is organized as follows:  We will briefly introduce the neutrinophilic 2HDMs in section~\ref{sec:intro_ch5}. In Section~\ref{sec:2hdm} we explain the theory of the neutrinophilic 2HDMs, specifically with the $\Z_2$ and a softly broken $U(1)$ symmetries. In Section~\ref{sec:constraints} we study the theoretical and experimental Electroweak data which constrain the neutrinophilic 2HDMs. Then in Section~\ref{sec:ana} we analyze the models. We summarize our conclusions in Section~\ref{sec:conc_chap5}.

\section{Introduction}\label{sec:intro_ch5}
The smallness of neutrino masses suggests a mass generating mechanism
distinct from the usual Higgs mechanism which reside in a scale
different from the electroweak one. From neutrino oscillation
experiments, we know that neutrinos are massive and that mass and
flavor eigenstates do not coincide. Besides, other
terrestrial~\cite{Aseev:2011dq, Kraus:2004zw} and
cosmological~\cite{Hinshaw:2012aka, Ade:2015xua} experiments indicate
that neutrino masses should be below the eV scale. Therefore, if the
same Higgs mechanism is responsible for the top and neutrino masses,
then the Yukawa couplings would span twelve orders of magnitude,
evincing an unpleasant hierarchy.

A well known alternative is the seesaw
mechanism~\cite{Minkowski:1977sc, Mohapatra:1979ia, Schechter:1980gr}.
In this scenario, the light neutrino masses are suppressed by some
heavy physics, for instance, right-handed neutrino Majorana
masses~\cite{Minkowski:1977sc, Mohapatra:1979ia, Yanagida:1979as,
  GellMann:1980vs}. What typically happens is that the scale at which
new physics can be found is extremely high, much above the TeV scale,
rendering the model intangible, except for the possible presence of
neutrinoless double beta decay\footnote{Nevertheless, there are
  alternative models which exhibit a low scale, as for instance the
  inverse seesaw scenario~\cite{Mohapatra:1986aw, Mohapatra:1986bd,
    Bernabeu:1987gr}.}. The latter could also originate from some
physics that do not comprise the main contribution to neutrino
masses~\cite{Schechter:1981bd, Duerr:2011zd}, and hence it does not
consist of a test of the seesaw mechanism by itself.

Another possibility is to generate neutrino masses by a copy of the
Higgs mechanism, having a second Higgs doublet, but with a much
smaller vacuum expectation value (vev). This can be achieved in a two
Higgs doublet model (2HDM) where one of the scalars gives mass to the
charged fermions, while the other one acquires a very small vev and
generates neutrino masses with $\O(1)$ Yukawa couplings, a
\emph{neutrinophilic} 2HDM. As a consequence, neutrino masses would
generically require new physics at the TeV scale (or even lower).  For
instance, by imposing lepton as a symmetry and adding three
right-handed neutrinos which carry no lepton number, a type I seesaw
mechanism can be realized below the TeV
scale~\cite{Ma:2000cc}. Moreover, lepton number could be conserved and
a $\Z_2$ symmetry~\cite{Gabriel:2006ns, Haba:2011nb} or a global
$U(1)$~\cite{Davidson:2009ha} could be used to prevent the SM Higgs
boson to couple to neutrinos, yielding Dirac neutrinos. Also, the 2HDM
could be augmented by a type III seesaw and a $\mu-\tau$ symmetry,
giving rise to interesting LHC
phenomenology~\cite{Bandyopadhyay:2009xa}; or by a singlet scalar and
a $\Z_3$ symmetry, possibly generating interesting lepton flavour
violating signals~\cite{Haba:2010zi}. It is also interesting to notice
that such models are stable against radiative
corrections~\cite{Morozumi:2011zu, Haba:2011fn}.

On general grounds, a new symmetry is typically invoked to prevent the
first scalar doublet from coupling to neutrinos as well as to enforce
the second one to interact only with them.  These models introduce a
minimal new field content which should materialize as particles below
the TeV scale.
The presence of such a low scale in the theory might have important
phenomenological consequences, like the presence of light scalar
particles (for instance, supernova energy loss strongly constrain such
scenarios~\cite{Zhou:2011rc}). After the discovery of a 125 GeV scalar
by the LHC experiments, new limits from electroweak precision data can
be derived on the allowed parameter space of such models.  The purpose
of this manuscript is to investigate to what extent these minimal
neutrinophilic 2HDMs can survive electroweak precision data
scrutiny.

\section{Neutrinophilic Two Higgs Doublet Models}\label{sec:2hdm}
We first start by making general considerations on the two Higgs doublet
models and the link to neutrino masses.  The most general scalar
potential for a two Higgs doublet model is
\begin{align}
\label{v2hdm}
V(\Phi_1,\Phi_2)&=m_{11}^2\Phi_1^\dagger\Phi_1 +m_{22}^2\Phi_2^\dagger\Phi_2-(m_{12}^2\Phi_1^\dagger\Phi_2+\text{h.c.})\nonumber\\
&+\dfrac{\lambda_1}{2}(\Phi_1^\dagger \Phi_1)^2+\dfrac{\lambda_2}{2}(\Phi_2^\dagger \Phi_2)^2
+\lambda_3 \Phi_1^\dagger\Phi_1 \Phi_2^\dagger\Phi_2+\lambda_4 \Phi_1^\dagger\Phi_2 \Phi_2^\dagger\Phi_1\\
&+\left[\dfrac{\lambda_5}{2}(\Phi_1^\dagger\Phi_2)^2 +\left(\lambda_6 \Phi_1^\dagger\Phi_1 + \lambda_7 \Phi_2^\dagger\Phi_2\right)\Phi_1^\dagger\Phi_2 + \text{h.c.}\right],\nonumber
\end{align}
where $\Phi_1$ and $\Phi_2$ are the two scalar doublets with
hypercharge $Y=+1$. For the vacuum expectation values (VEVs) of the
two scalars, we adopt the notation $\vev{\Phi_1}=v_1/\sqrt{2}$,
$\vev{\Phi_2}=v_2/\sqrt{2}$, and we pick $\Phi_2$ to be the one
responsible for neutrino masses. In order to have sizable Yukawa
coupling for neutrinos, it is required that $v_2 \ll v_1 \sim 246~\GeV
= v$, where $v^2=v_1^2+v_2^2$.  In principle, the parameters
$m_{12}^2$, $\lambda_5$, $\lambda_6$, and $\lambda_7$ can be
complex. Nevertheless, in all models we analyse, the symmetries will
forbid both $\lambda_6$ and $\lambda_7$, and only $m_{12}^2$ or
$\lambda_5$ will be allowed to be non-zero. A single phase of the
aforementioned parameters can always be absorbed in a redefinition of
the scalar fields, and therefore we can take all scalar potential
parameters to be real without loss of generality.

To forbid the coupling between neutrinos and $\Phi_1$, a symmetry is
called for.  In this minimal setup, there are two straightforward
examples. The first possibility is a $\Z_2$ symmetry under which only
$\Phi_2$ and the right handed neutrinos are charged, forcing
$m_{12}=\lambda_6=\lambda_7=0$~\footnote{We notice that, if instead a
  $\Z_N$ symmetry is postulated, the only difference is that
  $\lambda_5$ might be forcefully zero as well. Hence, the $\Z_2$ is
  the less restrictive case of $\Z_N$ symmetries.}.  An alternative is
to trade the $\Z_2$ by a global $U(1)$, yielding, in principle,
$m_{12}=\lambda_5=\lambda_6=\lambda_7=0$. In this case, to avoid the
presence of a massless Goldstone boson, a soft breaking is introduced
by having a nonzero but small $m_{12}$.  Anyhow, in all realizations
we will study here $\lambda_6=\lambda_7=0$, so these couplings will be
disregarded henceforth.

One last option that one could consider would be to gauge the $U(1)$
symmetry, avoiding the massless Goldstone boson. Nevertheless, in such
a scenario, the corresponding gauge boson as well as one of the
neutral scalars would be extremely light, with mass around the $v_2$
scale. This seems phenomenologically quite problematic. We do not
investigate this possibility here as it would require a completely
different study compared to the other two cases.
 
The two complex scalar $SU(2)$ doublets can be written as
\begin{equation}
\label{Higgs}
\Phi_a 	=
	\begin{pmatrix}
		\phi_a^+\\ (v_a+\rho_a+i\eta_a)/\sqrt{2} 
	\end{pmatrix},\qquad a=1,2.
\end{equation}

 We determine the mass eigenstates defining the following rotations:
 \begin{eqnarray}
 \begin{pmatrix}
 \phi_1^-\\
 \phi_2^-
 \end{pmatrix}&=&-
 \begin{pmatrix}
 c_\beta&-s_\beta\\
 s_\beta&c_\beta
 \end{pmatrix}
 \begin{pmatrix}
 G^-\\
 H^-
 \end{pmatrix},\\
 \begin{pmatrix}
 \eta_1\\
 \eta_2
 \end{pmatrix}&=&-
 \begin{pmatrix}
 c_\beta&-s_\beta\\
 s_\beta&c_\beta
 \end{pmatrix}
 \begin{pmatrix}
 G^0\\
 A
 \end{pmatrix},\\
 \begin{pmatrix}
 \rho_1\\
 \rho_2
 \end{pmatrix}&=&-
 \begin{pmatrix}
 c_\alpha&-s_\alpha\\
 s_\alpha&c_\alpha
 \end{pmatrix}
 \begin{pmatrix}
 H\\
 h
 \end{pmatrix},
 \end{eqnarray}
 where $c_{\alpha(\beta)}\equiv\cos{\alpha(\beta)}$ and
$s_{\alpha(\beta)}\equiv\sin{\alpha(\beta)}$ .
After electroweak symmetry breaking, three Goldstone bosons $G^\pm$ and $G^0$ 
become the longitudinal modes of the $W$ and $Z$ bosons. Then, the
remaining scalar spectrum is composed of two charged particles,
$H^\pm$, two CP-even neutral bosons, $h$ and $H$, and one CP-odd
neutral boson, $A$. The physical fields are given by
\begin{align}
H^+ =\phi_1^+\sin\beta -\phi_2^+\cos\beta, \qquad \qquad A =\eta_1\sin\beta-\eta_2\cos\beta,\\
H \;\;=-\rho_1\cos\alpha -\rho_2\sin\alpha, \qquad \qquad h =\rho_1\sin\alpha -\rho_2\cos\alpha,
\end{align}
where the angles $\alpha$ and $\beta$ are associated with the rotations
that diagonalize the mass matrices
\begin{align}\label{eq:alpha}
	\tan(2\alpha)&=\dfrac{2(-m_{12}^2+\lambda_{345}v_1 v_2)}{m_{12}^2(v_2/v_1-v_1/v_2)+\lambda_1 v_1^2-\lambda_2 v_2^2},\\\label{eq:beta}
	\tan\beta &= \dfrac{v_2}{v_1},
\end{align}
where $\lambda_{345}\equiv\lambda_3+\lambda_4+\lambda_5$. We will see
below that both $\alpha$ and $\beta$ are expected to be very
small. Hence, $H$ is essentially the SM Higgs and $h$ is the
neutrinophilic scalar.

As we will see in sec.~\ref{sec:ana}, the tree level stationary
conditions of the potential, $\partial V/\partial \Phi_i=0$, can be
used to write the diagonal mass parameters $m_{ii}$ as functions of
$m_{12}^2$, the quartic couplings and the vevs. With that in mind, we
can consider the quartic couplings as free parameters and express them
in terms of the physical masses, vevs and mixing
angles~\cite{Kanemura:2004mg}:
\begin{align}
\label{lambda1}
\lambda_1  &= \frac{1}{v^2}\left(-\tan^2\beta M^2+\frac{\sin^2\alpha}{\cos^2\beta} m_h^2+\frac{\cos^2\alpha}{\cos^2\beta}m_H^2\right), \\
\label{lambda2}
\lambda_2  &=\frac{1}{v^2}\left(-\cot^2\beta M^2+\frac{\cos^2\alpha}{\sin^2\beta} m_h^2+\frac{\sin^2\alpha}{\sin^2\beta}m_H^2\right), \\
\label{lambda3}
\lambda_3 &=\frac{1}{v^2}\left(-M^2+2 m_{H^\pm}^2+\dfrac{\sin(2\alpha)}{\sin(2\beta)}(m_H^2-m_h^2)\right),\\
\label{lambda4}
\lambda_4 &= \frac{1}{v^2} \left(M^2+m_A^2-2 m_{H^\pm}^2\right) \\
\label{lambda5}
\lambda_5 &= \frac{1}{v^2} \left(M^2- m_A^2\right),
\end{align}
where $M^2\equiv\dfrac{m_{12}^2}{\sin\beta \cos\beta}$. Inversely, we have 
\begin{align}
m_H^2 &= M^2 \sin^2(\alpha-\beta)\nonumber\\ \label{eq:mH}
&\quad +\left(\lambda_1 \cos^2 \alpha \cos^2\beta+\lambda_2 \sin^2 \alpha \sin^2\beta+\frac{\lambda_{345}}{2}\sin 2\alpha \sin 2\beta\right)v^2, \\ 
m_h^2 &= M^2 \cos^2(\alpha-\beta)\nonumber\\ \label{eq:mh}
&\quad+\left(\lambda_1 \sin^2 \alpha \cos^2\beta+\lambda_2 \cos^2 \alpha \sin^2\beta-\frac{\lambda_{345}}{2}\sin 2\alpha \sin 2\beta\right)v^2, \\ \label{eq:mA}
m_A^2 &= M^2 - \lambda_5 v^2, \\ \label{eq:mHpm}
m_{H^\pm}^2 &= M^2 - \frac{\lambda_{45}}{2} v^2.
\end{align}

Next we describe the two specific realizations of the neutrinophilic
scenarios that will be studied in this chapter.

\subsection{Neutrinophilic 2HDM: $\Z_2$ symmetry}
\label{subsec:z2}
The model to be studied was proposed by Gabriel and
Nandi~\cite{Gabriel:2006ns}\footnote{The same model was previously
  also proposed in ref.~\cite{Wang:2006jy} where the focus was on the
  origin of the second doublet from neutrino condensation.}.  It
consists of a 2HDM where both the right-handed neutrinos and one of
the scalar doublets, $\Phi_2$, are charged under a $\Z_2$
symmetry. The consequence is that the masses of the charged fermions
come solely from the $\Phi_1$ vev, and neutrinos, which are Dirac
fermions in this scenario as the authors impose lepton number
conservation, couple exclusively to $\Phi_2$.  This extra symmetry
can, in principle, be dropped allowing for Majorana neutrinos with a
low scale realization of the seesaw mechanism.  We will also investigate this
possibility in our analysis.

In the scalar potential (\ref{v2hdm}) of this model, the parameters
$m_{12}^2$ and $\lambda_{6,7}$ will vanish due to $\Z_2$ symmetry. The
smallness of neutrino masses is explained by the very low scale where
$\Z_2$ is broken, preferably $v_2\lesssim\O({\rm eV})$. A tiny
$v_2/v_1$ ratio and the absence of an explicit breaking $m_{12}^2$
term leads to almost no mixing between the doublets. The smallness of
$\tan\beta$ and $\tan\alpha$ can be seen from Eqs.~(\ref{eq:alpha})
and (\ref{eq:beta}) after imposing $v_2/v_1\rightarrow0$. Therefore,
apart from its couplings to neutrinos, $\Phi_1$ behaves almost
identically to the SM Higgs doublet, so we do not expect any
observable deviation from the Higgs couplings to the SM particles,
except possibly the loop induced couplings, e.g. $H\gamma\gamma$.

The second doublet displays some interesting features. Through the
Yukawa coupling, the neutral components couple almost only to
neutrinos (e.g. $c_\alpha\bar{\nu}\nu h$ or $c_\beta\bar{\nu}\nu A$),
while the charged scalars mediate interactions between neutrinos and
charged leptons (e.g. $c_\beta \bar{l}\nu H^+$). The Yukawas are
ideally expected to be of $\O(1)$. The neutral scalars couple to the
$W$ and $Z$ bosons, but notice that triple gauge couplings (TGCs)
involving only one scalar are highly suppressed by the small $v_2$
vev. Obviously, TGCs with two scalars and one gauge boson are present
and may provide a sizeable pair production cross section at colliders,
for instance $pp\to A^* \to H^+H^-$ at the LHC. This will be discussed
in the next section.

The scalar spectrum of this model is quite constrained. By setting
$m_{12}^2=0$ in Eqs.~(\ref{eq:mH}-\ref{eq:mA}), as well as
$\sin^2\alpha,\sin^2\beta\ll 1$, we notice that: (1) $H$ is identified
as the 125~GeV Higgs particle found at the LHC, and this essentially
fix $\lambda_1 \approx 0.26$ (see Eq.~(\ref{lambda1})); (2) the
neutrinophilic neutral scalar $h$ is extremely light, $m_h
\sim\O(v_2)\ll v$; and (3) For not so large values of the quartic
couplings, the charged scalars and the pseudoscalar masses are bounded
to be about or below the TeV scale.

When we analyse the viability of this model in sec.~\ref{sec:ana}, it
will turn out that oblique parameters will play a decisive role in
constraining it, due to the peculiar structure of the scalar spectrum.
The sensitivity of the $S$ parameter to the presence of a very light
neutral scalar, $m_h\sim\O(v_2)$, will essentially rule out the model.

\subsection{Neutrinophilic 2HDM: softly broken global $U(1)$ symmetry}
\label{sec:u1}
The second model we study was proposed by Davidson and
Logan~\cite{Davidson:2009ha}.  Analogously to the other scenario, both
$\Phi_2$ and right-handed neutrinos are charged under a new global
$U(1)$. The model spans $\lambda_{5,6,7}=0$ and a small $m_{12}^2$
which breaks the symmetry softly and generates neutrino masses. The
presence of the soft breaking mass term, is required in order to avoid
a massless Goldstone boson which might create problems with cosmology
and electroweak precision data.  Neutrinos are Dirac particles, as the
Majorana mass term is strictly forbidden by the new $U(1)$. From
Eq.~(\ref{eq:mA}), we write
\begin{equation}
  m_{12}^2 = \sin\beta\cos\beta m_A^2,
\end{equation}
and we observe that to obtain simultaneously $v_2\sim$~eV and $m_A
\sim \mathcal{O}(100~\GeV)$ one would need $m_{12}^2\sim
(200~\keV)^2$.  As said before, to avoid the issues of having a
massless Goldstone, instead of softly breaking the new $U(1)$
symmetry, one could also envisage to gauge it. Nonetheless, the theory
would contain a very light vector resonance as a consequence of the
small vev, and it is not clear if such a model can satisfy all
neutrino data and astrophysical constraints. We do not explore this
possibility here.

The presence of a non zero $m_{12}^2$ term makes this case fairly
different from the last one. From Eq.~(\ref{eq:mh}), we notice that
the mass of the neutrinophilic scalar, $m_h$, is proportional to
$M^2$, and therefore the $h$ mass in this scenario is not bounded by
$v_2$ as in the previous case.  As we will see later, this will ease
the constraints from the oblique parameters. Combining
Eq.~(\ref{lambda2}) and the definition
$M^2=m_{12}^2/\sin\beta\cos\beta$, and imposing $\tan\beta\sim\sin\beta=v_2/v\ll 1$, we obtain
\begin{align}
\label{eq:lambda2}
\lambda_2  &=\frac{1}{v^2}\left(-\cot^2\beta M^2+\frac{\cos^2\alpha}{\sin^2\beta} m_h^2+\frac{\sin^2\alpha}{\sin^2\beta}m_H^2\right) \simeq \frac{1}{v_2^2}\left(m_h^2-m_{12}^2 \frac{v}{v_2}\right)+\frac{\sin^2\alpha}{\sin^2\beta}\frac{m_H^2}{v^2},
\end{align}
which indicates that
\begin{equation}
  |m_h^2-m_{12}^2 v/v_2|\lesssim \O(v_2^2).
\end{equation}
To grasp the impact of this conclusion, assume that $m_{12}^2 = m_h^2
v_2/v$. Hence, from Eq.~(\ref{eq:mA}) we see that $m_A\approx m_h$, so
the CP-odd scalar is degenerate in mass with the neutrinophilic
scalar. We emphasize that this degeneracy by itself is not a fine
tuning of the model: the degenerate spectrum arises naturally given
the symmetries of the scalar potential and the hierarchy between the
vevs. As a last comment, we emphasize that since $m_{12}^2$ is the
only source of $U(1)$ breaking, it is natural in the t'Hooft sense --
$m_{12}^2$ only receives radiative corrections proportional to
itself~\cite{1107.1026, 1107.3203}.

\section{Theoretical and Experimental Electroweak Data Constraints}
\label{sec:constraints}

\subsection{Theoretical Constraints}

There are a number of conditions to be fulfilled by the scalar
potential. These will be used to constrain the parameter space,
ultimately restricting the range of physical scalar masses, having an
important impact on the phenomenology of the models.  To have
stability at tree level, the following constraints should be
fulfilled~\cite{hep-ph/0207010}
\begin{equation}
\label{contrainte}
\lambda_{1,2}>0,\qquad{}\lambda_3>-(\lambda_1\lambda_2)^{1/2}\qquad{}\text{and}\qquad{}\lambda_3+\lambda_4-|\lambda_5|>-(\lambda_1 \lambda_2)^{1/2}.
\end{equation}
In addition, the stationary conditions $\partial V/\partial \Phi_i=0$ read
\begin{equation}
\begin{aligned}\label{eq:stationary}
	&\dfrac{\lambda_1}{2}v_1^3+\dfrac{\lambda_{345}}{2}v_1v_2^2+
m_{11}^2v_1-m_{12}^2v_2=0,\\
	&\dfrac{\lambda_2}{2}v_2^3+\dfrac{\lambda_{345}}{2}v_2v_1^2+
m_{22}^2v_2-m_{12}^2v_1=0,
\end{aligned}
\end{equation}
which allow us to write $m_{ii}^2$ as functions of $m_{12}^2$, $v_1$ and $v_2$.
If $m_{12}^2=0$, it is easy to see that there are at least two
equivalent stable solutions, $(v,0)$ or $(0,v)$ (although they may not
be the global minima).
In this case, the vev is precisely the electroweak scale, one of the
scalars is exactly the Higgs and the other one is inert.  For
$m_{12}^2 \neq 0$, these equations cannot be solved
analytically. Nevertheless, if $m_{12}^2\ll v^2$ a perturbative
approach yields
\begin{equation}
   v_1\approx v, \qquad v_2\approx\dfrac{m_{12}^2}{\lambda_{345}\, v^2+m_{22}^2}v,
\end{equation}
and a symmetric solution interchanging the indices $1\leftrightarrow
2$, which reveals that the small vev necessary to satisfactorily
explain small neutrino masses might require a correspondingly small
$m_{12}^2$ parameter. This can be understood intuitively, as the
breaking of the $U(1)$ happens only through the soft breaking term
$m_{12}^2$.  In general, there can be more than one solution
satisfying the stationary conditions~(\ref{eq:stationary}), and hence
different non-trivial and non-degenerate minima $(v_1,v_2)$ and
$(v_1',v_2')$ might coexist. It is possible to check analytically if
the chosen vacuum is the deepest one in the potential for a 2HDM with
$\lambda_{6,7}=0$~\cite{1303.5098}. In this case, the potential
describes a $\Z_2$ symmetry softly broken by $m_{12}^2$. Both models
we deal here are special cases of such scenario. In the absence of an
explicit breaking, that is $m_{12}^2=0$, there can be multiple minima,
but they are degenerate and hence stability is not threatened. This is
the case of the $\Z_2$ model we analyse. For the softly broken $U(1)$
model, it can be shown that the chosen vacuum is the deepest one (at
tree level) if and only if the following condition is satisfied:

\begin{equation}
	D=m_{12}^2(m_{11}^2-\kappa^2 m_{22}^2)(\tan\beta-\kappa)>0,
\end{equation}
with $\kappa=\sqrt[4]{\lambda_1/\lambda_2}$. Although for a general
2HDM scenario this bound may be important, for the neutrinophilic case
we have checked that it does not lead to any significant effect on the
parameter space, after the other constraints are taken into account,
but we include it in the analysis of the softly broken $U(1)$ model
for completeness.

Another theoretical requirement is to have tree level pertubative
unitarity~\cite{Kanemura:1993hm, Arhrib:2000is}.
If the quartic couplings are too large, the lowest order amplitudes
for scalar--scalar scattering may violate unitarity at high enough
scales, requiring additional physics to mitigate this issue. To obtain
the constraint, the scalar--scalar $S$ matrix is computed and the
following conditions are imposed on its eigenvalues
\begin{equation}\label{eq:unitarity}
	|a_{\pm}|,|b_{\pm}|,|c_{\pm}|,|f_{\pm}|,|e_{1,2}|,|f_1|,|p_1|< 8 \pi,
\end{equation}
where 
\begin{align}
a_\pm &= \frac{3}{2}(\lambda_1+\lambda_2)\pm \sqrt{\frac{9}{4}(\lambda_1-\lambda_2)^2+(2\lambda_3+\lambda_4)^2}, \\
b_\pm &= \frac{1}{2}(\lambda_1+\lambda_2)\pm \frac{1}{2}\sqrt{(\lambda_1-\lambda_2)^2+4\lambda_4^2}, \\
c_\pm &= \frac{1}{2}(\lambda_1+\lambda_2)\pm \frac{1}{2}\sqrt{(\lambda_1-\lambda_2)^2+4\lambda_5^2}, \\
f_+ &= \lambda_3 + 2\lambda_4 + 3 \lambda_5, \\
f_- &= \lambda_3 + \lambda_5, \\
e_1 &= \lambda_3 + 2\lambda_4 - 3 \lambda_5, \\
e_2 &= \lambda_3- \lambda_5, \\
f_1 &= \lambda_3 + \lambda_4, \\
p_1 &= \lambda_3 - \lambda_4. 
\end{align}
To have an idea of the impact of these bounds, one can conservatively
assume that all $|\lambda_i|$ should be smaller than $8\pi$ (the
actual bound is always more stringent than that). 

Evidently, even if tree level unitarity is satisfied, loop corrections
could still play an important role leading to violation of unitarity
at some scale and thus demanding the presence of new physics below
such energies. This could be particularly relevant when some of the
tree level constraint is just barely satisfied, as the size of the
quartic couplings could enhance the loop contributions. Nevertheless,
we only take into account unitarity constraints at tree level, as a
full one loop evaluation of the parameter space is beyond the
scope of this work.

\subsection{Electroweak Data Constraints}
\label{sub:ew-cons}
{\bf Oblique Parameters.}  The impact of a second Higgs doublet in the
so-called electroweak precision tests (EWPT), encoded in the
Peskin-Takeuchi parameters $S$, $T$, and $U$~\cite{Peskin:1991sw}, has
been studied in the literature to a great extent (see for instance
refs.~\cite{0711.4022, 0802.4353, 1011.6188}). These are radiative
corrections to the gauge boson two point functions, known as oblique
corrections. 

The $S$ parameter encodes the running of the neutral gauge bosons two
point functions ($ZZ$, $Z\gamma$ and $\gamma\gamma$) between zero
momentum and the $Z$ pole. Therefore, it should be specially sensitive
to new physics at low scales, particularly below the $Z$ mass. Thus,
we expect it to be important in the presence of very light neutral
scalars, as is the case for the $\Z_2$ model.  The $T$ parameter
measures the breaking of custodial symmetry at zero momentum, that is,
the difference between the $WW$ and the $ZZ$ two point functions at
$q^2=0$. It usually plays a significant role in constraining the
parameter space of particles charged under $SU(2)_L$. Splitting the
masses of particles in a doublet breaks custodial symmetry and affects
$T$. As we will see later, in the softly broken $U(1)$ scenario, the
$T$ parameter will provide the major constraint on the mass splitting
$m_{H^\pm}-m_A$, forcing the scalar spectrum of this model to be
somewhat degenerate.  Last, and this time least, the $U$ parameter (or
better, the combination $S+U$) is somewhat similar to $S$ but for the
$W$ bosons, being sensitive to light charged particles in the loops.
Given the fact that light charged particles are excluded by LEP
data~\cite{Abbiendi:2013hk, Agashe:2014kda}, usually $U$ is the least
important of these three precision parameters, having a minor impact
on the model phenomenology, which we have checked that this is indeed the
case for all scenarios analyzed here.

To evaluate the impact on EWPT on the neutrinophilic 2HDM
scenarios, we calculate $S$, $T$, and $U$ using the results available
in ref.~\cite{1011.6188}, and we use the latest GFITTER values for the
best fit, errors and covariance matrix~\cite{Baak:2014ora},
\begin{equation}
  \begin{aligned}
    & \Delta S^{SM} = 0.05\pm0.11,\\
    & \Delta T^{SM} = 0.09\pm0.13,\\
    & \Delta U^{SM} = 0.01\pm0.11,\\
  \end{aligned}
\qquad\qquad
V = \left(\begin{array}{ccc}
1 & 0.90 & -0.59\\
0.90 & 1 & -0.83\\
-0.59 & -0.83 & 1
\end{array}\right),
\end{equation}
composing the $\chi^2$ function as
\begin{equation}
  \chi^2= \sum_{i,j}(X_i - X_i^{\rm SM})(\sigma^2)_{ij}^{-1}(X_j - X_j^{\rm SM}),
\end{equation}
with $X_i=\Delta S, \Delta T, \Delta U$ and the covariance matrix
$\sigma^2_{ij}\equiv\sigma_iV_{ij}\sigma_j$, in which
$(\sigma_1,\sigma_2,\sigma_3)=(0.11,0.13,0.11)$. As we are interested
in the goodness of fit of the model to the EWPT data, the 1, 2, and
3$\sigma$ regions are calculated using $\chi^2 = 3.5,8.0,14.2$,
respectively.

{\bf Higgs invisible width.} 
When the first doublet acquires a vev, triple scalar vertices like
$HSS$ ($S=h,A$) are induced. Therefore, light neutral scalars with $2
m_S < m_H$ could contribute to the Higgs invisible width $H\to S S$,
and sequentially $S\to\bar\nu\nu$. Because of the small $\tan\beta$ of
the model, the Higgs boson couplings to the Standard Model particles is
basically unchanged. Hence, the contribution to the Higgs total width
due to the invisible decay will suppress all Standard Model branching
fractions by the ratio $\Gamma_H^{\rm SM}/\Gamma_H^{\rm new}$. In this
scenario, as the only modification to the Higgs branching fractions is
the addition of an invisible channel, the LHC 8 TeV data bound is
${\rm BR}(H\to {\rm invisible})<0.13$ at 95\% CL~\cite{Ellis:2013lra}.

In our framework, the decay rate of such a process is given
by~\cite{Bernon:2014nxa}
\begin{align}
&\Gamma(H\to S S) = \frac{g_{HSS}^2}{32 \pi m_H} \sqrt{1-\frac{4 m_S^2}{m_H^2}}, 
\qquad {\rm with}\\
&g_{HAA}=\frac{1}{2v}\Bigg{[}(2 m_A^2-m_H^2)\frac{\sin(\alpha-3\beta)}{\sin 2\beta}
  \nonumber\\
&\qquad\qquad\qquad +(8m_{12}^2-\sin 2\beta(2m_A^2+3 m_H^2)) 
  \frac{\sin(\beta+\alpha)}{\sin^2 2\beta}\Bigg{]},\label{eq:HAA}\\
&g_{Hhh}=-\frac{1}{v} \cos(\beta-\alpha)\Bigg[\frac{2 m_{12}^2}{\sin 2\beta} 
  +\left(2 m_h^2+m_H^2-\frac{6 m_{12}^2}{\sin 2\beta}\right)
  \frac{\sin 2\alpha}{\sin 2\beta}\Bigg].\label{eq:Hhh}
\end{align}
We emphasize that the small terms in the denominator always cancel
out. While the couplings between the SM Higgs and the SM fermions,
$g_{Hff}=m_f/v$, are well below one due to the suppresion by the EW
scale (except for the top, to which the Higgs cannot decay), the
trilinear scalar couplings are typically much larger, $g_{hSS}\sim
m_h^2/v \sim 60\, \mathrm{GeV}$, unless there are some sort of
cancellations happening~\cite{Bernon:2014nxa}. Therefore, SM Higgs
decays to lighter scalars may have an important phenomenological
impact, specially because the total Higgs width in the Standard Model
is predicted to be very small, around $4.07~\MeV$~\cite{Denner:2011mq}



{\bf Higgs to diphoton.} The charged scalars will contribute to the
$H\to\gamma\gamma$ width, and thus we also analyse the impact on this
observable\footnote{The $H\to Z\gamma$ decay will also be modified,
  but due to the smaller branching ratio and subsequent suppression by
  requiring the $Z$ to decay leptonically, we do not expect it to
  provide any significant sensitivity in the near future.}. The $H$
diphoton width is a destructive interference effect mainly between $W$
and top loops, where the latter dominate. Charged scalars contribute
with the same sign as the $W$s, and their contribution usually do not
overcome the top one. Therefore, we expect $H\to\gamma\gamma$ to be
somewhat suppressed in most cases.  The expression for the
$H\to\gamma\gamma$ width at one loop can be found in many papers, see,
for instance ref.~\cite{Almeida:2012bq}.

{\bf $Z$ invisible width.}  We also have to consider possible extra
contributions to the $Z$ invisible width coming from the decays $Z \to
h \nu \bar{\nu}$ and $Z \to A \nu \bar{\nu}$. The $Z$ invisible decay
width reads
\begin{align}
\label{eq:zinv}
  \Gamma(Z\to S \nu\bar{\nu}) &= 
       \dfrac{1}{384 \pi^3 m_Z^5} \left(\frac{g}{2 \cos\theta_W}\right)^2 \frac{m^2_{\nu,\mathrm{tot}}}{v_2^2} \int_0^{q_\mathrm{max}^2}\mathrm{d}q^2\, \lambda^{1/2}(q^2,m_Z^2,m_S^2)\notag\\
       &\times\Bigg{[}q^2-4m_Z^2+\frac{4 m_Z^2(m_Z^2+m_S^2-q^2)}{\lambda^{1/2}(q^2,m_S^2,m_Z^2)}\coth^{-1}\left(\frac{m_Z^2+m_S^2-q^2}{\lambda^{1/2}(q^2,m_Z^2,m_S^2)}\right) \Bigg{]},
\end{align}
where $q_{\mathrm{max}}^2=(m_Z-m_S)^2$,
$\lambda(a^2,b^2,c^2)=(a^2-(b-c)^2)(a^2-(b+c)^2)$ and the scalar $S$
can be either $h$ or $A$. The ratio between $m^2_{\nu,{\rm
    tot}}\equiv\sum m_{\nu,i}^2$ and $v^2_2$ arrives from the neutrino
Yukawas. Clearly, if the Yukawas are small, both widthes vanish, so we
expect this bound to be more significant for lower $v_2$ and larger
neutrino masses.
To constrain extra contribution from new physics to the Z invisibles
width, we use LEP result
$\Gamma^{\rm{exp}}(Z\to\rm{invisible})=499.0(15)$~MeV and the Standard
Model prediction
$\Gamma^{\rm{SM}}(Z\to\rm{invisible})=501.69(6)$~MeV~\cite{Agashe:2014kda},
which yields $\Gamma^{\rm{NP}}(Z\to\rm{invisible})<1.8$ MeV at 3
$\sigma$ (notice that there is a mild $2\sigma$ discrepancy between
the data and the SM predicted value). In the case of the $\mathbb{Z}_2$ symmetry model,
one must take care while doing the computation, since $m_h\ll m_Z$,
the expression for the width in \eqref{eq:zinv} has an infrared divergence. We calculated one-loop corrections
for $\Gamma(Z\to\bar{\nu}\nu)$ coming from the $h$ scalar and added it to
$\Gamma(Z\to S\bar{\nu}\nu)$ in order to eliminate the IR divergence.

{\bf Collider bounds on charged scalars.} The charged scalars can be
pair produced directly at colliders via $s$-channel off shell photon
or $Z$ exchange. Due to the neutrinophilic character of the second
Higgs doublet and small admixture with the SM degrees of freedom, the
charged scalars decay almost only to $\ell\nu$. Therefore, we expect
the LEP bound to be $m_{H^\pm}>80~\GeV$~\cite{Abbiendi:2013hk,
  Agashe:2014kda}.

It is not clear how LHC data improves the
situation. There has been some studies on the LHC sensitivity to such
charged scalars, mainly focused on 14~TeV center of mass
energy~\cite{Davidson:2010sf, Morozumi:2012sg, Morozumi:2013rfa,
  Choi:2014tga}, but to the best of our knowledge, there has been no
dedicated experimental search for charged scalars in neutrinophilic
2HDMs.  As $v_2$ is very small, the main production modes of $H^\pm$
would be pair production through vector boson fusion or off-shell
$S$-channel photon and $Z$ exchange, and the tipical $t \to H^+ b$
would be absent due to small $\tan\beta$.  The LHC sensitivity then
would come mainly from opposite sign diletpton plus missing energy,
which has SM $W$ pair production as an irreducible
background. Moreover, the branching ratios of the charged scalar
depend on the neutrino masses and the mass ordering. If the $\tau\nu$
branching ratio is dominant, the sensitivity is expected to be
smaller. Therefore, to be conservative, we will scan the parameter
space considering only the LEP bound.

{\bf Anomalous magnetic moments and other constraints.} In principle,
the charged scalars could also contribute to charged lepton $g-2$
values, but the corresponding amplitude at one loop is suppressed by
$m_\ell^4/m_{H^\pm}^4$ (see ref.~\cite{1409.3199} for a recent
analysis on the impact of a second Higgs doublet on the muon
$g-2$). We have checked that the 1-loop contribution to both muon and
electron $g-2$ is negligible due to that suppression, while the tau
$g-2$ is not measured with enough precision to pose a bound. For a
general 2HDM, it has been noticed that two loop Barr-Zee
diagrams~\cite{hep-ph/0009292} can be more important than 1-loop
contributions, but this is not the case in the neutrinophilic 2HDM, as
the charged lepton couplings to $h$ and $A$ are suppressed by
$\tan\beta$~\footnote{There would be a small contribution due to
  modifications of the $H\to\gamma\gamma$ coupling, but Higgs data
  already constraint it to the level that there is no observable
  modification to the muon $g-2$.}. Therefore we conclude that the
electron, muon and tau $g-2$ measurements do not pose any bound on
this scenario.

Flavour physics constraints have also been studied in the
literature. The charged scalars will mediate lepton flavour violating
decays. In $\mu\to e\gamma$, for instance, the additional branching
ratio is proportional to $(m_{H^\pm}v_2)^{-4}$.  It has been argued
that for 100~GeV charged scalars and $v_2=2$~eV, this branching ratio
could be large enough to be excluded by current
data~\cite{Davidson:2009ha}. Nevertheless, if $v_2$ is above 10~eV,
the decay goes out of the reach of current and near future
experiments. Therefore, flavour physics make no difference in the
projections we will show of the parameter space, and it will be
disregarded.

\section{Analysis of the Models}
\label{sec:ana}
For each model $5.1\times 10^6$ points are generated, the corresponding
scalar potential parameters are calculated and the perturbative
unitarity and stability conditions, Eqs.~(\ref{eq:unitarity}) and
(\ref{contrainte}), respectively, are derived. We only show on our
plots points which fulfill these theoretical constraints, about
760000 and 510000 for the
$\Z_2$ and the global $U(1)$ model, respectively. Unless stated
otherwise, the points are color coded accordingly to the fit to EWPT
data: blue, green, and red correspond to the 1, 2, and 3$\sigma$
allowed regions, while gray points are excluded at $3\sigma$ or more.

\begin{figure}[t]
  \begin{center}
    \includegraphics[scale=0.71]{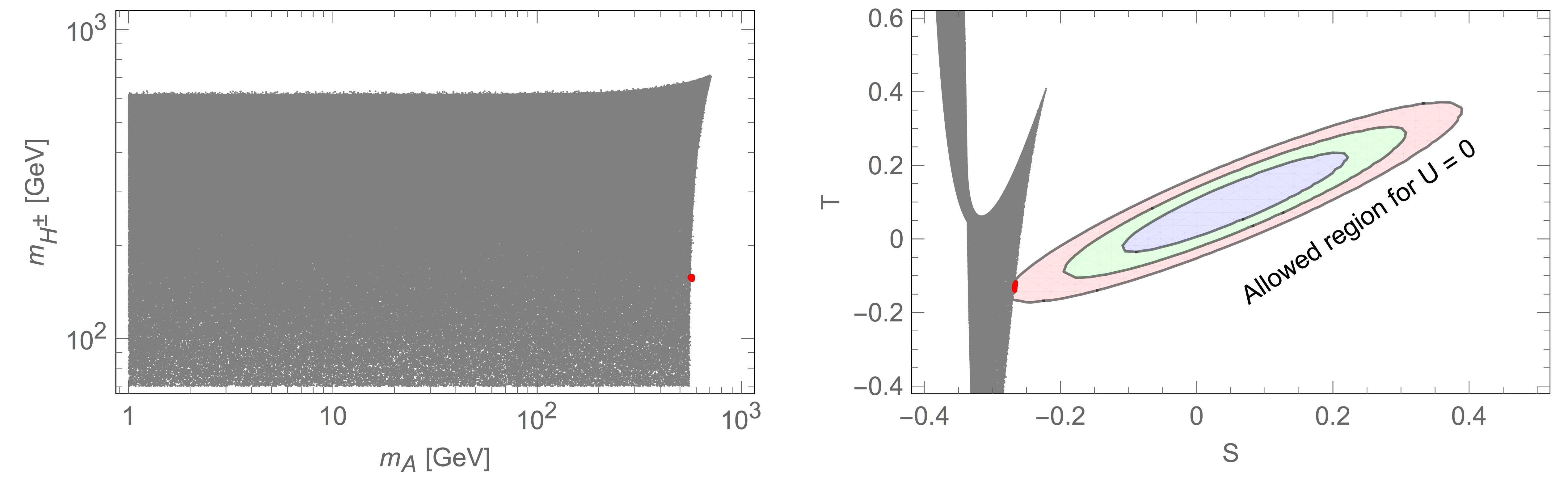}  
  \end{center}
  \caption{Neutrinophilic 2HDM with $\Z_2$ symmetry. The red points
    are allowed by electroweak precision data (oblique parameters) at
    $3\sigma$, while the gray points are ruled out at $3\sigma$ or more. No
    point was found within the $2\sigma$ region.  Left: points that
    satisfy perturbative unitarity and stability constraints in the
    $m_A\times m_{H^\pm}$ plane. Right: Projection of these points in
    the $S\times T$ plane.}
  \label{fig:gabriel-nandi-TH}
\end{figure}

{\bf 2HDM with a $\Z_2$ symmetry.}  Let's first discuss the results
for the 2HDM with a $\Z_2$ symmetry.  As discussed in
sec.~\ref{subsec:z2}, the model has a very light neutral scalar.  In
fact, we verified that Eq.~(\ref{eq:mh}) and the perturbative
unitarity conditions (\ref{eq:unitarity}) require $m_h\lesssim 10
\times v_2$.  Moreover, as the scalar potential parameters $\lambda_i$
and $m_{ij}^2$ can be written in terms of the physical masses and the
vevs, we perform a scan in the physical parameter space, imposing the
following conditions

\begin{align*}
  0.01~{\rm eV} < &m_h < 1~\GeV, \\
  124.85~\GeV < &m_H < 125.33~\GeV, \\
  70~\GeV < &m_{H^\pm}< 1~\TeV,\\
  1~\GeV < &m_A < 1~\TeV,\\
  -\pi/2<&\alpha<\pi/2,\\
  0.01~{\rm eV}<&v_2<1~\MeV.
\end{align*}

The Higgs mass range is taken from the CMS measurements in
ref.~\cite{Aad:2015zhl}

From the left panel of figure~\ref{fig:gabriel-nandi-TH}, the power of
perturbative unitarity constraints can be further appreciated: the
CP-odd and charged scalars are restricted to be below $\sim
600-700~\GeV$. This can be easily understood from Eqs.~(\ref{eq:mA})
and (\ref{eq:mHpm}). Since $M^2\propto m_{12}^2=0$ and $\lambda_{4,5}$
cannot be too large, the masses cannot go arbitrarily above the
electroweak vev.

\begin{figure}[t]
  \begin{center}
    \includegraphics[scale=0.71]{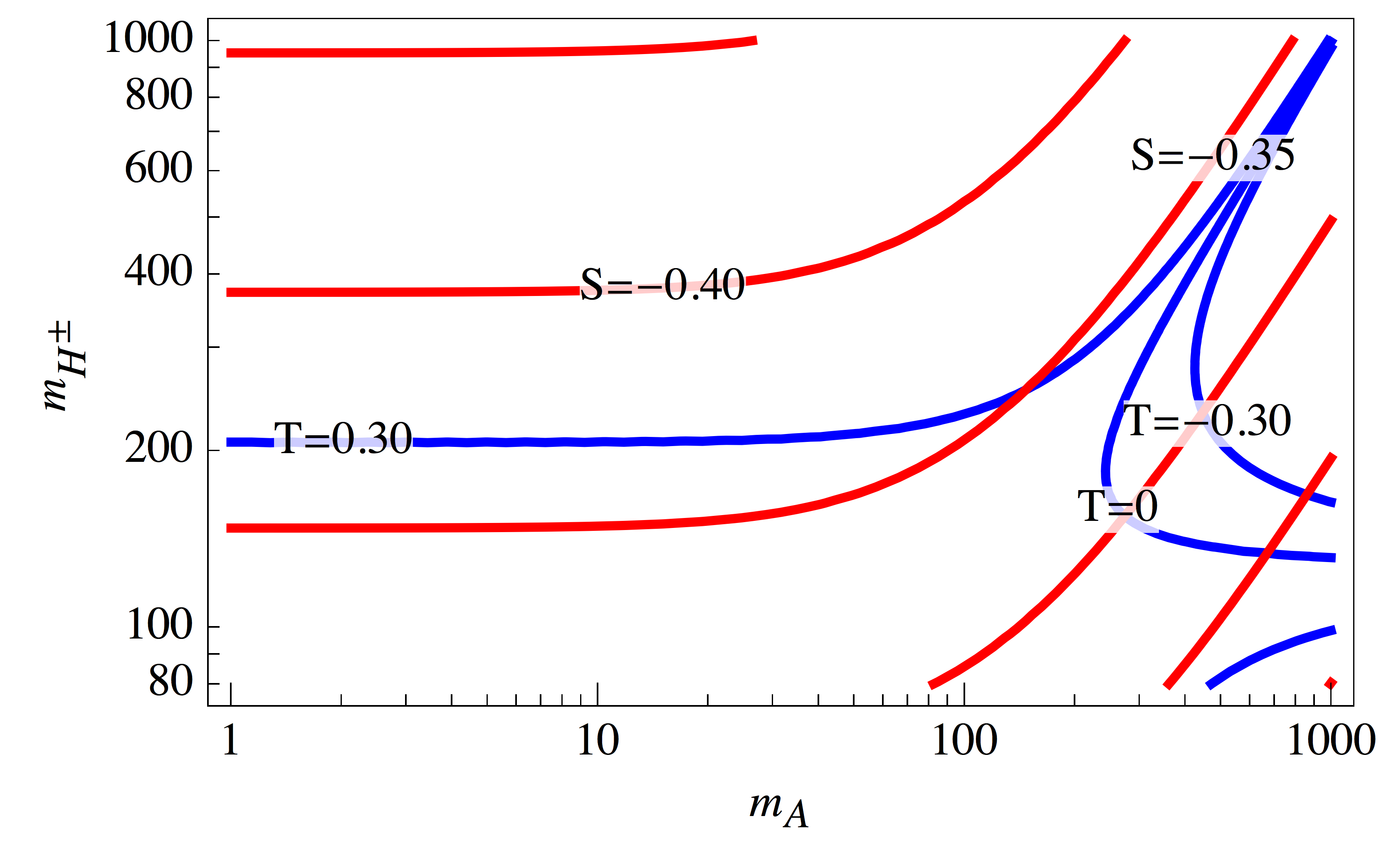}\includegraphics[scale=0.57]{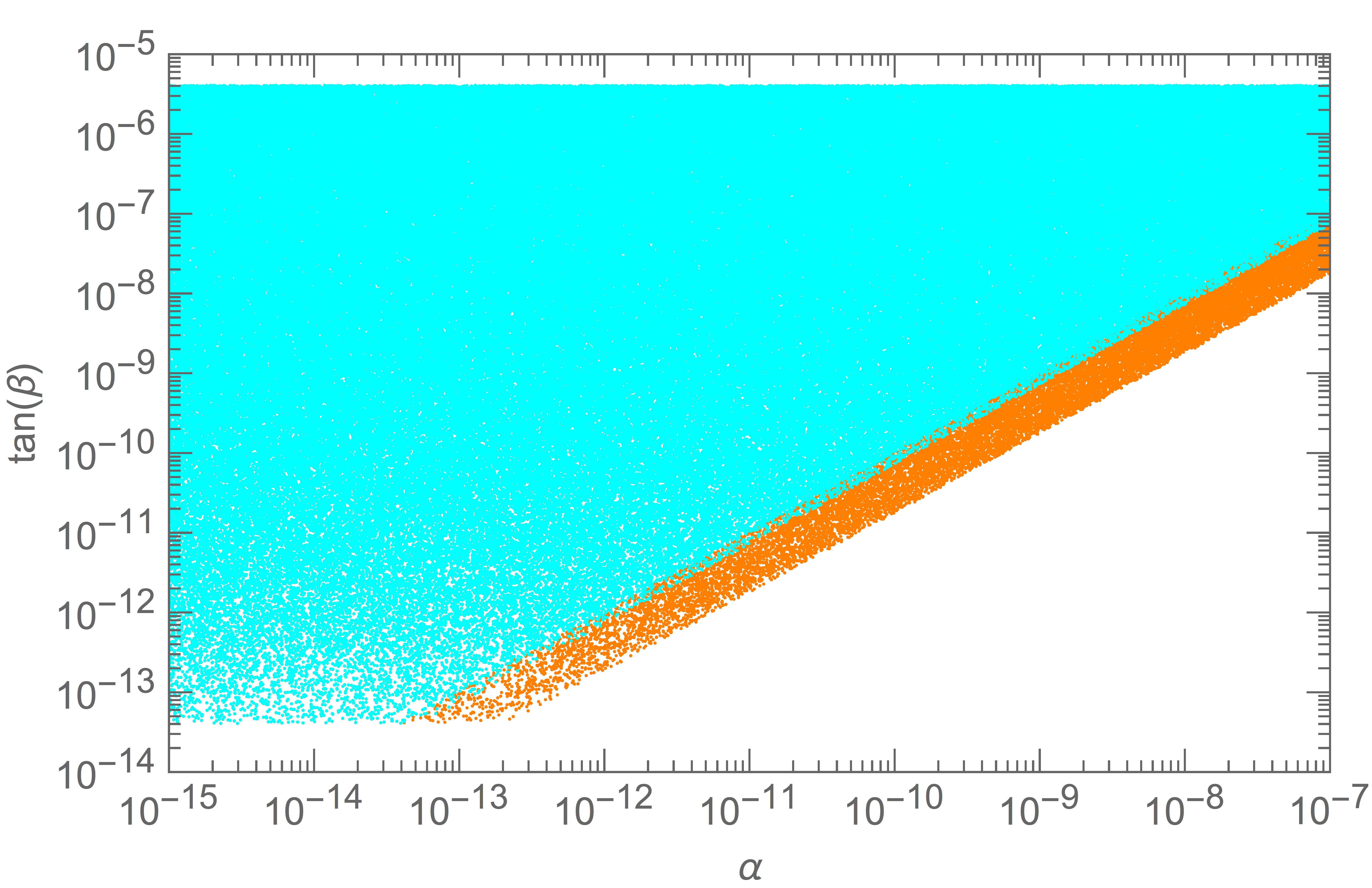}
  \end{center}
  \caption{Neutrinophilic 2HDM with $\Z_2$ symmetry. Left: Predicted
    values for $S$ and $T$ (left) and isolines of $S$ and $T$ values
    as a function of $m_A$ and $m_{H^\pm}$, for $m_h\ll m_Z$. Right:
    $\alpha\times\beta$ plane. Orange points are excluded by the Higgs
    invisible width, while cyan points are allowed. }
  \label{fig:gabriel-nandi-TH-2}
\end{figure}

Moreover, the presence of a very light scalar in the spectrum, below
the GeV scale, yields a substantial negative contribution to the $S$
parameter. The impact of the EWPT can be seen in the right panel of
figure~\ref{fig:gabriel-nandi-TH}, where all points scanned were
projected in the $S \times T$ plane and the allowed region by EWPT was
drawn.  Remarkably, only very few points (in red) were found which
provide a viable model, within the 3$\sigma$ allowed region for the
EWPT. To understand better why, we present the $S$ and $T$ parameters
dependency on $m_A$ and $m_{H^\pm}$ in the left panel of
figure~\ref{fig:gabriel-nandi-TH-2}, for $m_h\ll m_Z$.
It can be concluded that: the $T$ parameter strongly prefers $m_A
\approx m_{H^\pm}$ or a lighter $H^\pm$ with $m_{H^\pm}\sim 150~\GeV$
together with a $m_A > 300~\GeV$; while the $S$ parameter, although it
depends very mildy on the charged and pseudo scalar masses, exhibits a
slight preference to this latter region. All in all, the values of $S$
are always below $\sim -0.25$, revealing a tension with EWPT always
above the $2.97\sigma$ level\footnote{To be precise about such strong
  statements, we also included in our analysis the accepted points of
  a second scan centered on the red region, where the charged scalar
  mass range was changed to 150--160~GeV and the pseudoscalar mass
  range was changed to 500--580~GeV, with $10^5$ points.}.

From this analysis, we can conclude that the 2HDM with a $\Z_2$
symmetry is definitely very disfavored by data. It is not even clear
that the region found which is in the $3\sigma$ border of EWPT is
really viable. A closer look into this region of the parameter space
reveals that these points suffer from at least one of the following
worrisome situations: (i) the $e_1$ scattering amplitude, in
Eqs.~(\ref{eq:unitarity}), is on the verge of violating unitarity,
with at least about $\sim98\%$ of the bound saturated; (ii) the same
for $a_+$ scattering amplitude, with at least $\sim 98\%$ of the bound
saturated; (iii) the stability condition is very fragile, with the
third condition of Eq.~(\ref{contrainte}) satisfied with a margin of
less than $\sim 0.001$; and (iv) the same but for the second condition
of Eq.~(\ref{contrainte}), satisfied with a margin of less than
$\sim0.05$.  Therefore, given this delicate region of the parameter
space, it would be important to include radiative corrections to see
if the stability and unitarity of the theory still holds at one loop.

One could think that a possible way to evade these problems would be
to have a larger $v_2$ so that the mass spectrum, specially $m_h$,
becomes more flexible. Nevertheless, unless $v_2\gtrsim\O(\GeV)$, the
problem does not disappear, strongly disfavoring this minimal model as
an explanation for neutrino masses. A slightly non minimal scenario
that would be surely allowed by data consists of a $m_{12}^2$ term
which breaks $\Z_2$ softly. Such a model would be more general than
the two models considered here, as it would span non zero $m_{12}^2$
and $\lambda_5$ parameters. Therefore, its allowed parameter space
would be even larger than the softly broken $U(1)$ model, which we
will analyse shortly.

One could now be tempted to include a right-handed neutrino
contribution, dropping the lepton number conservation symmetry of the
model. In fact, as $v_2$ is small, it may be possible to have a low
energy realization of the type I seesaw scenario which leads to
observable sterile neutrino phenomenology, and hopefully could
increase a bit the value of the $S$ parameter to make the model
viable. As the effect on $S$ grows with the mass of the fermions in
the loop, we make a distinction between two regimes: the right handed
neutrinos can be below or above the GeV scale. In the first, what
happens is that the contribution to the $S$ parameter is suppressed by
the ratio between these small masses and the $Z$ mass and can be
neglected (for instance, the active neutrino contribution to $S$ is
virtually zero). In the second case, although the sterile neutrino
masses might be large, the coupling to the $Z$ is suppressed by the
active--sterile mixing which generically goes as the ratio between the
active to sterile neutrino masses, $m_\nu/m_N$. Therefore the impact
of right-handed neutrinos is never large enough to substantially
change the $S$ parameter\footnote{This fact has also been checked
  numerically using the expressions in ref.~\cite{Akhmedov:2013hec}.}

For completeness, we also show in the right panel of
figure~\ref{fig:gabriel-nandi-TH-2} the impact of the Higgs invisible
width measurement in the $\alpha\times\tan\beta$ plane. Given the
preference for heavier $S$, we will consider the case where only $H\to
hh$ is present. From Eq.~(\ref{eq:Hhh}), since $m_{12}^2=0$ in this
model, the $g_{Hhh}$ coupling can be rewritten in the limit of
small $\beta$ and $\alpha$ as
\begin{equation}
  g_{H h h} \approx -\frac{m_H^2}{v} \dfrac{\sin(2\alpha)}{\sin(2\beta)},
\end{equation}
which can be sizable only if $\alpha \gtrsim \beta$, explaining the
behavior of the excluded region (orange) in
fig.~(\ref{fig:gabriel-nandi-TH-2}).  Since the ratio $\alpha/\beta$
is already constrained by the theoretical limits (see
Eq.~(\ref{eq:lambda2})), this constraint turns out to be less
stringent than the others.  As a last comment, the charged scalars
could also have an impact in $H\to\gamma\gamma$. In the small
3$\sigma$ allowed region, the modifications to the diphoton width are
generically between $\pm10\%$, depending on the precise values of
$\lambda_3$. This quartic coupling only affects $m_h$, so it is only
weakly bounded by perturbative unitarity.

\begin{figure}[t]
  \begin{center}
    \includegraphics[scale=0.62]{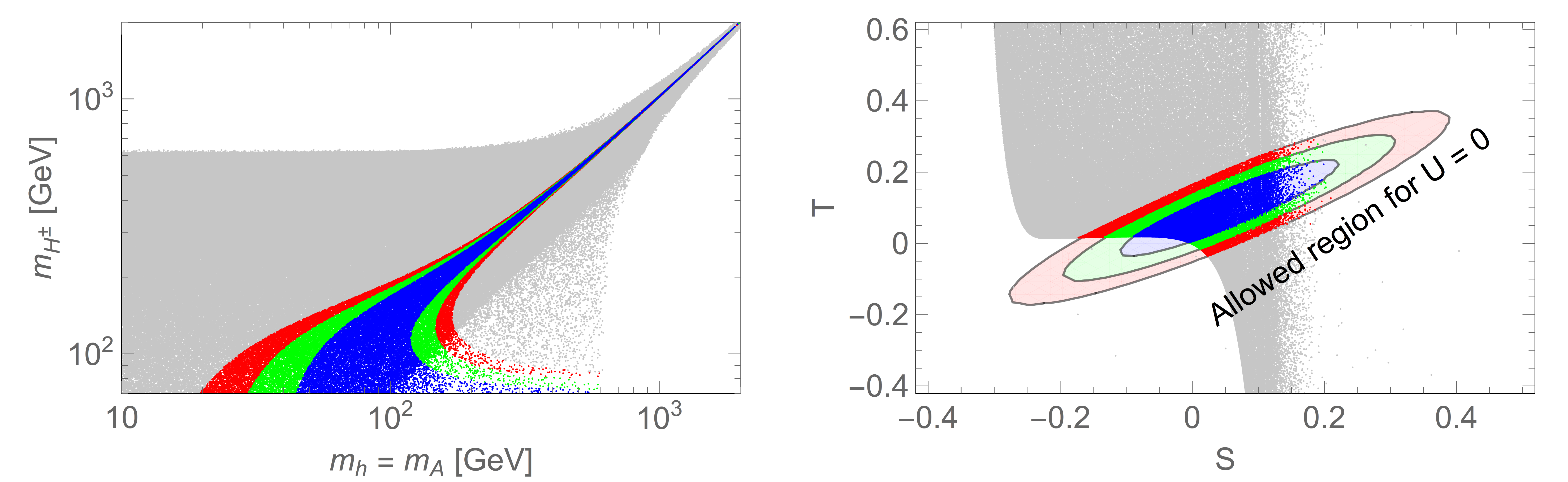}\\
    \includegraphics[scale=0.62]{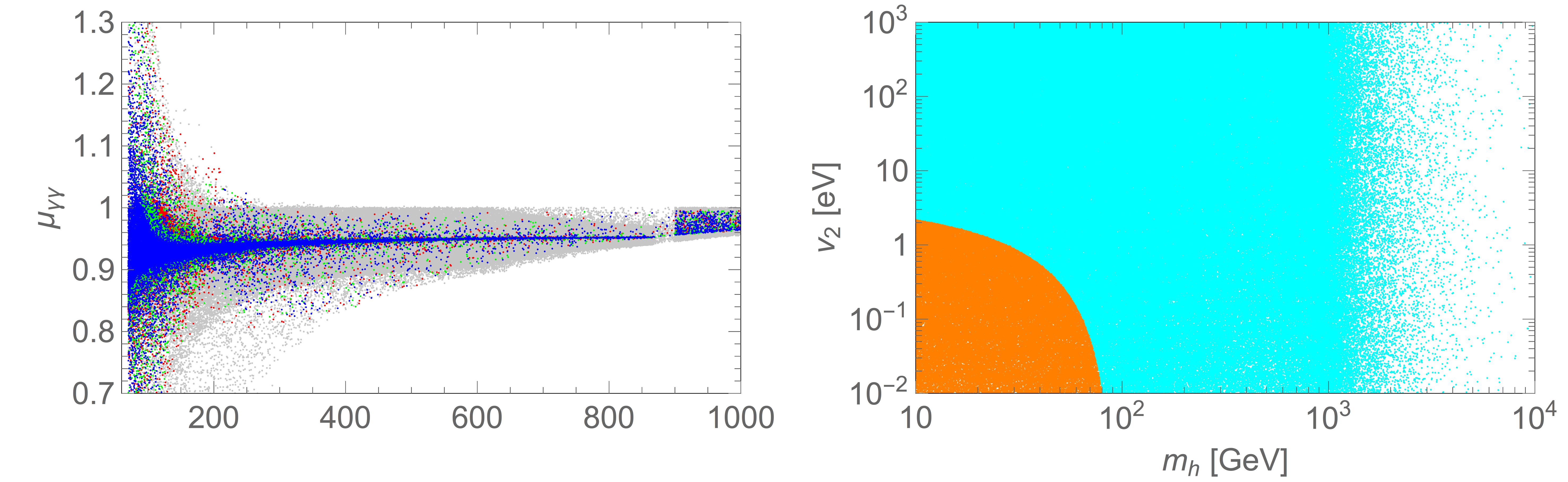}
  \end{center}
  \caption{Neutrinophilic 2HDM with softly broken global $U(1)$
    symmetry. The blue, green and red points are allowed by EWPT at
    $1\sigma$, $2\sigma$, and $3\sigma$, respectively, while the gray
    points are ruled out at $3\sigma$. Top left: parameter space in
    the plane $m_h \times m_{H^\pm}$ which satisfy perturbative
    unitarity and stability constraints. Top right: projection of
    these points in the $S\times T$ plane. Bottom left:
    $H\to\gamma\gamma$ signal strength as a function of
    $m_{H^\pm}$. Bottom right: region in the $m_h\times v_2$ plane
    that is excluded by the Z invisible width (orange
    points).}
  \label{fig:davidson-logan-TH}
\end{figure}

{\bf 2HDM with a global $U(1)$ symmetry.}  We now focus on the
phenomenology of the softly broken $U(1)$ model. A non zero $m_{12}^2$
term allows for heavier $h$, presenting a major change in the
phenomenology with respect to the previous model. Without the
requirement of a light scalar, we enlarge the scanned region
accordingly.  The absence of the $\lambda_5$ quartic coupling makes
the pseudoscalar degenerate in mass with $h$ (to first order in
$v_2$). Therefore we perform an initial scan of the spectrum parameter
space, this time in the region

\begin{align*}
  10~\GeV < &m_h < 1~\TeV, \\
  124.85~\GeV < &m_H < 125.33~\GeV, \\
  70~\GeV < &m_{H^\pm}< 1~\TeV,\\
  m_A &= m_h,\\
  -\pi/2<&\alpha<\pi/2,\\
  0.01~{\rm eV}<&v_2<1~\MeV,
\end{align*}

as well as a second scan with $m_{H^\pm}$ and $m_A$ heavier then 1 TeV
and almost degenerate.  We follow the same procedure as before,
showing only the points allowed by perturbative unitarity and
stability constraints.  The results are presented in
figure~\ref{fig:davidson-logan-TH}. 

In contrast to the previous case, due to the possibility of obtaining
a heavier $h$ in the mass spectrum, there is a region of the parameter
space of this model which passes the electroweak precision tests and
theoretical constraints. The behavior of the $T$ parameter is similar
to the previous scenario: or the mass splitting between $A$ and
$H^{\pm}$ is at most $\sim 80~\GeV$, or the charged scalar is around
100~GeV while $m_h=m_A>150~\GeV$, with negative values of $T$ for
larger $m_{H^\pm}$. This explains the strong correlation on the
allowed region in the upper left panel of
figure~\ref{fig:davidson-logan-TH}.
We also present the projection of these points in the $S\times T$
plane in the upper right panel of figure~\ref{fig:davidson-logan-TH}.

As discussed in the previous sections, this model can also accommodate
a pair of neutral scalars ($S=h,A$) satisfying $m_S< m_H/2$ if
$m_{12}^2$ is small enough. In this case, the constraints coming from
the Higgs invisible decays are similar to those described for the
model with a $\mathbb{Z}_2$ symmetry and turn out to be relatively
weak. On other hand, the $Z$ invisible width can provide us valuable
constraints when the channel $Z\to S\nu\nu$ is open. To perform this
analysis we conservatively assume the lightest neutrino to be massless
and the neutrino mass ordering to be normal, so that the sum of
neutrino masses is about 0.05~eV and $\sum_{i} m_{\nu,i}^2$ takes the
smallest possible value. We show on the bottom right panel of
figure~\ref{fig:davidson-logan-TH} the excluded region (orange points)
under these assumptions in the $m_h\times v_2$ plane. The shape of the
excluded region can be understood from Eq.~(\ref{eq:zinv}), since the
decay rate $Z\to S \nu \nu$ tends to be smaller when there is less
phase space available.

For heavy enough $H^{\pm}$, as can be seen in the lower left panel of
fig.~\ref{fig:davidson-logan-TH}, the $H\to\gamma\gamma$ signal
strength is diminished by about $\sim 5\%$. We can understand this non
decoupling feature by noticing that the $H\,H^+ H^-$ coupling is
$-i\lambda_3v$, which in turn has a correlation with $m_{H^\pm}$,
specially in the larger mass region.  This can be understood by
noticing that, in Eq.~(\ref{lambda3}), for large $m_{H^\pm}$, we have
\begin{equation}
  \lambda_3\approx\left(1-\dfrac{\sin 2\alpha}{\sin 2\beta}\right)\dfrac{m_{H^\pm}^2}{v^2}.
\end{equation}
Typically, $\alpha \ll \beta$, which corresponds to a strong
correlation between $\lambda_3$ and $m_{H^\pm}$, and this is the
denser region around $\mu=0.95$. However this is not always the case,
and the correlation is lost when the ratio of sines is closer to 1,
now corresponding to the sparser points with a much weaker
correlation. Nevertheless, we see that for heavy enough charged
scalar, the contribution to the Higgs diphoton width is always
negative.

\section{Conclusion}\label{sec:conc_chap5}
We performed an analysis of minimal neutrinophilic two Higgs doublet
models which can accommodate neutrino masses by means of a tiny vev of
the additional Higgs doublet. The models studied here differ among
themselves by the symmetry that forbids the couplings between
neutrinos and the scalar which gets an electroweak vev. The cases
studied here span a discrete $\Z_2$ and a global $U(1)$ symmetries.

The bounds considered come from both theory and experiment. The
requirement that the theory is unitary and perturbative at tree-level
strongly constrains the scalar mass spectrum of these models, either
with the presence of very light neutral scalars, in the $\Z_2$ model,
or with a degeneracy between the scalar and pseudoscalar particles, in
the global $U(1)$ scenario.

If there is no additional particle content, the $\Z_2$ symmetry model
was found to be in severe tension with the electroweak precision
tests, due to the presence of a very light neutral scalar, which
generates a large negative contribution to the $S$ parameter. There is
still a region of the parameter space that is barely allowed by EWPT
at the edge of the $3\sigma$ level, but it suffers from a fragile
stability or it saturates the perturbativity conditions. Thus, to
really evaluate the viability of that region, radiative corrections
should be taken into account. The inclusion of a Majorana mass term
for the right handed neutrinos, providing a low scale realization of
the seesaw type I mechanism, does not save the model, as the right
handed neutrino contribution to the $S$ parameter is always
negligible. Therefore, we conclude that \emph{the neutrinophilic 2HDM
  with a spontaneously broken $\Z_2$ symmetry is strongly disfavored
  by data}.

The analysis of the model with an explicit broken global $U(1)$
symmetry reveals a region of the parameter space which is allowed by
all bounds considered. Due to the set of constraints and the
symmetries of the model itself, the spectrum is quite constrained.
The $U(1)$ symmetry predicts that the neutrinophilic scalar is
degenerate in mass with the pseudoscalar, $m_h=m_A$.  Besides, the
electroweak precision tests play a very important role, specially the
$T$ parameter which is sensitive to the absolute mass splitting of the
pseudoscalar and the charged scalars, limiting it to be at most
$\sim80~\GeV$.  Therefore, an important consequence of the theoretical
and experimental constraints is that, if the new scalars are above
$\sim 400~\GeV$, all these particles should have similar masses.
Moreover, the $Z$ invisible width can be sensitive to the region
$v_2<\O({\rm eV})$ provided that $m_h<m_H/2$. Besides, the
$H\to\gamma\gamma$ branching fraction might be modified by about
$-30\%\sim 30\%$ for $m_{H^\pm}<200~\GeV$, while for heavier $H^\pm$,
above 500~GeV, it could be between $-5\%\sim 0\%$.  Finally, we stress
that this model can be well within the reach of LHC 13~TeV, by probing
the $H\to\gamma\gamma$ branching fraction of by direct pair production
of the charged scalars, if they are below $\O(300~\GeV)$.

\addcontentsline{toc}{chapter}{Summary}
\chapter*{Summary}\label{Summary}

The main goal of this thesis was studying the phenomenology of the sterile neutrinos. To do this, we used the data of various neutrino oscillation experiments to probe different models which contained sterile neutrinos and found constraints on the parameters of these models. 

In the first chapter of this thesis we briefly reviewed the framework of neutrino oscillations. In Section~\ref{Standard_Picture_of_Neutrinos} we introduced the standard picture of 3-neutrinos. Then in Section~\ref{Sterile_Neutrinos} we explained the experimental anomalies which led to the sterile neutrino hypothesis, and the simplest $3+1$ scenario which explains these anomalies. The neutrino experiments and neutrinos in cosmology were also briefed in Sections \ref{neutrino_experiments} and \ref{neutrinos_cosmology}, respectively.

Chapter~\ref{chap2} was devoted to the study of the light sterile neutrinos with $\Delta m_{41}^2\sim (10^{-3}-10^{-1})\,{\rm eV}^2$, using the data of the medium baseline reactor experiments. We reproduced the results of these experiments in the $3\nu$ framework in Section~\ref{sec:standard3nu}. We showed that our results are in a good agreement with the results of the collaborations. Then we performed the $3+1$ analysis, first for the Double Chooz experiment and then combining Double Chooz, RENO and Daya Bay. The Double Chooz experiment which consists of only a far detector provides the energy distribution of events, while the other 2 experiments only provide the information in total number of events. We showed that a discrepancy in the energy of the Double Chooz experiment leads to a better fit in the $(3+1)_{\rm light}$ than the $3\nu$ framework, in a way that these best-fit values were obtained: $\sin^2 2\theta_{13}=0.036$, $\sin^2 2\theta_{14}=0.129$ and $\Delta m^2_{41}=0.027~{\rm eV}^2$. Also, the $3\nu$ framework was excluded at $\sim 2.2\sigma$ C.L.. We found that for the case of the Double Chooz experiment, the best-fit value of $\theta_{13}$ angle is significantly different than the reported value in the experiment, and $\theta_{13}=0$ is allowed in less than $1\sigma$ C.L.. We showed these results in Section~\ref{analysis3p1dc}.

After combining the results of Double Chooz with the rate information of RENO and Daya Bay, we found these best fit values: $\sin^2 2\theta_{13}=0.074$, $\sin^2 2\theta_{14}=0.059$ and $\Delta m^2_{41}=0.027~{\rm eV}^2$~. With the combined data the $(3+1)_{\rm light}$ model was favored at $\sim1.2\sigma$ C.L.. The value of $\theta_{13}$ angle also became close to the reported value in the $3\nu$ framework and so the robustness of the determination of $\theta_{13}$ was claimed. Despite the preference for the $(3+1)_{\rm light}$ model, a large part of the parameter space of this model was excluded in our analysis, better than the constraints from the other analyses by a factor of 2. We presented these results in Section~\ref{combined}.\\

In Chapter~\ref{chap3} we studied the models with Large Extra Dimensions (LED) which were primarily introduced to explain the hierarchy problem in the Higgs sector, but can also explain the smallness of neutrino masses. In this model the Kaluza-Klein modes appear as towers of the sterile neutrino states on the brane. We studied the phenomenological consequences of this picture for the high energy atmospheric neutrinos. The existence of these KK modes can change the oscillation probability of neutrinos. In Section~\ref{sec:osc} we calculated the probabilities in the LED model in the presence of the matter effects of the earth. To understand and interpret the results of this work, we studied with details the KK modes in the LED model and found an equivalence between the LED model and a $(3+n)$ scenario consisting of 3 active and $n$ extra sterile neutrino states. This equivalence provided us a clear and intuitive picture of the oscillation pattern of the atmospheric neutrinos in the LED model and was used to explain the features obtained by the numerical calculations. We did this in Section~\ref{sec:3+n}. 

We used the high energy atmospheric data of the IceCube experiment to perform an analysis on the LED model in Section~\ref{sec:icecube}. We obtained the limits on the LED parameters (especially the radius of extra dimension $R_{\rm ED}$) by analyzing the zenith distributions of IC-40 and IC-79 data. For $m_1^D\lesssim0.1$~eV the upper limit $R_{\rm ED}\leq4\times10^{-5}$~cm (at $2\sigma$ level) have been set by the IceCube data and is independent of the value of $m_1^D$. For $m_1^D\gtrsim0.1$~eV the limit depends on the value of $m_1^D$ and is stronger: $R_{\rm ED}\lesssim3\times10^{-6}({\rm eV}/m_1^D)$~cm. 
We also discussed the prospect of improving the bounds by taking into account the energy distribution of the muon-track events in the IceCube. We showed that with a sample of data three times larger than the IC-79 data set, it would be possible to exclude the $2\sigma$ preferred region by the reactor and gallium anomalies. \\

Although most of the anomalies seen in the neutrino sector are in favor of the sterile models with the mass squared difference $\Delta m^2_{41}\sim1$ eV$^2$, there are conflicts between the sterile hypothesis and cosmology, since the light sterile states thermalize in the early universe through their mixing with the active neutrinos, while the Planck results $\Delta N_{\rm eff}<0.7$ with $90\%$ C.L.~\cite{Ade:2015xua}. This problem could be solved if the sterile states have interactions with a light extra gauge boson with mass $M_X\sim$ a few MeV. We studied in Chapter~\ref{chap4} the secret interaction (SI) of the sterile neutrinos. The SI model can change the oscillation probability of neutrinos drastically. We studied the probability in the SI model in Section~\ref{sec:formalism_cha4}. Using the data of the MINOS long-baseline experiment, we showed in Section~\ref{sec:analysis_cha4} that values above $\alpha=92.4$ are excluded at $2\sigma$ C.L., which means it is unlikely the sterile neutrinos can have very huge field strength with the new gauge boson. We also constrained the mass of the light gauge boson using the MINOS neutrino experiments. We showed that $M_X\lesssim10-24~$MeV is excluded with $2\sigma$ C.L.. \\

In Chapter~\ref{chap5} we studied the smallness of neutrino masses using the neutrinophilic 2 Higgs doublet models (2HDM). In such models, the vacuum expectation value (vev) of the first Higgs doublet is responsible for the masses of the particles in the SM, while the second Higgs doublet is the sole responsible for the masses of neutrinos through its small vev. We introduced 2 specific symmetries which prevent the neutrinos to couple to the first Higgs Doublet: the model with $\Z_2$ symmetry (Section~\ref{subsec:z2}) and the model with a softly broken $U(1)$ symmetry (Section~\ref{sec:u1}). We studied the theoretical and experimental Electroweak data which constrain the neutrinophilic 2HDMs in Section~\ref{sec:constraints}.

We found that if there is no additional particle content, the model with $\Z_2$ symmetry will be in severe tension with the electroweak precision tests due to a very light $h$ scalar. Therefore, the neutrinophilic 2HDM with a spontaneously broken $\Z_2$ symmetry is strongly disfavored by data. The analysis of the model with global $U(1)$ symmetry reveals a region of the parameter space which is allowed by all bounds considered, however this parameter space is considerably constrained by current data. Particularly, the mass of the new charged scalar has to be similar to the mass of the new neutral scalars.


\begin{thebibliography}{99}
\bibitem{CMS:2014ega} 
  CMS Collaboration [CMS Collaboration],
and studies of the compatibility of its couplings with the standard model,''
  CMS-PAS-HIG-14-009.
\end{thebibliography}
\end{document}